\documentclass[12pt,a4paper, epsfig]{report}
\usepackage[centertags]{amsmath}
\usepackage{amsfonts}
\usepackage{amssymb}
\usepackage{amsthm}

\usepackage{xthesis} %DAL Thesis Style
\usepackage{xtocinc} %Include Table of Contents as the first entry in TOC

\usepackage[dvips]{graphicx}

\begin{document}

\pagenumbering{roman}

\title{\bf{Electronic Properties of Quantum Wire Networks}}

%\author{Igor Kuzmenko}

\author{\bf{Thesis submitted in partial fulfillment}\\
        \bf{of the requirements for the degree of}\\
        \bf{DOCTOR OF PHILOSOPHY}\\
        \\
        \\
        \\
%        \vspace{39mm}
        {\bf\Large by}\\
        \\
        {\bf\Large Igor Kuzmenko}\\
        \\
        \\
        \\
        \bf{Submitted to the Senate of Ben-Gurion University}\\
        \bf{of the Negev}
        \\
        \\
        \date{\bf{\today}\\
        \vspace{15mm}
        \bf{Beer-Sheva}}}

\maketitle

\begin{center}
 {\Large Electronic Properties of Quantum Wire Networks}
\end{center}
\thispagestyle{empty}

\vspace{35mm}

\begin{center}
 Thesis submitted in partial fulfillment\\
 of the requirements for the degree of\\
 DOCTOR OF PHILOSOPHY\\
 \vspace{10mm}
 {\large by}\\
 \vspace{10mm}
 {\large Igor Kuzmenko}\\
 \vspace{20mm}
 Submitted to the Senate of Ben-Gurion University\\
 of the Negev
\end{center}

 \vspace{15mm}

\begin{flushleft}
 Approved by the advisor\underline{\hspace{40mm}}

 \vspace{10mm}

 Approved by the Dean of the Kreitman School of Advanced Graduate
 Studies\underline{\hspace{20mm}}
\end{flushleft}

 \vspace{30mm}

\begin{center}
 \date{\today}

 \vspace{10mm}

 Beer-Sheva
\end{center}

\newpage

\thispagestyle{empty}
 \vspace*{5cm}

 This work was carried out
under the supervision of Prof. Yshai
Avishai\\

\vspace{35mm}

In the Department \underline{ of Physics }

\vspace{10mm}

Faculty \underline{ of Natural Sciences }

\newpage

\thispagestyle{empty}

\thanks{\begin{center}{\Large\bf Acknowledgments}\end{center}
   I am grateful to my advisers Professor Sergey Gredeskul,
   Professor Konstantin Kikoin, and Professor Yshai Avishai
   for their guidance, help and many hours of fruitful
   discussions.}

\begin{abstract}

Quantum wire networks are novel artificial nano-objects that
represent a two dimensional ($2D$) grid formed by superimposed
crossing arrays of parallel conducting quantum wires, molecular
chains or metallic single-wall carbon nanotubes. Similar
structures arise naturally as crossed striped phases of doped
transition metaloxides. The mechanical flexibility of the
networks, the possibility of excitation of some of their
constituents (a nanotube or a single wire) by external electric
field, and the existence of bistable conformations in some of them
(molecular chain) make the networks one of the most attractive
architectures for designing molecular-electronic circuits for
computational application.

Since such networks have the geometry of crossbars, we call them
``quantum crossbars'' (QCB). Spectral properties of QCB cannot be
treated in terms of purely $1D$ or $2D$ electron liquid theory. A
constituent element of QCB (quantum wire or nanotube) possesses
the Luttinger liquid (LL) fixed point. A single array of parallel
quantum wires is still a LL-like system qualified as a sliding LL
phase provided only the density-density or/and current-current
interaction between adjacent wires is taken into account. Two
crossing arrays (QCB) coupled only by capacitive interaction in
the crosses have similar low-energy, long-wave properties
characterized as a crossed sliding LL phase. QCB with
electrostatic interaction in the crosses possess a cross-sliding
Luttinger liquid (CSLL) zero energy fixed point.

In this Thesis we develop a theory of interacting Bose excitations
(plasmons) in a superlattice formed by $m$ crossed arrays of
quantum wires. The subject of the theory is the spectrum of
excitations and response functions in the $2D$ Brillouin zone so
it goes far beyond the problem of stability of the CSLL fixed
point.

In the first part we analyze spectrum of boson fields and
two-point correlators in double ($m=2$) square QCB [double tilted
and triple ($m=3$) QCB are considered in Appendices
\ref{append-tilted} and \ref{subsec:Triple}, respectively]. We
show that the standard bosonization procedure is valid, and the
system behaves as a cross-sliding Luttinger liquid in the infrared
limit, but the high frequency spectral and correlation
characteristics have either $1D$ or $2D$ nature depending on the
direction of the wave vector in the $2D$ elementary cell of
reciprocal lattice. As a result, the crossover from 1D to 2D
regime may be experimentally observed. It manifests itself as
appearance of additional peaks of optical absorption, non-zero
transverse space correlators and periodic energy transfer between
arrays ("Rabi oscillations").

In the second part, the effectiveness of infrared (IR)
spectroscopy is studied. IR spectroscopy can be used as an
important and effective tool for probing QCB at finite frequencies
far from the LL fixed point. Plasmon excitations in QCB may be
involved in resonance diffraction of incident electromagnetic
waves and in optical absorption in the IR part of the spectrum.
The plasmon velocity is much smaller than the light velocity.
Therefore, an infrared radiation incident on an {\em isolated}
array, cannot excite plasmons at all. However in QCB geometry,
each array serves as a diffraction lattice for its partner, giving
rise to Umklapp processes of reciprocal super-lattice vectors. As
a result, excitation of plasmons in the center of the Brillouin
zone (BZ) occurs.

To excite QCB plasmons with non-zero wave vectors, an additional
diffraction lattice (DL) coplanar with the QCB can be used. Here
the diffraction field contains space harmonics with wave vectors
perpendicular to the DL that enable one to eliminate the wave
vector mismatch and to scan plasmon spectrum within the BZ. In the
general case, one can observe single absorption lines forming two
equidistant series. However, in case where the wave vector of the
diffraction field is oriented along some resonance directions,
additional absorption lines appear. As a result, an equidistant
series of split doublets can be observed in the main resonance
direction (BZ diagonal). This is the central concept of
dimensional crossover mentioned above with direction serving as a
control parameter. In higher resonance directions, absorption
lines form an alternating series of singlets and split doublets
demonstrating new type of dimensional crossover. The latter occurs
in a given direction with frequency as a control parameter.

In the third part, dielectric properties of QCB interacting with
semiconductor substrate are studied. It is shown that a capacitive
contact between QCB and a semiconductor substrate does not destroy
the Luttinger liquid character of the long wave QCB excitations.
However, the dielectric losses of a substrate surface are
drastically modified due to diffraction processes on the QCB
superlattice. QCB-substrate interaction results in additional
Landau damping regions of the substrate plasmons. Their existence,
form and the spectral density of dielectric losses are sensitive
to the QCB lattice constant and the direction of the wave vector
of the substrate plasmon. Thus, the dielectric losses in the
QCB-substrate system serve as a feasible tool for studying QCB
spectral properties.

In the fourth part we formulate the principles of ultraviolet (UV)
spectroscopy, i.e., Raman-like scattering. UV scattering on QCB is
an effective tool for probing QCB spectral properties, leading to
excitation of QCB plasmon(s). Experimentally, such a process
corresponds to sharp peaks in the frequency dependence of the
differential scattering cross section. The peak frequency strongly
depends on the direction of the scattered light.  As a result,
$1D\to 2D$ crossover can be observed in the scattering spectrum.
It manifests itself as a splitting of single lines into multiplets
(mostly doublets). The splitting magnitude increases with
interaction in the QCB crosses, while the peak amplitudes decrease
with electron-electron interaction within a QCB constituent.

The following novel results were obtained in the course of this
research:

\begin{itemize}
\item It is shown that the bosonization procedure may be applied
      to the Hamiltonian of $2D$ quantum networks. QCB plasmons
      may have either $1D$ or $2D$ character depending on the
      direction of the wave vector.
      The crossover from $1D$ to $2D$ regime may be experimentally
      observed. Indeed, due to inter-wire interaction, unperturbed states,
      propagating along the two arrays are always mixed, and transverse
      components of correlation functions do not vanish.
      For quasi-momenta near the resonant line of the BZ, such mixing
      is strong, and the transverse correlators possess specific
      dynamical properties. One of the main effects is the
      possibility of a periodic energy transfer between the
      two arrays of wires.
\item The principles of spectroscopic studies of the excitation
      spectrum of quantum crossbars are established, which possesses
      unique property of dimensional crossover. The plasmon excitations
      in QCB may be involved in resonance diffraction of incident
      light and in optical absorption in the IR part of the spectrum.
      One can observe $1D \to 2D$ crossover behavior of QCB by scanning
      an incident angle. The crossover manifests itself in the appearance
      of a set of absorption doublets instead of the set of single lines.
      At special directions, one can observe new type of crossover where
      doublets replace the single lines with changing frequency at a fixed
      direction of a wave vector.
\item It is shown that a capacitive contact between QCB and semiconductor
      substrate does not destroy the LL character of the long wave
      excitations. However, quite unexpectedly the interaction between
      the surface plasmons and plasmon-like excitations of QCB
      essentially influences the dielectric properties of a
      substrate. First, combined resonances manifest themselves in
      a complicated absorption spectra. Second, the QCB may be treated
      as the diffraction grid for a substrate surface, and an Umklapp
      diffraction processes radically change the plasmon dielectric
      losses. So the surface plasmons are more fragile against
      interaction with superlattice of quantum wires than the LL
      plasmons against interaction with $2D$ electron gas in a substrate.
\item The principles of inelastic UV Raman spectroscopy of QCB are
      formulated. An effective Hamiltonian for QCB-light interaction
      is expressed via the same boson fields as the Hamiltonian of the
      QCB themselves. One can observe $1D\to2D$ crossover of QCB by
      scanning an scattered angle. The crossover manifests itself in
      the appearance of multiplets (mostly doublets) instead of single
      lines.
\end{itemize}

%\newpage

\begin{center}
 {\large\bf Key Words}
\end{center}
Quantum crossbars, Luttinger liquid, strongly correlated
electrons, bosonization, plasmons, dimensional crossover, Rabi
oscillations, $ac$ conductivity, infrared absorption spectroscopy,
dielectric function, Dyson equation, Landau damping, Raman
spectroscopy.
\end{abstract}

\pagenumbering{arabic}

%%%%%%%%%%%%%%%%%%%%%%%%%%%%%%%%%%%%%%%%%%%

\tableofcontents %\listoffigures

%%%%%%%%%%%%%%%%%%%%%%%%%%%%%%%%%%%%%%%%%%%

%%%%%%%%%%%%%%%%%%%%%%%%%%%%%%%%%%%%%%%
\chapter{Background and Objectives}\label{intro}
%%%%%%%%%%%%%%%%%%%%%%%%%%%%%%%%%%%%%%%

The behavior of electrons in arrays of one-dimensional ($1D$)
quantum wires was recognized as a challenging problem soon after
the consistent theory of elementary excitations and correlations
in a Luttinger liquid (LL) of interacting electrons in one
dimension was formulated (see \cite{Voit} for a review). In
contrast to the Fermi liquid (FL) theory \cite{Landau,Landau1},
one dimensional electron liquids exhibits the following
properties:
\begin{itemize}
\item There is no elementary fermionic quasi-particles, the
      generic excitations are bosonic fluctuations.
\item Charge and spin quasi-particles are spatially separated
      and move with different velocities
      ({\em charge-spin separation}).
\item The correlations between these excitations show up as an
      interaction-dependent non-universal power law.
\end{itemize}
There are several possible regimes in which a $1D$ electron liquid
can exist \cite{Levitov}. First, there is an insulating regime,
where charge and spin excitations are gapped. Second, there is a
conducting (Tomonaga-Luttinger) regime where the charge sector is
gapless. In this case the spin sector is either gapped
(Luther-Emery regime) or gapless.

One of the fascinating challenges is a search for LL features in
higher dimensions \cite{Anders}. Although the Fermi liquid state
seems to be rather robust for $D>1$, a possible way to retain some
$1D$ excitation modes in $2D$ and even $3D$ systems is to consider
highly anisotropic objects, in which the electron motion is
spatially confined in the major part of real space (e.g., it is
confined to separate linear regions by potential relief). One may
hope that in this case, weak enough interaction does not violate
the generic long-wave properties of the LL state.

Recent achievements in material science and technology have led to
fabrication of an unprecedented variety of artificial structures
that possess properties never encountered in "natural" quantum
objects. One of the most exciting developments in this field is
fabrication of $2D$ networks by means of self-assembling, etching,
lithography and imprinting techniques \cite{Diehl,Wei}. Another
development is the construction of $2D$ molecular electronic
circuits \cite{Luo} where the network is formed by chemically
assembled molecular chains. Arrays of interacting quantum wires
may be formed in organic materials and in striped phases of doped
transition metal oxides. Artificially fabricated structures with
controllable configurations of arrays and variable interactions
are available now (see, e.g., Refs. \cite{Rueckes,Dai,Dalton}).
Such networks have the geometry of crossbars, and bistable
conformations of molecular chains may be used as logical elements
\cite{Tseng}. Especially remarkable is a recent experimental
proposal to fabricate $2D$ periodic grids from single-wall carbon
nanotubes (SWCNT) suspended above a dielectric substrate
\cite{Rueckes}. The possibility of excitation of a SWCNT by
external electric field together with its mechanical flexibility
makes such a grid formed by nanotubes an excellent candidate for
an element of random access memory for molecular computing.

From a theoretical point of view, such double $2D$ grid, i.e., two
superimposed crossing arrays of parallel conducting quantum wires
\cite{Avr,Avi,Guinea} or nanotubes \cite{Mukho1}, represents a
unique nano-object - quantum crossbars (QCB). Its spectral
properties cannot be treated in terms of purely $1D$ or $2D$
electron liquid theory. A constituent element of QCB (quantum wire
or nanotube) possesses the Luttinger liquid (LL)-like spectrum
\cite{Bockrath,Egger}. The inter-wire interaction may transform
the LL state existing in isolated quantum wires into various
phases of $2D$ quantum liquid. The most drastic transformation is
caused by {\it inter-wire} tunneling in arrays of quantum wires
with {\it intra-wire} Coulomb repulsion. The tunneling constant
rescales towards higher values for strong intra-wire interaction,
and the electrons in an array transform into $2D$ Fermi liquid
\cite{Wen,Schultz1}. The reason for this instability is the
orthogonality catastrophe, i.e., the infrared divergence in the
low-energy excitation spectrum that accompanies the inter-wire
hopping processes.

Unlike inter-wire tunneling, density-density or current-current
inter-wire interaction does not modify the low-energy behavior of
quantum arrays under certain conditions. In particular, it was
shown recently \cite{Luba00,Vica01,Luba01,Mukho1} that
``vertical'' interaction which depends only on the distance
between the wires, imparts the properties of a {\it sliding phase}
to $2D$ array of $1D$ quantum wires. Such LL structure can be
interpreted as a quantum analog of classical sliding phases of
coupled $XY$ chains \cite{Hern}. Recently, it was found
\cite{Sond,Sond1} that a hierarchy of quantum Hall states emerges
in sliding phases when a quantizing magnetic field is applied to
an array.

Similar low-energy, long-wave properties are characteristic of QCB
as well. Its phase diagram inherits some properties of sliding
phases in case when the wires and arrays are coupled only by
capacitive interaction \cite{Mukho1}. When the inter-array
electron tunneling is possible, say, in crosses, dimensional
crossover from LL to $2D$ FL occurs \cite{Guinea2,Mukho1}. If
tunneling is suppressed and the two arrays are coupled only by
electrostatic interaction in the crosses, the system possesses the
LL zero energy fixed point \cite{Luba01}.

The physics of dimensional crossover is quite well studied, e.g.,
in thin semiconductor or superconductor films where the film
thickness serves as a control parameter that governs the crossover
(see e.g. Ref. \cite{Buch,Buch2}). It occurs in strongly
anisotropic systems like quasi-one-dimensional organic conductors
\cite{sault} or layered metals
\cite{metals,metals2,metals3,metals4,metals5}. In the latter
cases, temperature serves as a control parameter and crossover
manifests itself in inter-layer transport. In metals, the layers
appear ``isolated'' at high temperature, but become connected at
low temperatures to manifest $3D$ conducting properties.

The most promising type of artificial structures where dimensional
crossover is expected is a periodic $2D$ system of $m$ crossing
arrays of parallel quantum wires or carbon nanotubes. We call it
``quantum crossbars'' (QCB). Square grids of this type consisting
of two arrays were considered in various physical contexts in
Refs. \cite{Avr,Avi,Guinea,Guinea2}. In Refs.
\cite{Guinea,Guinea2} the fragility of the LL state against
inter-wire tunnelling in the crossing areas of QCB was studied. It
was found that a new periodicity imposed by the inter-wire hopping
term results in the appearance of a low-energy cutoff
$\Delta_l\sim\hbar{v/a}$ where $v$ is the Fermi velocity and $a$
is the period of the quantum grid. Below this energy, the system
is ``frozen'' in its lowest one-electron state. As a result, the
LL state remains robust against orthogonality catastrophe, and the
Fermi surface conserves its $1D$ character in the corresponding
parts of the $2D$ Brillouin zone (BZ). This cutoff energy tends to
zero at the points where the one-electron energies for two
perpendicular arrays $\epsilon_{k_1}$ and $\epsilon_{k_2}$ become
degenerate. As a result, a dimensional crossover from $1D$ to $2D$
Fermi surface (or from LL to FL behavior) arises around the points
$\epsilon_{F_1}=\epsilon_{F_2}$.

Unlike inter-wire tunneling, the density-density or
current-current inter-wire interaction does not modify the
low-energy behavior of quantum arrays under certain conditions. In
particular, it was shown recently \cite{Luba00,Vica01,Luba01} that
``vertical'' interaction which depends only on the distance
between the wires, imparts the properties of a {\it sliding
Luttinger liquid phase} to $2D$ array of $1D$ quantum wires. Such
LL structure can be interpreted as a quantum analog of classical
sliding phases of coupled $XY$ chains \cite{Hern}. Recently, it
was found \cite{Sond} that a hierarchy of quantum Hall states
emerges in sliding phases when a quantizing magnetic field is
applied to an array. Similar low-energy, long-wave properties are
characteristic of QCB as well. Its phase diagram inherits some
properties of sliding phases in case when the wires and arrays
are coupled only by capacitive interaction \cite{Mukho1}. If
tunneling is suppressed and the two arrays are coupled only by
electrostatic interaction in the crosses, the system possesses a
{\em cross-sliding Luttinger liquid} (CSLL) zero energy fixed
point.

In this Thesis we develop a theory of interacting Bose excitations
(plasmons) in a superlattice formed by crossed interacting arrays
of quantum wires. This theory goes far beyond the problem of
stability of the CSLL fixed point. We do not confine ourselves
with the studying the conditions under which the LL behavior is
preserved in spite of inter-wire interaction. We consider
situations where the {\em dimensional crossover} from $1D$ to $2D$
occurs. It turns out that the standard bosonization procedure is
valid in a $2D$ reciprocal space under certain conditions. The QCB
behaves as a sliding Luttinger liquid in the infrared limit, and
exhibits a rich Bose-type excitation spectrum (plasmon modes)
arising at finite energies in $2D$ BZ. We derive the Hamiltonian
of the QCB, analyze the spectrum of boson fields away from the LL
fixed point and compute two-point correlation functions in QCB
with short range inter-array capacitive interaction. We study new
type of dimensional crossover, i.e., a {\it geometrical} crossover
where the quasimomentum serves as a control parameter, and the
excitations in a system of quantum arrays demonstrate either $1D$
or $2D$ behavior in different parts of reciprocal space. A rather
pronounced manifestation of this kind of dimensional crossover is
related to the QCB response to an external ac electromagnetic
field. We formulate the principles of spectroscopy for the QCB. We
consider an infrared (IR) absorption spectroscopy of the QCB and
an ultraviolet (UV) scattering on the QCB, and study the main
characteristics of IR absorption spectra and UV scattering
observables.

The structure of the Thesis is as follows. In the second Chapter
the progress in the theory of interacting fermions in
low-dimensional systems (such as quantum wires, metallic carbon
nanotubes, array of quantum wires, and QCB) exhibiting LL-like
behavior is briefly reviewed. The bosonization procedure is
introduced for a simple model of $1D$ spinless interacting
electrons. The LL theory is applied for describing the low-energy
behavior of interacting electrons in real systems such as quasi
one-dimensional quantum wires and single-wall carbon nanotubes.
The existence of sliding LL phase is established for an array of
weakly coupled parallel quantum wires. This analysis is extended
to a system of two crossed arrays of $1D$ quantum wires (QCB) with
a capacitive inter-wire coupling. Such a system exhibits a
crossed-sliding LL phase. We also consider QCB with virtual
wire-to-wire electron tunneling, and find the necessary condition
under which the one-electron tunneling is suppressed and the
cross-sliding LL phase is stable.

In the third Chapter the spectrum of boson fields and two-point
correlation functions are analyzed in a double square QCB. We show
that the standard bosonization procedure is valid, and that the
system behaves as a sliding Luttinger liquid in the infrared
limit, but the high frequency spectral and correlation
characteristics have either $1D$ or $2D$ nature depending on the
direction of the wave vector in the $2D$ elementary cell of the
reciprocal lattice. As a result, the crossover from $1D$ to $2D$
regime may be experimentally observed. It manifests itself as
appearance of additional peaks of optical absorption, non-zero
transverse space correlators and periodic energy transfer between
arrays (``Rabi oscillations'').

In the fourth Chapter the effectiveness of infrared spectroscopy
is studied. It is shown that plasmon excitations in the QCB may be
involved in resonance diffraction of incident electromagnetic
waves and in optical absorption in the IR part of the spectrum.
The plasmon velocity is much smaller than the light velocity.
Therefore, an infrared radiation incident on an {\em isolated}
array, cannot excite plasmons at all. However in QCB geometry,
each array serves as a diffraction lattice for its partner, giving
rise to Umklapp processes of reciprocal super-lattices vectors. As
a result, plasmons may be excited in the BZ center. To excite QCB
plasmons with non-zero wave vectors, an additional diffraction
lattice (DL) coplanar with the QCB can be used. Here the
diffraction field contains space harmonics with wave vectors
perpendicular to the DL that enable one to eliminate the wave
vector mismatch and to scan plasmon spectrum within the BZ. In the
general case, one can observe single absorption lines forming two
equidistant series. However, in case where the wave vector of the
diffraction field is oriented along some resonance directions,
additional absorption lines appear. As a result, an equidistant
series of split doublets can be observed in the main resonance
direction (BZ diagonal). This is the central concept of
dimensional crossover mentioned above with direction serving as a
control parameter. In higher resonance directions, absorption
lines form an alternating series of singlets and split doublets
demonstrating new type of dimensional crossover. The latter occurs
in a given direction with a frequency as a control parameter.

The fifth Chapter is devoted to the study of dielectric properties
of a semiconductor substrate with the imposed $2D$ QCB. We
demonstrate that a capacitive contact between the QCB and
semiconductor substrate does not destroy the Luttinger liquid
character of the long wave QCB excitations. However, dielectric
losses of a substrate surface are drastically modified due to
diffraction processes on the QCB superlattice. QCB-substrate
interaction results in additional Landau damping regions of the
substrate plasmons. Their existence, form and the density of
dielectric losses are strongly sensitive to the QCB lattice
constant and the direction of the wave vector of the substrate
plasmon. Thus, dielectric losses in the QCB-substrate system serve
as a good tool for studying QCB spectral properties.

In the sixth Chapter, the principles of UV spectroscopy for QCB
are formulated and the main characteristics of scattering spectra
are described. We study inelastic scattering of an incident photon
leading to the creation of a QCB plasmon. Experimentally, such a
process corresponds to sharp peaks in the frequency dependence of
the differential scattering cross section. We show that the peak
frequency strongly depends on the direction of the scattered
light. As a result, the $1D \to 2D$ crossover can be observed in
the scattering spectrum. It manifests itself as a splitting of
single lines into multiplets (mostly doublets).

All technical details are contained in Appendices
\ref{append:Empty}, \ref{append:DoublSpectr&Corr}, and
\ref{append:Inter}. Double tilted and triple QCB are considered in
Appendices \ref{append-tilted} and \ref{subsec:Triple},
respectively.

This work was partially presented by posters and lectures in
scientific conferences and schools (see List of Presentations).
The first part of the results was published in Refs. 1-4 (see List
of Publications). The second and third parts were published in
Refs. 5-8. The fourth part was published in Refs. 9, 10.

The author is grateful to V. Liubin, M. Klebanov, and Y. Imry for
discussions of the effectiveness of the infrared absorption and
ultraviolet scattering in probing spectral properties of QCB.

\newpage

\begin{center}
 {\LARGE\bf List of Presentations}
\end{center}
\begin{enumerate}
\item \underline{I. Kuzmenko}. {\sl Ultraviolet Scattering on
      Quantum Crossbars} (poster). Third Windsor School on
      Condensed Matter Theory ``Field Theory on Quantum Coherence,
      Correlations and Mesoscopics'', Windsor, UK, August 9-22, 2004.
\item \underline{I. Kuzmenko}, S. Gredeskul, K. Kikoin,
      Y. Avishai. {\sl Optical Properties of Quantum
      Crossbars} (poster). SCES'04, the International
      Conference, on Strongly Correlated Electron Systems,
      Karlsruhe, Germany, July 26-30, 2004.
\item \underline{I. Kuzmenko}. {\sl Spectrum and Optical
      Properties of Quantum Crossbars} (lecture). Condensed Matter
      Seminar, Department of Physics, Ben-Gurion University of the
      Negev, Beer Sheva, Israel, June 7, 2004.
\item \underline{I. Kuzmenko}, S. Gredeskul, K. Kikoin,
      Y. Avishai. {\sl Dielectric Properties of Quantum
      Crossbars} (lecture). International School and
      Workshop on Nanotubes $\&$ Nanostructures,
      Frascati, Italy, September 15-19, 2003.
\item \underline{I. Kuzmenko}, S. Gredeskul, K. Kikoin,
      Y. Avishai. {\sl Optical Absorption and Dimensional
      Crossover in Quantum Crossbars} (poster). International
      Seminar and Workshop on Quantum transport and Correlations
      in Mesoscopic Systems and Quantum Hall Effect, Dresden,
      Germany, July 28 - August 22, 2003.
\item \underline{I. Kuzmenko}, S. Gredeskul, K. Kikoin,
      Y. Avishai. {\sl Electronic Excitations in $2D$ Crossbars} (poster).
      International School of Physics "Enrico Fermi", Varenna, July
      9-19, 2002.
\item \underline{I. Kuzmenko}, S. Gredeskul, K. Kikoin,
      Y. Avishai. {\sl Electronic Properties of Quantum Wire Networks} (poster).
      19th Winter School for Theoretical Physics, Jerusalem, December
      30, 2001 - January 8, 2002.
\item \underline{I. Kuzmenko}, S. Gredeskul, K. Kikoin,
      Y. Avishai. {\sl Electronic Properties of Quantum Bars} (poster).
      Meeting of Israel Physical Society, Tel-Aviv,
      December 17, 2001.
\item \underline{I. Kuzmenko}, S. Gredeskul, K. Kikoin,
      Y. Avishai. {\sl Energy Spectrum of Quantum Bars} (poster).
      NATO ASI at Windsor, UK, August 13-26, 2001.
\end{enumerate}

\newpage

\begin{center}
 {\LARGE\bf List of Publications}
\end{center}
\begin{enumerate}
\item I. Kuzmenko, S. Gredeskul, K. Kikoin, Y. Avishai.
      {\it Electronic Excitations and Correlations in Quantum Bars}.
      Low Temperature Physics, {\bf 28}, 539 (2002);
      [Fizika Nizkikh Temperatur, {\bf 28}, 752 (2002)].
\item K. Kikoin, I. Kuzmenko, S. Gredeskul, Y. Avishai.
      {\it Dimensional Crossover in 2D Crossbars}.
      Proceedings of NATO Advanced Research Workshop
      ``Recent Trends in Theory of Physical Phenomena in High
      Magnetic Fields'' (Les Houches, France, February 25 -
      March 1, 2002), p. 89; {\tt cond-mat/0205120}.
\item I. Kuzmenko, S. Gredeskul, K. Kikoin, Y. Avishai.
      {\it Plasmon Excitations and One to Two Dimensional
      Crossover in Quantum Crossbars}. Phys. Rev. B {\bf 67},
      115331 (2003); {\tt cond-mat/0208211}.
\item S. Gredeskul, I. Kuzmenko, K. Kikoin, Y. Avishai.
      {\it Spectrum, Correlations and Dimensional Crossover in
      Triple $2D$ Quantum Crossbars}.
      Physica E {\bf 17}, 187 (2003).

\item S. Gredeskul, I. Kuzmenko, K. Kikoin, and Y. Avishai.
      {\it Quantum Crossbars: Spectra and Spectroscopy}.
      Proceeding of NATO Conference MQO, Bled, Slovenia,
      September 7-10, 2003, p.219.
\item I. Kuzmenko, S. Gredeskul.
      {\it Infrared Absorption in Quantum Crossbars}.
      HAIT. Journal of Science and Engineering {\bf 1},
      130 (2004).
\item I. Kuzmenko, S. Gredeskul, K. Kikoin, and Y. Avishai,
      {\it Infrared Spectroscopy of Quantum Crossbars}.
      Phys. Rev. B {\bf 69}, 165402 (2004);
      {\tt cond-mat 0306409}.
\item I. Kuzmenko. {\it Landau Damping in a $2D$ Electron Gas
      with Imposed Quantum Grid.}
      Nanotechnology {\bf 15}, 441 (2004); {\tt cond-mat 0309546}.

\item I. Kuzmenko. {\it X-ray Scattering on Quantum Crossbars.}
      Physica B {\bf 359-361}, 1421 (2005).
\item I. Kuzmenko, S. Gredeskul, K. Kikoin, and Y. Avishai,
      {\it Ultraviolet Probing of Quantum Crossbars}.
      Phys. Rev. B {\bf 71}, 045421 (2005);
      {\tt cond-mat 0411184}.
\end{enumerate}

%\newpage

%%%%%%%%%%%%%%%%%%%%%%%%%%%%%%%%%%%%%%%
\chapter{From Quantum Wires to Quantum Crossbars}
         \label{LL}
%%%%%%%%%%%%%%%%%%%%%%%%%%%%%%%%%%%%%%%

%%%%%%%%%%%%%%%%%%%%%%%%%%%%%%%%%%%%%%%
\section{Introduction}\label{sub-LL-intro}
%%%%%%%%%%%%%%%%%%%%%%%%%%%%%%%%%%%%%%%

In this Chapter, a brief review of electron properties of
low-dimensional systems is presented. In Section \ref{sub-LL},
Luttinger-liquid (LL) theory is introduced. In Sections
\ref{sub-wire} and \ref{subsec:LL-SWCN}, the LL theory is applied
to describe low-energy properties of an electron liquid in a quasi
one dimensional quantum wire and a carbon nanotube. In Section
\ref{subsec:sliding}, an array of quantum wires or carbon
nanotubes with local density-density and/or current-current
inter-wire interactions is considered which result in generalized
Luttinger-liquid theory. Similar Luttinger-liquid behavior of
electron liquid in crossed arrays is considered in Sections
\ref{subsec:cross-sliding} and \ref{subsec:QCB-with-tunneling}.

%%%%%%%%%%%%%%%%%%%%%%%%%%%%%%%%%%%%%%%
\section{Luttinger-Liquid Theory}\label{sub-LL}
%%%%%%%%%%%%%%%%%%%%%%%%%%%%%%%%%%%%%%%

Following conventional Luttinger liquid (LL) theory
\cite{Tomonaga,Luttinger,Delft,kane-fisher}, we consider in this
Section a simple model: a $1D$ conductor containing spinless
right- and left-moving electrons (spin sector is assumed to be
gapped). In the Tomonaga-Luttinger model, the free-electron
dispersion is assumed to be linear $\varepsilon_{Lk}=\pm\hbar
v_Fk$ around two Fermi points $\pm k_F$, and a local
electron-electron interaction is parameterized by the
dimensionless coupling constants $g_2$ and $g_4$. The model
Hamiltonian $H_{TL}=H_{\rm kin}+H_{\rm int}$ is
\begin{eqnarray}
  H_{\rm kin} &=& i\hbar v_F
       \int\limits_{-L/2}^{L/2}{dx}
       \left(
            \psi_{L}^{\dag}(x)\partial_{x}\psi_{L}(x)-
            \psi_{R}^{\dag}(x)\partial_{x}\psi_{R}(x)
       \right),
  \label{H-kinetic-0}
  \\
  H_{\rm int} &=&
  \pi\hbar v_F
  \int\limits_{-L/2}^{L/2}{dx}
  \left[
       g_4
       \left(
            \rho_L^2(x)+
            \rho_R^2(x)
       \right)+
       2g_2\rho_L(x)\rho_R(x)
  \right].
  \label{H-inter-0}
\end{eqnarray}
Here $\psi_{L}(x)$ ($\psi_{R}(x)$) is the field operator for
left-moving (right-moving) fermions satisfying the
anti-commutation relations $\{\psi_{\alpha}(x),
\psi_{\alpha'}^{\dagger}(x')\}= \delta_{\alpha\alpha'}
\delta(x-x')$ ($\alpha,\alpha'=L,R$); $\rho_{\alpha}(x)=
\psi_{\alpha}^{\dag}(x) \psi_{\alpha}(x)$ are the density
operators for left- and right-movers. The total number of left
(right) moving electrons is a good quantum number. Therefore all
excitations are electron-hole-like and hence have bosonic
character.

It is convenient to write the Hamiltonian in terms of bosonic
fields. The electron density $\rho_{L/R}(x)$ can be expressed in
terms of derivative fields $\partial_x\varphi_{L/R}(x)$:
\begin{eqnarray}
  :\rho_L(x):= \frac{1}{2\pi}\partial_x\varphi_L(x),
  \ \ \ \ \
  :\rho_R(x):=-\frac{1}{2\pi}\partial_x\varphi_R(x).
  \label{bosonization-rho-varphi}
\end{eqnarray}
Here $:\rho_{L/R}:=\rho_{L,R}-\langle0|\rho_{L,R}|0\rangle$
denotes the fermion-normal-ordering with respect to the Fermi sea
$|0\rangle$ defined as following \cite{Delft}, to normal-order a
function of operators of creation and annihilation of fermions,
operators of creation of fermions above the Fermi level and
operators of annihilation of fermions below the Fermi level are to
be moved to the left of all other operators (namely operators of
creation of fermions below the Fermi level and operators of
annihilation of fermions above Fermi level). The fields
$\varphi_{L,R}(x)$ satisfy the following commutation relations
\cite{Delft}
\begin{eqnarray}
 [\varphi_{L/R}(x),\varphi_{L/R}(x')]=
 \mp i\pi{\rm{sign}}(x-x'),
 \ \ \ \ \
 [\varphi_L(x),\varphi_R(x')]=0.
 \label{commut-phi}
\end{eqnarray}

It is convenient to define
\begin{eqnarray}
  \theta(x) =
  \frac{1}{\sqrt{4\pi}}
  \left[\varphi_{L}(x)-\varphi_{R}(x)\right],
  \ \ \ \ \
  \phi(x) =
  \frac{1}{\sqrt{4\pi}}
  \left[\varphi_{L}(x)+\varphi_{R}(x)\right],
  \label{boson-theta-phi}
\end{eqnarray}
where $\theta(x)$ is a density variable and $\phi(x)$ is the
conjugate phase variable \cite{kane-fisher}. Then one obtains
\begin{eqnarray}
  H_{TL} &=&
  \frac{\hbar v}{2}
  \int\limits_{-L/2}^{L/2}dx
  \left(
       \frac{1}{g}\pi^2(x)+
       g\left(\partial_x\theta(x)\right)^2
  \right).
  \label{h-boson}
\end{eqnarray}
The Hamiltonian (\ref{h-boson}) describes a set of harmonic
oscillators, where $\theta(x)$ and $\pi(x)=\partial_x\phi(x)$
satisfy the commutation relations of conventional canonically
conjugate operators of a ``coordinate'' and a ``momentum'':
$[\theta(x),\pi(x')]=i\delta(x-x')$. The renormalized Fermi
velocity $v$ and the LL parameter $g$ are given by the equations
\begin{eqnarray}
  v=v_F\sqrt{(1+g_4)^2-g_2^2}~,
  \ \ \ \ \
  g=\sqrt{\frac{1+g_4-g_2}{1+g_4+g_2}}~.
  \label{LL-v-g}
\end{eqnarray}
The dimensionless parameter $g$ is a measure of the strength of
the electron-electron interactions. It plays a central role in the
LL theory. The noninteracting value of $g$ (i.e., for $g_2=0$) is
$1$, and for repulsive interactions ($g_2>0$) $g$ is less than
$1$.

%%%%%%%%%%%%%%%%%%%%%%%%%%%%%%%%%%%%%%%
 \section{Quasi One Dimensional Quantum Wire}
 \label{sub-wire}
%%%%%%%%%%%%%%%%%%%%%%%%%%%%%%%%%%%%%%%

\begin{figure}[htb]
\centering
\includegraphics[width=80mm,height=56mm,angle=0,]{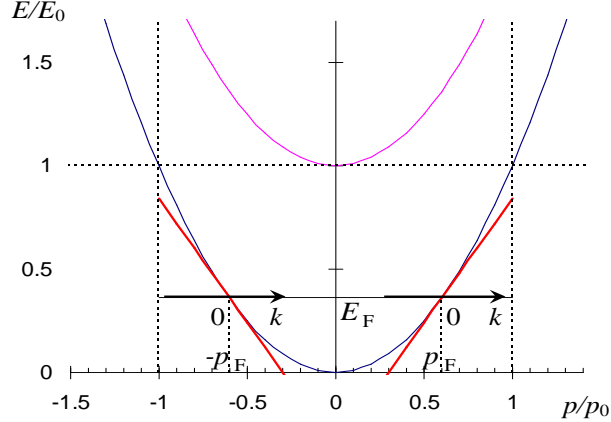}
\caption{The low energy part of the free-electron spectrum. Here
$p_0=\pi/R_0$ and $E_0=\hbar^2p_0^2/(2m_e)$. $k$ is measured from
$p_F$ ($-p_F$) for right-moving (left-moving) fermions. Tilted
lines describe the linear approximation $E_F\pm\hbar{v_F}k$ for
right/left moving electrons.}
 \label{SP-QW}
\end{figure}

In this Section, the LL theory is applied to a conductor slab of
length $L$, width $R_0$ and thickness $r_0$ ($L{\gg}R_0{\gg}r_0$)
containing free spinless electrons (spin sector is still assumed
to be gapped). In experimentally realizable setups, such a
structure can be created by cleverly gating $2D$ electron gas in
{Ga}{As} inversion layers \cite{meirav,timp,zotov1,zotov2}, and
doped helical polyacetilene nanofibres \cite{aleshin}. The
position of an electron is described by a $2D$ vector
${\bf{r}}=(x,y)$, where the $x$-axis is taken along the wire
direction $(0<x<L)$, and $y$ is taken along transverse direction
($0<y<R_0$). The $2D$ momentum of an electron is
${\bf{p}}=(p,\kappa)$ and its dispersion is
$E({\bf{p}})=\hbar^2{\bf{p}}^2/(2m_e)$ (see Fig. \ref{SP-QW}).
Here $p$ ($\kappa$) is the wave number in the direction of the
wire axes (in the transverse direction), $m_e$ is an effective
electron mass. Up to scales $|p|<p_0=\pi/R_0$ and
$E<E_0=\hbar^2p_0^2/(2m_e)$, all the excitations are one
dimensional. In the subsequent discussion we assume that $p_F<p_0$
and $E_F<E_0$. Then we have ``left''- and ``right''-moving
quasi-particles with energies near the Fermi level and momenta
near $p_F$ ($-p_F$) for right-moving (left-moving) fermions. We
introduce the momentum index $k=p-p_F$ ($k=p+p_F$) for
right-moving (left-moving) fermions. It is seen that
$-p_F<k<\infty$ ($-\infty<k<p_F$) for right-moving (left-moving)
fermions.

%%%%%%%%%%%%%%%%%%%%%%%%%%%%%%%%%%%%%
%\begin{figure}[htb]
%\centering
%\includegraphics[width=60mm,height=15mm,angle=0,]{stripe.eps}
%%\epsfxsize=70mm
%\caption{Conductor stripe of length $L$ width $R_0$ and thickness
%$r_0$.} \label{stripe}
%\end{figure}
%%%%%%%%%%%%%%%%%%%%%%%%%%%%%%%%%%%%%%

Following Ref. \cite{Delft}, we extend the range of $k$ to be
unbounded by introducing additional unphysical ``positron states''
at the bottom of the Fermi sea. Next, we factor out the rapidly
fluctuating $e^{\pm{ip_Fx}}$ phase factors and express the
physical fermion field $\Psi(x)$ in terms of two fields
$\psi_{L/R}(x)$ that vary slowly on the scale of $1/p_F$:
\begin{equation}
  \Psi(x)=
  e^{ ip_Fx}\psi_{R}(x)+
  e^{-ip_Fx}\psi_{L}(x).
  \label{Psi-LR}
\end{equation}
The energies near the Fermi level can be written as
$E_k=E_F+\hbar{v_Fk}$ ($E_k=E_F-\hbar{v_Fk}$) for right-moving
(left-moving) fermions, where $v_F=\hbar p_F/m_e$ is Fermi
velocity. Then in this approximation the kinetic energy
Hamiltonian has the form of the Hamiltonian (\ref{H-kinetic-0}).

Electron-electron interaction is a Coulomb interaction screened in
the long-wave limit \cite{Schulz1}. Indeed, quantum wires are not
pure $1D$ objects and screening arise due to their finite
transverse size $R_0$ which is the characteristic screening length
\cite{Sasaki}. It is described by the following Hamiltonian
\begin{eqnarray}
 H_{\rm int} &=& \frac{1}{2}
     \int\frac{dx  dy }{R_0}
     \int\frac{dx' dy'}{R_0}
        U({\bf{r-r}}')
        \rho(x, y)
        \rho(x',y').
 \label{H-C}
\end{eqnarray}
Here
\begin{eqnarray}
  U({\bf{r}})=
  \frac{
       e^2
       \zeta
       \displaystyle{
     \left(
          \frac{x}{R_0}
     \right)}}
       {\sqrt{|{\bf{r}}|^2+r_0^2}},
  \label{U-stripe}
\end{eqnarray}
where $r_0$ is the stripe thickness, the screening function
$\zeta(\xi)$ (introduced phenomenologically) is of order unity for
$|\xi|<1$ and vanishes for $|\xi|>1$. In the long-wavelength
limit, the electron density operator
$\rho(x,y)\equiv\rho(x)=\Psi^{\dag}(x)\Psi(x)$ can be written as
\begin{eqnarray}
  &&
  \rho(x)=
    \rho_{L}(x)+
    \rho_{R}(x)+
    \psi_{L}^{\dag}(x)
    \psi_{R}(x)
    e^{2ip_Fx}+
    \psi_{R}^{\dag}(x)
    \psi_{L}(x)
    e^{-2ip_Fx},
  \label{rho}
\end{eqnarray}
where $\rho_{\alpha}(x)=\psi_{\alpha}^{\dag}(x)\psi_{\alpha}(x)$
($\alpha=L,R$) are density operators for left- and right-moving
fermions.

%%%%%%%%%%%%%%%%%%%%%%%%%%%%%%%%%%%%
%%%%%%%%%%%%%%%%%%%%%%%%%%%%%%%%%%%%%

Then one can write the interaction (\ref{H-C}) in the form
(\ref{H-inter-0}), where
\begin{eqnarray*}
  g_4=\frac{1}{\pi}
      \int\frac{dx dy}{R_0}
      \frac{U({\bf{r}})}{\hbar v_F}
      \approx\frac{2e^2}{\hbar v_F},
  \ \ \ \ \
  g_2=\frac{1}{\pi}
      \int\frac{dx dy}{R_0}
      \frac{U({\bf{r}})}{\hbar v_F}
      \left(1-\cos(2p_Fx)\right)
      \approx\frac{2e^2}{3\hbar v_F}\left(p_FR_0\right)^2.
  &&
\end{eqnarray*}

With Eqs. (\ref{bosonization-rho-varphi}) and
(\ref{boson-theta-phi}), the Hamiltonian $H=H_{\rm{kin}}+
H_{\rm{int}}$ acquires the form (\ref{h-boson}), where
renormalized Fermi velocity $v$ and the dimensionless interaction
parameter $g$ are given by Eqs. (\ref{LL-v-g}). In the {Ga}{As}
slab with a density of one electron per $10$~nm and the width
$R_0\sim1$~nm, $m_e=0.068m_0$ ($m_0$ is the free electron mass),
$v_F\sim10^7$~cm/sec and then $g\sim0.97$.

%\newpage

%%%%%%%%%%%%%%%%%%%%%%%%%%%%%%%%%%%%%%%%%%%%%%%%%%%%%%%%%%%%%%%%%%%%%%%
 \section{Luttinger Liquid Behavior  in Single-Wall Carbon Nanotubes (SWCNT)}
 \label{subsec:LL-SWCN}
%%%%%%%%%%%%%%%%%%%%%%%%%%%%%%%%%%%%%%%%%%%%%%%%%%%%%%%%%%%%%%%%%%%%%%%

\begin{figure}[htb]
\centering
\includegraphics[width=85mm,height=65mm,angle=0,]{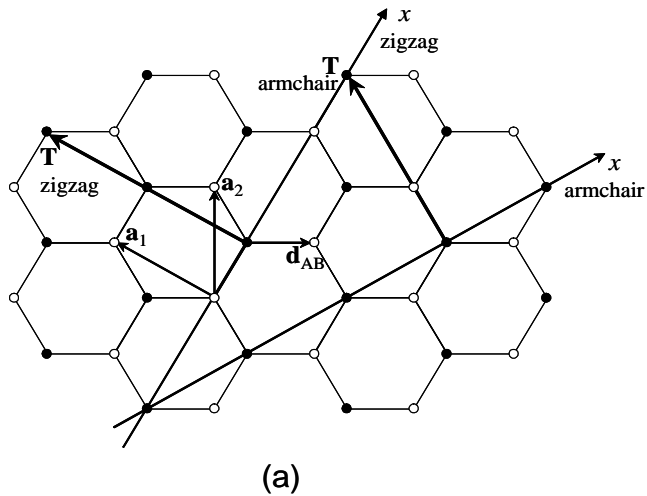}
\caption{Honeycomb lattice of $2D$ carbon sheet and the coordinate
system. Hexagonal sublattices $A$ and $B$ are labelled by
$\bullet$ and $\circ$, respectively. Two primitive translation
vectors are ${\bf{a}}_1$ and ${\bf{a}}_2$, where
$|{\bf{a}}_1|=|{\bf{a}}_2|=a_0=2.5$~{\AA} \cite{Ando}. The vector
directed from an $A$ site to a nearest neighbor $B$ site is
${\bf{d}}_{AB}$. A nanotube is specified by a chiral vector
${\bf{T}}$ corresponding to the circumference of the nanotube
whereas the $x$-axes is oriented along the nanotube axes.}
 \label{CN-Lattice}
\end{figure}

\begin{figure}[htb]
 \centering
 \includegraphics[width=70mm,height=65mm,angle=0,]{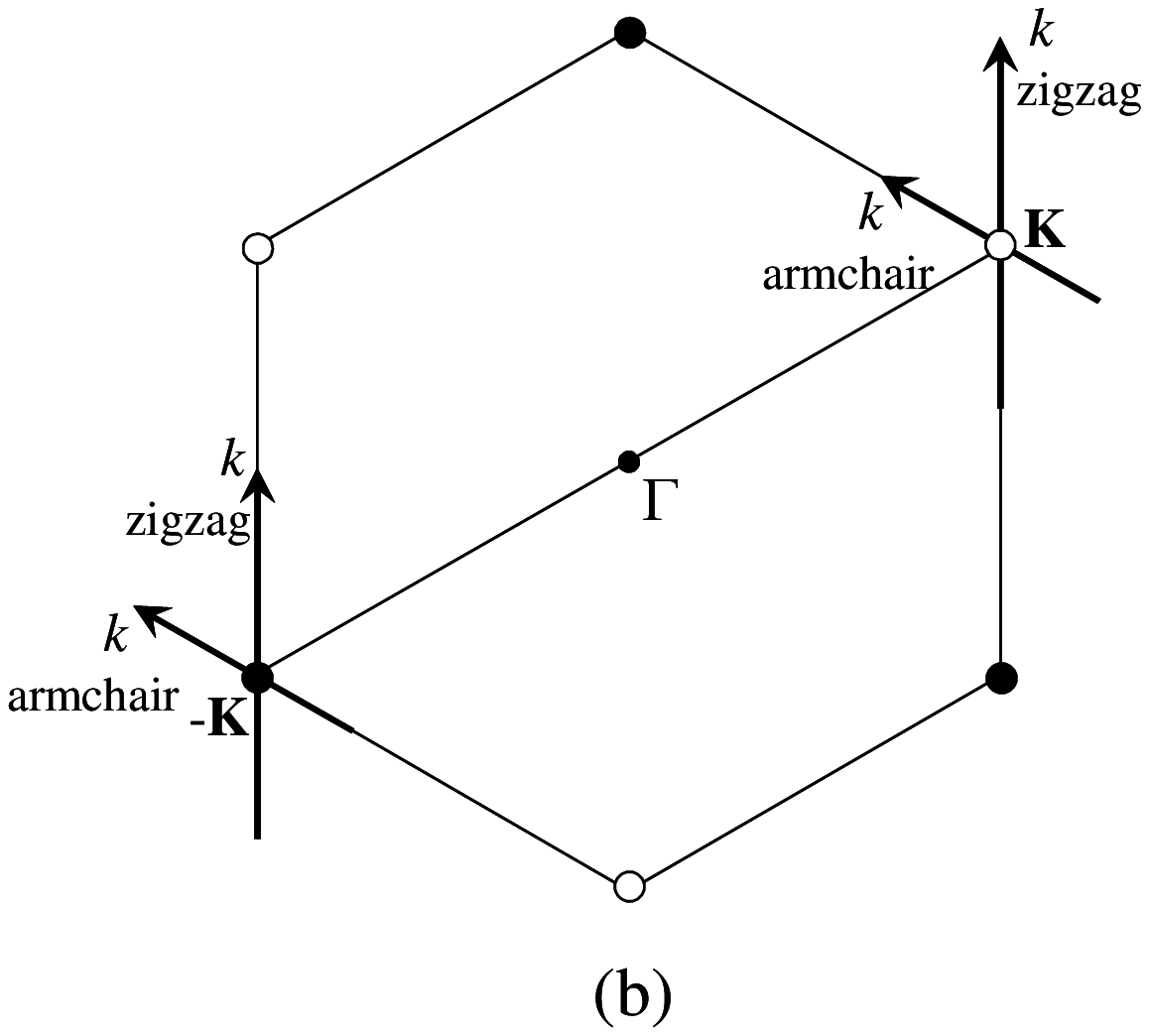}
\caption{First BZ of $2D$ carbon sheet. Here $\circ$ and $\bullet$
are two vertices of the BZ corresponding to the vectors ${\bf{K}}$
and $-{\bf{K}}$ respectively, $|{\bf{K}}|=4\pi/(\sqrt{3}a_0)$.
Wave vectors of low-energy quasi-particles lie at the axes $k$ in
the vicinity of the points $\pm{\bf{K}}$.}
 \label{CN-BZ}
\end{figure}

Nanotubes are tubular nanoscale objects which can be thought as a
graphite sheet wrapped into a cylinder \cite{Iijima1,Iijima2}. The
arrangement of carbon atoms on the tube surface is determined by
the integer indices $0{\le}m{\le}n$ of the wrapping superlattice
vector ${\bf{T}}=n{\bf{a}}_1+m{\bf{a}}_2$ \cite{Ando,Ando2}, where
${\bf{a}}_1$ and ${\bf{a}}_2$ are the primitive Bravais
translation vectors of the honeycomb lattice (see Fig.
\ref{CN-Lattice}). The first Brillouin zone of the honeycomb
lattice is a hexagon (Fig. \ref{CN-BZ}), and there are two
independent Fermi points denoted by ${\bf{K}}$ and $-{\bf{K}}$,
with two linearly dispersing bands around each of these points.
The necessary condition of metallicity of SWCNT is $2n+m=3I$ for
an integer $I$ \cite{Ando}. If this condition is not fulfilled,
the nanotube exhibits the band gap $\Delta E \sim 2\hbar
v_F/(3R_0) \sim1$~eV \cite{Ando,egger98}, where $v_F$ is the Fermi
velocity and $R_0$ is the nanotube radius. Even if the necessary
condition is fulfilled, the rearrangement of local bonds due to
the curvature of the cylinder can introduce a gap,
$\Delta{E}\sim10$~meV, which implies narrow-gap semiconducting
behavior. For very small tube diameter ($1$~nm or less), due to
the strong curvature-induced hybridization of $\sigma$ and $\pi$
orbitals, this effect can be quite pronounced \cite{blase}. In the
cases of {\em ``armchair''} ($n=m$) and {\em ``zigzag''} ($m=0$)
nanotubes, however, the formation of a secondary gap is prevented
by the high symmetry, and therefore armchair and zigzag nanotubes
are metallic \cite{yevtush}.

In this Section, a single metallic SWCNT is under consideration.
The effective low-energy description of SWCNT is derived. Coulomb
interaction between electrons induce a breakdown of Fermi liquid
theory. It is shown that the bosonization procedure (similar to
the procedure derived in Section \ref{sub-LL-intro}) is valid. As
a result, interacting electrons in a metallic SWCNT exhibit
Luttinger liquid behavior.

%%%%%%%%%%%%%%%%%%%%%%%%%%%%%%%%%%%%
%%%%%%%%%%%%%%%%%%%%%%%%%%%%%%%%%%%%

The electronic properties of carbon nanotubes are due to special
band-structure of the $\pi$-electrons in graphite \cite{Louie}.
The band structure exhibits two Fermi points $\kappa{\bf{K}}$
($\kappa=\pm$) with a right- and left-moving ($\alpha=R/L$) branch
around each Fermi point. These branches are highly linear with
Fermi velocity $v_F\approx 8\times10^7$~cm/s. The R- and L-movers
arise as linear combinations of the $\tau=A,B$ sublattice states
reflecting the two C atoms in the basis of the honeycomb lattice.
The dispersion relation is linear for energy scale $E < D$, with
the bandwidth cutoff scale $D\approx\hbar{v_F/R_0}$ for tube
radius $R_0$. For typical SWCNT, $D$ is of the order $1$~eV. The
large overall energy scale together with the structural stability
of SWCNTs explain their unique potential for revealing Luttinger
liquid (LL) physics. The fermionic quasi-particles in the vicinity
of the Fermi level of graphite are described by the $2D$ massless
Dirac Hamiltonian \cite{Ando}. This result can also be derived in
terms of ${\bf{k}}\cdot{\bf{p}}$ theory \cite{Ando2}.

%%%%%%%%%%%%%%%%%%%%%%%%%%%%%%%%%%%%
%%%%%%%%%%%%%%%%%%%%%%%%%%%%%%%%%%%%%

%The effective low-energy theory of graphite is the $2D$ massless
%Dirac Hamiltonian \cite{Ando}. This result can also be derived in
%terms of ${\bf{k}}\cdot{\bf{p}}$ theory.

%%%%%%%%%%%%%%%%%%%%%%%%%%%%%%%%%%%%
\begin{figure}[htb]
\centering
\includegraphics[width=70mm,height=60mm,angle=0,]{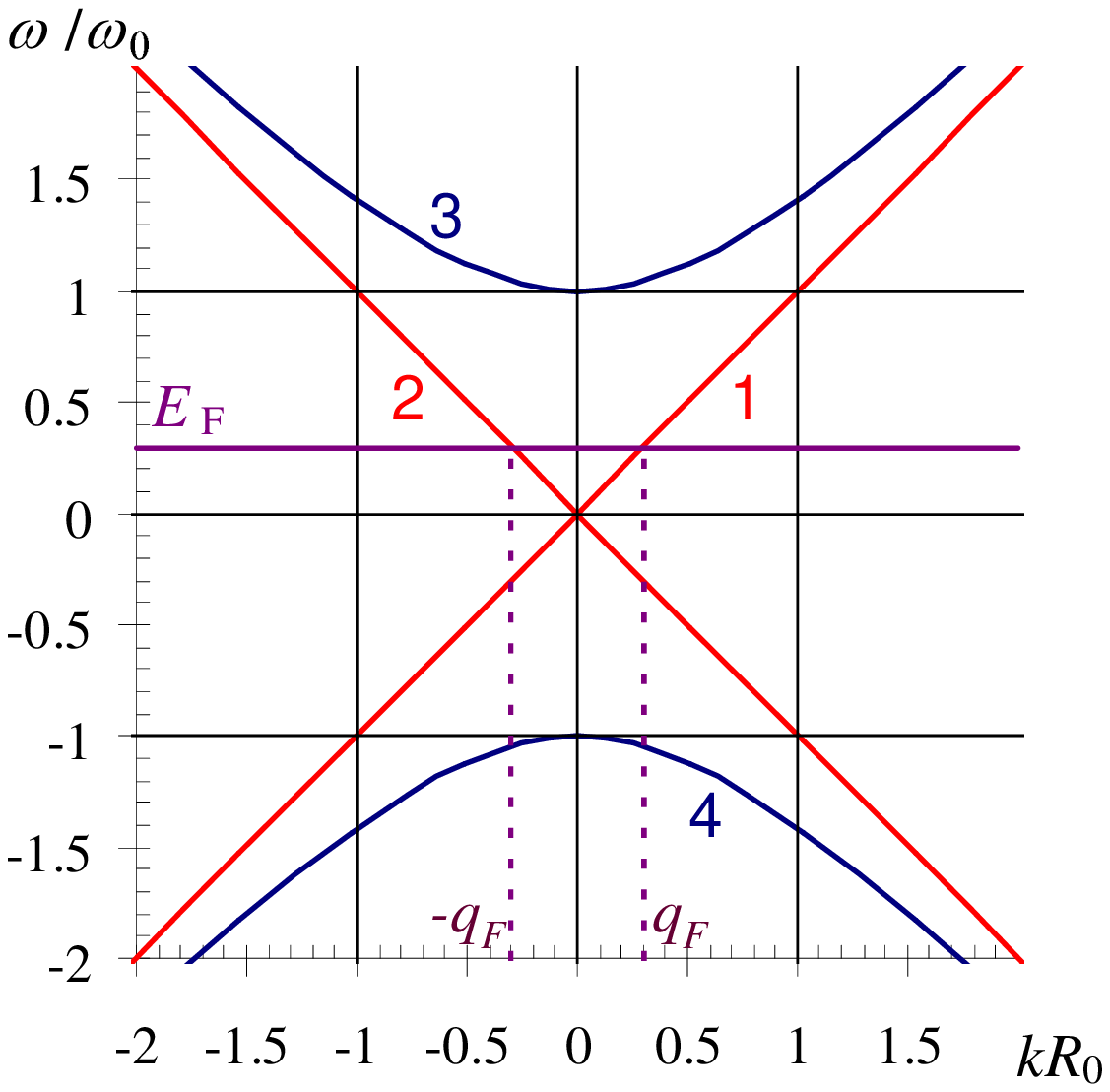}
\caption{The low-energy part of the band structure of a metallic
 nanotube in the vicinity of the vertices ${\bf{K}}$ and ${\bf{K}}'$
 of the BZ (see Fig. \ref{CN-BZ}). Lines $1,2$ ($3,4$)
 correspond to excitations with orbital quantum number $m=0$
 ($m=\pm1$). The bandwidth cutoff scale is $\omega_0={v_F/R_0}$.
 The Fermi level $E_F$ (and momentum $q_F=E_F/v_F$) can be tuned by an
 external gate.}
 \label{CN-sp-2D}
\end{figure}
%%%%%%%%%%%%%%%%%%%%%%%%%%%%%%%%%%%%%

Wrapping the graphite sheet onto a cylinder then leads to the
generic band-structure of a metallic SWCNT shown in
Fig.~\ref{CN-sp-2D}. Quantization of transverse (azimuthal) motion
now allows for a contribution $\propto \exp(i m y/R_0)$ to the
wave function. Here the $x$-axis is taken along the tube
direction, the circumferential variable is $0<y<2\pi R_0$ ($y$ is
really the azimuthal angle multiplied by the nanotube radius
$R_0$). However, excitation of angular momentum states other than
$m=0$ costs a huge energy of order $D\approx1$~eV. In an effective
low-energy theory, we may thus omit all transport bands except
$m=0$ (assuming that the SWCNT is not excessively doped).
Evidently, the nanotube forms a $1D$ quantum wire with only two
transport bands intersecting the Fermi energy. This strict
one-dimensionality is fulfilled up to a remarkably high energy
scales (eV) here, in contrast to conventional $1D$ conductors. The
Hamiltonian of kinetic energy is:
\begin{equation}
  H_{\rm kin}=
  -i\hbar v_F
  \sum_{\kappa\sigma}
  \int{dx}
  \left(
       \psi_{A\kappa\sigma}^{\dagger}(x)
       \partial_x
       \psi_{B\kappa\sigma}(x)+
       \psi_{B\kappa\sigma}^{\dagger}(x)
       \partial_x
       \psi_{A\kappa\sigma}(x)
  \right),
  \label{h0}
\end{equation}
where $\psi_{\tau\kappa\sigma}(x)$ is the ``smooth'' field
operator of fermions in the sublattice $\tau=A,B$ in the vicinity
of the Fermi point $\kappa{\bf{K}}$ ($\kappa=\pm$) with spin
$\sigma=\uparrow,\downarrow$ \cite{egger98}. Switching from the
sublattice ($\tau=A,B$) description to the right- and left-movers
($\alpha=R,L$),
$$
  \psi_{R/L\kappa\sigma}(x)=
  \frac{1}{\sqrt{2}}
  \left(
       \psi_{A\kappa\sigma}(x)
       \mp
       \psi_{B\kappa\sigma}(x)
  \right),
$$
implies two copies ($\kappa=\pm$) of  massless $1D$ Dirac
Hamiltonians (similar to the Hamiltonian (\ref{H-kinetic-0})) for
each spin direction,
$$
  H_{\rm kin}=
  i\hbar v_F
  \sum_{\kappa\sigma}
  \int{dx}
  \left(
       \psi_{L\kappa\sigma}^{\dagger}(x)
       \partial_x
       \psi_{L\kappa\sigma}(x)-
       \psi_{R\kappa\sigma}^{\dagger}(x)
       \partial_x
       \psi_{R\kappa\sigma}(x)
  \right).
$$

The electron-electron interaction is the screened Coulomb
interaction between charge fluctuations \cite{Sasaki}. The
re-distribution of a charge induced by ``external'' charge can be
described by the envelope function (introduced phenomenologically)
$\zeta(\xi)$, $\xi={x}/{R_0}$, $\zeta(\xi)=\zeta(-\xi)$,
$\zeta(0)\sim1$. This function is of order of unity for
$|\xi|\sim1$ and vanishes outside this region. Thus similarly to
Eq. (\ref{U-stripe}), the interaction is introduced as
\begin{equation}
  U({\bf{r}})=
  \frac{e^2
        \zeta
        \left(
             \displaystyle{\frac{x}{R_0}}
        \right)}
       {\left[
             x^2+
             4R_0^2
             \sin^2
             \left(
                  \displaystyle{\frac{y}{2R_0}}
             \right)+
             r_B^2
       \right]^{1/2}}\ ,
  \label{unsc}
\end{equation}
where $r_B$ denotes the average distance between a $2p_z$ electron
and the nucleus, i.e., the ``thickness'' of the graphite sheet. In
the long-wavelength limit, electron-electron interactions are then
described by the second-quantized Hamiltonian \cite{egger98}
\begin{equation}
  H_I=
  \frac{1}{2}
  \sum_{\tau\tau'\sigma\sigma'}
  \sum_{\{\kappa_i\}}
  \int{dxdx'}
  V^{\tau\tau'}_{\{\kappa_i\}}(x-x')
  \psi^{\dagger}_{\tau\kappa_1\sigma}(x)
  \psi^{\dagger}_{\tau'\kappa_2\sigma'}(x')
  \psi_{\tau'\kappa_3\sigma'}(x')
  \psi_{\tau\kappa_4\sigma}(x),
  \label{general}
\end{equation}
with the $1D$ interaction potentials
\begin{equation}
  V^{\tau\tau'}_{\{\kappa_i\}}(x-x')=
  \int\frac{dydy'}{(2\pi R_0)^2}~
  e^{i{\bf{K}}
   \left(
        (\kappa_1-\kappa_4){\bf{r}}+
        (\kappa_2-\kappa_3){\bf{r}}'
   \right)}
  U({\bf{r}}-{\bf{r}}'+{\bf{d}}_{pp'}).
  \label{intpot}
\end{equation}
These potentials depend only on $x-x'$ and on the $1D$ fermion
quantum numbers. For interactions involving different sublattices
$\tau\neq\tau'$ for ${\bf{r}}$ and ${\bf{r}}'$ in
Eq.~(\ref{intpot}), one needs to take into account the shift
vector ${\bf{d}}_{pp'}$
(${\bf{d}}_{AB}=-{\bf{d}}_{BA}$, $|{\bf{d}}_{AB}|=1.44$~\AA~%
\cite{Ando}) between sublattices (see Fig. \ref{CN-Lattice}).

To simplify the resulting $1D$ interaction (\ref{general}), we now
exploit momentum conservation, assuming $E_F\neq 0$ (see Fig.
\ref{CN-sp-2D}) so that Umklapp electron-electron scattering can
be ignored, and only the processes conserving the number of
electrons for each channel $\tau\kappa$ will be considered, i.e.,
$\kappa_1+\kappa_2=\kappa_3+\kappa_4$. We then have ``forward
scattering'' processes and ``exchange interaction'', where
$\kappa_1=\kappa_4$ and $\kappa_2=\kappa_3$. In addition,
``backscattering'' processes may be important, where
$\kappa_1=-\kappa_2=\kappa_3=-\kappa_4$.

In the next step, we employ the fact that the potentials
$V_{\{\kappa_i\}}(x-x')$ are screened with the radius of screening
being of order of the nanotube radius $R_0$, whereas the field
operators $\psi$'s are slowly varying on this distance scale. As a
result, one can approximate the short-range potentials
$V_{\{\kappa_i\}}(x-x')$ by delta-like potentials. Then, switching
to the right- and left-mover representation, the Hamiltonian
(\ref{general}) can be written in the form
\begin{eqnarray}
  H_I &=& H_f+H_x+H_b,
  \nonumber\\
  H_f &=&
  \frac{e^2}{2}
  \int{dx}
  \left[
       \gamma_0\rho^2(x)+
       \frac{\gamma_1}{4}
       \sum_{\kappa}
        \rho_{L\kappa}(x)
        \rho_{R\kappa}(x)
  \right],
  \label{H-int-f}\\
  H_x &=&
  \frac{e^2\gamma_1}{8}
  \int{dx}
  \left[
       \sum_{\kappa}
       \rho_{L\kappa\overline{\kappa}}(x)
       \rho_{R\overline{\kappa}\kappa}(x)+
       2\sum_{\kappa\kappa'}
       {\bf{S}}_{L\kappa\kappa'}(x)
       {\bf{S}}_{R\kappa'\kappa}(x)
  \right],
  \label{H-int-x}\\
  H_b &=&
  \frac{e^2\gamma_b}{2}
  \sum_{\alpha\alpha'\kappa}
  \int{dx}
  \rho_{\alpha\kappa\overline{\kappa}}(x)
  \rho_{\alpha'\overline{\kappa}\kappa}(x),
  \ \ \
  \alpha,\alpha'=L,R,
  \ \
  \kappa=\pm,
  \ \
  \overline{\kappa}=\mp.
  \label{H-int-b}
\end{eqnarray}
Here $H_f$ describes the ``forward scattering'' processes, $H_x$
is the exchange interaction, and $H_b$ corresponds to
``backscattering'',
\begin{eqnarray*}
 &&
  \rho_{\alpha\kappa\kappa'}(x)=
  \sum_{\sigma}
  \psi^{\dag}_{\alpha\kappa\sigma}(x)
  \psi_{\alpha\kappa'\sigma}(x),
  \ \ \ \ \
  \rho_{\alpha\kappa}(x)
  =
  \rho_{\alpha\kappa\kappa}(x),
  \ \ \ \ \
  \rho(x)=
  \sum_{\alpha\kappa}
  \rho_{\alpha\kappa}(x),
 \\
 &&
  {\bf{S}}_{\alpha\kappa\kappa'}(x)=
  \sum_{\sigma\sigma'}
  \psi_{\alpha\kappa\sigma}^{\dag}(x)
  {\mbox{${\boldsymbol{\tau}}$}}_{\sigma\sigma'}
  \psi_{\alpha\kappa'\sigma'}(x),
\end{eqnarray*}
${\boldsymbol\tau}_{\sigma\sigma'}$ is the vector of Pauli
matrices. The effective coupling constants $\gamma_0$, $\gamma_1$,
and $\gamma_b$ are given by the equations:
\begin{eqnarray}
  e^2\gamma_0
  =\int\frac{d^2{\bf{r}}}{2\pi R_0}
    U({\bf{r}}),
  \ \ \ \ \
  \gamma_1
  \approx
    \left(2q_Fd_x^{AB}\right)^2
    \gamma_0,
  \ \ \ \ \
  e^2\gamma_b
  =
    \int\frac{d^2{\bf{r}}}{2\pi R_0}
    e^{i{\bf{K}}{\bf{r}}}
    U({\bf{r}}),
  \label{gamma-01b}
\end{eqnarray}
where $d_x^{pp'}$ is the $x$-component of the vector
${\bf{d}}_{pp'}$, $q_F=E_F/({\hbar}v_F)$. For carbon nanotube,
$d^{AB}=1.44$~\AA,~$q_F\approx1/(3R_0)$,
$R_0\approx4$~\AA~\cite{Egger}, $K=4\pi/(a_0\sqrt{3})$,
$a_0=2.5$~\AA~\cite{Ando}, then $\gamma_0=1.3$,
$\gamma_1\approx0.09$, $\gamma_b\approx0.08$, i.e.,
$\gamma_1,\gamma_b\ll\gamma_0$.

The electron density $\rho_{\alpha\kappa\sigma}$ can be expressed
in terms of the derivative fields
$\partial_x\varphi_{\alpha\kappa\sigma}$ similar to
(\ref{bosonization-rho-varphi})
\begin{eqnarray*}
  &&:\rho_{L\kappa\sigma} (x):=
    \frac{1}{2\pi}\partial_x \varphi_{L\kappa\sigma}(x) \;,
    \ \ \ \ \
    :\rho_{R\kappa\sigma} (x):=
    -\frac{1}{2\pi}\partial_x \varphi_{R\kappa\sigma}(x) \;.
\end{eqnarray*}
Here $:\rho_{\alpha\kappa\sigma}:= \rho_{\alpha\kappa\sigma}-
\langle0|\rho_{\alpha\kappa\sigma}|0\rangle$ denotes fermion
normal ordering with respect to the Fermi sea $|0\rangle$
\cite{Delft}. The fields $\varphi_{\alpha\kappa\sigma}(x)$ satisfy
the following commutation relations
\begin{eqnarray*}
  &&
  \left[
       \varphi_{L/R~\kappa\sigma}(x),
       \varphi_{L/R~\kappa'\sigma'}(x')
  \right]=
  \mp i\pi\delta_{\kappa\kappa'}\delta_{\sigma\sigma'}{\rm{sign}}(x-x'),
  \ \ \ \ \
  \left[
       \varphi_{L\kappa\sigma}(x),
       \varphi_{R\kappa'\sigma'}(x')
  \right]=
  0.
\end{eqnarray*}
It is natural to introduce the standard linear combinations
$\theta_{\lambda\nu}(x)$ and their dual fields
$\phi_{\lambda\nu}(x)$ subject to the algebra
\begin{equation}
  \label{alg2}
  \left[
       \theta_{\lambda\nu}(x),
       \phi_{\lambda'\nu'}(x')
  \right]=
  -\frac{i}{2}
   \delta_{\lambda\lambda'}
   \delta_{\nu\nu'}{\rm{sign}}(x-x') \;.
\end{equation}
The bosonic density fields $\theta_{\lambda\nu}(x)$ for the total
($\nu=g$) and relative ($\nu=u$) charge ($\lambda=c$) and spin
($\lambda=s$) channels are constructed as
\begin{eqnarray*}
  \theta_{c,g/u} &=&
  \frac{1}{4\sqrt{\pi}}
  \left[
         \varphi_{L+\uparrow}
        -\varphi_{R+\uparrow}
       \pm\varphi_{L-\uparrow}
       \mp\varphi_{R-\uparrow}
        +\varphi_{L+\downarrow}
        -\varphi_{R+\downarrow}
       \pm\varphi_{L-\downarrow}
       \mp\varphi_{R-\downarrow}
  \right],
  \\
  \theta_{s,g/u} &=&
  \frac{1}{4\sqrt{\pi}}
  \left[
         \varphi_{L+\uparrow}
        -\varphi_{R+\uparrow}
       \pm\varphi_{L-\uparrow}
       \mp\varphi_{R-\uparrow}
        -\varphi_{L+\downarrow}
        +\varphi_{R+\downarrow}
       \mp\varphi_{L-\downarrow}
       \pm\varphi_{R-\downarrow}
  \right].
\end{eqnarray*}
Their dual phase fields $\phi_{\lambda\nu}$ are defined similarly
\begin{eqnarray*}
  \phi_{c,g/u} &=&
  \frac{1}{4\sqrt{\pi}}
  \left[
         \varphi_{L+\uparrow}
        +\varphi_{R+\uparrow}
       \pm\varphi_{L-\uparrow}
       \pm\varphi_{R-\uparrow}
        +\varphi_{L+\downarrow}
        +\varphi_{R+\downarrow}
       \pm\varphi_{L-\downarrow}
       \pm\varphi_{R-\downarrow}
  \right],
  \\
  \phi_{s,g/u} &=&
  \frac{1}{4\sqrt{\pi}}
  \left[
         \varphi_{L+\uparrow}
        +\varphi_{R+\uparrow}
       \pm\varphi_{L-\uparrow}
       \pm\varphi_{R-\uparrow}
        -\varphi_{L+\downarrow}
        -\varphi_{R+\downarrow}
       \mp\varphi_{L-\downarrow}
       \mp\varphi_{R-\downarrow}
  \right].
\end{eqnarray*}
The Hamiltonian $H_f$ can be written purely in terms of charge
bosonic field operators $\theta_{c,g/u}(x)$ and
$\pi_{c,g/u}(x)=\partial_x\phi_{c,g/u}(x)$,
\begin{eqnarray*}
  H_f &=&
  e^2
  \int{dx}
  \left\{
       \left(
            2\gamma_0-
            \frac{\gamma_1}{8}
       \right)
       \left(
            \partial_x\theta_{cg}(x)
       \right)^2-
       \frac{\gamma_1}{8}
       \left[
            \pi_{cg}^2(x)+
            \pi_{cu}^2(x)-
            \left(
                 \partial_x\theta_{cu}(x)
            \right)^2
       \right]
  \right\}.
\end{eqnarray*}
The Hamiltonian $H_x$ leads to nonlinearities in the $\theta_{cu}$
charge field and the $\theta_{s,g/u}$ spin fields. The four
channels are obtained by combining charge and spin degrees of
freedom as well as symmetric and antisymmetric linear combinations
of the two Fermi points, $\kappa=\pm$. The bosonized expression
for $H_x$ reads \cite{egger}
\begin{eqnarray}
  H_x &=&
  -\frac{4\gamma_1}{L^2}
  \int dx \;
  :\left[
       \cos\left(\sqrt{4\pi} \, \theta_{cu}(x) \right)
       \cos\left(\sqrt{4\pi} \, \theta_{su}(x) \right)+
  \right.\label{H-f2}\\&&\left.+
       \cos\left(\sqrt{4\pi} \, \theta_{cu}(x) \right)
       \cos\left(\sqrt{4\pi} \, \theta_{sg}(x) \right)-
       \cos\left(\sqrt{4\pi} \, \theta_{sg}(x) \right)
       \cos\left(\sqrt{4\pi} \, \theta_{su}(x) \right)
  \right]:.
  \nonumber
\end{eqnarray}
Here $:\ldots:$ denotes boson-normal-ordering defined as follows:
to boson-normal-order a function of operators of creation and
annihilation of bosons, all creation operators are to be moved to
the left of all annihilation operators.

Similar to $H_x$, the backscattering Hamiltonian leads to
nonlinearities in the $\theta_{cu}$ and $\theta_{su}$ fields. The
bosonized expression for $H_{b}$ takes the form \cite{egger}
\begin{eqnarray}
  H_{b} &=&
  \frac{4\gamma_b}{L^2} \int dx \,
  :\left[
       \cos\left(\sqrt{4\pi} \, \theta_{cu}(x) \right)
       \cos\left(\sqrt{4\pi} \, \theta_{su}(x) \right)+
  \right.\label{H-b} \\ \nonumber &&\left.+
       \cos\left(\sqrt{4\pi} \, \theta_{cu}(x) \right)
       \cos\left(\sqrt{4\pi} \, \phi_{su}(x) \right)+
       \cos\left(\sqrt{4\pi} \, \theta_{su}(x) \right)
       \cos\left(\sqrt{4\pi} \,\phi_{su}(x) \right)
  \right]:.
\end{eqnarray}

Writing the non-interacting Hamiltonian $H_0$ (\ref{h0}) in terms
of bosonic field operators, one obtains
\begin{eqnarray}
  H=\sum_{\lambda\nu}H_{\lambda\nu}+H_x+H_b\ ,
  \ \ \
  H_{\lambda\nu} =
  \frac{\hbar v_{\lambda\nu}}{2}
  \int dx
  \left[
       g_{\lambda\nu}\pi_{\lambda\nu}^2(x)+
       \frac{1}{g_{\lambda\nu}}(\partial_x\theta_{\lambda\nu}(x))^2
  \right],
  \label{H-a-boson}
\end{eqnarray}
where $\lambda=c,s$; $\nu=g,u$; $H_x$ is given by Eq.
(\ref{H-f2}), $H_b$ is given by Eq. (\ref{H-b}),
\begin{eqnarray*}
  && v_{\lambda\nu}=\frac{v_F}{g_{\lambda\nu}},
     \ \ \ \ \
     g_{cg}\equiv g=
     \left[
          \frac{1-\widetilde{g}_1}
               {1+\widetilde{g}_0+\widetilde{g}_1}
     \right]^{1/2}
     \approx0.25,
     \ \ \ \ \
     g_{cu}\approx g_{s,g/u}\approx1.
\end{eqnarray*}
Here
$$
  \widetilde{g}_1=\frac{\gamma_1e^2}{4\hbar v_F},
  \ \ \ \ \
  \widetilde{g}_0=\frac{4\gamma_0e^2}{\hbar v_F}.
$$
Clearly, the charged $(cg)$ mode propagates with significantly
higher velocity than the three neutral modes. There is a further
renormalization of the values $v_{cu}$ and $v_{s,g/u}$, however,
this effect is very small and can be neglected. Renormalization
group analysis \cite{egger98} shows that the contribution $H_{x}$
is {\em marginally irrelevant}, whereas the backscattering part
$H_{b}$ is {\em marginally relevant}.

There are several possible regimes in which a nanotube can exist
\cite{Levitov}. First, there is an insulating regime with the
density at half filling ($q_F=0$), where all excitations are
gapped. Second, there are conducting states which can be realized
by applying various external fields. These fields may close some
gaps or even all of them, provided their magnitudes exceed certain
critical values. For example, by varying the chemical potential
(or changing $q_F$), one can close all the gaps. This leads to a
transition into a metallic regime.

%\newpage

%%%%%%%%%%%%%%%%%%%%%%%%%%%%%%%%%%%%%%%%%%%%%%%%%%%
 \section{Sliding Luttinger Liquid Phase}
 \label{subsec:sliding}
%%%%%%%%%%%%%%%%%%%%%%%%%%%%%%%%%%%%%%%%%%%%%%%%%%%

The simplest $2D$ ensemble of $1D$ nanoobjects is an array of
parallel quantum wires or nanotubes. The inter-wire interaction
may transform the LL state existing in isolated quantum wires into
various phases of $2D$ quantum liquid. However, the
density-density or/and current-current inter-wire interactions do
not modify the low-energy behavior of quantum arrays under certain
conditions. In particular, it was shown recently \cite{Vica01}
that ``vertical'' interaction which depends only on the distance
between the wires, imparts the properties of a {\it sliding phase}
to $2D$ array of $1D$ quantum wires.

In this Section, a $2D$ array of coupled $1D$ quantum wires is
considered and the question of existence of a stable $2D$ phase
that retains some of the properties of $1D$ Luttinger liquid is
addressed. Following Ref. \cite{Vica01} this phase will be called
as {\em sliding Luttinger liquid} (SLL). An anisotropic $2D$
system composed of parallel chains with spinless Luttinger liquids
(LL) in each chain (LL in the spin-gapped phase) is considered. It
will be shown that the long-wavelength density-density and/or
current-current interactions between neighboring Luttinger liquids
are marginal operators which result in the sliding
Luttinger-liquid phase.

Let us consider an array with $N$ chains, each labelled by an
integer $n_2=1,2,\ldots,N$. The conventional LL regime in a single
$1D$ quantum wire is characterized by bosonic fields describing
charge modes (LL in the spin-gapped phase). It is assumed that all
wires of the array are identical. They have the same length $L,$
Fermi velocity $v$ and Luttinger liquid parameter $g.$ The period
of the array is $a$. The axis $x_1$ is chosen along the array
direction, whereas the $x_2$ axis is perpendicular to the array.
The excitation motion in QCB is one-dimensional in major part of
the $2D$ plane. The anisotropy in real space imposes restrictions
on the possible values of the coordinate $x_2$. It should be an
integer multiple of the array period $a$, so that the vector
${\bf{r}}=(x_1,n_2a)$ characterizes the point with the $1D$
coordinate $x_1$ lying at the $n_2$-th wire of the array. The
low-energy Luttinger liquid of each wire with spinless interacting
fermions is described by the Hamiltonian (\ref{h-boson}). The
Hamiltonian of the array without inter-wire interaction reads
\begin{equation}
  H_1^0 =
        \frac{\hbar v}{2}
        \sum_{n_2}
        \int\limits_{-L/2}^{L/2}{dx_1}
        \left\{
             g\pi^2(x_1,n_2a)+
             \frac{1}{g}({\partial}_{x_1}\theta(x_1,n_2a))^2
        \right\},
  \label{1D-Hamilt}
\end{equation}
where $\theta,\pi$ are the conventional canonically conjugate
boson fields.

The interactions between the chains correspond to couplings
between the long wavelength components of the densities
$\rho(x_1,n_2a)$ and of the currents $j(x_1,n_2a)$ \cite{Luba01}:
\begin{eqnarray}
  H_{\rm int} &=&
              \frac{\pi\hbar v}{2}\sum_{{n_2}\neq{n'_2}}
              \int\limits_{-L/2}^{L/2}dx_1
              \Big[
                   W_J(n_2-n'_2)j(x_1,n_2a)j(x_1,n'_2a)+
              \label{intra-array-int}\\&&+
                   W_{\rho}(n_2-n'_2)\rho(x_1,n_2a)\rho(x_1,n'_2a)
              \Big].
  \nonumber
\end{eqnarray}
With Eqs. (\ref{bosonization-rho-varphi}) and
(\ref{boson-theta-phi}), the electron density $\rho=\rho_L+\rho_R$
and current $j=\rho_L-\rho_R$ can be expressed in terms of
$\partial_{x_1}\theta$ and $\pi$ respectively.
%These formulas reads
%$:\rho(x_1,n_2a):=\partial_{x_1}\theta(x_1,n_2a)/\sqrt\pi$,
%$:j(x_1,n_2a):=\pi(x_1,n_2a)/\sqrt\pi$.
%where $:\ldots:$ denotes fermion-normal-ordering.
Then the bosonized form of the Hamiltonian $H_1=H_1^0+H_{\rm int}$
for the interacting liquids has the form
\begin{eqnarray*}
  H_1 &=&
  \frac{\hbar v}{2}
  \sum\limits_{n_2n'_2}\int\limits_{-L/2}^{L/2}dx_1
  \big\{
       K_J(n_2-n'_2)
       \pi(x_1,n_2a)\pi(x_1,n'_2a)+
  \nonumber\\&&+
       K_{\rho}(n_2-n'_2)
       (\partial_{x_1}\theta(x_1,n_2a))
       (\partial_{x_1}\theta(x_1,n'_2a))
  \big\},
\end{eqnarray*}
where the coupling matrices are
\begin{eqnarray*}
  K_J(n_2-n'_2)=
  \left(
       g\delta_{n_2n'_2}+
       W_J(n_2-n'_2)
  \right),
  \ \ \ \ \
  K_{\rho}(n_2-n'_2)=
  \left(
       \frac{\delta_{n_2n'_2}}{g}+
       W_{\rho}(n_2-n'_2)
  \right).
\end{eqnarray*}
The Hamiltonian $H_1$ describes coupled harmonic oscillators and
can be diagonalized. By introducing Fourier transforms in the
direction perpendicular to the wires,
$$
  \theta(x_1,n_2a)=\frac{1}{\sqrt{N}}\sum_{q_2}
  e^{iq_2n_2a}\theta_{q_2}(x_1),
  \ \ \ \ \
  \pi(x_1,n_2a)=\frac{1}{\sqrt{N}}\sum_{q_2}
  e^{iq_2n_2a}\pi_{q_2}(x_1),
$$
($|q_2|<Q/2$, $Q=2\pi/a$, $\theta_{q_2}^{\dag}=\theta_{-q_2}$,
$\pi_{q_2}^{\dag}=\pi_{-q_2}$) the Hamiltonian $H_1$ can be
rewritten in the form similar to the Hamiltonian
(\ref{1D-Hamilt}):
\begin{eqnarray}
  H_1 =
  \sum\limits_{q_2}
  \frac{\hbar v_{q_2}}{2}
  \int\limits_{-L/2}^{L/2}dx_1
  \left\{
       g_{q_2}
       \pi_{q_2}^{\dag}(x_1)\pi_{q_2}(x_1)+
       \frac{1}{g_{q_2}}
       \left(\partial_{x_1}\theta_{q_2}^{\dag}(x_1)\right)
       \left(\partial_{x_1}\theta_{q_2}(x_1)\right)
  \right\},
  \label{H-first-array}
\end{eqnarray}
where the Luttinger liquid parameters $g_{q_2}$ and the
velocities $v_{q_2}$ are defined as \cite{Luba01}
\begin{eqnarray*}
  v_{q_2} =
  v\sqrt{K_J(q_2)K_{\rho}(q_2)},
  \ \ \ \ \
  g_{q_2} =
  \sqrt{\frac{K_J(q_2)}{K_{\rho}(q_2)}},
  \ \ \ \ \
  K_{J/\rho}(q_2)=\sum_{n_2}K_{J/\rho}(n_2)e^{iq_2n_2a}.
\end{eqnarray*}

The Hamiltonian (\ref{intra-array-int}) is invariant under the
transformations $\phi(x_1,n_2a)\to\phi(x_1,n_2a)+C_{n_2}$ and
$\theta(x_1,n_2a)\to\theta(x_1,n_2a)+D_{n_2}$, where
$\pi(x_1,n_2a)=\partial_{x_1}\phi(x_1,n_2a)$, $C_{n_2}$ and
$D_{n_2}$ are constants on each wire. The corresponding phase is
called as a SLL one \cite{Vica01}. In this phase, the total
numbers of left (right) moving electrons on each chain are good
quantum numbers and expectation values
$\langle\psi_{\alpha}(x_1,n_2a;t)
\psi_{\alpha'}^{\dag}(x'_1,n'_2a;0) \rangle$
$(\alpha,\alpha'=L,R)$ for ${n_2}\neq{n'_2}$ are {\em necessarily
zero} in this phase. This corresponds to a perfect charge
insulator in the transverse direction. Density correlations in the
transverse direction are, however, nontrivial. For short ranged
density-density and current-current interactions, they decay
exponentially with increasing distance between the wires. The low
energy modes are sets of $1D$ density oscillations (sound)
propagating along {\it each} wire of the array with a wave number
$k_1$ and a phase shift $q_2a$ between adjacent wires. The
dispersion of the modes is linear with respect to $1D$ wave number
$k_1$, $E(k_1,q_2)=v_{q_2}|k_1|$. These modes can, for instance,
transport heat perpendicular to the chains although the system is
a perfect charge insulator in this direction.

%\newpage

%%%%%%%%%%%%%%%%%%%%%%%%%%%%%%%%%%%%%%%%%%%%%%%%%%%
\section{Cross-Sliding Luttinger Liquid Phase}
\label{subsec:cross-sliding}
%%%%%%%%%%%%%%%%%%%%%%%%%%%%%%%%%%%%%%%%%%%%%%%%%%%

\begin{figure}[htb]
\centering
\includegraphics[width=70mm,height=30mm,angle=0,]{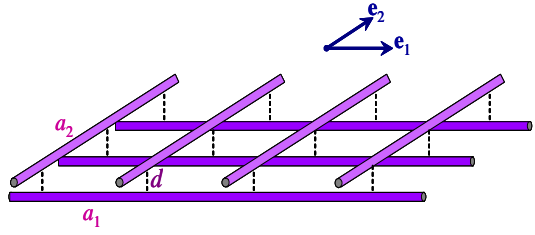}
\caption{$2D$ crossbars formed by two interacting arrays of
parallel quantum wires. Here ${\bf e}_{1}, {\bf e}_{2}$ are the
unit vectors of the superlattice, $a_1=a_2{\equiv}a$ is the
superlattice period (the case $a_1{\ne}a_2$ is considered in
Appendix \ref{append-tilted}) and $d$ is the vertical inter-array
distance.} \label{Bar3}
\end{figure}

Next, a square network of $1D$ wires formed by coupling two
perpendicular arrays of chains is considered. In experimentally
realizable setups \cite{Rueckes} these are cross-structures of
suspended single-wall carbon nanotubes placed in two parallel
planes separated by an inter-plane distance $d$. However, some
generic properties of QCB may be described in assumption that QCB
is a genuine $2D$ system. The system consists of two periodically
crossed arrays of $1D$ quantum wires. It is assumed that all wires
are identical. They have the same length $L$, Fermi velocity $v$
and Luttinger parameter $g$. A coordinate system is chosen so that
the axes $x_j$ and the corresponding basic unit vectors
${\bf{e}}_j$ are oriented along the $j$-th array ($j=1,2$ is the
array index). The period of crossbars is $a$, and the basic
vectors are ${\bf{a}}_j=a{\bf{e}}_j$ (Fig. \ref{Bar3}). The
interaction between the excitations in different wires includes
both interaction between wires from the same array (intra-array
interaction) and wires from different arrays (inter-array
interaction). The former is given by Eq.(\ref{intra-array-int}),
the latter is assumed to be concentrated around the crossing
points with coordinates
$n_1{\bf{a}}_1+n_2{\bf{a}}_2\equiv(n_1a,n_2a)$. The integers $n_j$
enumerate the wires within the $j$-th array. Following Refs.
\cite{Luba00,Luba01} it will be shown that it exhibits a new {\em
crossed sliding Luttinger liquid} (CSLL) phase.

%%%%%%%%%%%%%%%%%%%%%%%%%%%%%%%%%%%%
%%%%%%%%%%%%%%%%%%%%%%%%%%%%%%%%%%%%%

The LL of the first array is described by the Hamiltonian $H_1$
(\ref{H-first-array}). The Hamiltonian $H_{2}$ of the second array
is obtained from $H_{1}$ after permutation $1\leftrightarrow2$ in
the arguments of the fields. The density-density interactions
between electrons on intersecting wires gives rise to a term in
the Hamiltonian in the form
\begin{eqnarray}
  H_{12} &=&
  \sum_{n_1n_2}\int\limits_{-L/2}^{L/2}dx_1dx_2
  V(x_1-n_1a,x_2-n_2a)\rho_1(x_1,n_2a)\rho_2(n_1a,x_2),
  \label{interaction-interarray}
\end{eqnarray}
where $V(x_1-n_1a,x_2-n_2a)$ is a short-range inter-array
interaction. Using the definitions
$\rho_1(x_1,n_2a)=\partial_{x_1}\theta_1(x_1,n_2a)/{\sqrt\pi}$ and
$\rho_2(n_1a,x_2)=\partial_{x_2}\theta_2(n_1a,x_2)/{\sqrt\pi}$,
one obtains the bosonized form of the Hamiltonian $H_{12}$
\cite{Luba01}:
\begin{eqnarray}
  H_{12} &=&
  \frac{1}{\pi}
  \sum_{n_1n_2}\int\limits_{-L/2}^{L/2}dx_1dx_2
  V(x_1-n_1a,x_2-n_2a)
  \partial_{x_1}\theta_1(x_1,n_2a)
  \partial_{x_2}\theta_2(n_1a,x_2).
  \label{H-12}
\end{eqnarray}

Let us introduce Fourier transforms according to \cite{KGKA2}
\begin{eqnarray}
  \theta_1(x_1,n_2a) &=&
  \frac{1}{\sqrt{NL}}\sum_{{\bf qm}_1}
  e^{i(q_1+m_1Q)x_1+iq_2n_2a}\theta_{1{\bf q+m}_1},
  \label{Fourier-theta1}\\
  \theta_2(n_1a,x_2) &=&
  \frac{1}{\sqrt{NL}}\sum_{{\bf qm}_2}
  e^{iq_1n_1a+i(q_2+m_2Q)x_2}\theta_{2{\bf q+m}_2},
  \label{Fourier-theta2}
\end{eqnarray}
where wave vector ${\bf{q}}$ belongs to the first Brillouin zone,
$|q_{1,2}|<Q/2$, ${\bf{m}}_{1,2}=m_{1,2}Q{\bf{e}}_{1,2}$ are
reciprocal superlattice vectors \cite{KGKA3,KGKA4} ($Q=2\pi/a$ is
reciprocal superlattice constant and $m_{1,2}$ are integers). Then
removing degrees of freedom with wavelengthes smaller than $2a$
(that is, considering only the modes with ${\bf{m}}_{1,2}=0$), one
obtains the total Hamiltonian $H=H_1+H_2+H_{12}$ \cite{Luba01}
\begin{eqnarray}
  H &=&
  \sum_{\bf{q}}
  \left(
  \frac{\hbar v_{q_2}g_{q_2}}{2}
  \pi_{1{\bf{q}}}^{\dag}\pi_{1{\bf{q}}}+
  \frac{\hbar v_{q_1}g_{q_1}}{2}
  \pi_{2{\bf{q}}}^{\dag}\pi_{2{\bf{q}}}
  \right)+
  \frac{\hbar v}{2g}\sum_{ij{\bf{q}}}
       \theta_{i{\bf{q}}}^{\dag}
       \Omega^{ij}_{{\bf{q}}}
       \theta_{j{\bf{q}}},
  \label{H-cross-sliding}
\end{eqnarray}
where $i,j=1,2$ denote the array number,
\begin{eqnarray*}
  &&
  \Omega^{ij}_{{\bf{q}}} =
  q_iq_j
  \left[
  \left(
       \frac{v_{q_2}}{v}\frac{g}{g_{q_2}}\delta_{j1}+
       \frac{v_{q_1}}{v}\frac{g}{g_{q_1}}\delta_{j2}
  \right)
  \delta_{ij}+
  g\widetilde{V}({\bf{q}})
  (1-\delta_{ij})
  \right],
  \\
  &&
  \widetilde{V}({\bf{q}}) =
  \frac{1}{\pi\hbar va}\int\limits_{-L/2}^{L/2}dx_1dx_2
  V(x_1,x_2)e^{iq_1x_1+iq_2x_2}.
\end{eqnarray*}
The Hamiltonian (\ref{H-cross-sliding}) exhibits a CSLL phase.
Renormalization group analysis \cite{Luba01} shows that additional
interactions between the two arrays, such as the Josephson and
charge-density-wave couplings, are irrelevant and the CSLL phase
is stable.

It should be noted that the Hamiltonian (\ref{H-cross-sliding})
does not include interactions of the long-wavelength plasmons with
quasi-particles whose wavelengthes smaller than $2a$, i.e.,
plasmons in higher energy bands. Being weak, these interactions
renormalize the plasmon velocity (see Eq. (\ref{omega-12})). When
the interaction strength increases, the lowest QCB modes soften
and their frequencies vanish {\em in a whole BZ} at a certain
critical of the interaction strength (see Section
\ref{subsubsec:Approx}).

%\newpage

%%%%%%%%%%%%%%%%%%%%%%%%%%%%%%%%%%%%%%%%%%%%%%%%%%%
\section{Quantum Crossbars with Virtual Wire-to-Wire Tunneling}
\label{subsec:QCB-with-tunneling}
%%%%%%%%%%%%%%%%%%%%%%%%%%%%%%%%%%%%%%%%%%%%%%%%%%%

{To finalize the substantiation of the CSLL family, consider the
condition when the tunneling of electrons in quantum crossbars is
suppressed and the Luttinger-liquid-like phase is stable. For the
case when the charge degrees of freedom are quenched, we derive
the effective spin Hamiltonian of QCB.}

Let us consider two non-parallel metallic nanotubes. One of them
belongs to the first array and has the number $n_2$, and another
one belongs to the second array and has the number $n_1$. The
intersecting point has the coordinates $(n_1a,n_2a)$. The electron
hopping between the wires gives rise to a term in the Hamiltonian
of the form: $\sum_{\sigma}t [\Psi_{1\sigma}^{\dag}(n_1a,n_2a)
\Psi_{2\sigma}(n_1a,n_2a)+h.c.]$, where $t$ is an effective
hopping constant, and $\sigma= \uparrow, \downarrow$ is a spin
index. Next we introduce ``slowly varying'' field operators
$\psi_{L\sigma}$ and $\psi_{R\sigma}$ (similar to (\ref{Psi-LR})).
Assuming that the Fermi vector $k_F$ is not commensurate with the
reciprocal superlattice vector $Q=2\pi/a$, we can write
$k_F=m_FQ+q_0$, with $m_F$ being integer and $|q_0|<Q/2$. Then we
represent the hopping between the two arrays as
\begin{eqnarray}
 H_t =
     t\sum_{n_1n_2}\sum_{\alpha,\alpha'}\sum_{\sigma}
     \left[
          e^{-iq_{\alpha}n_1a+iq_{\alpha'}n_2a}
          \psi_{1\alpha\sigma}^{\dag}(n_1a,n_2a)
          \psi_{2\alpha'\sigma}(n_1a,n_2a)
          +h.c.
     \right],&&
 \label{H-t-lr}
\end{eqnarray}
where $q_L=q_0$ and $q_R=-q_0$. Here and below we assume that $t$
is real and positive.

The energy cost of the electron wire-to-wire tunneling is the
energy $2E_C$ necessary to charge both wires,
\begin{eqnarray}
  E_C &=& \frac{2e^2}{L}\ln\left(\frac{L}{2R_0}\right),
  \label{Coulomb-Blocade}
\end{eqnarray}
where $L$ is the nanotube length and $R_0$ is the nanotube radius.
The tunneling is suppressed if ${t}\ll{E_C}$. The tunneling
constant $t$ can be estimated from the transport experiment
through crossed nanotubes. The elastic and van der-Waals
interactions between crossed nanotubes determine two equilibrium
positions \cite{Rueckes} with inter-wire distances $d=1$~nm and
2~nm. It is shown that for $d=2$~nm, the resistance is
$R\sim{10}^{10}$~$\Omega$. On the other hand, resistance can be
evaluated from the well known Landauer formula
\begin{equation}
  R=\frac{1}{G},
  \ \ \ \
  G=\frac{2e^2}{h}\frac{\nu_0t^2}{E_Ca^2},
  \ \ \ \
  \nu_0=\frac{L}{2\pi\hbar v_F},
  \label{Resist}
\end{equation}
where $\nu_0$ is the density of states in the quantum wires. For
real nanotubes, the length of a ballistic transport is
$L_{\rm{exp}}\sim1\mu$m and $E_C\sim20$~meV \cite{Egger}. Taking
$v=8\cdot10^7$~cm/sec, $R_0=0.4$~nm, and $a=20$~nm, we have
$t/a\sim1~\mu$eV, i.e., ${t}\ll{E_C}$. Then, single-particle
hopping between nanotubes is suppressed, and only backward
scattering interaction can be relevant. However, if $k_F$ is not
commensurate with the reciprocal lattice constant $Q$, the back
scattering is also suppressed.

In this case the effective interaction Hamiltonian can be obtained
from the initial one (\ref{H-t-lr}) by means of elimination the
states from adjacent charge sectors by using the second order
perturbation theory with respect to $t/(aE_C)\ll1$:
\begin{eqnarray}
  H_{\rm{int}} = \frac{t^2}{2E_C}
          \sum_{n_1n_2}
          \big\{
            \rho_{1}(n_1a,n_2a)\rho_{2}(n_1a,n_2a)+
%          \nonumber\\&+&
            2{\bf{S}}_{1}(n_1a,n_2a)\cdot{\bf{S}}_{2}(n_1a,n_2a)
          \big\}.
  \label{H-eff3}
\end{eqnarray}
Here $\rho_j$ is the electron density operator and ${\bf{S}}_j$ is
the spin operator,
$$
  \rho_{j}=
  \sum_{\alpha\sigma}
  \psi_{j\alpha\sigma}^{\dag}
  \psi_{j\alpha\sigma},
  \ \ \ \ \
  {\bf{S}}_{j}=
  \frac{1}{2}
  \sum_{\alpha\sigma\sigma'}
  \psi_{j\alpha\sigma}^{\dag}
  {\boldsymbol{\tau}}_{\sigma\sigma'}
  \psi_{j\alpha\sigma'},
$$
where ${\boldsymbol{\tau}}$ is the vector of Pauli matrices.

The first term under the sum in the right-hand side of
Eq.(\ref{H-eff3}) describes the potential interaction between
charge fluctuations and simply renormalizes the coupling strength
$V(x_1,x_2)$ of the Coulomb inter-wire interaction
(\ref{interaction-interarray}). The second term under the sum in
the right-hand side of Eq.(\ref{H-eff3}) is spin-spin inter-wire
interaction.

The Hamiltonian (\ref{H-eff3}) shows that the virtual wire to wire
hopping results in only a slight renormalization of the strength
of the capacitive inter-wire interaction. However, the virtual
hoping results in the effective spin-spin interaction.

\textbf{Short summary and future outlook:} \textit{It was
demonstrated in this Chapter that the LL fixed point is conserved
in a quasi $2D$ periodic network (quantum crossbars). In the
following chapters we concentrate on another aspect of the
problem: we consider the properties of QCB at finite wave vectors
${\bf{q}}$ and frequencies $\omega$ in the whole $2D$ Brillouin
zone, study the spectrum of Bose excitations $\omega(q_1,q_2)$,
consider situations where dimensional crossover
$1D\leftrightarrow2D$ occurs, formulate the principles of infrared
and ultraviolet spectroscopy for QCB and study the basic
characteristics of the spectra.}

%\newpage

%%%%%%%%%%%%%%%%%%%%%%%%%%%%%%%%%%%%%%%%%%%%%%%%%%%
 \chapter{Plasmon Excitations and One to Two Dimensional
          Crossover in Quantum Crossbars}
 \label{sec:spectrum}
%%%%%%%%%%%%%%%%%%%%%%%%%%%%%%%%%%%%%%%%%%%%%%%%%%%

{In this Chapter we analyze the spectrum of a specific nano-object
- square double quantum crossbars (QCB). We show that the standard
bosonization procedure is valid and the system behaves as
cross-sliding Luttinger liquid in the infrared limit. However
plasmon excitations in QCB demonstrate both $1D$ and $2D$ behavior
depending on the direction of the plasmon wave vector. We discuss
several crossover effects such as appearance of non-zero
transverse space correlators and the periodic energy transfer
between arrays (``Rabi oscillations''). The spectra of a tilted
double and triple QCB are considered in Appendices
\ref{append-tilted} and \ref{subsec:Triple}.}

%%%%%%%%%%%%%%%%%%%%%%%%%%%%%%%%%%%%%%%%%%%%%%%%%%%
 \section{Introduction}
 \label{subsec:Spectr-intro}
%%%%%%%%%%%%%%%%%%%%%%%%%%%%%%%%%%%%%%%%%%%%%%%%%%%

A double $2D$ grid, i.e., two superimposed crossing arrays of
parallel conducting quantum wires or nanotubes, represents a
specific nano-object which is called quantum crossbars (QCB). Its
spectral properties cannot be treated in terms of purely $1D$ or
$2D$ electron liquid theory. A constituent element of QCB (quantum
wire or nanotube) possesses the LL-like spectrum
\cite{Bockrath,Egger}. A single array of parallel quantum wires is
still a LL-like system qualified as a sliding phase \cite{Mukho1}
provided only the electrostatic interaction between adjacent wires
is taken into account. If tunnelling is suppressed and the two
arrays are coupled only by electrostatic interaction in the
crosses, the system possesses the LL zero energy fixed point.

In the present Chapter we concentrate on another aspect of the
problem of capacitively interacting arrays of quantum wires.
Instead of studying the conditions under which the LL behavior is
preserved in spite of inter-wire interaction, we show that a rich
Bose-type excitation spectrum (plasmon modes) arises at finite
energies in the $2D$ Brillouin zone (BZ) \cite{Kuzm,KKGA} and
consider situations where the {\it dimensional crossover} from
$1D$ to $2D$ occurs \cite{KGKA2,GKKA}. We start our studies of QCB
with a double square QCB, namely, $m=2$. In the first two Sections
\ref{subsubsec:Notions} and \ref{subsubsec:Hamilt} we introduce
basic notations and construct the Hamiltonian of the QCB. The main
approximations are discussed in Section \ref{subsubsec:Approx}.
Here we substantiate the method used (separable interaction
approximation) and show that interaction between arrays in QCB is
weak.  The energy spectra for square QCB are described in detail
in Section \ref{subsubsec:Spectr}. Various correlation functions
and related experimentally observable quantities (optical
absorption, space correlators) are discussed in the last Section
\ref{subsubsec:Observ2}. We predict here the effect of peculiar
``Rabi oscillations'' - periodic energy transfer from one of the
QCB array to another. All technical details are placed in
Appendices \ref{append:Empty} and \ref{append:DoublSpectr&Corr}.

Double tilted QCB is considered in Appendix \ref{append-tilted}.
Triple QCB ($m=3$) formed by three arrays lying in parallel planes
are studied in Appendix \ref{subsec:Triple}. Such hexagonal grids
may be useful for three-terminal nanoelectronic devices
\cite{Luo}. The plasmon spectra of triple QCB possess some
specific features in comparison with double QCB. We introduce the
main notions and construct the Hamiltonian of symmetric triple QCB
(section \ref{subsubsec:NotHam}), analyze the peculiarities of the
frequency spectrum (section \ref{subsubsec:SpecTriple}), and
illustrate them by description of triple Rabi oscillations -
periodic energy transfer between all three arrays (part
\ref{subsubsec:Observ3}). The results are summarized in the
Conclusions.

%%%%%%%%%%%%%%%%%%%%%%%%%%%%%%%%%%%%%%%%%%%%%%%%%%%
\section{Basic Notions}
 \label{subsubsec:Notions}
%%%%%%%%%%%%%%%%%%%%%%%%%%%%%%%%%%%%%%%%%%%%%%%%%%%

Double square QCB is a $2D$ periodic grid, which is formed by two
periodically crossed arrays of $1D$ quantum wires. In
experimentally realizable setups \cite{Rueckes} these are
cross-structures of suspended single-wall carbon nanotubes placed
in two parallel planes separated by an inter-plane distance $d$.
However, some generic properties of QCB may be described in the
assumption that QCB is a genuine $2D$ system.  We assume that all
wires of QCB are identical. They have the same length $L,$ Fermi
velocity $v$ and Luttinger parameter $g.$ The periods of a
crossbars is $a.$ The arrays are oriented along the unit vectors
${\bf e}_{1,2}$, and corresponding basic vectors are
${\bf{a}}_j=a{\bf{e}}_j$ (Fig. \ref{Bar3}).

The interaction between the excitations in different wires is
assumed to be concentrated around the crossing points with
coordinates $n_1{\bf a}_1+n_2{\bf a}_2\equiv(n_1a,n_2a)$. The
integers $n_j$ enumerate the wires within the $j$-th array. Such
an interaction imposes a super-periodicity on the energy spectrum
of initially one dimensional quantum wires, and the eigenstates of
this superlattice are characterized by a $2D$ quasimomentum
${\bf{q}}=q_1{\bf{e}}_1+q_2{\bf{e}}_2\equiv(q_1,q_2)$. The
corresponding basic vectors of the reciprocal superlattice have
the form ${\bf{m}}={\bf{m}}_1+{\bf{m}}_2$, where
${\bf{m}}_{1,2}=m_{1,2}Q{\bf{e}}_{1,2}$, $Q=2\pi/a$ and $m_{1,2}$
are integers.

\begin{figure}[htb]
\centering
\includegraphics[width=140mm,height=60mm,angle=0,]{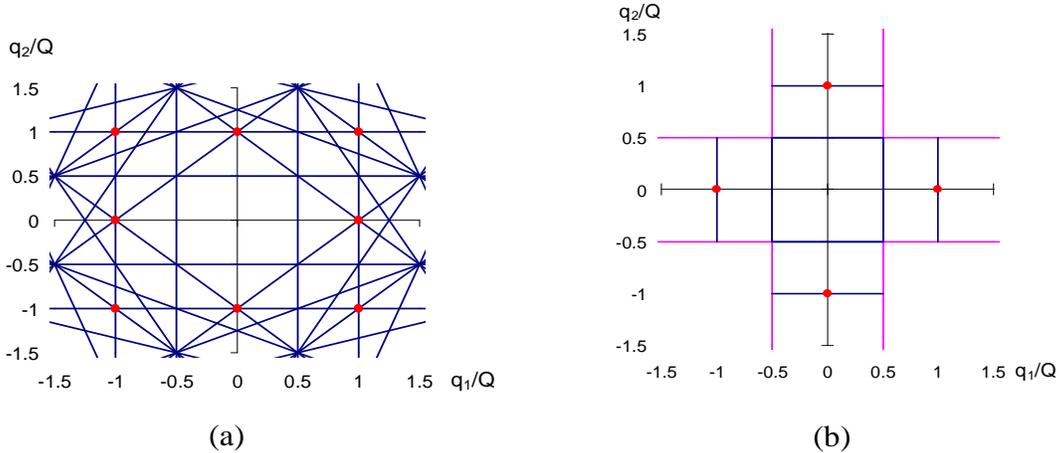}
\caption{The Bragg lines in conventional $2D$ square lattice
{\bf(a)} and that of $2D$ square crossbars {\bf(b)}. In both
panels, solid lines are the Bragg lines, whereas points correspond
to reciprocal superlattice vectors.} \label{Bragg-lines}
\end{figure}

However, the crossbars kinematics differs from that of a standard
$2D$ periodic system. In conventional $2D$ systems, forbidden
states in the inverse space arise due to Bragg diffraction in a
$2D$ periodic potential, whereas the whole plane is allowed for
wave propagation in real space, at least when the periodic
potential is weak enough. A set of Bragg lines correspond to
different reciprocal lattice vectors ${\bf{m}}$. A Brillouin zone
is bounded by the Bragg lines. It coincides with a Wigner-Seitz
cell of reciprocal lattice (Fig. \ref{Bragg-lines}a). In contrast,
most of the real space is forbidden for electron and plasmon
propagation in sharply anisotropic QCB. The Bragg conditions for
the wave vectors of a quasi-particle in an array are modulated by
a periodic potential created by another array, unlike those in
conventional $2D$ plane. There are two sets of Bragg lines
corresponding to reciprocal lattice vectors ${\bf{m}}_1$ and
${\bf{m}}_2$. These conditions are essentially one-dimensional.
The corresponding BZ is a Wigner-Seitz cell of a reciprocal
lattice shown in Fig. \ref{Bragg-lines}b.

%%%%%%%%%%%%%%%%%%%%%%%%%%%%%%%%%%%%
%%%%%%%%%%%%%%%%%%%%%%%%%%%%%%%%%%%%%

\begin{figure}[htb]
\centering
\includegraphics[width=60mm,height=60mm,angle=0,]{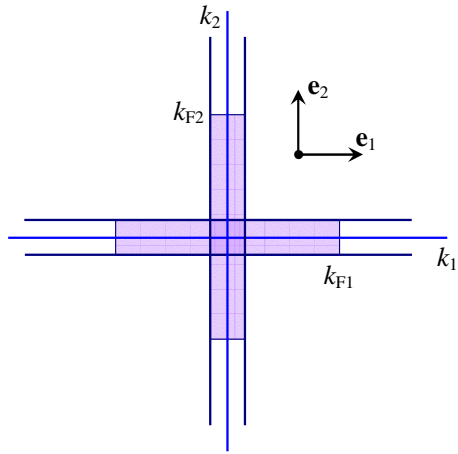}
\caption{Fermi surface of $2D$ metallic quantum bars in the
absence of charge transfer between wires.} \label{FS3}
\end{figure}

Indeed, the excitation motion in QCB is one-dimensional in major
part of the $2D$ plane.  The anisotropy in real space imposes
restrictions on the possible values of $2D$ coordinates
$x_{1},x_{2}$ (${\bf r}=x_{1}{\bf e}_{1}+x_{2}{\bf e}_{2}$).  At
least one of them, e.g., $x_2$ ($x_{1}$) should be an integer
multiple of the corresponding array period $a$, so that the vector
${\bf r}=(x_1,n_2a)$ (${\bf r}=(n_1a,x_2)$) characterizes the
point with the $1D$ coordinate $x_1$ ($x_2$) lying at the $n_2$-th
($n_1$-th) wire of the first (second) array. As a result, one
cannot resort to the standard basis of $2D$ plane waves when
constructing the eigenstate with a given wave vector ${\bf k}$.
Even in {\it non-interacting} arrays of quantum wires (empty
superlattice) the $2D$ basis is formed as a superposition of two
sets of $1D$ waves. The first of them is a set of $1D$ excitations
propagating along {\it each} wire of the first array characterized
by a unit vector $k_1{\bf e}_1$ with a phase shift $ak_2$ between
adjacent wires. The second set is the similar manifold of
excitations propagating along the wires of the second array with
the wave vector $k_2{\bf e}_2$ and the phase shift $ak_1$. The
dispersion law of these excitations has the form
$\omega^{0}({\bf{k}})=\omega(k_1)+\omega(k_2)$. The states of
equal energy obtained by means of this procedure form straight
lines in the $2D$ reciprocal space.  For example, the Fermi
surface of QCB developed from the points $\pm k_{F1,2}$ for
individual quantum wire consists of two sets of lines
$|k_{1,2}|=k_{F1,2}$. Respectively, the Fermi sea is not a circle
with radius $k_F$ like in the case of free $2D$ gas, but a cross
in the $k$ plane bounded by these four lines \cite{Guinea} (see
Fig. \ref{FS3}). Finally, the Bragg conditions read
\begin{eqnarray*}
    &&
    \omega(k_1)-\omega(k_1+ m_1Q)
    +\omega(k_2)-\omega(k_2+ m_2Q)=0.
%   \label{Bragg}
\end{eqnarray*}
and the lines $|k_1|=Q/2$ and $|k_2|=Q/2$ satisfying these
conditions, form a $2D$ BZ of double QCB.

%%%%%%%%%%%%%%%%%%%%%%%%%%%%%%%%%%%%
%%%%%%%%%%%%%%%%%%%%%%%%%%%%%%%%%%%%%%

Due to the inter-array interaction, the plasmons in QCB (see Figs.
\ref{BZ2}, \ref{BZ3} below) exist in a $2D$ Brillouin zone. These
excitations are characterized by the quasimomentum
${\bf{q}}=(q_1,q_2)$.  However, in case of weak interaction, the
$2D$ waves are constructed from the $1D$ plane waves in accordance
with the above procedure. Two sets of $1D$ plane waves form an
appropriate basis for the description of elementary excitations in
QCB in close analogy with the nearly free electron approximation
in conventional crystalline lattices. It is easily foreknown that
a weak inter-array interaction does not completely destroy the
above quasimomentum classification of eigenstates, and the $2D$
reconstruction of the spectrum may be described in terms of wave
mixing similarly to the standard Bragg diffraction in a weak
periodic potential. Moreover, the classification of eigenstates of
an empty superlattice may be effectively employed for the
classification of energy bands in a real QCB superlattice where
the super-periodicity is imposed by interaction.

Complete kinematics of an empty super-chain (wave functions,
dispersion laws, relations between quasiparticle second
quantization operators) is developed in Appendix
\ref{append:Empty}. In terms of these $1D$ Bloch functions (see
Eqs.  (\ref{WaveFunc}), (\ref{WaveFunc1}) of Appendix
\ref{append:Empty}) we construct the $2D$ basis of Bloch functions
for an empty superlattice
\begin{equation}
  \Psi_{ss'{\bf q}}({\bf r})=
   \psi_{sq_1}(x_1)\psi_{s'q_{2}}(x_{2}).
   \label{Psi}
\end{equation}
Here $s,s'=1,2,\ldots,$ are the band numbers, and the $2D$
quasimomentum ${\bf q}=(q_1, q_2)$ belongs to the first BZ,
$|q_j|\leq Q/2.$ The corresponding eigenfrequencies
$$
    \omega_{ss'}({\bf q})=
    \omega_{1s}(q_1)+\omega_{2s'}(q_2).
$$
Here $\omega_{js}(q_j)\equiv \omega_{s}(q_j),$ and
$\omega_{s}(q_j)$ are formed by two $1D$ acoustic branches
propagating along the $j$-th array and belonging to the band $s.$
Their explicit form is defined in Appendix \ref{append:Empty}
(Eq.(\ref{WaveFunc1})). Alternatively, each mode with
quasimomentum ${\bf{q}}$ in the energy band $s$ (reduced BZ
description) propagating along the $j$-th array can be described
by the wave vector ${\bf{q}}+{\bf{m}}_j$ (extended BZ
description), where ${\bf{m}}_j=m_jQ{\bf{e}}_j$ and
$m_j=(-1)^{s}{\rm{sign}}(q_j)[s/2]$. We will use both these bases
in the next Section when constructing the excitation spectrum of
QCB within the reduced band scheme and in the next Chapters where
we formulate the spectroscopic principles of QCB within the
extended band scheme.

%%%%%%%%%%%%%%%%%%%%%%%%%%%%%%%%%%%%%%%%%%%%%%%%%%%%%%%%%%%%%%%
\section{Hamiltonian}\label{subsubsec:Hamilt}
%%%%%%%%%%%%%%%%%%%%%%%%%%%%%%%%%%%%%%%%%%%%%%%%%%%%%%%%%%%%%%%

When turning to the description of interaction in a QCB, one
should refer to a real geometry of crossbars, and recollect the
important fact that the equilibrium distance between two arrays is
finite and large enough to suppress direct electron tunnelling
\cite{Rueckes}.  We neglect also the elastic and van der-Waals
components of the interaction between real nanotubes, because
these interactions are not involved in the formation of collective
excitations in QCB. Then the full Hamiltonian of the QCB is
\begin{equation}
  H_{QCB} = H_1 + H_2 + H_{12},
  \label{TotHam1}
\end{equation}
where the Hamiltonian $H_1$ (\ref{1D-Hamilt}) describes the $1D$
boson field in the first array, the Hamiltonian $H_{2}$ is
obtained from $H_{1}$ after permutation $1\leftrightarrow 2$
in the arguments of the fields.

The inter-wire interaction includes both interactions between
wires from the same array (intra-array interaction) and  wires
from different arrays (inter-array interaction). The latter is the
Hamiltonian (\ref{H-12}) which results from a contact capacitive
coupling in the crosses of the bars. Physically, the short-range
inter-array interaction $V(x_1-n_1a,n_2a-x_2)$ represents the
screened Coulomb interaction between charge fluctuations around
the crossing point $(n_1a,n_2a).$ We assume that the crossed
nanotubes are suspended in an unpolarized medium, and  screening
arises due to intra-wire interaction. The nanotube diameter is the
only physical parameter which determines the screening length
$r_0$ in a tube (see e.g. Ref.\cite{Sasaki}). We describe the
re-distribution of a charge in tube $j$ induced by the interaction
with tube $i$ by the envelope function (introduced
phenomenologically)
\begin{equation}
    \zeta(\xi_j), \phantom{aa}
    \xi_j=\frac{x_j-n_ja}{r_0},
    \phantom{aa}
    \zeta(\xi)=\zeta(-\xi),\phantom{aa}
    \zeta(0)\sim 1.
    \label{zeta}
\end{equation}
This function is of order unity for $|\xi|\sim 1$ and vanishes
outside this region. Thus the on-cross interaction is introduced
as
\begin{equation}
    V({\bf r}_{12})=\frac{V_{0}}{2}
        \Phi\left
        (\frac{x_1-n_1a}{r_0},\frac{n_2a-x_2}{r_0}
        \right),
    \ \ \ \ \
    \Phi(\xi_{1},\xi_{2})=
        \frac
    {
    \displaystyle{
    \zeta(\xi_{1})
    \zeta(\xi_{2})
    }}
    {\displaystyle{
    \sqrt{1+\frac{|{\bf r}_{12}|^2}{d^2}
    }}
    },
        \label{F}
\end{equation}
where ${\bf r}_{12}=r_0\xi_{1}{\bf e}_{1}-r_0\xi_{2}{\bf e}_{2}$.
It is seen from these equations that $\Phi(\xi_1,\xi_2)$ vanishes
for $|\xi_{1,2}|\ge1$ and satisfies the condition
${\Phi}(0,0)\sim1$. The effective coupling strength is
\begin{equation}
    V_0=\frac{2e^{2}}{d}.
    \label{strength}
\end{equation}
In terms of boson field operators ${\theta}_j$, the inter-array
interaction is written in the form similar to (\ref{H-12})
\begin{eqnarray}
    H_{{int}} = V_{0}\sum\limits_{{n}_{1},{n}_{2}}
            \int dx_1 dx_2
            \Phi
            \left(
                 \frac{x_1-n_1a}{r_0},
                 \frac{n_2a-x_2}{r_0}
            \right)
            \partial_{x_1}\theta_1(x_1,n_2a)
            \partial_{x_2}\theta_2(n_1a,x_2).
     \label{Interaction}
\end{eqnarray}

As for the inter-wire interaction within each array, one can
neglect it for a couple of reasons.  First, the inter-wire
distance within the same array is much larger than the inter-array
distance. Second, this interaction is irrelevant in the long-wave
limit \cite{Luba01}. Thus Eq. (\ref{Interaction}) is the full
interaction Hamiltonian.

In the quasimomentum representation (\ref{Psi}) the full
Hamiltonian (\ref{TotHam1}) acquires the form,
\begin{eqnarray}
H_{QCB} & = & \frac{{\hbar}{v}{g}}{2}\displaystyle{
                      \sum_{j=1}^{2}
                      \sum_{s,{\bf q}}
                      }
                      {\pi}_{js{\bf q}}^{\dagger}
                      {\pi}_{js{\bf q}}+
          \frac{\hbar}{2vg}\displaystyle{
          \sum_{jj'=1}^{2}
                 \sum_{ss'{\bf q}}
                 }
                 {\Omega}_{jj'ss'{\bf q}}
                 {\theta}_{j s {\bf q}}^{\dagger}
                 {\theta}_{j's'{\bf q}},
                 \label{TotHam2}
\end{eqnarray}
where $\theta_{js{\bf{q}}}$ and $\pi_{js{\bf{q}}}$ are the Fourier
components of the boson fields $\theta_j$ and $\pi_j$.

The matrix elements for inter-array coupling are given by:
\begin{eqnarray}
  \Omega_{jj'ss'{\bf{q}}} & = &
    {\omega}_{s }(q_j)
    {\omega}_{s'}(q_{j'})
    \left[
         {\delta}_{jj'}{\delta}_{ss'}+
         \phi{\Phi}_{ss'{\bf{q}}}
         \left(
              1-{\delta}_{jj'}
         \right)
    \right],
  \ \ \ \ \
  \phi =
  \frac{gV_{0}r_{0}^{2}}{{\hbar}va}.
  \label{eq-phi}
\end{eqnarray}
Here $\omega_{s}(q_j)=v\left([s/2]Q+(-1)^{s-1}|q_j|\right)$ are
eigenfrequencies of the ``unperturbed'' $1D$ mode (see Eq.
(\ref{WaveFunc1}) of Appendix \ref{append:Empty}), pertaining to
an array $j$, band $s$ and quasimomentum ${\bf q}=q_j{\bf{e}}_j.$
The coefficients
\begin{eqnarray}
  {\Phi}_{ss'{\bf{q}}}=
  (-1)^{s+s'}
  {\rm{sign}}(q_1q_2)
   \int d{\xi}_{1}d{\xi}_{2}{\Phi}({\xi}_{1},{\xi}_{2})
   e^{-ir_0(q_1\xi_1+q_2\xi_2)}
   u_{s q_1}^{*}(r_0\xi_1)
   u_{s'q_2}^{*}(r_0\xi_2),
  \label{Phi-Fourier}
\end{eqnarray}
(${\Phi}_{ss'{\bf{q}}}={\Phi}^{*}_{s's{\bf{q}}}$) are proportional
to the dimensionless Fourier component of the interaction
strengths, where the Bloch amplitudes $u_{sq_j}(x_j)$ are given by
Eq. (\ref{WaveFunc1}) of Appendix \ref{append:Empty}.

The Hamiltonian (\ref{TotHam2}) describes a system of coupled
harmonic oscillators, which can be {\em exactly} diagonalized with
the help of a certain canonical linear transformation (note that
it is already diagonal with respect to the quasimomentum ${\bf
q}$).  The diagonalization procedure is, nevertheless, rather
cumbersome due to the mixing of states belonging to different
bands and arrays. However, it will be shown below that provided
$d\gg{r}_0$, a separable potential approximation is applicable,
that significantly simplifies the calculations.

%%%%%%%%%%%%%%%%%%%%%%%%%%%%%%%%%%%%%%%%%%%%%%%%%%%%%%%%%%%%%%%%%
\section{Approximations}
 \label{subsubsec:Approx}
%%%%%%%%%%%%%%%%%%%%%%%%%%%%%%%%%%%%%%%%%%%%%%%%%%%%%%%%%%%%%%%%%

As it has already been mentioned, we consider the rarefied QCB
with short range capacitive interaction.  In the case of QCB
formed by nanotubes, this is a screened Coulomb interaction at a
distance of the order of the nanotube radius $R_{0}$
\cite{Sasaki}, therefore $r_{0}\sim R_{0}.$ The minimal radius of
a single-wall carbon nanotube is about $R_{0}=0.35\div 0.4 nm$
(see Ref. \cite{Louie}). The inter-tube vertical distance $d$ in
artificially fabricated nanotube networks is estimated as
$d\approx 2$nm (see Ref. \cite{Rueckes}). Therefore the ratio
$r_0^{2}/d^{2}\approx{0.04}$ is really small and {\it the
dimensionless interaction} $\Phi(\xi_1,\xi_2)$ (\ref{F}) {\it in
the main approximation is separable}
\begin{equation}
  \Phi(\xi_{1},\xi_{2})\approx\Phi_{0}(\xi_{1},\xi_{2})=
  \zeta(\xi_1)\zeta(\xi_2).
  \label{separ}
\end{equation}

To diagonalize the Hamiltonian (\ref{TotHam2}), one should solve
the system of equations of motion for the field operators. The
generalized coordinates $\theta$ satisfy the equations
\begin{eqnarray}
  \left[\omega_{s}^{2}(q_1)-\omega^{2}\right]
  \theta_{1s{\bf{q}}}+
  \sqrt{\varepsilon}\zeta_{s}(q_1)\omega_{s}(q_1)
  \frac{r_0}{a}\sum\limits_{s'}
  \zeta_{s'}(q_2)\omega_{s'}(q_2)
  \theta_{2s'{\bf{q}}}=0,
  \ \ \
  s=1,2,\ldots,
  \label{Euler-Lagr}
\end{eqnarray}
and the similar equations obtained by permutation
$1\leftrightarrow 2$.  Here
\begin{equation}
  \zeta_{s}(q)=(-1)^s\mbox{sign}(q)\int d\xi
  \zeta(\xi) e^{ir_0q\xi}u_{sq}(r_0\xi),
  \label{FC-2}
\end{equation}
the Bloch amplitudes $u_{sq}(r_{0}\xi)$ are defined by Eq.
(\ref{WaveFunc1}) of Appendix \ref{append:Empty}, and
\begin{equation}
    \varepsilon=\left(\phi\frac{a}{r_0}\right)^{2}=
    \left(\frac{gV_0r_0}{{\hbar}v}\right)^{2}.
    \label{epsilon}
\end{equation}
Due to separability of the interaction, the equations of motion
(\ref{Euler-Lagr}) can be solved exactly. The corresponding
eigenfrequencies are determined by the characteristic equation
\begin{equation}
 F_{q_1}(\omega^2)F_{q_2}(\omega^2)=\frac{1}{\varepsilon},
 \label{secul-eq}
\end{equation}
where
\begin{equation}
 F_{q_j}(\omega^2)=\frac{r_0}{a}\sum\limits_{s}
 \frac{\zeta_{s}^{2}(q_j)\omega_{s}^{2}(q_j)}
      {\omega_{s}^{2}(q_j)-\omega^2}.
 \label{F_j}
\end{equation}
The function $F_{q_j}(\omega^2)$ has a set of poles at
$\omega^2=\omega_{s}^{2}(q_j)$ ($s=1,2,3,\ldots$). For squared
frequency smaller than all squared initial eigenfrequencies
$\omega_{s}^{2}(q_j)$, i.e., within the interval
$[0,\omega_{1}^{2}(q_j)]$, this is a positive and monotonically
increasing function. Its minimal value $F_0$ on the interval is
reached at $\omega^2=0,$ and it does not depend on quasimomentum
$q_j$
\begin{equation}
F_{q_j}(0)=\frac{r_0}{a}\sum\limits_{s}
 \zeta_{s}^{2}(q_j)=\int d\xi \zeta^2(\xi)\equiv F_0
\label{F_j0}
\end{equation}
(here Eqs.  (\ref{F_j}) and (\ref{FC-2}) are used). If the
parameter $\varepsilon$ is smaller than its critical value
$\varepsilon_c=1/F_0^2$, then all solutions $\omega^{2}$ of the
characteristic equation are positive. When $\varepsilon$
increases, the lowest QCB mode softens and its frequency vanishes
\textit{in the entire BZ} at $\varepsilon=\varepsilon_{c}$. For
exponential charge density distribution $\zeta(\xi)=\exp(-|\xi|),$
one obtains $\varepsilon_c\approx1$.

In our model, the dimensionless interaction $\varepsilon$ in
Eq.(\ref{epsilon}) can be written as
\begin{equation}
    \varepsilon=\left(\frac{2R_{0}}{d}\frac{ge^{2}}{\hbar v}\right)^{2}.
    \label{epsilon1}
\end{equation}
For nanotube QCB, the first factor within parentheses is about
$0.35.$ The second one which is nothing but the corresponding QCB
``fine structure'' constant, can be estimated as $0.9$ (we used
the values of $g=1/3$ and $v=8\times 10^{7}$cm/sec, see Ref.
\cite{Egger}). Therefore $\varepsilon$ approximately equals $0.1,$
so this parameter is really small. Thus the considered system is
stable, its spectrum is described by Eqs.(\ref{secul-eq}),
(\ref{F_j}) with a \emph{small} parameter $\varepsilon$.

The general Eq.(\ref{secul-eq}) reduces in the infrared limit
${\bf{q}},\omega\to0$ to an equation describing the spectrum of
two coupled sliding phases, i.e., $1:1$ arrays in accordance with
classification scheme, offered in Ref. \cite{Luba01}. Equation
(3.13) of this paper is the long wave limit of our equations
(\ref{RenFr}) and (\ref{freq_a_1}) derived in Appendix
\ref{append:DoublSpectr&Corr}. Therefore, the general analysis of
stability of the LL fixed point is applicable in our approach.

%%%%%%%%%%%%%%%%%%%%%%%%%%%%%%%%%%%%%%%%%%%%%%%%%%%%%%%%%%%%%%%
\section{Spectrum}
 \label{subsubsec:Spectr}
%%%%%%%%%%%%%%%%%%%%%%%%%%%%%%%%%%%%%%%%%%%%%%%%%%%%%%%%%%%%%%%

Due to the weakness of the interaction, the systematics of
unperturbed levels and states is grossly conserved, at least in
the low energy region corresponding to the first few energy bands.
This means that perturbed eigenstates could be described by the
same quantum numbers (array number, band number and quasimomentum)
as the unperturbed ones. Such a description fails in two specific
regions of reciprocal space. The first of them is the vicinity of
the high-symmetry lines $q_j=nQ_j/2$ with $n$ integer. Indeed, as
it follows from the equations of motion (\ref{Euler-Lagr}), around
these lines the inter-band mixing is significant. These lines with
$n=\pm1$ include BZ boundaries. The second region is the vicinity
of the lines where the resonance conditions are fulfilled
\begin{equation}
    \omega^{2}_{s}(q_{1})=\omega^{2}_{s}(q_{2}).
    \label{res}
\end{equation}
Here inter-array mixing within the same energy band $s$ is
significant. In what follows we will pay attention first of all to
these two regions because in the rest of the BZ the initial
systematics of the energy spectrum can be successfully used.

Equations (\ref{Euler-Lagr}), (\ref{secul-eq}), describing the
wave functions and the dispersion laws  are analyzed in Appendix
\ref{append:DoublSpectr&Corr}. We describe below some of these
dispersion curves for a square QCB based on this analysis (the
case of tilted QCB is described in Appendix \ref{append-tilted}).

\begin{figure}[htb]
 \centering
 \includegraphics[width=60mm,height=55mm,angle=0,]{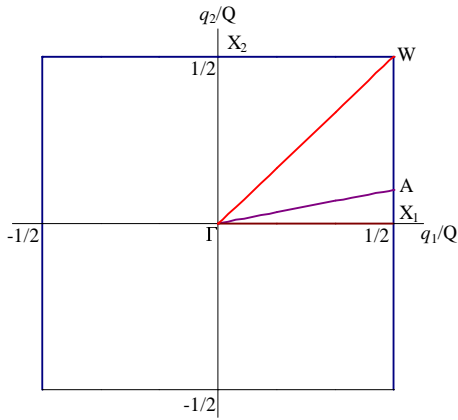}
 \caption{Two dimensional BZ of square QCB.}
 \label{BZ2}
\end{figure}

We start with the simplest case of square QCB formed by identical
wires. This means that all parameters (wire length, space period,
Fermi velocity, LL parameter, screening radius) are the same for
both arrays. The corresponding BZ is also a square (see Fig.
\ref{BZ2}). The resonant lines are the diagonals of BZ.

%%%%%%%%%%%%%%%%%%%%%%%%%%%%%%%%%%%%%%%%%%%%%%%%%%%%%%%%%%%%%%
%%%%%%%%%%%%%%%%%%%%%%%%%%%%%%%%%%%%%%%%%%%%%%%%%%%%%%%%%%%%%%

In the major part of the BZ, for quasimomenta ${\bf{q}}$ lying far
from the diagonals, each eigenstate mostly conserves its initial
systematics, i.e. belongs to a given array, and mostly depends on
a given quasimomentum component.  Corresponding dispersion laws
remain linear being slightly modified near the BZ boundaries only.
The main change is therefore the renormalization of the plasmon
velocity.

%%%%%%%%%%%%%%%%%%
\begin{figure}[htb]
\centering
\includegraphics[width=90mm,height=60mm,angle=0,]{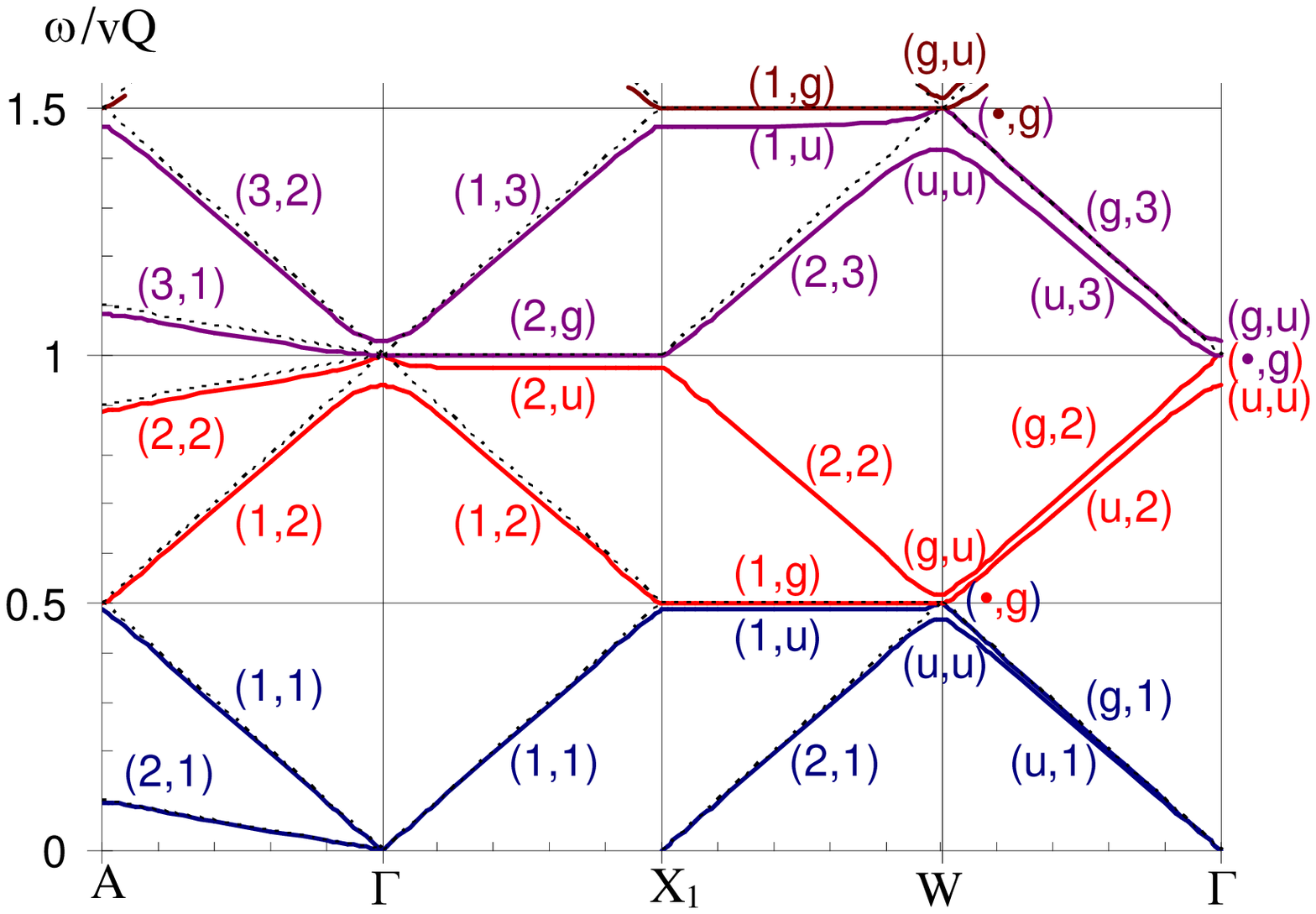}
 \caption{The energy spectrum of QCB (solid lines) and
 noninteracting arrays (dashed lines) for quasimomenta at the lines
 $A\Gamma,$ $\Gamma X_1,$ $X_1W$, and $W\Gamma$ of BZ.}
 \label{sp-sq}
\end{figure}
%%%%%%%%%%%%%

{\bf i. General case: The point ${\bf{q}}$ lies away from both the
high symmetry lines and the resonant lines.} This case is
illustrated in the left part of Fig. \ref{sp-sq}, where we display
dispersion curves corresponding to quasimomenta belonging to a
generic $\Gamma A$ line in the BZ. In what follows we use $(j,s)$
notations for the unperturbed boson propagating along the $j$-th
array in the $s$-th band. Then the lowest two curves in this part
of Fig. \ref{sp-sq} are, in fact, the slightly renormalized
dispersion of $(2,1)$ and $(1,1)$ bosons. The middle curves
describe $(1,2)$ and $(2,2)$ bosons, and the upper curves are the
dispersion of $(2,3)$ and $(1,3)$ bosons. It is seen that the
dispersion remains linear along the whole line $\Gamma A$ except
at the nearest vicinity of the BZ boundary (point $A$ in Fig.
\ref{sp-sq} and point $\Gamma$ for highest energy bands).

{\bf ii. Inter-band resonance in one of the arrays: the point
${\bf{q}}$ lies on a high symmetry line of only one array.} This
case is illustrated by the lines $\Gamma X_1$ and $X_1W$ in Fig.
\ref{BZ2}. Dispersion curves corresponding to quasi momenta lying
at the line $q_2=0$, $0\leq q_1\leq Q/2$ (line $\Gamma X_1$ in
Fig. \ref{sp-sq}) and the BZ boundary $q_{1}=Q/2,$ $0\leq
q_{2}\leq Q/2$ (line $X_1W$ in Fig. \ref{BZ2}) are displayed in
the central parts in Fig. \ref{sp-sq}). The characteristic feature
of these lines is the intra-band degeneracy in one of two arrays.
Indeed, in zero approximation, two modes $(2,s),$ $s=2,3$,
propagating along the second array with a quasimomentum lying in
the line $\Gamma X_1$ are degenerate with unperturbed frequency
$\omega=1$. The interaction lifts the degeneracy. This interaction
appears to be repulsive. As a result, the lowest of the two middle
curves in Fig. \ref{sp-sq} corresponds to $(2,u)$ boson, and the
upper among them describes $(2,g)$ boson. Here the indices $g,u$
denote a boson parity with respect to the transposition of the
band numbers. Note that the $(2,g)$ boson exactly conserves its
unperturbed frequency $\omega=1.$ The latter fact is related to
the square symmetry of the QCB.

Other curves correspond to almost non perturbed bosons of the
first array. The lowest two curves describe the dispersion of the
$(1,1)$ and $(1,2)$ waves. Plasmons in the third band $(1,3)$ are
described by the uppermost curve in the figure. Their dispersion
laws are nearly linear, and deviations from linearity are observed
only near the boundary of the BZ (point $X_1$ in Fig. \ref{BZ2})
and near the point $\Gamma$ for the highest bands.

Similarly, in zero approximation, two modes $(1,s),$ $s=1,2$
($3,4$), propagating along the first array with a quasimomentum
lying in the line $X_1W$ are degenerate with unperturbed frequency
$\omega=0.5$ ($1.5$). The interaction lifts the degeneracy. As a
result, the lowest of two middle (highest) curves in Fig.
\ref{sp-sq} corresponds to $(2,u)$ boson, and the upper of them
describes $(2,g)$ boson. As in the previous case, $(2,g)$ boson
exactly conserves its unperturbed frequency $\omega=0.5$ ($1.5$).

Other curves correspond to almost non perturbed bosons of the
second array. The lowest curve describes the dispersion of the
$(2,1)$ wave. The two middle curves describe the dispersion of
$(2,2)$ and $(2,3)$ plasmons. Their dispersion laws are nearly
linear, and deviations from linearity are observed only near the
corner of the BZ (point $W$ in Fig. \ref{BZ2}) as well as in the
vicinity of the point $X_1$ of the BZ.

{\bf iii Inter-band resonance in both arrays: the point $X_1$
($X_2$) is a crossing point of the two high symmetry lines away
from all resonant lines.} This case is illustrated by the points
$X_1$ and $X_2$ in Fig. \ref{BZ2}. Consider for example point
$X_1$. Here $q_1=Q/2$, $Q_2=0$. In zero approximation, two modes
$(1,s),$ $s=1,2$ ($s=3,4$), propagating along the first array are
degenerate with unperturbed frequency $\omega=0.5$ ($\omega=1.5$).
The lower (higher) two lines correspond to even $(1,g)$ and odd
$(1,u)$ superpositions of the $1$-st array states of the first and
second (third and fourth) bands. Similarly, two modes $(2,s)$ with
$s=2,3$ are degenerate in zero approximation with unperturbed
frequency $\omega=1$. Therefore the middle two lines describe the
same superpositions of the $2$-d array states from the second and
third bands.

{\bf iv.  Inter-array resonance: The point ${\bf{q}}$ lies only on
one of the resonant lines away from the high symmetry lines.} This
case is illustrated by the diagonal $\Gamma W$ of BZ. Consider now
dispersion relations of modes with quasi-momenta on the diagonal
$\Gamma W$ of BZ and start with ${\bf q}$ not too close to the BZ
corner $W$ ($q_{1}=q_{2}=Q/2$) and the $\Gamma$ point. This
diagonal is actually one of the resonance lines. Two modes in the
first band corresponding to different arrays are strongly mixed.
They mostly have a definite $j$-parity with respect to
transposition of array numbers $j=1,2$. Interaction between these
modes appears to be attractive (repulsive) for $q_{1}q_{2}>0$
($q_{1}q_{2}<0$). Therefore the odd modes $(u,s),$ at the BZ
diagonal $\Gamma W$ $s=1,2,$ correspond to lower frequencies and
the even modes $(g,s)$ correspond to higher ones. The
corresponding dispersion curves are displayed in the right part of
Fig. \ref{sp-sq}.

{\bf v. Inter-array and inter-band resonance: The point ${\bf{q}}$
lies at the cross of two resonant lines.} There are two points,
the BZ corner $W$ and the BZ center $\Gamma$ in Fig. \ref{BZ2}. At
the BZ corner $q_{1}=q_{2}=Q/2$ (point $W$ in Fig. \ref{BZ2}) all
four initial modes $j=1,2$ and $s=1,2$ ($s=3,4$) are degenerate in
the lowest approximation. This four-fold degeneracy results from
the square symmetry of BZ (the resonant lines are diagonals of the
BZ). Weak inter-array interaction partially lifts the degeneracy.
However, the split modes have a definite $s$-parity with respect
to transposition of band numbers $s=1,2$ ($s=3,4$). The highest
frequency corresponds mostly to $(g,u)$ boson, symmetric with
respect to transposition of array numbers, but antisymmetric with
respect to the transposition of band numbers. The lower curve
describes a $(u,u)$ boson with both $j$-parity and $s$-parity odd.
The two middle modes with even band parity, $(g,g)$ and $(u,g)$
bosons, remain degenerate and their frequencies conserve the
unperturbed value $\omega=0.5$ ($\omega=1.5$). This also results
from the square symmetry of QCB (\ref{F}).

Similar behavior is observed in the BZ center $\Gamma$. All four
initial modes $j=1,2$ and $s=2,3$ are degenerate in the lowest
approximation. Weak inter-array interaction partially lifts the
degeneracy. The highest frequency corresponds mostly to $(g,u)$
boson,the lower curve describes a $(u,u)$ boson, and the two
middle bosons, $(g,g)$ and $(u,g)$, remain degenerate and their
frequencies conserve the unperturbed value $\omega=1$.

%%%%%%%%%%%%%%%%%%%%%%%%%%%%%%%%%%%%%%%%%%%%%%%%%%%%%%%%%%%%%%%
\begin{figure}[htb]
\centering
\includegraphics[width=65mm,height=65mm,angle=0,]{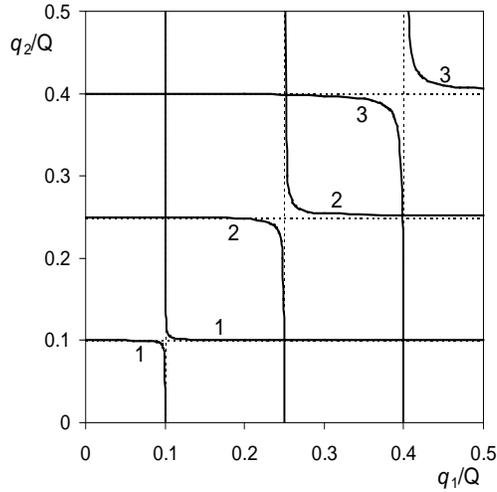}
\caption{Lines of equal frequency of the lowest mode for QCB
(solid lines) and for noninteracting arrays (dashed lines).  The
lines $1,2,3$ correspond to the frequencies $\omega_{1}=0.1,$
$\omega_{2}=0.25,$ $\omega_{3}=0.4.$} \label{IsoEn}
\end{figure}
%%%%%%%%%%%%%%%%%%%%%%%%%%%%%%%%%%%%%%%%%%%%%%%%%%%%%%%%%%%%%%
All these results show that the quantum states of the $2D$ QCB
conserve the quasi $1D$ character of the Luttinger--like liquid in
major part of momentum space, and that $2D$ effects can be
successfully calculated within the framework of perturbation
theory. However, bosons with quasimomenta close to the resonant
line (diagonal $OC$) of the BZ are strongly mixed bare $1D$
bosons.  These excitations are essentially two-dimensional, and
therefore the lines of equal energy in this part of the BZ are
modified by the $2D$ interaction (see Fig. \ref{IsoEn}).  It is
clearly seen that deviations from linearity occur only in a small
part of the BZ. The crossover from LL to FL behavior around
isolated points of the BZ due to a single-particle hybridization
(tunnelling) for Fermi excitations was noticed in Refs.
\cite{Guinea,Guinea2}, where a mesh of horizontal and vertical
stripes in superconducting cuprates was studied.

%%%%%%%%%%%%%%%%%%%%%%%%%%%%%%%%%%%%%%%%%%%%%%%%%%%%%%%%%%%%%%%%%%%
\section{Correlations and Observables}
 \label{subsubsec:Observ2}
%%%%%%%%%%%%%%%%%%%%%%%%%%%%%%%%%%%%%%%%%%%%%%%%%%%%%%%%%%%%%%%%%%%

The structure of the energy spectrum analyzed above predetermines
optical and transport properties of QCB. We consider here three
types of correlation functions manifesting dimensional crossover
in QCB.

%%%%%%%%%%%%%%%%%%%%%%%%%%%%%%%%%%%%%%%%%%%%%%%%%%%%%%%%%%%%%%%%
\subsection{Optical Absorption}
 \label{subsec-OptAbs}
%%%%%%%%%%%%%%%%%%%%%%%%%%%%%%%%%%%%%%%%%%%%%%%%%%%%%%%%%%%

We start with {\it ac} conductivity
\begin{eqnarray*}
 {\sigma}_{jj'}({\bf{q}},\omega)=
 {\sigma}'_{jj'}({\bf{q}},\omega)+i{\sigma}''_{jj'}({\bf{q}},\omega).
\end{eqnarray*}
The real part ${\sigma}'_{jj'}({\bf{q}},\omega)$ determines an
optical absorption.  The spectral properties of \emph{ac}
conductivity are given by a current--current correlator
\begin{equation}
      {\sigma}_{jj'}({\bf{q}},\omega)=\frac{1}{\omega}
      \int\limits_{0}^{\infty}dt{e}^{i{\omega}t}
      \left\langle \left[
          {J}_{j1{\bf{q}}}(t),{J}_{j'1{\bf{q}}}^{\dag}(0)
      \right] \right\rangle.
    \label{CurrCorr}
\end{equation}
Here ${J}_{js{\bf{q}}}=\sqrt{2}vg{\pi}_{js{\bf{q}}}$ is a current
operator for the $j$-th array (we restrict ourselves to the first
band, for the sake of simplicity).

The current-current correlator for non-interacting wires is
reduced to the conventional LL expression \cite{Voit},
\begin{eqnarray*}
\left\langle \left[
     {J}_{j1{\bf{q}}}(t),{J}_{j'1{\bf{q}}}^{\dag}(0)
\right] \right\rangle_0=
  -2ivg{\omega}_{1{\bf{q}}}
  \sin({\omega}_{1{\bf{q}}}t)
  {\delta}_{jj'}
\end{eqnarray*}
with metallic-like peak
\begin{equation}
{\sigma}'_{jj'}({\bf{q}},\omega>0)= {\pi}vg
\delta({\omega}-{\omega}_{1{\bf{q}}}) {\delta}_{jj'}.
\label{Drude_peak}
\end{equation}
For QCB this correlator is calculated in Appendix
\ref{append:DoublSpectr&Corr}. Its analysis leads to the following
results. The longitudinal absorption
\begin{equation}
    \sigma'_{11}({\bf q},\omega)\propto
    (1-\phi^{2}_{1{\bf q}})
    \delta(\omega-\tilde\omega_{1{\bf q}})+
    \phi^{2}_{1{\bf q}}
    \delta(\omega-\tilde\omega_{2{\bf q}})\nonumber
%   \label{eq:LongOut}
\end{equation}
contains well pronounced peak on the modified first array
frequency and weak peak at the second array frequency (the
parameter $\phi_{1{\bf{q}}}= \sqrt{\varepsilon}
 \zeta_1(q_1)\zeta_1(q_2) \omega_1(q_1)\omega_1(q_2)$ is small).
The modified frequencies $\tilde\omega_{1{\bf q}}$ and
$\tilde\omega_{2{\bf q}}$ coincide with the eigenfrequencies
$\omega_{+1{\bf q}}$ and $\omega_{-2{\bf q}}$ respectively, if
$\omega_{1{\bf q}}>\omega_{2{\bf q}}.$ In the opposite case the
signs $+,-$ should be changed to the opposite ones.

The transverse absorption component contains two weak peaks
\begin{equation}
    \sigma'_{12}({\bf q},\omega)\propto
    \phi_{1{\bf q}}\left[
    \delta(\omega-\tilde\omega_{1{\bf q}})+
    \delta(\omega-\tilde\omega_{2{\bf q}})
    \right].\nonumber
\end{equation}

At the resonant line, the results are drastically modified. Both
longitudinal and transverse components of the optical absorption
contain two well pronounced peaks corresponding to slightly split
modified frequencies
\begin{equation}
    \sigma'_{11}({\bf q},\omega)\propto\frac{1}{2}
    \left[
    \delta(\omega-\tilde\omega_{1{\bf q}})+
    \delta(\omega-\tilde\omega_{2{\bf q}})
    \right].\nonumber
%   \label{eq:LongAt}
\end{equation}

%%%%%%%%%%%%%%%%%%%%%%%%%%%%%%%%%%%%%%%%%%%%%%%%%%%%%%%%%%%%
\subsection{Space Perturbation}
 \label{subsec-SpaceCorr}
%%%%%%%%%%%%%%%%%%%%%%%%%%%%%%%%%%%%%%%%%%%%%%%%%%%%%%%%%%%%

One of the main effects specific for QCB is the appearance of
non-zero transverse momentum--momentum correlation function. In
space-time coordinates $({\bf{x}},t)$ its representation reads,
\begin{eqnarray*}
  G_{12}({\bf x};t) =
            i\left\langle \left[
                              {\pi}_{1}(x_1,0;t),
                              {\pi}_{2}(0,x_2;0)
            \right] \right\rangle,
  \ \ \ \ \
  {\bf{x}}=(x_1,x_2).
\end{eqnarray*}
%%%%%%%%%%%%%%%%%%%%%%%%%%%%%%%%%%%%%%%%%%%%%%%%%
\begin{figure}[htb]
\centering
\includegraphics[width=75mm,height=50mm,angle=0,]{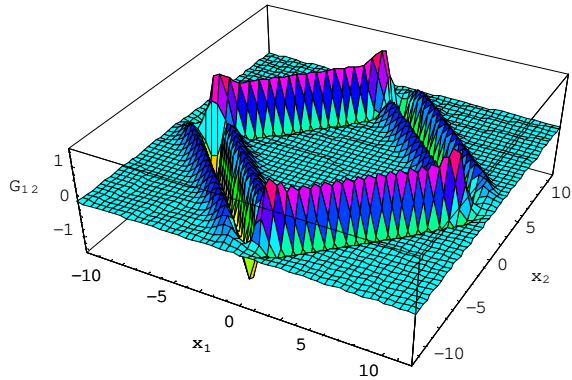}
\caption{The transverse correlation function $G_{12}(x_1,x_2;t)$
for $r_0=1$ and $vt=10$.} \label{GF12}
\end{figure}
%%%%%%%%%%%%%%%%%%%%%%%%%%%%%%%%%%%%%%%%%%%%%%%%%%
This function describes the momentum response at the point
$(0,x_{2})$ of the second array for time $t$ caused by an initial
($t=0$) perturbation localized in coordinate space at the point
$(x_{1},0)$ of the first array.  Standard calculations similar to
those described above, lead to the following expression,
\begin{eqnarray*}
 {G}_{12}({\bf{x}};t) =
                     \displaystyle{
                     \frac{V_0r_0^2}{4{\pi}^2{\hbar}}
                     \int\limits_{-\infty}^{\infty} dk_1 dk_2
                     }
                     {\phi}_1({k_1}){\phi}_2({k_2})k_1k_2
                     \sin(k_1x_1)
                     \sin(k_2x_2)
% \times\nonumber\\&&\times
                     \displaystyle{
                     \frac{vk_2\sin(vk_2t)-
                           vk_1\sin(vk_1t)}
                          {v^2(k_2^2-k_1^2)},
                     }
\end{eqnarray*}
where ${\phi}_{j}(k)$ is the form-factor (\ref{FC-2}) written in
the extended BZ. This correlator is shown in Fig. \ref{GF12}.  It
is mostly localized at the line determined by the obvious
kinematic condition $|x_1|+|x_2|=vt$ (``horizon of events"). The
time $t$ in the r.h.s. is the total time of plasmon propagation
from the starting point $(x_{1},0)$ to the final point $(0,x_{2})$
or vice versa, along any of the shortest ways compatible with a
restricted geometry of the $2D$ grid. The finiteness of the
interaction radius slightly spreads this peak and modifies its
profile.

%%%%%%%%%%%%%%%%%%%%%%%%%%%%%%%%%%%%%
\subsection{Rabi Oscillations}
 \label{subsec-Rabi}
%%%%%%%%%%%%%%%%%%%%%%%%%%%%%%%%%%%%%%%%%%%%%%%%%

Further manifestation of the 2D character of QCB system is related
to the possibility of periodic energy transfer between the two
arrays. Consider an initial perturbation which excites a plane
wave with amplitude $\theta_{0}$ within the first array in the
system of {\it non}-interacting arrays,
\begin{eqnarray*}
  {\theta}_{1}(x_1,n_2a;t)
  & = &
  \theta_{0}
  \sin(q_1x_1+q_2n_2a-v|q_1|t).
\end{eqnarray*}
If the wave vector ${\bf q},$ satisfying the condition
$|{\bf{q}}|<<Q/2,$ is not close to the resonant line of the BZ,
weak inter-array interaction $\phi$ (\ref{eq-phi}) slightly
changes the $\theta_{1}$ component and leads to the appearance of
a small $\theta_{2}\sim\phi$ component. But for ${\bf q}$ lying on
the resonant line ($v|q_1|=v|q_2|\equiv \omega_{\bf{q}}$), both
components within the main approximation have the same order of
magnitude,
\begin{eqnarray*}
   {\theta}_{1}(x_1,n_2a;t) & = &
   {\theta}_{0}
   \cos\left(
            \frac{1}{2}
            {\phi}_{1{\bf{q}}}
            {\omega}_{{\bf{q}}}t
       \right)
   \sin(q_1x_1+q_2n_2a-{\omega}_{{\bf{q}}}t),
   \\
   {\theta}_{2}(n_1a,x_2;t) & = &
   {\theta}_{0}
   \sin\left(
            \frac{1}{2}
            {\phi}_{1{\bf{q}}}
            {\omega}_{{\bf{q}}}t
       \right)
   \cos(q_1n_1a+q_2x_2-{\omega}_{{\bf{q}}}t).
\end{eqnarray*}
%%%%%%%%%%%%%%%%%%%%%%%%%%%%%%
\begin{figure}[htb]
\centering
\includegraphics[width=65mm,height=55mm,angle=0,]{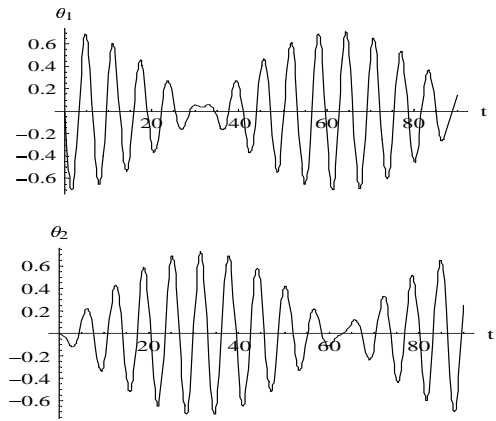}
\caption{Periodic energy exchange between arrays (Rabi
oscillations).} \label{RO3}
\end{figure}
%%%%%%%%%%%%%%%%%%%%%%%%%%%%%%
This corresponds to 2D propagation of a plane wave with wave
vector ${\bf q },$ {\it modulated} by a ``slow'' frequency
$\sim\phi\omega.$ As a result, beating arises due to periodic
energy transfer from one array to another during a long period
$T\sim (\phi\omega)^{-1}$ (see Fig. \ref{RO3}).  These peculiar
``Rabi oscillations'' may be considered as one of the fingerprints
of the physics exposed in QCB systems. Similar periodic energy
transfer between three arrays can be observed in triple QCB (see
Section \ref{subsubsec:Observ3} of Appendix \ref{subsec:Triple}).

%%%%%%%%%%%%%%%%%%%%%%%%%%%%%%%%%%%%%%%%%%%%%%%%%%%%%%%%%%%%%%
\section{Conclusions}
 \label{subsec3:Concl}
%%%%%%%%%%%%%%%%%%%%%%%%%%%%%%%%%%%%%%%%%%%%%%%%%%%%%%%%%%%%%%

We discussed in this Chapter the kinematics and dynamics of
plasmon spectrum in QCB. These nanostructures may be fabricated
from single-wall carbon nanotubes \cite{Rueckes,Diehl}.  On the
one hand, QCB is promised to become an important component of
future molecular electronics \cite{Rueckes,Tseng}.  On the other
hand, the spectrum of elementary excitations (plasmons) in these
grids possesses the features of both $1D$ and $2D$ electron
liquid. As is shown in Refs. \cite{Luba00,Luba01} and confirmed in
the present study, the energy spectrum of QCB preserves the
characteristic properties of LL at $|{\bf q}|,\omega\to 0,$.  At
finite ${\bf q},\omega$ the density and momentum waves in QCB may
have either $1D$ or $2D$ character depending on the direction of
the wave vector.  Due to inter-array interaction, unperturbed
states, propagating along arrays are always mixed, and transverse
components of correlation functions do not vanish. For
quasi-momentum lying on the resonant lines of the BZ, such mixing
is strong and transverse correlators have the same order of
magnitude as the longitudinal ones.  Periodic energy transfer
between arrays (``Rabi oscillations'') is predicted.

The crossover from $1D$ to $2D$ regime may be experimentally
observed. One of the experimental manifestations, i.e. the
crossover from isotropic to anisotropic (spatially nonuniform)
conductivity was pointed out in Ref. \cite{Luba01}. The current
may be inserted in QCB at a point on an array $j$ and extracted
from another array $i$ at a distance $r$. Then a temperature
dependent length scale $l(T)$ arises, so that for $r\gg l$ the
resistance is dominated by small $q$ and therefore, the current is
isotropic. In the opposite limit $r<l$ the dependence of the
current on the points of injection/extraction may be detected.  At
$T=0$ the length $l$ becomes infinite, and current can only be
carried along the wires. These effects are in fact manifestations
of the LL behavior of the QCB in the infrared limit.

To observe the crossover at finite $\{\omega, {\bf q}\}$, one
should find a way of exciting the corresponding plasmon modes.
Then, scanning the $\omega(q_1,q_2)$ surfaces, one may in
principle detect the crossover from quasi $1D$ to $2D$ behavior in
accordance with the properties of the energy spectra presented in
Section \ref{subsubsec:Spectr} and Appendices \ref{append-tilted}
and \ref{subsec:Triple}. Plasmons in QCB may be excited either by
means of microwave resonators or by means of interaction with
surface plasmons. In the latter case one should prepare the grid
on a corresponding semiconductor substrate and measure, e.g., the
plasmon loss spectra. The theory of these plasmon losses will be
presented in Chapter \ref{sec:Damp}.

%\newpage

%%%%%%%%%%%%%%%%%%%%%%%%%%%%%%%%%%%%%%%%%%%%%%%%%%%
\chapter{Infrared Spectroscopy of Quantum Crossbars}
\label{sec:Infrared}
%%%%%%%%%%%%%%%%%%%%%%%%%%%%%%%%%%%%%%%%%%%%%%%%%%%
%
%In this Chapter we study the possibilities of direct observations
%of dimensional crossover outside the LL fixed point by the method
%of infrared (IR) spectroscopy. QCB plasmons may be involved in the
%resonant diffraction of incident electro-magnetic waves and in
%optical absorption in the IR part of spectrum. The absorption of
%external electromagnetic field in QCB strongly depends on the
%direction of the wave vector an incident wave. As a result two
%types of one- to two-dimensional crossover with varying angle of
%an incident wave of its frequency can be observed.
%
%%%%%%%%%%%%%%%%%%%%%%%%%%%%%%%%%%%%%%%%%%%%%%%%%%%
\section{Introduction}
\label{subsec:IR-Intro}
%%%%%%%%%%%%%%%%%%%%%%%%%%%%%%%%%%%%%%%%%%%%%%%%%%%

In this Chapter we consider various possibilities of direct
observation of plasmon spectra at high frequencies and wave
vectors by the methods of infrared (IR) spectroscopy. The QCB
plasmons can be treated as a set of dipoles distributed within QCB
constituents. In a single wire, the density of the dipole momenta
is proportional to the LL boson field $\theta(x)$ ($x$ is the
coordinate along the wire). Few sets of coupled $1D$ dipoles form
a unique system which possesses the properties of $1D$ and $2D$
liquid depending on the type of experimental probe. Some
possibilities of observation of $1D\to 2D$ crossover in transport
measurements were discussed in Ref. \cite{Mukho1}.

In transport measurements, the geometric factors regulate the
crossover from anisotropic to isotropic resistivity of QCB: one
may study the dc response for a field applied either parallel to
one of the constituent arrays or in arbitrary direction.  One may
also study spatially nonuniform response by means of two probes
inserted at different points of QCB and regulate the length scale,
i.e., the distance between the two probes in comparison with the
periods of the crossbar superlattice. These methods give
information about the nearest vicinity of LL fixed point at
$({\bf{q}},\omega,T)\to 0.$

Several crossover effects such as appearance of non-zero
transverse space correlators and periodic energy transfer between
arrays ("Rabi oscillations") were discussed in the previous
Chapter. Unlike transport measurements, the methods of infrared
spectroscopy provide an effective tool for investigating the
excitation spectrum in a rather wide $({\bf{q}},\omega)$ area well
beyond the sliding phase region \cite{KGKA3}. We will show that
the IR spectroscopy allows scanning of the $2D$ Brillouin zone in
various directions and thereby elucidates dimensional crossover in
the high symmetry points of the BZ.

The direct manifestation of dimensional crossover is through the
response to an external ac electromagnetic field
\cite{GKKA2,KGKA3,KG}. To estimate this response one should note
that the two main parameters characterizing the plasmon spectrum
in QCB are the Fermi velocity $v$ of electrons in a wire and the
QCB period $a$ (we assume both periods to be equal). These
parameters define both the typical QCB plasmon wave numbers
$q=|{\bf q}|\sim Q=2\pi/a$ and the typical plasmon frequencies
$\omega\sim \omega_{Q}=vQ$. Choosing according to Refs.
\cite{Rueckes,Egger} $v\approx 0.8\cdot 10^{6}$~m/sec and
$a\approx 20$~nm, one finds that characteristic plasmon
frequencies lie in the far infrared region $\omega\sim
10^{14}$~sec$^{-1}$, while characteristic wave vectors are
estimated as $q\sim 10^{6}$~cm$^{-1}$.

Here we study high frequency properties of the simplest double
square QCB (generalization to more complicated geometries is
straightforward).  We start from QCB interacting with an external
infrared radiation.  The plasmon velocity $v$ is much smaller than
the light velocity $c$ and the light wave vector $k$ is three
orders of magnitude smaller than the characteristic plasmon wave
vector $Q$ corresponding to the same frequency. Therefore,
infrared radiation incident directly on an {\em isolated} array,
can not excite plasmons at all (it could excite plasmon with
$\omega\neq 0$).  However in QCB geometry, each array serves as a
diffraction lattice for its partner, giving rise to Umklapp
processes of wave vectors $nQ,$ $n$ integer. As a result,
excitation of plasmons in the BZ center $q=0$ with frequencies
$nvQ$ occurs.

To excite QCB plasmons with $q\neq 0$ one may use an additional
diffraction lattice (DL) with period $A>a$ coplanar to the QCB.
Here the diffraction field contains space harmonics with wave
vectors $2\pi M/A,$ $M$ integer, that enables one to scan plasmon
spectrum within the BZ. Dimensional crossover manifests itself in
the appearance of additional absorption lines when the wave vector
of the diffraction field is oriented along specific directions.
In the general case one observes the single absorption lines
forming two sets of equidistant series.  Instead of that, in the
main resonance direction (QCB diagonal) an equidistant series of
split doublets can be observed.  In the case of higher resonance
direction, absorption lines form an alternating series of singlets
and split doublets demonstrating new type of dimensional crossover
related to the frequency change with direction fixed.

The structure of the present Chapter is as follows: In Sections
\ref{subsec:Direct} and \ref{subsec:Scan} we study interaction of
QCB with an external field. In Section \ref{subsec:Direct} we
consider the case when the incident infrared radiation falls
directly on the QCB. In Section \ref{subsec:Scan} we study
possible scanning of QCB spectrum with the help of an external DL.
In the Conclusions we summarize the results obtained.

%%%%%%%%%%%%%%%%%%%%%%%%%%%%%%%%%%%%%%%%%%%%%%%%%%%%%%%%%%%%%%%%%%
%\section{Infrared Light Absorption by QCB}\label{subsec:Absorption}
%%%%%%%%%%%%%%%%%%%%%%%%%%%%%%%%%%%%%%%%%%%%%%%%%%%%%%%%%%%%%%%%%%%
\section{Long Wave Absorption}\label{subsec:Direct}
\begin{figure}[htb]
\centering
\includegraphics[width=35mm,height=52mm,angle=0,]{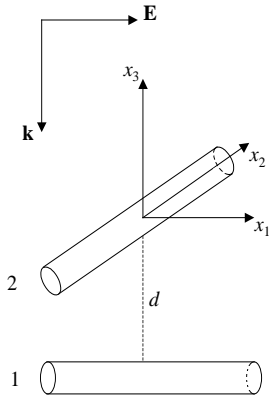}
\caption{The incident field orientation with
respect to QCB. The axes $x_{1}$ and $x_{2}$ are directed along
the corresponding arrays, and $d$ is the inter-array vertical
distance (along the $x_{3}$ axis).} \label{Cross1}
\end{figure}

%%%%%%%%%%%%%%%%%%%%%%%%%%%%%%%%%%%%%%%%%%%%%%%%%%%%%%%%%%%%%%%%%%%
In the case of a dielectric substrate transparent in the infrared
region, one can treat QCB as an isolated grid (without substrate)
interacting directly with the incident radiation.  Consider the
simplest geometry where an external wave falls normally onto QCB
plane, and its electrical field
$${\bf E}=E_{0}{\bf e}_{1}\cos{({\bf k}{\bf r}-\omega t)}$$
is parallel to the lower (first) array (see Fig. \ref{Cross1} for
details). In this geometry the field ${\bf E}$ is {\it
longitudinal} for array $1$ and {\it transverse} for the array
$2$. The eigenfrequencies of transverse modes in array $2$
substantially exceed the IR frequency of the incident wave and
even the standard LL ultraviolet cutoff frequency.  Thus, the
incident wave can be treated as a static polarization field for
this array, and the factor $\cos{\omega t}$ can be omitted.  Then,
the polarization waves in array $2$ form a longitudinal
diffraction field for array $1$ with quasi wave vectors $nQ$ ($n$
integer). Further, the characteristic order of magnitude $Q$ of a
QCB plasmon wave vector is much larger than the wave vector
${\bf{k}}$ of the incident light, so one can assume the latter to
be equal to zero from the very beginning. Then, the light
wavelength is much larger than a nanotube diameter and the
geometrical shadow effect can be neglected. As a result the total
field which affects array 1 consists of an external field and a
diffraction field produced by a static charge induced in array 2.

To calculate the diffraction field, we consider first the field
${\bf E}^{0}$ produced by the quantum wire of array 2 which is
located at $x_{1}=x_{3}=0$ and labelled by $n_{1}=0$.  The large
distance between the wire under consideration and its neighbor
partners from the same array allows us to neglect the influence of
the charges induced on them.  The static potential on the surface
of the wire includes external potential of an incident field and
the potential $\Phi^{0}$ of the charge induced on the wire. On the
other hand, this static potential should be equal to a constant
which we choose to be zero. In cylindrical coordinates
$r,\vartheta, x_{2},$ $x_{1}=r\cos\vartheta,
x_{3}=r\sin\vartheta,$ this condition reads
\begin{eqnarray}
    \Phi^{0}(R_0,\vartheta,x_{2}) = E_0r_0\cos\vartheta.
    \label{Pot}
\end{eqnarray}
Outside the wire, the induced potential $\Phi^{0}$ satisfies the
Laplace equation $\Delta\Phi^{0}=0.$ Solving this equation with
boundary condition (\ref{Pot}) we obtain the static part of the
induced potential
\begin{eqnarray*}
    \Phi^{0}(r,\vartheta,x_{2})=\frac{E_0r_0^2}{r}\cos\vartheta
    \label{IndPhi}
\end{eqnarray*}
and the corresponding static part of the induced field along the
$x_{1}$ direction,
\begin{eqnarray*}
    E_{1}^{0}(x_{1},x_{3})=-E_{0}
    \frac{r_0^2\left(x_{3}^2-x_{1}^2\right)}
           {\left(x_{3}^2+x_{1}^2\right)^2}.
    \label{IndField}
\end{eqnarray*}
The first component of the diffraction field is the sum of the
fields induced by all wires of the upper array,
\begin{eqnarray}
        E_{1}(x_{1};t)=\cos\omega t\sum_{n_{1}}
        E_{1}^{0}(x_{1}-n_{1}a,-d)
        =
       -E_{0}\cos\omega t
       \sum_{n_{1}}
       \frac{r_0^2\left(d^2-(x_{1}-n_{1}a)^2\right)}
            {\left(d^2+(x_{1}-n_{1}a)^2\right)^2}.
    \label{Field}
\end{eqnarray}
This field is a periodic function of $x_1$ with a period $a.$
Therefore, its Fourier expansion contains only wave vectors
$k_{1n}=nQ$ ($n$ is the order of diffraction).  This means that
only frequencies $\omega_{n}=nvQ$ can be excited.  In this case it
is more convenient to expand the field over Bloch eigenfunctions
of an ``empty'' wire \cite{KGKA2}. These functions are labelled by
quasimomentum $q_{1},$ $|q_{1}|\leq Q/2,$ and the band number $s$.
The expansion includes only $q_{1}=0$ components and has the form
\begin{eqnarray*}
    E_{1}(x_{1};t)=\cos\omega t\sum_{s}
    E_{[s/2]}
    u_{s}(x_{1}),
    \label{Expansion1}
\end{eqnarray*}
where $u_s(x)$ is the $q_1=0^{+}$ Bloch amplitude $u_{sq_1}(x)$
(\ref{WaveFunc1}) within the $s$-th band and
\begin{equation}
    E_{n}=-E_{0}
    \frac{\pi r_{0}^{2}}{ad}
    nQde^{-nQd}.
    \label{Expansion2}
\end{equation}
The excited eigenfrequency $\omega_{n}=\omega_{[s/2]}$ belongs
simultaneously to the top of the lower even band with number
$s=2n$ and to the bottom of the upper odd band with number
$s=2n+1$ (this is the result of ${\bf{E}}(x)={\bf{E}}(-x)$
parity). The incident field cannot excite plasmons at all and we
do not take it into account.

Turning to the $({\bf q},s)$ representation with the help of the
expansion
\begin{eqnarray}
 \theta_1(x_1,n_2a)=\frac{\sqrt{a}}{L}
 \sum\limits_{s{\bf{q}}}\theta_{1s{\bf{q}}}
 e^{i(q_1x_1+q_2n_2a)}
 u_{sq_1}(x),
 \label{Fourier-IR}
\end{eqnarray}
and similarly for $\theta_{2}$ and $\pi_{1,2},$ one easily sees
that only the ${\bf q}={\bf 0}$ components are involved in the
interaction of plasmons with the incident radiation.

Consider an initial frequency $\omega$ close to $\omega_{n}.$ In a
resonant approximation, only four equations of motion for the
``coordinate'' operators $\theta_{s}$ with $s=2n,2n+1$ are
relevant
\begin{eqnarray}
  \ddot{\theta}_{1,2n}+\omega_n^2{\theta}_{1,2n}+
  \phi\omega_n^2
  \left({\theta}_{2,2n}-{\theta}_{2,2n+1}\right)
  &=& Lf_n\cos\omega t,
  \nonumber\\
%%%%%%%%%%%%%%%%%%%%
  \ddot{\theta}_{1,2n+1}+\omega_n^2{\theta}_{1,2n+1}-
  \phi\omega_n^2
  \left({\theta}_{2,2n}-{\theta}_{2,2n+1}\right)
  &=& Lf_{n}\cos\omega t,
  \nonumber\\
%%%%%%%%%%%%%%%%%%%%%%%%%%%%%%%%%%%%%%%%
  \ddot{\theta}_{2,2n}+\omega_n^2{\theta}_{2,2n}+
  \phi\omega_n^2
  \left({\theta}_{1,2n}-{\theta}_{1,2n+1}\right)
  &=& 0,
  \label{EqMot1}\\
%%%%%%%%%%%%%%%%%%%%
  \ddot{\theta}_{2,2n+1}+\omega_n^2{\theta}_{2,2n+1}
  -\phi\omega_n^2
  \left({\theta}_{1,2n}-{\theta}_{1,2n+1}\right)
  &=& 0,
  \nonumber
\end{eqnarray}
where
$$f_n=\frac{\sqrt{2}{v}g{e}}{\hbar\sqrt{a}}E_{n},$$
and we assume that $\zeta_s(q_{1,2})$ (\ref{FC-2}) is equal to $1$
for the first few bands.

The homogeneous part of this system defines four eigenfrequencies:
$\omega_{g,g}=\omega_{u,g}=\omega_{n}$,
$\omega_{g/u,u}\approx\omega_{n}(1\pm\phi)$. The corresponding
eigenvectors are symmetrized combinations of the four operators
which enter Eq.(\ref{EqMot1}).  They have a fixed parity with
respect to permutation of arrays (the first index) and neighboring
bands (the second index).  Only two modes (even with respect to
band index)
\begin{eqnarray*}
  \theta_{g/u,g} &=& \frac{1}{2}
  \left({\theta}_{1,2n}+
        {\theta}_{1,2n+1}\pm
        {\theta}_{2,2n}\pm
        {\theta}_{2,2n+1}
  \right),
\end{eqnarray*}
interact with an external field.  Therefore only the unperturbed
frequency $\omega_{n}=\omega_{gg}=\omega_{ug}$ will be absorbed.
The two equations of motion for the operators $\theta_{gg,ug}$
have the same form
\begin{eqnarray*}
  \ddot{\theta}_{\alpha}
  +2\gamma\dot{\theta}_{\alpha}
  +\omega_{n}^2{\theta}_{\alpha}
  = Lf_{n}\cos\omega t,
  \label{EqMotDiag}
\end{eqnarray*}
where $\alpha=gg,ug.$ Employing standard procedure in the vicinity
of the resonance $|\omega-\omega_n|\ll\omega_n$ immediately yields
the relative absorption of the Lorentz type
\begin{eqnarray}
  \frac{\Delta I_{n}}{I_{0}}&=&
           2g\frac{e^{2}}{\hbar c}
           \left(\frac{{\pi}r_0^2}{{a}{d}}\right)^2
           \frac{\gamma vQ}{\left(\omega-\omega_n\right)^2+\gamma^2}
           \left[
                nQde^{-nQd}
           \right]^{2},
  \label{EnerAbsorp}
\end{eqnarray}
where
$$
 I_0=\frac{cL^2}{4\pi}E_0^2
$$
is the intensity of light that falls on the QCB.

Due to the exponential term in the r.h.s of Eq.(\ref{Expansion2}),
$E_{n}$ decreases fast with $n$ and only the first few terms
contribute to absorption.  The characteristic dimensionless scale
of the induced field ${{r}_0^2}/{(ad)}$ for typical values of QCB
parameters equals $0.004$.  We tabulate below the lowest
dimensionless Fourier components of the induced field.
\begin{center}
\begin{tabular}{| c | c | c | c | c | c |}
 \hline
 n & 1 & 2 & 3 & 4 & 5 \\ \hline
 $-\displaystyle{\frac{ad}{r_0^2}\frac{{E}_{n}}{E_0}}$
 & 1.05306 & 1.12359 & 0.89914 & 0.63957 & 0.42650 \\
 \hline
\end{tabular}
\end{center}
The results show that one can hope to probe at least the first
five spectral lines corresponding to $\omega_{n}$ with
$n=1,2,\ldots,5.$

The width of the absorption line (\ref{EnerAbsorp}) is governed by
an attenuation coefficient $\gamma.$ We introduce it
phenomenologically, but one can (at least qualitatively) estimate
its value. The attenuation is caused by decay of plasmon into
phonons. The one phonon decay of the plasmon with wave number $k$
and frequency $\omega=v|k|$ into a single phonon with the same
$\omega$ and $k$ occurs at a single point in $1D$ and does not
yield finite attenuation at all. Multi-phonon decay is weak
because of the small anharmonic coupling within the wire. As a
result, the form of the absorption lines should be determined
mainly by the instrumental line-width.
%%%%%%%%%%%%%%%%%%%%%%%%%%%%%%%%%%%%%%%%%%%%%%%%%%%%%%%%%%%%%%%%%
\section{Scanning of the QCB Spectrum within the BZ}
\label{subsec:Scan}
\begin{figure}[htb]
\centering
\includegraphics[width=60mm,height=42mm,angle=0,]{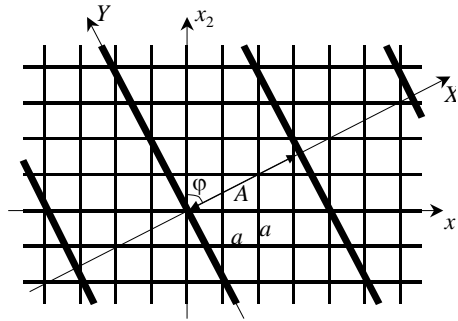}
\caption{QCB and DL. The $X,Y$ axes are oriented
along the DL stripes and the wave vector ${\bf K}$ of the
diffraction field respectively. The DL (QCB) period is $A$ ($a$).}
\label{QCB-DLFig}
\end{figure}

%%%%%%%%%%%%%%%%%%%%%%%%%%%%%%%%%%%%%%%%%%%%%%%%%%%%%%%%%%%%%%%%%%%
Within a geometry considered in the previous subsection, one can
probe plasmon spectrum only at the BZ center.  To study plasmons
with nonzero wave vectors one should add to the system an external
diffraction lattice (DL), namely, a periodic array of metallic
stripes parallel to the $Y$ axis (see Fig. \ref{QCB-DLFig}).  The
DL plane $Z=0$ is parallel to the QCB planes $Z=-D$ for the upper
second array and $Z=-(D+d)$ for the lower first array (the $Z$
axis is parallel to the $x_{3}$ axis).  The distance $D$ between
DL and second array is of the same order as the inter-array
distance $d=2$~nm. The angle between the DL wires and the {\em
second} array is $\varphi$ ($0<\varphi<\pi/2$). To get a wave
number $K$ of a diffraction field much smaller than $Q$ one needs
a DL with a period $A$ much larger than the QCB period $a$.  In
the following numerical estimations we choose $A\approx200$~nm.

Consider an incident field with electric vector ${\bf E}=E_{0}{\bf
e}_{X} \cos({\bf k}{\bf r}-\omega t)$ oriented along the $X$ axis
(perpendicular to the DL wires).  The radius $R_{0}$ of a DL wire
is assumed to be not much larger than the nanotube radius $r_{0}.$
In this case, light scattering on the DL is similar to that
considered in subsection \ref{subsec:Direct}. Then the diffraction
field is concentrated along the $X$ direction and has the form
(compare with Eq.(\ref{Field}))
\begin{eqnarray*}
 E_X(X,Z,t) =
 -E_0\cos\omega t
 \sum_{N}
      \frac{R_0^2(Z^2-(X-NA)^2)}
           {\left(Z^2+(X-NA)^2\right)^2}.
 \label{FieldDL}
\end{eqnarray*}
The Fourier transform of the diffraction field is
\begin{eqnarray*}
 {E_{X}}({\bf K},Z)=-E_0
      \frac{\pi R_0^2}{A|Z|}
      |KZ|e^{-|KZ|},
 \label{Fourier1}
\end{eqnarray*}
where ${\bf{K}}(M)= K{\bf{e}}_{X}= K(\sin\varphi,\cos\varphi)=
(K_{1},K_{2})$, with $K={2\pi M}/{A}$, $M$ being a positive
integer. This means that all the points ${\bf K}$ lie on the same
ray oriented along the positive direction of the $X$ axis.  The
vector ${\bf K}(M)$ for a fixed $M$ can be uniquely represented as
a sum of quasimomenta lying in the first BZ and two reciprocal
lattice vectors
\begin{eqnarray*}
    {\bf K}(M)={\bf q}(M)+{\bf m}_{1}(M)+{\bf m}_{2}(M).
    \label{direction}
\end{eqnarray*}
The field components
\begin{eqnarray}
 E_{1{\bf{K}}}={E_{X}}({\bf K},D+d)\sin\varphi,
 \ \ \ \ \ \ \ \ \ \
 E_{2{\bf{K}}}={E_{X}}({\bf K},D)\cos\varphi
 \label{Components}
\end{eqnarray}
parallel to the quantum wires, can excite plasmons and contribute
to the absorption process.

The Hamiltonian describing the interaction of QCB with an external
field reads,
\begin{eqnarray}
      H_{E}&=& \frac{\hbar{L}}{2vg}
      \sum_{{\bf K}}\Big[
      f_{1,{\bf{K}}}\left(
      \theta_{1,{\bf K}}+
      \theta_{1,{\bf K}}^{\dag}
      \right)+
 %     \nonumber\\&&+
      f_{2,{\bf{K}}}\left(
      \theta_{2,{\bf K}}+\theta_{2,{\bf K}}^{\dag}
      \right)%\nonumber
      \Big],
      \label{Hams}
\end{eqnarray}
where
\begin{eqnarray}
      f_{j,{\bf{K}}}&=&\frac{\sqrt{2}vge}{\hbar\sqrt{a}}E_{j,{\bf{K}}},
      \phantom{aaaa}
      {\bf{m}}={\bf{m}}_1+{\bf{m}}_2.
\label{Explan}
\end{eqnarray}
In this Section we are interested not in the form of the
absorption line but only in the resonant frequencies. Therefore,
we do not introduce any phenomenological attenuation. The
equations of motion for boson fields have the form
\begin{eqnarray}
  &&\ddot{\theta}_{1,{\bf{q}}+{\bf{m}}_1}+
  \omega_{q_1+m_1Q}^2{\theta}_{1,{\bf{q}}+{\bf{m}}_1}+
%  \nonumber\\+
  \phi\sum_{m_2}\Phi_{{\bf{q}}+{\bf{m}}}
  {\theta}_{2,{\bf{q}}+{\bf{m}}_2}
  =
  L\sum_{M,m_2}f_{1,{\bf{K}}}\delta_{{\bf{K}},{\bf{q}}+{\bf{m}}},
  \nonumber\\
%%%%%%%%%%%%%%%%%%%%
  &&\ddot{\theta}_{2,{\bf{q}}+{\bf{m}}_2}+
  \omega_{q_2+m_2Q}^2{\theta}_{2,{\bf{q}}+{\bf{m}}_2}+
%  \nonumber\\+
  \phi\sum_{m_1}\Phi_{{\bf{q}}+{\bf{m}}}
  {\theta}_{1,{\bf{q}}+{\bf{m}}_1}
  =
  L\sum_{M,m_1}f_{2,{\bf{K}}}\delta_{{\bf{K}},{\bf{q}}+{\bf{m}}},
  \label{EqMot2}
\end{eqnarray}
where $\phi$ is the dimensionless coupling strength
(\ref{eq-phi}), $\Phi_{{\bf{k}}}$ is given by Eq.
(\ref{Phi-Fourier}).

Only the first few terms in the sum over $K$ in the r.h.s. of Eq.
(\ref{EqMot2}) really excite the QCB plasmons. Indeed, the
diffraction field (\ref{Fourier-IR}) is proportional to the same
dimensionless function of the type $te^{-t}$ ($t=|KZ_{j}|$) as in
the previous subsection (see Eq.(\ref{Expansion2})).  This
function has its maximum at $t=1$ and differs significantly from
zero for $0.2<t<2.7$.  For $a=20$~nm, $D=2$~nm, it is of order
unity within the interval $0.18Q<|K|<2.13Q$ for the first array
($Z_{1}=D+d$), and within the interval $0.36Q<|K|<4.26Q$ for the
second array ($Z_{1}=D$). This means that one can excite the modes
of the four lower bands ($K<2Q$) of the first array and the modes
of eighth lower bands ($K<4Q$) of the second array.

According to Eqs. (\ref{Components}) the field $E_{j{\bf K}(M)}$
is coupled with plasmons of wave vectors ${\bf q}+{\bf m}_{j}={\bf
q}(M)+{\bf m}_{j}(M)$ within the $j$-th array. The nature of the
excited plasmons as well as their frequencies depend on the
direction of the vector ${\bf K}(M).$ For simplicity we restrict
ourselves by acute angles $0<\varphi<\pi/2$ describing orientation
of both the DL and the vector ${\bf K}(M)$. There are four kinds
of dimensional crossover events depending on the specific
directions in the BZ. Each type of crossover is characterized by
its own set of absorption lines. The first one takes place in a
common case when ${\bf K}(M)$ for any $M$ never reaches neither a
resonant direction nor the BZ boundary. The second case
corresponds to the bisectorial direction $\varphi=\pi/4$ where the
main resonant condition $\omega(K_{1})=\omega(K_{2})$ is
fulfilled. The third set of directions is determined by another
resonant condition $\omega(K_{1})=\omega(nQ\mp K_{2}).$ Finally,
the fourth set is formed by directions intersecting with the BZ
boundaries for some values of $M.$ In what follows we consider
these four cases separately.

{\bf 1.} In the general case, the points ${\bf K}(M)$ for all $M$
are far from the BZ diagonals and boundaries. Therefore each of
them corresponds to a couple of plasmons mostly propagating along
the $j$-th array, $j=1,2,$ with unperturbed frequencies
$\omega_{K_{j}(M)}=vK_{j}(M)$. The inter-array interaction
slightly renormalizes the eigenfrequencies
\begin{eqnarray*}
 \omega_{1{\bf{K}}}^2 = \omega_{K_1}^2+
            \phi^2\sum_{m_2}
            \frac{\omega_{K_1}^{2}
                  \omega_{K_2+m_2Q}^{2}}
                 {\omega_{K_1}^2-\omega_{K_2+m_2Q}^2},
 \ \ \ \ \
 \omega_{2{\bf{K}}}^2 = \omega_{K_2}^2+
            \phi^2\sum_{m_1}
            \frac{\omega_{K_2}^{2}
                  \omega_{K_1+m_1Q}^{2}}
                 {\omega_{K_2}^2-\omega_{K_1+m_1Q}^2}.
\end{eqnarray*}
Thus, increasing the frequency of an incident light one observes a
set of single absorption lines that consists of two almost
equidistant subsets with frequencies corresponding to excitation
of plasmons in the first or second arrays. The frequency spacing
between adjacent lines within each subset are
\begin{eqnarray*}
  \Delta\omega_{1}=v\Delta K_{1}=2\pi v \sin\varphi/A,
  \ \ \ \ \
  \Delta\omega_{2}=v\Delta K_{2}=2\pi v \cos\varphi/A,
\end{eqnarray*}
and their ratio depends only on the DL orientation $\varphi$
$$\frac{\Delta\omega_{1}}{\Delta\omega_{2}}=\tan\varphi.$$

{\bf 2.} In the resonant case $\varphi=\pi/4,$ the relation
$K_1(M)=K_2(M)$ is satisfied for all $M$.  Therefore modes
propagating along the two arrays are always degenerate.
Inter-array interaction lifts the degeneracy.  Indeed, in the
resonant approximation, the coupled equations of motion for the
field operators read
\begin{eqnarray*}
  \ddot{\theta}_{1{\bf{K}}}+\omega_{K_1}^2\theta_{1{\bf{K}}}+
  \phi\omega_{K_1}^2\theta_{2{\bf{K}}}
  =f_{1{\bf{K}}},
  \ \ \ \ \
  \ddot{\theta}_{2{\bf{K}}}+\omega_{K_1}^2\theta_{2{\bf{K}}}+
  \phi\omega_{K_1}^2\theta_{1{\bf{K}}}
  =f_{2{\bf{K}}}.
\end{eqnarray*}
After symmetrization
$\theta_{g,u}=(\theta_{1{\bf{K}}}\pm\theta_{2{\bf{K}}})/\sqrt{2}$,
they have the same form
$$
  \ddot{\theta}_{\alpha}+\omega_{\alpha}^2\theta_{\alpha}
  =f_{\alpha},
$$
where $\omega_{g/u}=\omega_{K}\sqrt{1\pm\phi}$ are the
renormalized frequencies,
$f_{g/u}=({f}_{1{\bf{K}}}\pm{f}_{2{\bf{K}}})/\sqrt{2}$, and
$\alpha=g,u$.  The amplitudes $f_{g,u}$ are of the same order of
magnitude because the distances $D$ and $d$ are different but have
the same order of magnitude. As a result, increasing the frequency
of an incident light one observes an equidistant set of absorption
doublets with distance $\pi\sqrt{2}v/A$ between adjacent doublets.

{\bf 3.} Consider now the directions $\varphi$ determined by the
equation
\begin{eqnarray*}
    \sin\left(\varphi\pm\frac{\pi}{4}\right)=
    \frac{nA}{\sqrt{2}M_{0}a},
\end{eqnarray*}
where $n$ and $M_{0}$ are mutually prime integers.  For this
direction, two components of the first $M_{0}-1$ points ${\bf
K}(M)$ do not satisfy any resonant condition while the $M_{0}$-th
one does
\begin{equation}
     K_1(M_{0})\pm K_2(M_{0})=nQ.
     \label{resonant}
 \end{equation}
With increasing $M$ this situation is reproduced periodically so
that all the points ${\bf K}(pM_{0})$ with $p$ integer satisfy a
similar condition with $pn$ standing instead of $n,$ while all
intermediate points are out of resonance.

In zero order approximation with respect to the inter-array
interaction we expect to observe two set of absorption lines with
frequencies $p\omega_{j}=vK_{j}(pM_{0}),$ $j=1,2,$ corresponding
to excitation of plasmons within the $pm_{j}(M_{0})$-th band of
the $j$-th array.  The ratio of the frequencies $\omega_{j}$ is
defined by the DL orientation
\begin{eqnarray*}
    \frac{\omega_{1}}{\omega_{2}}=\tan\varphi.
    \label{relation}
\end{eqnarray*}
However, due to the resonance condition (\ref{resonant}), a
plasmon in the first array with wave vector $K_{1}(pM_{0})$ and
frequency $\omega_{1}=vK_{1}(pM_{0})$ is coupled with a plasmon in
the second array with the same frequency and wave vector
$K'_{2}=\mp(npQ- K_{1}(pM_{0}))$ (inter-array degeneracy).
Similarly, a plasmon in the second array with wave vector
$K_{2}(pM_{0})$ and frequency $\omega_{2}=vK_{2}(pM_{0})$ is
coupled with a plasmon in the first array with the same frequency
and wave vector $K'_{1}=npQ\mp K_{2}(pM_{0}).$ This degeneracy of
two modes corresponding to the same band but to different arrays
is lifted by the inter-array interaction.  As a result one has two
sets of doublets instead of two sets of single lines.

Thus, for such orientation of the DL, increasing the frequency of
an incident wave one should observe two equidistant sets of single
absorption lines with two sets of equidistant doublets built in
these series
\begin{eqnarray*}
 \omega_{1{\bf{K}}} =
  \omega_{K_1(pM_{0})}\left(1\pm\frac{1}{2}\phi\right),
  \ \ \ \ \
 \omega_{2{\bf{K}}} =
  \omega_{K_2(pM_{0})}\left(1\pm\frac{1}{2}\phi\right).
 \label{omega-res}
\end{eqnarray*}
In the case $n=1$ the lower doublet lies in the first energy band,
whereas the upper one lies in the second band.  For $A/a=10$ (that
corresponds to the realistic values of the parameters $a=20$~nm
and $A=200$~nm) the lowest doublet ($p=1$) will be observed for
example for integer $M_{0}=8,$ at the angle $\varphi(8)\approx
17^{\circ}$, around frequencies $\omega_{1}(8)=0.76vQ,$
$\omega_{2}(8)=0.24vQ.$

{\bf 4.} A similar behavior will be manifested in the case when
the points ${\bf K}_{pM}$ lie at one of the BZ boundaries, i.e.,
they satisfy the relation
\begin{eqnarray*}
    K_{j}(pM_{j})=\frac{npQ}{2}
    \label{boundaries}
\end{eqnarray*}
with some specific values $j,$ $M_{j}$ and $n.$ Such situation is
realized at specific angles that depend on the integers
$j,n,M_{j}.$ In the vicinity of the points ${\bf K}(pM_{j})$ two
frequencies corresponding to the unperturbed modes of the $j$-th
array from the $np$-th and $(np+1)$-th bands coincide.  This is
the case of inter-band degeneracy that is also lifted by
inter-array interaction. Due to the square symmetry (invariance
with respect to ${x_j}\to{-x_j}$ inversion), only one of the two
components with frequency $\omega=v|K_{j}(pM_{j})|$ may be excited
by a diffraction field.  Therefore, this case is not distinct from
case {\bf 1} considered above and two sets of equidistant
single lines can be observed.

We emphasize that by studying absorption of light by QCB one can
expose, beyond the studied above \cite{KGKA2} dimensional
crossover with respect to an angle (direction), also the
occurrence of a new type of crossover with {\em an external
frequency} as a control parameter. This occurs for special
directions of type ${\bf 3}$ where, with increasing frequency, the
set of single lines is periodically intermitted by doublets.

%%%%%%%%%%%%%%%%%%%%%%%%%%%%%%%%%%%%%%%%%%%%%%%%%%%%%%%%%%%%%
\section{Conclusions}\label{subsec:IRConclu}
%%%%%%%%%%%%%%%%%%%%%%%%%%%%%%%%%%%%%%%%%%%%%%%%%%%%%%%%%%%%%
In conclusion, we investigated the possibility of spectroscopic
studies of the excitation spectrum of quantum crossbars, which
possesses unique property of dimensional crossover both in spatial
coordinates and in $({\bf q},\omega)$ coordinates.  It follows
from our studies that the plasmon excitations in QCB may be
involved in resonance diffraction of incident electromagnetic
waves and in optical absorption in the IR part of spectrum.

In the case of direct interaction of external electric field with
QCB, infrared absorption strongly depends on the direction of the
wave vector ${\bf q}$.  One can observe dimensional crossover from
$1D\to 2D$ behavior of QCB by scanning an incident angle.  The
crossover manifests itself in the appearance of a set of
absorption doublets instead of the set of single lines.  At
special directions, one can observe new type of crossover where
doublets replace the single lines with changing frequency at a
fixed ${\bf q}$ direction.

Dimensional crossover in QCB plays a significant role in all the
above phenomena.

%\newpage

%%%%%%%%%%%%%%%%%%%%%%%%%%%%%%%%%%%%%%%%%%%%%%%%%%%%%%%%%%%%%%%%%%%%
\chapter{Landau Damping in a $2D$ Electron Gas
with Imposed Quantum Grid} \label{sec:Damp}
%%%%%%%%%%%%%%%%%%%%%%%%%%%%%%%%%%%%%%%%%%%%%%%%%%%%%%%%%%%%%%%%%%%%

%%%%%%%%%%%%%%%%%%%%%%%%%%%%%%%%%%%%%%%%%%%%%%%%%%%%%%%%%
\section{Introduction}\label{sect:Damp-intro}
%%%%%%%%%%%%%%%%%%%%%%%%%%%%%%%%%%%%%%%%%%%%%%%%%%%%%%%%%

In this Chapter, QCB interaction with semiconductor substrate is
studied. Any surface wave excited in the substrate is coupled with
QCB-plasmon modes due to the substrate-QCB interaction. This
interaction might be strong enough compared with the frequency
spacing of surface waves and QCB plasmons because surface plasmon
waves exist in the same frequency and wave vector area as plasmons
in QCB (see subsection \ref{subsect-H-Int} for details). Therefore
by exciting the substrate plasmons one can probe the QCB
characteristics \cite{KGKA3,KG,K}. Indeed, substrate-QCB
interaction substantially changes the conventional picture of
substrate dielectric losses. Due to such interaction, new regions
of Landau damping appear. The existence of these regions
themselves, as well as their structure and the density of losses
are sensitive both to the QCB period $a$ and the direction of the
wave vector ${\bf{k}}$ of the initial wave. Thus, dielectric
losses in QCB-substrate system serve as a good tool for studying
QCB spectral properties.

The structure of this Chapter is as follows. In Section
\ref{sect-H}, we briefly describe double square QCB interacting
with the dielectric substrate and introduce the necessary
definitions. Dielectric properties of the system considered are
studied in Section \ref{sect-DF}, where Dyson-type equations for
the polarization operator are obtained and analyzed. The detailed
description of new regions of Landau damping is presented in
Section \ref{sect-Damping}. In the Conclusion Section we summarize
the results obtained.

%%%%%%%%%%%%%%%%%%%%%%%%%%%%%%%%%%%%%%%%%%%%%%%%%%%%%%%%%
\section{Quantum Crossbars on Semiconductor Substrate}\label{sect-H}
%%%%%%%%%%%%%%%%%%%%%%%%%%%%%%%%%%%%%%%%%%%%%%%%%%%%%%%%%

\begin{figure}[htb]
\centering
\includegraphics[width=60mm,height=20mm,angle=0,]{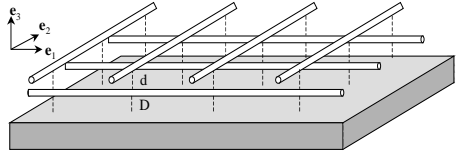}
 \caption{QCB on a substrate.  ${\bf{e}}_{\mu}$
($\mu=1,2,3$) are basic vectors of the coordinate system.  The
vector ${\bf{e}}_{1}$ (${\bf{e}}_{2}$) is oriented along the first
(second) array.  The inter-array distance is $d$ and the distance
between the substrate and the first (lower) array is $D.$}
\label{Substrate}
\end{figure}

Let us consider a square QCB on a semiconductor substrate (see
Fig. \ref{Substrate}). We choose coordinate system so that {\bf
1)} the axes $x_{j}$ and the corresponding basic unit vectors
${\bf e}_{j}$ are oriented along the $j$-th array ($j=1,2$); {\bf
2)} the $x_{3}$ axis is perpendicular to the QCB plane; {\bf 3)}
the $x_{3}$ coordinate is zero for the second array, $-d$ for the
first one, and $-(d+D)$ for the substrate. The basic vectors of
the reciprocal superlattice for a square QCB are $Q{\bf e}_{1,2},$
$Q=2\pi/a$ so that an arbitrary reciprocal superlattice vector
${\bf{m}}$ is a sum ${\bf{m}}={\bf{m}}_{1}+ {\bf{m}}_{2},$ where
${\bf{m}}_{j}=m_{j}Q {\bf{e}}_{j}$ ($m_j$ integer). An arbitrary
vector ${\bf k}=k_{1}{\bf e}_{1}+k_{2}{\bf e}_{2}$ of reciprocal
space can be written as ${\bf q}+{\bf m}$ where ${\bf q}$ belongs
to the first BZ $|q_{1,2}|\leq Q/2.$

%%%%%%%%%%%%%%%%%%%%%%%%%%%%%%%%%%%%%%%%%%%%%%%%%%%%%%%%%%%%%%%%%
%%%%%%%%%%%%%%%%%%%%%%%%%%%%%%%%%%%%%%%%%%%%%%%%%%%%%%%%%%%%%%%%%

%%%%%%%%%%%%%%%%%%%%%%%%%%%%%%%%%%%%%%%%%%%%%%%%%%%
\subsection{Substrate Characteristics}\label{subsect-H-Sub}
%%%%%%%%%%%%%%%%%%%%%%%%%%%%%%%%%%%%%%%%%%%%%%%%%%%

The substrate is described by the Hamiltonian
\begin{equation}
 H_s=H_K+H_C,
 \label{Hsub}
\end{equation}
where
\begin{eqnarray*}
  H_{K} =
  \sum_{\bf{qm}}\varepsilon_{\bf{q+m}}c_{\bf{q+m}}^{\dag}c_{\bf{q+m}},
  \ \ \ \
  \varepsilon_{{\bf{k}}}=\displaystyle{\frac{\hbar^{2}k^{2}}{2m}},
%  \label{Kinetic}
\end{eqnarray*}
is a kinetic energy of the substrate electrons with effective mass
$m$ and quadratic dispersion law (we omit the irrelevant spin
variables), and
\begin{eqnarray*}
  H_C = \frac{1}{2}\sum_{\bf{qm}}U_{\bf{q+m}}
          \rho_{\bf{q+m}}^{\dag}\rho_{\bf{q+m}},
  \ \ \ \ \
  \rho_{\bf{k}}=\frac{1}{L}\sum_{\bf{k'}}c_{\bf{k'}}^{\dag}c_{\bf{k+k'}},
  \ \ \ \ \
  U_{\bf{k}}=\frac{2\pi e^2}{k},
  \ \ \ \ \
  {\bf{k}}={\bf{q}}+{\bf{m}},
\end{eqnarray*}
is Coulomb interaction within the substrate.

Dielectric properties of the substrate {\it per se} are described
by its dielectric function $\epsilon_{s}({\bf{k}},\omega)$,
\begin{equation}
 \frac{1}{\epsilon_{s}({\bf{k}},\omega)}=
 1+U_{\bf{k}}\Pi_{s}({\bf k},\omega).
 \label{DielFunSub}
\end{equation}
Within the RPA approach, the polarization operator
$\Pi_{s}({\bf{k}},\omega)$ is approximated by the Lindhard
expression
\begin{eqnarray}
  \left(\Pi_{s}({\bf k},\omega)\right)^{-1} &=&
    \left(\Pi_{0}({\bf k},\omega)\right)^{-1}-U_{\bf k}
    \label{Pi_s},
\end{eqnarray}
with
\begin{eqnarray}
  \Pi_{0}({\bf k},\omega) &=&
    \frac{1}{L^2}\sum\limits_{\bf k'}
    \frac{\vartheta(\varepsilon_F-\varepsilon_{\bf k'})
         -\vartheta(\varepsilon_F-\varepsilon_{\bf k+k'})}
         {\hbar\omega-(\varepsilon_{\bf k+k'}-
                     \varepsilon_{\bf k'})+i0}
 \label{Pi_0}
\end{eqnarray}
at $T=0$.

\begin{figure}[ht]
\centering
\includegraphics[width=60mm,height=45mm,angle=0,]{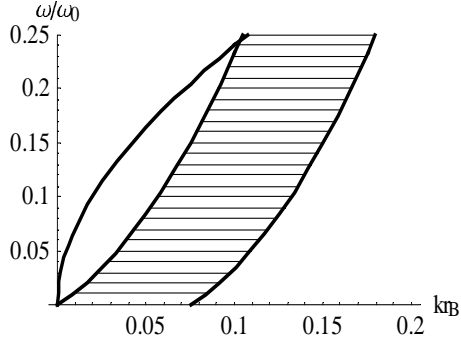}
\caption{Dispersion of the substrate plasmons (upper line) and
quasi-continuum spectrum of electron-hole excitations, (dashed
area). Frequency is measured in $\omega_0=v_F/r_B$ units.}
 \label{Disp0}
\end{figure}

Active branches of substrate excitations are the surface density
fluctuations which consist of $2D$ electron-hole pair continuum
and surface plasmon mode with dispersion law \cite{Stern}
\begin{eqnarray*}
 \omega_{s}({\bf{k}})=v_F k\sqrt{1+\frac{1}{2 kr_B}},
 \ \ \
 r_B=\displaystyle{\frac{\hbar^2}{me^2}},
 \ \ \
 k=|{\bf{k}}|.
\end{eqnarray*}
The {R}{P}{A} spectrum of surface excitations is shown in Fig.
\ref{Disp0}.

%%%%%%%%%%%%%%%%%%%%%%%%%%%%%%%%%%%%%%%%%%%%%%%%%%%%%%%%%%%%%%%%%
%%%%%%%%%%%%%%%%%%%%%%%%%%%%%%%%%%%%%%%%%%%%%%%%%%%%%%%%%%%%%%%%%

In the case of {Ga}{As}, the substrate parameters are
$m=0.068m_0$, $m_0$ is the free electron mass,
$v_F=8.2\cdot{10}^6$~cm/sec, $r_B=0.78$~nm,
$\omega_0=v_F/r_B={1.05}\times{10}^{14}$~sec$^{-1}$. For
$k<k^{*}\approx 0.104r_B^{-1}$, the plasmon frequency lies above
the continuum spectrum of electron-hole pairs and the substrate
plasmons are stable. Besides, one may easily satisfy the resonance
condition for the collective plasmon mode near the stability
threshold $k\sim k^{*}$ and QCB excitations with frequency
$\omega\sim\omega^{*}=2.6\cdot{10}^{13}$~sec$^{-1}$. For large
enough $k>k^{*}$ the plasmon dispersion curve lies within the
quasi-continuum spectrum and plasmons become unstable with respect
to decay into electron-hole pairs (Landau damping of the substrate
plasmons). Dielectric losses of an isolated substrate are
described by an imaginary part $\Im\epsilon_{s}({\bf{k}},\omega)$
of its dielectric function (\ref{DielFunSub}). This imaginary part
is non-zero within the dashed region in Fig. \ref{Disp0} due to
the appearance of an imaginary part
\begin{eqnarray*}
  &&\Im\Pi_0({\bf k},\omega) =
  \frac{m}{2\pi\hbar^2}
  \frac{1}{\kappa^2}
  \left[
      \vartheta(\kappa^2-\nu_{+}^2)
       \sqrt{\kappa^2-\nu_{+}^2}-
      \vartheta(\kappa^2-\nu_{-}^2)
       \sqrt{\kappa^2-\nu_{-}^2}
  \right],
  \\&&
  \kappa=\frac{k}{k_{F}}, \ \ \ \nu_{\pm}=\nu\pm\kappa^{2}/2,
  \ \ \ \nu=\frac{\omega}{v_{F}k_{F}}
\end{eqnarray*}
of the bare polarization operator $\Pi_0({\bf k},\omega)$
(\ref{Pi_0}).

%%%%%%%%%%%%%%%%%%%%%%%%%%%%%%%%%%%%%%%%%%%%%%%%%%%
\subsection{Interaction}\label{subsect-H-Int}
%%%%%%%%%%%%%%%%%%%%%%%%%%%%%%%%%%%%%%%%%%%%%%%%%%%

Interaction between QCB and the substrate is a capacitive coupling
of charge fluctuations in the substrate with collective modes in
the quantum wires.  Assuming the distance $D$ between the first
array and the substrate to be much smaller than the distance $d$
between arrays, one can keep only the interaction $H_{s1}$ between
the substrate and the first array.  The interaction between the
substrate density fluctuation at the point ${\bf r}\equiv
(x_1,x_2)$ and the density fluctuations located in the vicinity of
the point $x'_1$ which belongs to the $n_2$-th wire of the first
array is described by its amplitude $W(x_1-x'_1,x_2-n_2a)$, where
\begin{eqnarray*}
    W({\bf{r}}) &=&
    \frac{\sqrt{2}e^2
          \zeta
          \left(
               {\displaystyle{{x_1}/{r_0}}}
          \right)}
         {\sqrt{\left|{\bf{r}}\right|^2+D^2}}.
\end{eqnarray*}
The function $\zeta(x_1/r_0)$ in the numerator describes the
screening of Coulomb interaction within a wire or nanotube
\cite{Sasaki}. In the momentum representation the interaction
Hamiltonian between substrate and array has the form:
\begin{equation}
  H_{s1}= \sqrt{\frac{\hbar}{vg}}\sum_{\bf{mq}}W_{{\bf{q}}+{\bf{m}}}
             \rho_{\bf{q+m}}\theta_{1,{\bf{q+m}}_1}^{\dag},
  \label{Hs1}
\end{equation}
where
\begin{eqnarray}
  W_{{\bf{k}}} = ik_1\sqrt{\frac{v g}{\hbar a}}\int d{\bf r}
  W({\bf{r}})e^{i{\bf{kr}}}
  \label{W_k}
\end{eqnarray}
is proportional to the Fourier component of the interaction
amplitude $W({\bf{r}})$.

Finally, the Hamiltonian of QCB interacting with a semiconductor
substrate is the sum of Hamiltonians (\ref{TotHam2}), (\ref{Hsub})
and (\ref{Hs1}),
\begin{equation}
  H = H_{QCB}+H_s+H_{s1}.
  \label{H}
\end{equation}

%%%%%%%%%%%%%%%%%%%%%%%%%%%%%%%%%%%%%%%%%%%%%%%%%%%%%%%%%
\section{Dielectric Function}\label{sect-DF}
%%%%%%%%%%%%%%%%%%%%%%%%%%%%%%%%%%%%%%%%%%%%%%%%%%%%%%%%%

High frequency properties of the system at zero temperature are
determined by zeroes of its dielectric function
\begin{equation}
 \frac{1}{\epsilon({\bf{k}},\omega)}=
 1+U_{\bf{k}}\Pi({\bf k},\omega).
 \label{DielFun}
\end{equation}
Here
\begin{eqnarray*}
  \Pi({\bf{k}},\omega)=-\frac{i}{\hbar}\int\limits_{0}^{\infty}dt
  e^{i\omega t}
         \left\langle\left[
              \rho_{\bf{k}}(t),
              \rho_{\bf{k}}^{\dag}(0)
         \right]\right\rangle,
%  \label{PolarCorrFun}
\end{eqnarray*}
is the polarization of the {\it substrate interacting with QCB},
$\rho_{\bf{k}}(t)=e^{iHt/\hbar}\rho_{\bf{k}}e^{-iHt/\hbar}$ is the
density of the {\it substrate} electrons in the Heisenberg
representation, and averaging is performed over the ground state
of the Hamiltonian (\ref{H}).

The Umklapp processes stimulated by the interaction between the
substrate and the first array (\ref{Hs1}) as well as the
interaction between arrays (\ref{interaction-interarray}), produce
modes with wave vectors ${\bf{q}}+{\bf{m}}$ with various inverse
lattice vectors ${\bf{m}}$.  This necessarily leads to appearance
of non-diagonal polarization operators
\begin{eqnarray*}
  &&\Pi({\bf q} +{\bf m},{\bf q} +{\bf m}';\omega) =
  %\nonumber\\&&=
  -\frac{i}{\hbar}\int\limits_{0}^{\infty}dt
  e^{i\omega t}
  \big\langle\big[
             \rho_{\bf{q+m}}(t)
             \rho_{\bf{q+m'}}^{\dag}(0)
  \big]\big\rangle.
\end{eqnarray*}
In what follows we always consider a fixed frequency $\omega$ and
a fixed wave vector ${\bf q}$ from the BZ. So the variables
${\bf{q}}$ and $\omega$ are omitted below for simplicity. In the
framework of {R}{P}{A} approach, $\Pi({\bf{m}},{\bf{m}}')$
satisfies the Dyson-type equation
\begin{eqnarray}
  \Pi({\bf{m,m}}') &=& \Pi_{s}({\bf{m}})\delta_{{\bf{m,m}}'}+
  \Pi_{s}({\bf{m}})W_{\bf{m}}\Xi_1(m_1,{\bf{m}}').
  \label{EqDyson1}
\end{eqnarray}
The first term $\Pi_s({\bf{m}})$ in the right hand side is the
substrate polarization (\ref{Pi_s}) of the isolated substrate
itself, ${W}_{\bf{m}}\equiv{W}_{{\bf{q}}+{\bf{m}}}$ is a bare
vertex (\ref{W_k}) which describes substrate - (first) array
interaction, and
\begin{eqnarray}
  \Xi_j(m_j,{\bf{m}}') &=&
   -\frac{i}{\hbar}\int\limits_{0}^{\infty}dt
  e^{i\omega t}
  \big\langle\big[
       \theta_{j,{\bf{q+m}}_j}(t),
       \rho_{{\bf{q}}+{\bf{m}}'}^{\dag}(0)
  \big]\big\rangle
  \label{Dj}
\end{eqnarray}
is the correlation function of the $j$-{t}{h} array mode and the
substrate plasmon.

The Dyson equation (\ref{EqDyson1}) should be completed by two
equations for the correlation functions (\ref{Dj}) ($j=1,2$)
\begin{eqnarray}
  \Xi_1(m_1,{\bf{m}}') &=&
   D_1^0(m_1)\sum_{m_2}W_{\bf{m}}\Pi({\bf{m,m}}')+
  %\nonumber\\&+&
   D_1^0(m_1)\sum_{m_2}\Phi_{\bf{m}}
  \Xi_2(m_2,{\bf{m}}'),
  \label{EqDyson2}\\
  \Xi_2(m_2,{\bf{m}}') &=&
   D_2^0(m_2)\sum_{m_1}\Phi_{\bf{m}}
   \Xi_1(m_1,{\bf{m}}').
  \label{EqDyson3}
\end{eqnarray}
Here $D_{j}^{0}(m_j)$ ($j=1,2$) is the bare correlation function
of the $j$-{t}{h} array modes
\begin{eqnarray*}
  D_{j}^{0}(m_j) &=&
  -\frac{i}{v g}\int\limits_{0}^{\infty}dt
  e^{i\omega t}\left\langle\left[
                    \theta_{j,{\bf{q+m}}_j}(t),
                    \theta_{j,{\bf{q+m}}_j}^{\dag}(0)
               \right]\right\rangle_{0}
  =%\nonumber\\&=&
  \frac{1}{\omega^2-v^2(q_j+m_j)^2},
%  \label{D-j0}
\end{eqnarray*}
and another bare vertex $\Phi_{\bf{m}}$ describes the separable
inter-array interaction (\ref{Phi-Fourier}).

Solving the system of equations (\ref{EqDyson1}), (\ref{EqDyson2})
and (\ref{EqDyson3}) one obtains the diagonal element $\Pi({\bf
m})\equiv \Pi({\bf m,m})$ of the polarization operator
\begin{equation}
  \left[\Pi({\bf m})\right]^{-1}=
  \left[\Pi_s({\bf m})\right]^{-1}-
        |W_{\bf{m}}|^2D({\bf m}).
  \label{Polar-Solution}
\end{equation}
The second term on the right-hand side of this equation describes
renormalization of the substrate polarization operator
$\Pi_s({\bf{m}})$ by interaction between the substrate and QCB.
The factor $D({\bf m})$ is a renormalized correlation function of
modes of the first array
\begin{eqnarray}
  \left[D({\bf m})\right]^{-1}&=&
  \left[D_1^0(m_1)\right]^{-1}
  -\left(w({\bf m})+\varphi(m_1)\right).
  \label{Dm}
\end{eqnarray}
The first term $w({\bf{m}})$ describes the effective interaction
between the first array and the substrate
\begin{eqnarray}
  w({\bf{m}}) &=&F(m_{1})-
  |W_{\bf{m}}|^2\Pi_s({\bf m}),\ \ \ %\nonumber\\
  F(m_{1})=\sum_{m_{2}}|W_{\bf{m}}|^2\Pi_s({\bf m}).
       \label{w}
\end{eqnarray}
The second one $\varphi(m_1)$ is the effective interaction between
arrays
\begin{eqnarray}
  \frac{\omega_{m_{1}}^{2}}{\varphi(m_1)} &=&
   \left[
    \phi^{2}\sum\limits_{m_2}\omega_{m_{2}}^{2}D_2^0(m_2)
   \right]^{-1}-\Psi_{m_{1}},
  \ \ \ %\label{phi} \\
  \omega_{m_{j}}=v|q_{j}+m_{j}Q|,
  \label{phi} %\nonumber
\end{eqnarray}
renormalized by Coulomb interaction of array modes with the
substrate plasmons,
\begin{eqnarray}
   \Psi_{m_1} &=&
         \sum_{{m'_1}\ne{m_1}}
         \frac{\omega_{m'_{1}}^{2}}
         {\left(D_1^0(m'_1)
         \right)^{-1}-F(m_1')}.
  \label{M}
\end{eqnarray}
Equations (\ref{Polar-Solution}) - (\ref{M}) together with
definition (\ref{DielFun}) solve the problem of elucidation of the
dielectric properties of the combined system QCB-substrate.

The spectrum of collective excitations in QCB-substrate system is
determined by zeros of the dielectric function
$\epsilon({\bf{q}},\omega)=0.$ The key question here is the
robustness of the QCB spectrum against interaction with $2D$
substrate excitations. Detailed analysis shows that in the long
wave limit $q\ll Q$ the interaction just renormalizes the bare
dispersion laws of the arrays, conserving its LL linearity. This
result verifies stability of QCB plasmons with respect to
substrate-QCB interaction.

The QCB-substrate interaction also results in occurrence of some
special lines in the BZ. These lines correspond to resonant
interaction of the substrate with the first or the second array.
The resonance condition $\omega_{s}({\bf{k}})=\omega_j({\bf{k}})$
is fulfilled along the line $LJIN$ for $j=1$ and along the line
$KBM$ for $j=2$ in Fig. \ref{30nm}, below.

%%%%%%%%%%%%%%%%%%%%%%%%%%%%%%%%%%%%%%%%%%%%%%%%%%%%%%%%%
\section{Landau Damping}\label{sect-Damping}
%%%%%%%%%%%%%%%%%%%%%%%%%%%%%%%%%%%%%%%%%%%%%%%%%%%%%%%%%

As was mentioned in subsection \ref{subsect-H-Sub} above,
dielectric losses of an isolated substrate are related to the
Landau damping due to decay of substrate plasmons with momentum
$k>k^{*}\approx 0.104r_B^{-1}$ into electron-hole pairs. The
substrate-QCB interaction remarkably modifies the conventional
picture of substrate plasmon dielectric losses.  Due to
QCB-substrate interaction, new domains of Landau damping appear in
addition to the dashed region in Fig. \ref{Disp0}. Indeed, outside
the initial instability region where
$\Im\epsilon_{s}({\bf{k}},\omega)=0$, nonzero imaginary part
$\Im\epsilon({\bf{k}},\omega)$ (\ref{DielFun}) exists if the
imaginary part of the bare polarization operator
$\Im\Pi_0({\bf{k}}+{\bf{m}},\omega)$ differs from zero at least
for one of the reciprocal lattice vectors ${\bf{m}}$.  The main
contribution to $\Im\epsilon({\bf{k}},\omega)$ is related to the
renormalization term $w({\bf m})$ in Eq.(\ref{w}) due to Umklapp
processes along the $x_{2}$ axis (summation over $m_{2}$ in the
expression for the function $F(m_{1})$ is implied).  It is
proportional to the fourth power of QCB - substrate interaction
$W^{4}$.  The Umklapp processes along both directions $x_{1,2}$
contribute also to the renormalization term $\varphi(m_{1})$ in
Eq.(\ref{phi}).  However, they contain an additional small
parameter $\phi^{4}$ related to inter-array interaction within
QCB. These terms are not taken into account. Thus, the possible
Umklapp vectors have the form ${\bf m}_{2}=m_{2}Q{\bf e}_{2},$
$m_{2}=\pm 1,\pm 2, \ldots$ and in what follows we will label them
by an integer number $m_{2}.$

%%%%%%%%%%%%%%%%%%%%%%%%%%%%%%%%%%%%%%%%%%%
\begin{figure}[ht]
 \centering
 \includegraphics[width=70mm,height=60mm,angle=0,]{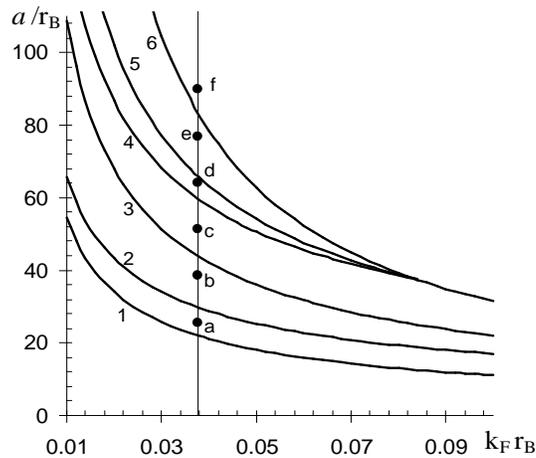}
 \caption{Phase diagram describing appearance and
structure of new regions of Landau damping. Lines $1-6$ separate
different types of new damping regions. Points $a-f$ correspond to
the structures displayed below in Figs. \ref{20nm}-\ref{70nm}.}
\label{diagram}
\end{figure}
%%%%%%%%%%%%%%%%%%%%%%%%%%%%%%%%%%%%%%%%%%%

The structure of the new Landau damping regions and their
existence themselves is governed by interplay between the Fermi
momentum of the substrate $k_{F}$ and the QCB period $a$. The
first of these parameters defines the width of the two-particle
excitation band (dashed region in Fig. \ref{Disp0}) while the
second determines the minimal reciprocal vector $Q$. In the case
of sufficiently thick QCB superlattice (small $a$) and
sufficiently low electron density within the substrate (small
$k_F$), Umklapp processes are always ineffective because they
change an initial plasmon wave-vector into the outer part of the
instability region. This means that only plasmons with momenta
$|k|>k^{*}$ decay into electron-hole pairs.

Increasing the QCB period or the Fermi momentum turn the Umklapp
processes effective, and additional Landau damping regions appear
within the circle $|k|\leq k^{*}$.  The first factors involved are
the smallest Umklapp vectors $\pm 1,$ then new damping regions
appear corresponding to the Umklapp vectors $\pm 2$ and so on.  As
a result one gets a rich variety of possible damping scenarios. We
describe them with the help of a ``phase diagram'' in the
$a-k_{F}$ plane displayed in Fig. \ref{diagram} (actually
dimensionless coordinates $a/r_{B}$ and $k_{F}r_{B}$ are used).
Here the set of curves labelled by numbers $1-6$ separate the
regions of parameters corresponding to the different Umklapp
vectors and different structures of the new damping regions. There
is no additional Landau damping regions below the first line.
Above the sixth line Landau damping takes place within the whole
circle $k\leq k^{*}.$ Above lines with numbers $2n-1,$ the Umklapp
vector $\pm n$ becomes effective. The corresponding additional
damping region has the form of a tail touching the initial Landau
damping region $|k|\geq k^{*}$. This tail turns to the additional
damping band well separated from the initial one above the line
number $2n$ ($n<3$) within some sector of directions in $k$-space
(in what follows, these directions will be labelled by
corresponding arcs of the circle).

%%%%%%%%%%%%%%%%%%%%%%%%%%%%%%%%%%%%%%%%%%%
\begin{figure}[htb]
 \centering
 \includegraphics[width=62mm,height=54mm,angle=0,]{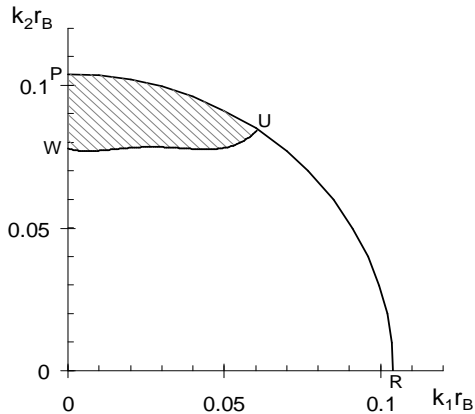}
 \caption{New Landau damping region $PUW$ for QCB with
period $a=20$~nm (point $a$ in Fig. \ref{diagram}) corresponds to
the Umklapp vector $-1.$ Other details of this figure are
explained in the text.} \label{20nm}
\end{figure}
%%%%%%%%%%%%%%%%%%%%%%%%%%%%%%%%%%%%%%%%%%%
%%%%%%%%%%%%%%%%%%%%%%%%%%%%%%%%%%%%%%%%%%%
\begin{figure}[htb]
 \centering
 \includegraphics[width=62mm,height=54mm,angle=0,]{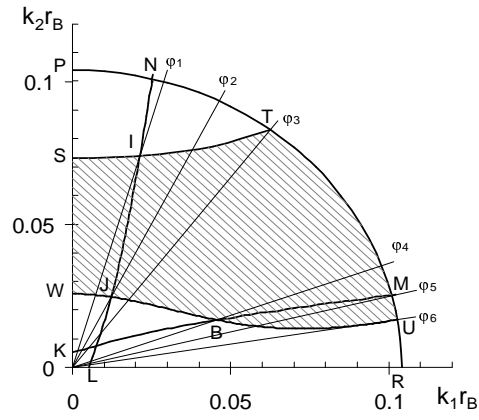}
 \caption{New damping region
$STUW$ for $a=30$~nm (point $b$ in Fig. \ref{diagram}) corresponds
to the Umklapp vector $-1$ and describes Landau damping tail
(within the arc $TU$) or separate landau band (within the arc
$TP$).} \label{30nm}
\end{figure}
%%%%%%%%%%%%%%%%%%%%%%%%%%%%%%%%%%%%%%%%%%%

\begin{figure}[ht]
\centering
 \includegraphics[width=62mm,height=54mm,angle=0,]{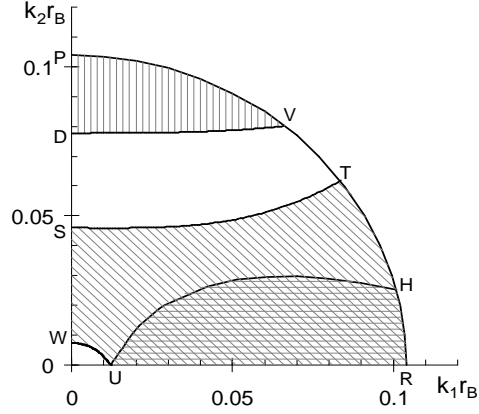}
 \caption{In the case $a=40$~nm (point $c$ in
Fig. \ref{diagram}), new Landau damping regions $PVD,$ $STRUW,$
and $UHR$ correspond to the Umklapp vectors $-2,$ $-1,$ and
$+1,$.} \label{40nm}
\end{figure}

\begin{figure}[ht]
\centering
 \includegraphics[width=62mm,height=54mm,angle=0,]{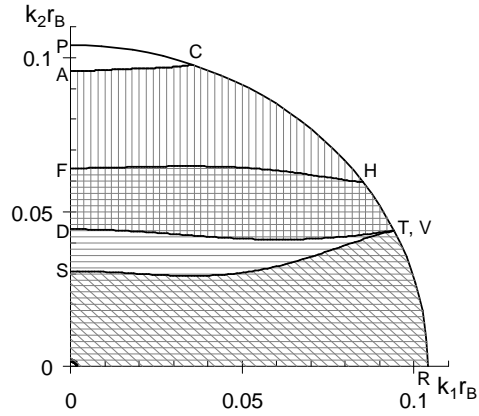}
 \caption{New regions of Landau damping for $a=50$~nm
(point $d$ in Fig. \ref{diagram}). Regions $ACVD$, $STR,$ and
$FHR$ correspond to the Umklapp vectors $-2,$ $-1,$ and $+1$}
\label{50nm}
\end{figure}

\begin{figure}[htb]
\centering
 \includegraphics[width=62mm,height=54mm,angle=0,]{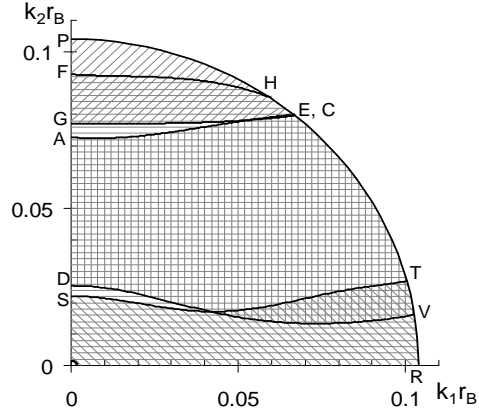}
 \caption{New regions of Landau damping for $a=60$~nm
(point $e$ in Fig. \ref{diagram}). Besides the regions $ACVD,$
$STR,$ and $FHR$ corresponding, as in the case $a=50nm,$ to the
Umklapp vectors $-2,$ $-1,$ and $+1,$ new Umklapp vector $-3$
appears (region $PEG$).}
 \label{60nm}
\end{figure}

\begin{figure}[htb]
\centering
 \includegraphics[width=62mm,height=54mm,angle=0,]{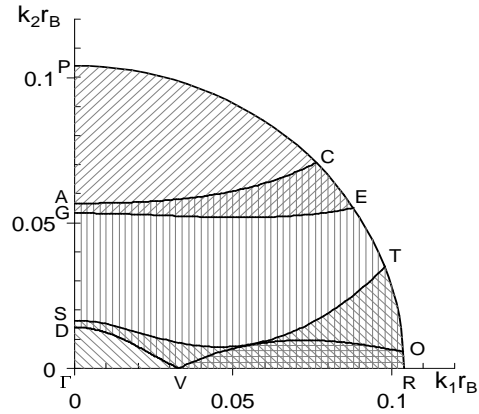}
 \caption{The case $a=70$~nm
(point $f$ in Fig. \ref{diagram}). New regions of Landau damping
$PEG,$ $ACVD,$ $STR,$ whole sector $\Gamma PR,$ and $VOR$
correspond to the Umklapp vectors $-3,$ $-2,$ $-1,$ $+1,$ and
$+2.$}
 \label{70nm}
\end{figure}

Possible structures of new damping regions corresponding to some
representative points $a-f$ in the $a-k_{F}$ plane (see Fig.
\ref{diagram}) are displayed in detail in Figs.
\ref{20nm}-\ref{70nm}. All these figures correspond to the
{Ga}{As} value of $k_{F}r_{B}\approx 0.038.$ Generally speaking we
should display new damping region within the whole circle $|k|\leq
k^{*}$ in the plane $k_{1},k_{2}.$ The circle center $\Gamma$ is
placed at the origin (we did not put the letter $\Gamma$ in Figs.
\ref{20nm}-\ref{70nm}). But this region is always symmetric with
respect to reflection $k_{1}\to -k_{1}$ and with respect to the
combined reflection $k_{2}\to -k_{2}, m_{2}\to -m_{2}.$ This
enables us to describe the damping scenarios only within the
quarter $k_{1,2}\geq 0$ (complete picture of the new damping
region can be easily obtained from the displayed one with the help
of the reflection symmetries mentioned above).

Damping of the substrate plasmon occurs inside the arc $PR$ in
Figs. \ref{20nm}-\ref{70nm} when at least one of the points in the
phase space with coordinates $({\bf{k}}+{\bf{m}}_2,\omega_s(k))$
lies within the quasi-continuum spectrum of the electron-hole
excitations, whereas the ``mother'' point $({\bf{k}},\omega_s(k))$
lies above the continuum (above the dashed area in Fig.
\ref{Disp0}). As was mentioned above, for small enough {Q}{C}{B}
period $a<a_1=17.3$~nm, the basic reciprocal lattice vector
$Q{\bf{e}}_2$ is too large, the points
$({\bf{k}}+{\bf{m}}_2,\omega_s(k))$ lie outside the
quasi-continuum for all $m_2$ and additional Landau damping region
does not exist. It appears only for $a>a_1$ ($Q<2k^{*}+2k_F$). For
$a_1<a<a_2=23.6$~nm ($2k^{*}+2k_F>Q>2k^{*}$), this is the region
$PUW$ (see Fig. \ref{20nm}) corresponding to the $m_{2}=-1$ (in
all Figs. \ref{20nm}-\ref{70nm}, the regions related to this
Umklapp vector are always hatched by the hatching tilted to left).
As a result, the damping tails touching the initial Landau damping
region appear in certain directions of the ${\bf{k}}$ plane.

For $a_2<a<a_3=34.6$~nm ($2k^{*}>Q>k^{*}+k_F$), the new damping
region is related to the same Umklapp vector $-1$, but now it has
a strip-like structure bounded by the line $STUW$ in Fig.
\ref{30nm}. Note that the damping is absent within the region
$PTS$. As a result, it is possible to divide the angular region
$0\le\varphi\le\pi/2$ into three sectors.  Within the first one
$PT$, $0\le\varphi\le\varphi_3$, a new damping region is separated
from the initial one.  The second sector $TU$,
$\varphi_3\le\varphi\le\varphi_6$, corresponds to a new damping
tail. Finally, within the third sector $UR$ ,
$\varphi_6\le\varphi\le{\pi}/{2}$, new damping region does not
exist at all.

%%%%%%%%%%%%%%%%%%%%%%%%%%%%%%%%%%%%%%%%%%%
%%%%%%%%%%%%%%%%%%%%%%%%%%%%%%%%%%%%%%%%%%%
%%%%%%%%%%%%%%%%%%%%%%%%%%%%%%%%%%%%%%%%%%%
%%%%%%%%%%%%%%%%%%%%%%%%%%%%%%%%%%%%%%%%%%%

For larger QCB period, $a_3<a<a_4=47.1$~nm ($k^{*}+k_F>Q>k^{*}$),
the new damping regions have a more complicated structure. In fact
the damping area consists of three parts (see Fig. \ref{40nm}).
The first one, $STRUW$ corresponds to the Umklapp vector $-1.$
Note that it is shifted to the bottom with respect to the previous
case $a=30$~nm. This part overlaps with the second part $HRU$. The
latter corresponds to the Umklapp vector $+1$ and is hatched in
Figs. \ref{40nm}-\ref{70nm} by horizontal hatching. The second
region $PDV$ corresponds to the Umklapp vector $-2$ (in Figs.
\ref{40nm}-\ref{70nm} such a regions are always hatched by
vertical hatching). As a result in the direction close enough to
the $k_{2}$ axis, one gets a new damping tail with $m_{2}=-2$) and
well separated new damping band with Umklapp vector $-1.$

Further increase of QCB period $a_4<a<a_5=52.3$~nm,
$k^{*}>Q>2(k^{*}+k_F)/3$ ($k^{*}\approx 2.8 k_{F}$ for {Ga}{As})
leads to further extension of new damping regions.  The region
$ADVC$ corresponding to the Umklapp vector $-2$ is partially
separated from the initial damping region.  It overlaps with the
region $FHR\Gamma$ ($m_{2}=+1$) which in its turn overlaps with
the region $STR\Gamma$ ($m_{2}=-1$).  Actually the two latter
regions do not include an extremely small vicinity of the origin
$\Gamma$ which is not shown in Fig. \ref{50nm}. Visible
coincidence of the points $V$ and $T$ in Fig. \ref{50nm} is an
artefact of the accuracy of the figure.

%%%%%%%%%%%%%%%%%%%%%%%%%%%%%%%%%%%%%%%%%%%
%%%%%%%%%%%%%%%%%%%%%%%%%%%%%%%%%%%%%%%%%%%
%%%%%%%%%%%%%%%%%%%%%%%%%%%%%%%%%%%%%%%%%%%
%%%%%%%%%%%%%%%%%%%%%%%%%%%%%%%%%%%%%%%%%%%

Within the next interval of QCB periods $a_5<a<a_6=65.2$~nm
($2(k^{*}+k_F)/3>Q>2k_F)$) new damping region $GPE$ corresponding
to the Umklapp vector $-3$ appears.  This region is hatched in
Figs. \ref{60nm} and \ref{70nm} by the hatching tilted to the
right. Beside that, the regions $ADVC$ ($m_{2}=-2$), $FHR\Gamma$
($+1$), and $STR\Gamma$ ($-1$) are present.  As in the previous
figure, visible coincidence of the points $E$ and $C$ in Fig.
\ref{60nm} is an artefact of the accuracy of the figure.

Finally for $a>a_6$ ($Q<2k_F$), Landau damping emerges in the
whole circle $|k|\leq k^{*}$ (Fig. \ref{70nm}) This occurs due to
processes with Umklapp vector $+1.$ The corresponding damping
region covers the whole quarter. Therefore we did not hatch it at
all, and used the same horizontal hatching for the new region
$VOR$ corresponding to the Umklapp vector $+2.$ The region $DACRV$
is related to the Umklapp vector $-2.$ We emphasize that at the
same time the vertex $V$ of this region is the vertex of the
region $VOR$. This is not an accidental approximate coincidence as
in the two previous figures. The regions $GPE$ and $\Gamma STR $
are related to the Umklapp vectors $-3$ and $-1$ respectively.

%%%%%%%%%%%%%%%%%%%%%%%%%%%%%%%%%%%%%%
\begin{figure}[htb]
\centering
 \includegraphics[width=55mm,height=40mm,angle=0,]{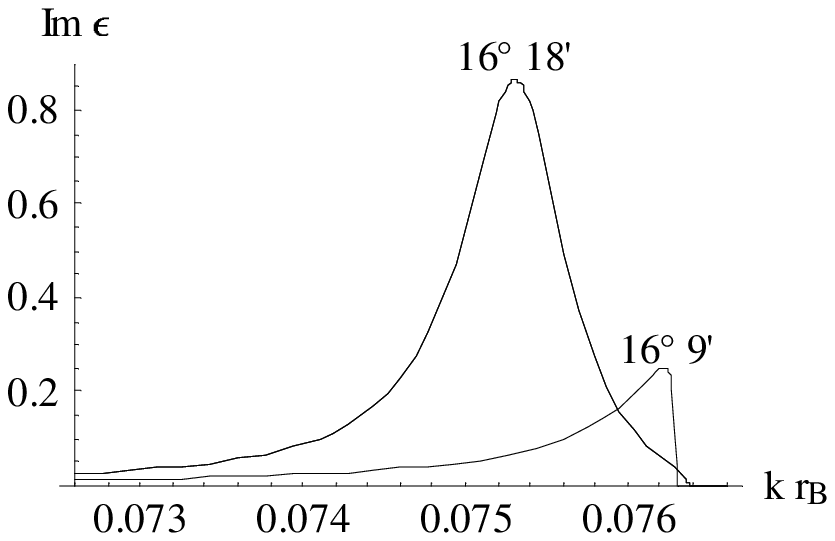}
\caption{Damping tail for $\varphi\approx16^{\circ}$. Precursor of
the resonant peak and the resonant peak are resolved quite well.}
 \label{ImDFAngl-16}
\end{figure}
%%%%%%%%%%%%%%%%%%%%%%%%%%%%%%%%%%%%%%%%%%%%%%%%%%%%%%%%%%%%%%%%%

Thus, the general structure of the additional damping regions is
described in Figs. \ref{20nm}-\ref{70nm}. However, there is an
additional structure of these regions. This fine structure is
related to possible resonance interaction between the substrate
plasmons and the QCB plasmons of the first or second array.  The
resonance condition for the first (second) array is written as
$\omega_{s}({\bf{k}})=\omega_{1}({\bf{k}})$
($\omega_{s}({\bf{k}})=\omega_{2}({\bf{k}})$).

%%%%%%%%%%%%%%%%%%%%%%%%%%%%%%%%%%%%%%%%%%%%%%%%%%%%%%%%%%%%%%%%%%%
\begin{figure}[htb]
\centering
 \includegraphics[width=70mm,height=50mm,angle=0,]{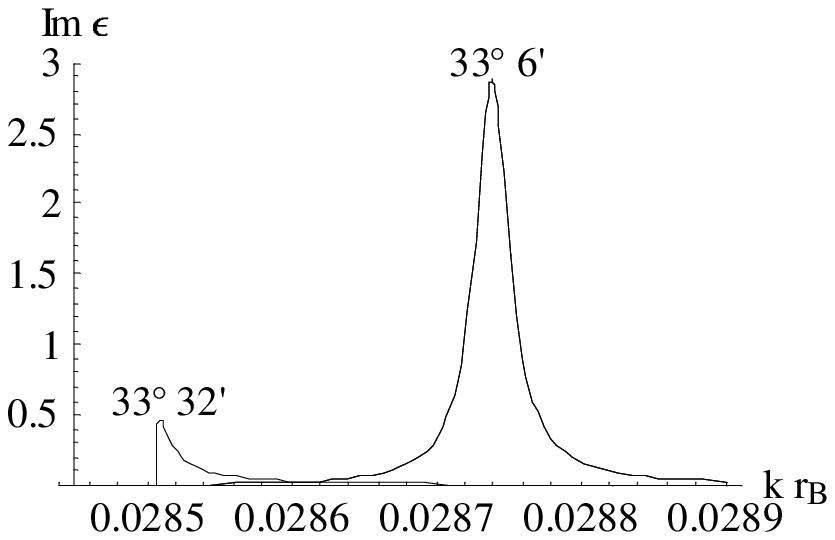}
 \caption{Damping tail for $\varphi\approx33^{\circ}$.
 Precursor of the resonant peak and the resonant peak are resolved
 quite well.} \label{ImDFAngl-33}
\end{figure}
%%%%%%%%%%%%%%%%%%%%%%%%%%%%%%%%%%%

Consider such fine structure of new Landau damping region in
details for QCB with period $a=30$~nm.  Here the resonance
conditions are satisfied along the lines $LJIN$ and $KBM$ (see
Fig. \ref{30nm}). These lines intersect with the damping region
boundaries at the points $J,I,B$ and $MB$ which define four rays
$OI$, $OJ$, $OB$, and $OU$ and four corresponding angles
$\varphi_{1}\approx16^{\circ}11',$
$\varphi_{2}\approx33^{\circ}19',$ $\varphi_{4}\approx69^{\circ},$
$\varphi_{5}\approx76^{\circ}.$ The resonance interaction takes
place within two sectors $\varphi_{1}<\varphi<\varphi_{2}$ and
$\varphi_{4}<\varphi<\varphi_{5}.$

For each $\varphi<\varphi_{1}$ the new damping region is a well
separated damping band.  The damping amplitude is small because of
the small factor of order $W^{4}$ mentioned above.  When
$\varphi\to \varphi_{1}-0$, small peak appears near the ``blue''
boundary of this damping region.  This peak is a precursor of the
resonance between the substrate plasmon and the first array QCB
plasmon (Fig. \ref{ImDFAngl-16}). The same happens from the
opposite side of the sector $(\varphi_{1},\varphi_{2}$ when
$\varphi\to \varphi_{2}+0$ (Fig. \ref{ImDFAngl-33}).

Within the sector $\varphi_{1}<\varphi<\varphi_{2},$ the damping
band contains a well pronounced peak corresponding to resonant
interaction between the substrate plasmon and the first array
plasmons (see Figs. \ref{ImDFAngl-16} and  \ref{ImDFAngl-33}). The
peak amplitude is of order of the damping amplitude within the
initial damping region. It has a Lorentz form placed on a wide and
low pedestal. The peak is especially sensitive to the strength of
the QCB-substrate interaction which is governed by the distance
$D$ between QCB and substrate.

\begin{figure}[ht]
\centering
 \includegraphics[width=80mm,height=50mm,angle=0,]{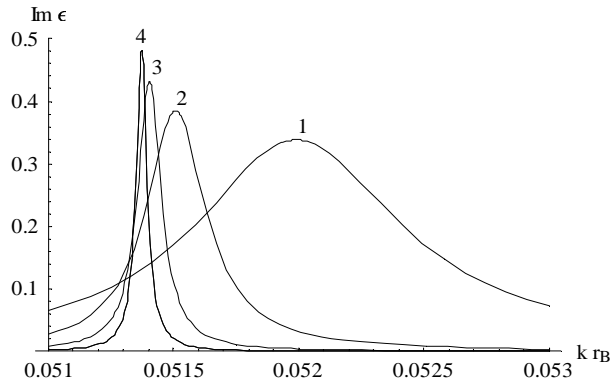}
 \caption{Damping tail for $\varphi=20^{\circ}$ for different
distances $D$ between QCB and substrate.  The curves ${\bf{1}}$,
${\bf{2}}$, ${\bf{3}}$, and ${\bf{4}}$ correspond to $D=1$~nm,
$1.5$~nm, $2$~nm, and $2.5$~nm, respectively.  With increasing
$D$, the resonant peak narrows and slowly increases, whereas the
area under the curve decreases.} \label{ImDFW}
\end{figure}

To study this $D$ dependence, let us consider the imaginary part
$\Im\epsilon({\bf k},\omega)$ of the dielectric function within
the considered sector $\varphi_1<\varphi<\varphi_2.$ In the
vicinity of the plasmon frequency $\omega\approx\omega_s(k)$ this
imaginary part is written as
\begin{eqnarray*}
 \Im\epsilon({\bf k},\omega)=
  \frac{|W_{\bf{k}}|^2}{U_{\bf{k}}}
  \frac{-\Im{w}({\bf{k}},\omega_s(k))}
       {\left(\omega^2-v^2k_1^2\right)^2+
        \left(\Im{w}({\bf{k}},\omega_s(k))\right)^2},
\end{eqnarray*}
with ${w}({\bf{k}},\omega_s(k))$ being of order $|W|^2$.  So the
resonance peak indeed has the Lorentz like shape with height of
order unity,
\begin{eqnarray*}
  \Im\epsilon_{max}\sim
     \frac{|W_{\bf{k}}|^2}
          {|W_{{\bf{k}}-Q{\bf{e}}_2}|^2}\sim1,
\end{eqnarray*}
whereas its half-width
\begin{eqnarray*}
  \Gamma=\Im{w}({\bf{k}},v|k_1|)\sim|W_{{\bf{k}}-Q{\bf{e}}_2}|^2,
\end{eqnarray*}
is of order $W^2$. The peak is displayed in Fig. \ref{ImDFW} for
different values of the distance $D$ between the substrate and the
nearest (first) array. It is seen that the amplitude changes
slowly with increasing distance $D$ while its width squeezes
sharply, $W^2\sim1/D^2$.

%%%%%%%%%%%%%%%%%%
%%%%%%%%%%%%%%%%%%

There is no resonance interaction within the sector
$\varphi_{2}<\varphi<\varphi_{4}$ but further increase of the
angle $\varphi_{4}<\varphi<\varphi_{5}$ leads to re-appearance of
the resonant peak within the damping tail.  In this case one deals
with a resonance between the substrate plasmon and the QCB plasmon
in the second array.  Existence of this resonance is caused by
inter-array interaction that brings additional small parameter to
the imaginary part of the dielectric function.  As a result, the
width of the peak is much smaller in the second sector while
surprisingly, the peak amplitude has the same order of magnitude
as in the case of resonance with the first array (closest to the
substrate).

Existence of the additional QCB bands (tails) of Landau damping
and appearance of the resonant peaks within the bands (tails) is a
clear manifestation of interplay between real $2D$ surface
plasmons and quasi-$2D$ QCB plasmons.

\section{Conclusions}\label{sec:Conclu}
%%%%%%%%%%%%%%%%%%%%%%%%%%%%%%%%%%%%%%%%%%%%%%%%%%%%%%%%%%%%%

In conclusion, the possibility of spectroscopic studies of the
excitation spectrum of quantum crossbars interacting with
semiconductor substrate is investigated. A capacitive contact
between QCB and substrate does not destroy the LL character of the
long wave excitations. However, the interaction between the
surface plasmons and plasmon-like excitations of QCB essentially
influences the dielectric properties of the substrate.  The QCB
may be treated as a diffraction grid for the substrate surface,
and Umklapp diffraction processes radically change the plasmon
dielectric losses.  Due to QCB-substrate interaction, additional
Landau damping regions of the substrate plasmons appear.  The
structure of these regions and the spectral density of dielectric
losses are strongly sensitive to the QCB period. The surface
plasmons are more fragile against interaction with superlattice of
quantum wires than the LL plasmons against interaction with $2D$
electron gas in a substrate.

%\newpage

%%%%%%%%%%%%%%%%%%%%%%%%%%%%%%%%%%%%%%%%%%%%%%%%%%%%%%%%%%%%%%%
\chapter{Ultraviolet Probing of Quantum Crossbars}
 \label{sec:UV}
%%%%%%%%%%%%%%%%%%%%%%%%%%%%%%%%%%%%%%%%%%%%%%%%%%%%%%%%%%%%%%%

%%%%%%%%%%%%%%%%%%%%%%%%%%%%%%%%%%%%%%%%%%%%
\section{Introduction}
 \label{subsec:Intro}
%%%%%%%%%%%%%%%%%%%%%%%%%%%%%%%%%%%%%%%%%%%%

The IR based methods mentioned above are not very convenient from
two points of view. First, as it was mentioned, one needs an
additional diffraction lattice to tune the light wave vector and
that of the QCB plasmon. Second, they probe QCB spectrum only in
some discrete points. The alternative method of studying QCB
spectrum by {\em ultraviolet} (UV) {\em light scattering} is the
subject of the present Chapter. The advantages of this method are
evident. It does not require any additional diffraction lattice.
It probes QCB spectrum in a continuous region of wave vectors.
Finally, its selection rules differ from those for IR absorption.
This gives rise to the observation of additional spectral lines
not visible in IR experiments \cite{KGKA4,K2}.

In this Chapter we formulate the principles of UV spectroscopy for
QCB and study the main characteristics of scattering spectra.  The
Chapter is organized as follows.  In Section \ref{subsec:scatt},
we discuss light scattering on QCB and present basic equations
describing this process. The main results of the Chapter are
contained in subsection \ref{subsec:scatt} where we classify the
basic types of the scattering indicatrices (angular diagrams of
differential cross section) corresponding to various detector
orientations. The results obtained are summarized in the
Conclusion Section. Technical details are concentrated in Appendix
\ref{append:Inter}. This Appendix is devoted derivations of an
effective QCB-light interaction and the basic formula for
differential cross section of light scattering.

%%%%%%%%%%%%%%%%%%%%%%%%%%%%%%%%%%%%%%%%%%%%%%%%%%%%%%%%%%%%%
\section{Light Scattering on QCB}
 \label{subsec:scatt}
%%%%%%%%%%%%%%%%%%%%%%%%%%%%%%%%%%%%%%%%%%%%%%%%%%%%%%%%%%%%%

%%%%%%%%%%%%%%%%%%%%%%%%%%%%%%%%%%%%%%%%%%%%%%%%%%%%%%%%%%%%%%%%%%%%
%\subsection{Light scattering}
% \label{subsubsec:Light}
%%%%%%%%%%%%%%%%%%%%%%%%%%%%%%%%%%%%%%%%%%%%%%%%%%%%%%%%%%%%%

\begin{figure}[htb]
\begin{center}
\includegraphics[width=100mm,height=45mm,angle=0,]{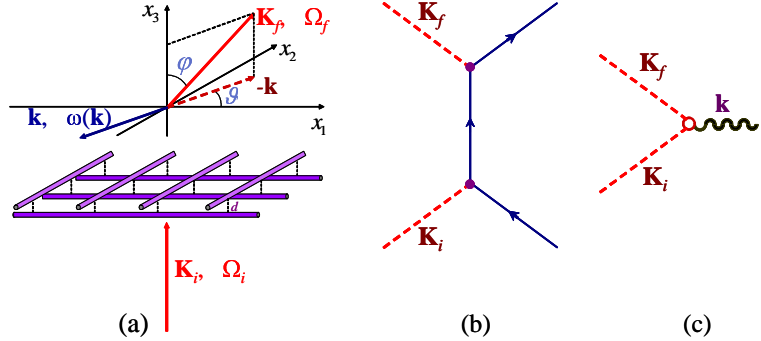}
\caption{(a) The scattering process geometry (all notations are
explained in the text). (b) Second order diagram describing light
scattering on QCB. Solid lines correspond to fermions, whereas
dashed lines are related to photons.  The vertices are described
by the interaction Hamiltonian (\ref{H-int-initial}). (c)
Effective photon-plasmon vertex. QCB excitation is denoted by wavy
line.} \label{X-Scatt}
\end{center}
\end{figure}

The simplest process contributing to the Raman-like light
scattering is an annihilation of an incident photon and creation
of a scattered photon together with a QCB plasmon (Fig.
\ref{X-Scatt}b). In terms of initial electrons, this is in fact
the second order process. Since the energies of incident and
scattered photons significantly exceed the electron excitation
energy in nanotube, one may consider the emission/absorption
process as an instantaneous act (see Appendix \ref{append:Inter}).
This process may be treated as the inelastic photon scattering
accompanied by emission or absorption of a plasmon (Fig.
\ref{X-Scatt}c).

%\vspace{20mm}

%%%%%%%%%%%%%%%%%%%%%%%%%%%%%%%%%%%%%%%%%%%%%%%%%%%%%%%%%%%%%%
%%%%%%%%%%%%%%%%%%%%%%%%%%%%%%%%%%%%%%%%%%%%%%%%%%%%%%%%%%%%%%

Let ${\bf{K}}_i$ (${\bf{K}}_f$) and $\Omega_i=cK_i\equiv
c|{\bf{K}}_i|,$ ($\Omega_f=cK_f$) be momentum and frequency of the
incident (scattered) photon. For simplicity we restrict ourselves
by the case of normal incidence ${\bf{K}}_{i}=(0,0,K_i).$
Direction of the scattered wave vector is characterized by the
unit vector ${\bf{n}}(\varphi,\vartheta),$ where $\varphi$ and
$\vartheta$ are polar and azimuthal angles of the spherical
coordinate system with polar axis oriented in the ${\bf{e}}_3$
direction. The momentum of the excited QCB plasmon is ${\bf{k}},$
the projection of the scattered photon momentum onto QCB plane is
$-{\bf{k}}$. This process is displayed in Fig. \ref{X-Scatt}a. If
the QCB plasmon frequency $\omega({\bf k})$ coincides with the
photon frequency loss $\omega=\Omega_i-\Omega_f,$ then a detector
oriented in a scattered direction ${\bf n}$ will register a sharp
well pronounced peak at the frequency loss $\omega=\omega({\bf
k})$. The frequency loss is much smaller than the incident and
scattered photon frequencies.  Therefore in what follows we use
the same notation $K$ for both $K_i$ and $K_f$ where it is
possible. Scanning the frequency loss $\omega,$ (or, that is the
same, modulus of the scattered wave vector $K_f$) for a fixed
detector orientation ${\bf n},$ two or more (up to six for a
square QCB) such peaks can be observed. The number of peaks and
their location strongly depend on the azimuthal angle $\vartheta$
in the QCB plane. Scanning this angle, one can change the number
of observable peaks. This is yet another manifestation of
dimensional crossover mentioned above.

An arbitrary vector ${\bf{k}}=(k_1,k_2)$ of reciprocal space can
be written as ${\bf{k}}={\bf{q}}+{\bf{m}},$ where
${\bf{q}}=(q_1,q_2)$ belongs to the first BZ. QCB eigenstates are
classified by quasi-momentum ${\bf{q}}$ and $2D$ band number.
However, the specific QCB geometry makes its spectral properties
rather unusual. Consider an {\em isolated} array 1. As was
mentioned in Section \ref{subsubsec:Notions}, within the
($x_1,x_2$) plane, its excitations are characterized by a pair of
$2D$ coordinates $(x_1,n_2a)$, i.e. a continuous longitudinal
coordinate $x_1$ parallel to ${\bf e}_1$ direction, and its
discrete transverse partner $n_2a$ parallel to ${\bf e}_2.$ As a
result, the longitudinal component $k_1=q_1+m_1Q$ of the
excitation momentum changes on the entire axis
$-\infty<k_1<\infty$ while its transverse momentum $q_2$ is
restricted to the region $|q_2|<Q/2$. Thus an eigenstate (plasmon)
of the first array is characterized by the vector
${\bf{k}}_1={\bf{q}}+{\bf{m}}_1=(k_1,q_2)$ and the frequency
$\omega({\bf{k}}_1)=v|k_1|$ which depends only on the longitudinal
component $k_1$ of the momentum ${\bf{k}}_1$. Similar description
of the second array is obtained by replacing $1\leftrightarrow2.$

The exact differential cross section of the scattering in a given
direction, considered as a function of the scattered frequency,
has almost equidistant sets of peaks. However, because of weak
inter-array interaction (\ref{Interaction}), only few peaks are
well pronounced, whereas other peaks are very low and will be
omitted below. Therefore in the approximation adopted here the
scattering cross section is characterized by a number of peaks,
their positions and intensities. Calculation of these quantities
as functions of the frequency loss $\omega$ and azimuthal angle
$\vartheta$ for various fixed values of the polar angle $\varphi,$
is the goal of our study.

To perform quantitative analysis, we should derive an expression
for the scattering cross section. All the details of our
calculations are contained in Appendix \ref{append:Inter}. Here we
present only the main steps of the derivation. We start with the
Hamiltonian
\begin{equation}
 h_{nl} = \frac{ev_F}{c}
     \int\frac{dx_1d\gamma}{2\pi}
     \psi_{\alpha}^{\dag}
     {\bf{A}}\cdot
     {\mbox{$\boldsymbol\sigma$}}_{\alpha,\alpha'}
     \psi_{\alpha'},
       \label{H-int-initial}
\end{equation}
which describes interaction between a single nanotube oriented,
e.g., along the ${\bf e}_1$ direction (the first array nanotube)
and an external electromagnetic field. The field is described by
its vector potential in the Landau gauge
\begin{equation}
    {\bf{A}}=A_1{\bf{e}}_1+A_2{\bf{e}}_2+A_3{\bf{e}}_3.
    \label{VecPot}
\end{equation}
The indices $\alpha=A,B$ enumerate sublattices in a honeycomb
carbon sheet, $(r,\gamma)$ are polar coordinates in the
$(x_{2},x_{3})$ plane,
$\psi_{\alpha}(x_{1},\gamma)\equiv\psi_{\alpha}(x_{1},r_{0},\gamma)$
and $\psi_{\alpha}^{\dag}$ are slowly varying electron field
operators at the nanotube surface $r=r_0,$ and the vector of Pauli
matrices ${\boldsymbol\sigma}$ is
$\boldsymbol\sigma={\bf{e}}_1\sigma_x+{\bf{e}}_\gamma\sigma_y$,
${\bf{e}}_{\gamma}=-{\bf{e}}_2\sin\gamma +{\bf{e}}_3\cos\gamma$.
%\begin{eqnarray*}
% {\mbox{$\boldsymbol\sigma$}}=
% {\bf{e}}_1\sigma_x+
% {\bf{e}}_\gamma\sigma_y,
% \ \ \ \ \ \ \ \ \ \
% {\bf{e}}_{\gamma}=-{\bf{e}}_2\sin\gamma +{\bf{e}}_3\cos\gamma.
%\end{eqnarray*}
The light wavelength is much longer than the nanotube radius, so
the vector potential can be taken at its axis, ${\bf A}(x_1,0,0)$.

Such form of a nanotube-light interaction leads to the following
expression for an effective QCB-light interaction Hamiltonian
\begin{eqnarray}
 H_{int} &=&
 \frac{\sqrt{2}}{4k}
  \frac{e^2}{\hbar c}
  \left(\frac{v_F}{c}\right)^2
 \sum_{n_2} \int dx_1
 \partial_{x_1}\theta_1(x_1,n_2a)
 {\bf{A}}_1^2(x_1,n_2a,0)
 +\nonumber\\&+&
 \frac{\sqrt{2}}{4k}
  \frac{e^2}{\hbar c}
  \left(\frac{v_F}{c}\right)^2
 \sum_{n_1} \int dx_2
 \partial_{x_2}\theta_2(n_1a,x_2)
 {\bf{A}}_2^2(n_1a,x_2,d).
 \label{Eff_Int_QCB}
\end{eqnarray}
Here ${\bf{A}}_j({\bf r})= {\bf{A}}({\bf{r}})+
\left(\sqrt{2}-1\right)A_j({\bf r}){\bf e}_j$ ($j=1,2$)
%\begin{equation}
%  {\bf{A}}_j({\bf r})={\bf{A}}({\bf r})+\left(\sqrt{2}-1
%  \right)A_j({\bf r}){\bf e}_j, \ \ j=1,2,
%  \nonumber
%  \label{A-J}
%\end{equation}
are two effective vector potentials affecting two arrays and
$A_{j}({\bf r})$ are two (of three) cartesian components of the
full vector potential (\ref{VecPot}). The main subject of our
interest is the scaled scattering differential cross-section
$\sigma(\omega,{\bf n})$ defined by equation
\begin{equation}
    {d\sigma}=
    \lambda L^{2}
    \sigma(\omega,{\bf n}){d\omega do},
    \ \ \ \ \
    \lambda=\frac{g}{4\pi ka}
    \left(\frac{e^{2}}{\hbar c}\right)^{2}
    \left(\frac{v_{F}}{c}\right)^{4},
    \label{eq:cross1}
\end{equation}
where $L^{2}$ is the QCB area and $do=\sin\varphi d\varphi
d\vartheta.$ A standard procedure applied to the Hamiltonian
(\ref{Eff_Int_QCB}), leads to the following form of the scaled
cross section
\begin{equation}
    \label{Fin_Cross}
    \sigma(\omega,{\bf n})=\frac{1}{4k}
    \sum_{P}
    \overline{|\langle P|\textrm{h}|0 \rangle_p|^2}
      \delta(\omega-\omega_P).
\end{equation}
Here $\omega=\Omega_i-\Omega_f$ is the frequency loss and
\begin{eqnarray}
    \label{Fin_Int}
    \textrm{h}=
    -\sum_j
    \frac{k_j}{\sqrt{|k_j|}}
    P_{j;\lambda_f,\lambda_i}(\varphi,\vartheta)
    \left(a_{j,{\bf k}_{j}}+
    a^{\dag}_{j,-{\bf k}_{j}}
    \right)
\end{eqnarray}
is the interaction Hamiltonian reduced to the subspace of QCB
states and summation is performed over all one-plasmon states
$|P\rangle.$ $P_{j;\lambda_f,\lambda_i}(\varphi,\vartheta),$
$j=1,2,$ are polarization matrices.  In the basis $(\|,\bot)$ they
are given by
\begin{eqnarray}
    \label{Pol_Matr1}
    P_1(\varphi,\vartheta)=
    -\left(
    \begin{array}{cc}
           2\sin\vartheta
           &
           i\cos\vartheta \\
           2i\cos\vartheta \cos\varphi
           &
           \sin\vartheta \cos\varphi
    \end{array}
    \right),
    \ \ \ \ \
    P_2(\varphi,\vartheta)=P_1\Big(\frac{\pi}{2}+\vartheta,\varphi\Big).
\end{eqnarray}
Equations (\ref{Fin_Cross}) - (\ref{Pol_Matr1}) serve as a basis
for the subsequent analysis.

%%%%%%%%%%%%%%%%%%%%%%%%%%%%%%%%%%%%%%%%%%%%%%%%%%%%%%%%%%%%%
\section{Scattering Cross Section}
 \label{sec:Scat}
%%%%%%%%%%%%%%%%%%%%%%%%%%%%%%%%%%%%%%%%%%%%%%%%%%%%%%%%%%%%%

%%%%%%%%%%%%%%%%%%%%%%%%%%%%%%%%%%%%%%%%%%%%%%%%%%%%%%%%%%%%%
\subsection{Cross Section: Basic Types}
 \label{subsubsec:Types}
%%%%%%%%%%%%%%%%%%%%%%%%%%%%%%%%%%%%%%%%%%%%%%%%%%%%%%%%%%%%%

According to Eqs.  (\ref{Fin_Cross}), (\ref{Fin_Int}), in order to
contribute to the cross-section (\ref{Fin_Cross}) for a fixed
detector orientation at ${\bf n},$ an excited QCB plasmon
$|P\rangle $ must contain at least one of two single-array states
$|1,-{\bf{k}}_1\rangle$ or $|2,-{\bf{k}}_2\rangle$. Analysis shows
that there are five basic types of excited QCB plasmons depending
on the location of the point ${\bf{k}}.$ Here we describe these
types of plasmons and the corresponding structure of the
differential scattering cross section. In this description beside
the polar angle $\varphi$ of the scattered photon wave vector
${\bf{K}}_f,$ we use in a sense a mixed representation. It is
based on the excited plasmon momentum ${\bf{k}}$ and the azimuthal
angle $\vartheta$ of the transverse component of the scattered
photon wave vector ${\bf{K}}_f$.

%%%%%%%%%%%%%%%%%%%%%%%%%%%%%%%%%%%%%%%%%%%%%%%%%%%%%%%%%%%%%
\begin{figure}[htb]
\begin{center}
\includegraphics[width=65mm,height=50mm,angle=0,]{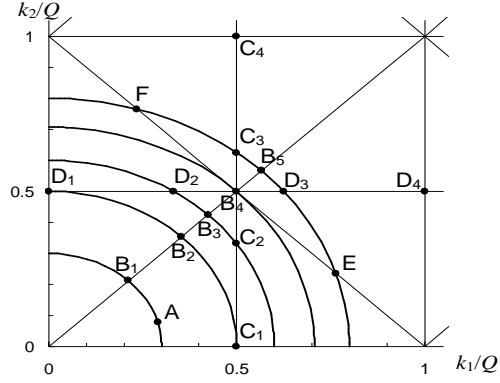}
\caption{Part of inverse space. The small square $k_j/Q\leq0.5$ is
the quarter of the first QCB BZ. High symmetry lines are parallel
to the coordinate axes. Resonant lines are parallel to BZ
diagonals. The arcs coming over the point $B_1,$ $B_2,$ $B_3,$
$B_4,$ $B_5,$ correspond to different wave numbers of excited
plasmons: $|k/Q|=0.3;$ $0.5;$ $0.6;$ $0.7;$ $0.8$.}\label{BZNew}
\end{center}
\end{figure}
%%%%%%%%%%%%%%%%%%%%%%%%%%%%%%%%%%%%%%%%%%%%%%%%%%%%%%%%%%%%%%

{\bf i. The general case: The point ${\bf{k}}$ lies away from both
the high symmetry lines and the resonant lines.} This case is
illustrated by the point $A$ in Fig. \ref{BZNew}. There are two
QCB plasmons $|P\rangle,$ mostly propagating along the first or
second array with the frequencies (\ref{RenFr}), which contribute
to the scattering. The differential cross-section is a sum of two
peaks centered at these frequencies. After averaging over initial
polarizations and summing over final polarizations, it has the
form $\sigma(\omega,{\bf{n}})=F_1({\bf n})\delta(\omega-\omega_1)+
F_{2}({\bf n})\delta(\omega-\omega_{2}),$
%\begin{eqnarray*}
%    \sigma(\omega,{\bf n}) =
%       F_1({\bf n})
%       \delta(\omega-\omega_{1})+
%       F_{2}({\bf n})
%       \delta(\omega-\omega_{2}),
%\end{eqnarray*}
where the functions $F_1({\bf{n}}){\equiv}F(\varphi,\vartheta)$
and $F_2({\bf{n}}){=}F(\varphi,\vartheta+{\pi}/{2})$ describe
universal angle dependence of the peak amplitudes,
\begin{eqnarray*}
  F(\varphi,\vartheta)
  =
  |\sin\varphi\cos\vartheta|
  \Big[
       1-
       \frac{3}{4}\sin^2\varphi\cos^2\vartheta+
       \frac{1}{4}\cos^2\varphi
  \Big].
\end{eqnarray*}
The functions $F_{1,2}$ are related to the corresponding array
plasmons. Each one of them vanishes when the scattered photon is
perpendicular to the corresponding array. However, these functions
describe the scaled cross section. The absolute value amplitude of
each peak has an additional factor $\lambda$ (see Eq.
(\ref{eq:cross1})). Strong electron-electron interaction in a
nanotube corresponds to small values of the Luttinger parameter
$g$ and therefore suppresses the scattering cross-section.

{\bf ii.  Inter-band resonance in one of the arrays: the point
${\bf{k}}$ lies on a high symmetry line of only one array.} This
case is illustrated by the points $C_{2,3}$ ($1$-st array) and
$D_{2,3}$ ($2$-d array) in Fig.  \ref{BZNew}.  Consider for
example point $C_2$ where $k_1=Q/2$, $k_2\neq nQ/2.$ Here
\emph{three} QCB plasmons contribute to the scattering. The first
one is $|2,k_2\rangle $ plasmon propagating along the second array
with quasimomentum $k_2$ and frequency $\omega_2,$ Eq.
(\ref{RenFr}). The two others are even or odd superpositions of
the $1$-st array states (Eq. (\ref{states_1}) with $j=1$) of the
two first zones with eigenfrequencies (\ref{freq_b_1}).  Due to
weakness of the inter-array interaction, three peaks of the
scattering cross section form a singlet $\omega_2$ and doublet
$\omega_{1g},\omega_{1u}$.  After averaging over initial and final
polarizations, the cross section has the form
$\sigma(\omega,{\bf{n}})=2^{-1}\sum_{j}F_j({\bf{n}})
 [\delta(\omega-\omega_{jg})+\delta(\omega-\omega_{ju}))].$
%\begin{eqnarray*}
%    \sigma(\omega,{\bf n}) &=&\frac{1}{2}
%       F_{1}({\bf n})
%       [\delta(\omega-\omega_{1g})+
%       \delta(\omega-\omega_{1u}))]+
%       F_{2}({\bf n})
%       \delta(\omega-\omega_{2}).
%\end{eqnarray*}

{\bf iii.  Inter-band resonance in both arrays: The point
${\bf{k}}$ is a crossing point of two high symmetry lines away
from all resonant lines.} This case is illustrated by the points
$C_4$ and $D_4$ in Fig. \ref{BZNew}.  Consider for example point
$C_4$. Here $k_1=Q/2$, $k_2=2Q/2,$ and \emph{four} QCB plasmons
contribute to the scattering. The first pair consists of even and
odd superpositions of the $1$-st array states of the first and the
second bands. These states and their frequencies are described by
Eqs. (\ref{high_cross}), (\ref{freq_b}) with $j=1.$ The second
pair consists of the same superpositions of the $2$-d array states
from the second and third bands and is described by the same
equations with $j=2.$ As a result, \emph{four} peaks, which form
two doublets (\ref{freq_b}), $j=1,2$, can be observed.  After
averaging over initial and final polarizations the cross section
is $\sigma(\omega,{\bf{n}})=2^{-1}\sum_{j}F_j({\bf{n}})
 [\delta(\omega-\omega_{jg})+\delta(\omega-\omega_{ju}))].$
%\begin{eqnarray*}
%    \sigma(\omega,{\bf n})=
%  \frac{1}{2}F_{1}({\bf n})
%       [\delta(\omega-\omega_{1g})+
%       \delta (\omega-\omega_{1u}))]+
%  \frac{1}{2}F_{2}({\bf n})
%       [\delta(\omega-\omega_{2g})+
%       \delta\big(\omega-\omega_{2u}))].
%\end{eqnarray*}

{\bf iv.  Inter-array resonance: The point ${\bf{k}}$ lies only on
one of the resonant lines away from the high symmetry lines.} This
case is illustrated by the points $B_{1-3,5},\ \ E,$ and $F$ in
Fig. \ref{BZNew}.  Here the QCB plasmons which contribute to the
scattering are two even and two odd superpositions of the first
and second array states (\ref{first_1}) whose eigenfrequencies
form two doublets (\ref{freq_a_1}). As in the previous case, the
scattering cross section contains \emph{four} peaks which form two
doublets.  After averaging over initial and final polarizations
the cross section is
$\sigma(\omega,{\bf{n}})=2^{-1}\sum_{j}F_j({\bf n})
 [\delta(\omega-\omega_g({\bf{k}}_j))+
  \delta(\omega-\omega_{u}({\bf{k}}_j))].$
%\begin{eqnarray*}
%    \sigma(\omega,{\bf n}) &=&
%  \frac{1}{2}
%  \sum_{j}
%  F_j({\bf n})
%       [\delta(\omega-\omega_{g}({\bf{k}}_j))+
%        \delta(\omega-\omega_{u}({\bf{k}}_j))].
%\end{eqnarray*}

Thus, the inter-array splitting is proportional to the main small
parameter (\ref{eq-phi}) of the theory, $\phi\approx 0.007$
\cite{KGKA2}. For the set of parameters described in the beginning
of subsection \ref{subsubsec:Spectr}, the inter-band splitting
defined by Eq. (\ref{freq_b_1}), is five times smaller because it
contains an additional factor $\phi a/r_0.$

{\bf v.  Inter-array and inter-band resonance: The point
${\bf{k}}$ lies at the intersection of two resonant lines.} There
is only one such point $B_4$ in Fig. \ref{BZNew}. In the general
case, where the parameter $n\neq 0$ for both crossing resonant
lines (the point $B_4$ is {\em not} the case), the QCB plasmons
involved in Raman scattering form two quartets. The first quartet
consists of four symmetrized combinations (\ref{eigenfun-quartet})
of the single-array states. QCB eigenstates for the second quartet
are obtained from these equations by replacing $1\leftrightarrow
2.$ The corresponding eigen-frequencies are described by Eqs.
(\ref{evar}), (\ref{odar}). The scattering cross section in this
case contains six peaks, and two of them are two-fold degenerate
\begin{eqnarray*}
  \sigma(\omega,{\bf n})
  &=&
  \frac{1}{4}
  \sum_{j}
  F_j({\bf n})[2\delta(\omega-\omega_{g,u/g}({\bf{k}}_j))+
       \delta(\omega-\omega_{uu}({\bf{k}}_j))+
       \delta(\omega-\omega_{ug}({\bf{k}}_j)].
\end{eqnarray*}
The point $B_4$ in Fig.  \ref{BZNew} lies on the main resonance
line with $n=0.$ Here $|k_1|=|k_2|,$ the frequencies of both
quartets coincide, and the scattering cross section contains one
four-fold degenerate peak and a symmetric pair of its two-fold
degenerate satellites.

This classification of all types of excited plasmons enables us to
describe completely the UV scattering on QCB.

%%%%%%%%%%%%%%%%%%%%%%%%%%%%%%%%%%%%%%%%%%%%%%%%%%%%%%%%%%%%%
\subsection{Scattering Indicatrices}
 \label{subsubsec:Rosettes}
%%%%%%%%%%%%%%%%%%%%%%%%%%%%%%%%%%%%%%%%%%%%%%%%%%%%%%%%%%%%%

In this part we describe the results of the scattering process
with the help of a family of scattering indicatrices.

To explain the indicatrix structure, we start with some
preliminary arguments. Consider the case where the detector is
tuned on the frequency $\Omega_f$ and is oriented in direction
${\bf{n}}.$ These parameters determine a point ${\bf{k}}$ by a
unique way. There are two ways of scanning QCB plasmons. The first
way is to scan the polar angle $\vartheta$. The corresponding
points ${\bf{k}}$ in Fig. \ref{BZNew} form a circle with radius
$k=(\Omega_f/c)\sin\varphi$. The second way is to tune the
detector frequency as $\Omega_f$. We are interested in frequency
loss of order of the plasmon frequencies within the two - three
lowest bands. This loss is of order of $\Omega_f(v/c)\ll\Omega_f.$
Therefore in this case the point ${\bf{k}}$ remains on its place
with a very good accuracy.

Each scattering indicatrix corresponds to a circular arc in Fig.
\ref{BZNew} and the structure of the indicatrix is completely
determined by the arc radius $k.$ The indicatrix represents a set
of curves displayed in polar coordinates with polar angle, which
coincides with the polar angle $\vartheta$ used above, and the
(dimensionless) radius $\omega/(vQ),$ where $\omega$ is the
frequency loss. Each point of the indicatrix corresponds to an
excitation of a QCB plasmon and therefore to a sharp peak in the
scattering cross section. The number of peaks depends on the polar
angle. Scanning the polar angle $\vartheta$ results in changing
the number of peaks. This is one more example of dimensional
crossover in QCB (see Chapter \ref{sec:Infrared} for similar
effects in IR spectroscopy).

\begin{figure}[htb]
\begin{center}
\includegraphics[width=120mm,height=45mm,angle=0,]{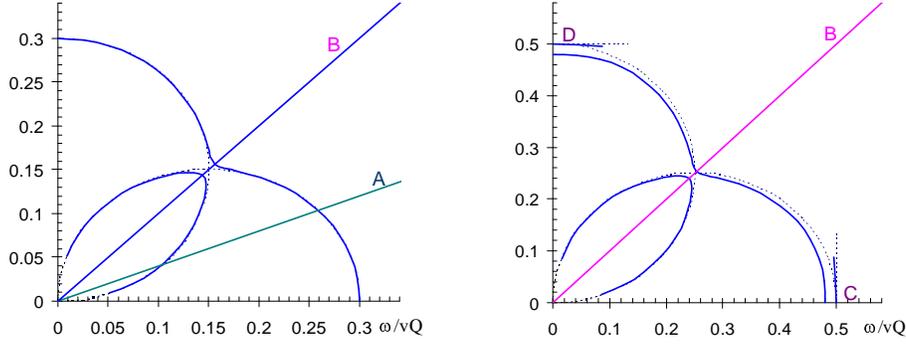}
\caption{{\bf{Left panel}}: Positions of the scattering peaks for
$|k/Q|=0.3.$ The doublet in the resonance direction (point $B_1$
in Fig. \ref{BZNew}) is well pronounced. {\bf{Right panel}}:
Positions of the scattering peaks for $|k/Q|=0.5.$ Two doublets
appear at the BZ boundaries (points $C_1$ and $D_1$ in Fig.
\ref{BZNew}).}
 \label{Kll_03}
\end{center}
\end{figure}

We start with the case of the smallest radius $K\sin\varphi=-0.3Q$
(arc $AB_1$ in Fig.  \ref{BZNew}). Here all points beside the
point $B_1$ are points of general type ({\bf{i}}). Each one of
them, e.g., point $A$, corresponds to excitation of two plasmons
in the two arrays and therefore leads to two separate peaks in the
scattering spectrum (see Fig. \ref{Kll_03}, left panel). The peaks
corresponding to the point $A$ in Fig. \ref{BZNew} lie on the ray
defined by the angle $\theta_{A}$. The point $B_1$ corresponds to
an inter-array resonance ({\bf iv}) and in this direction a split
doublet can be observed.

%%%%%%%%%%%%%%%%%%%%%%%%%%%%%%%%%%%%%%%%%%%%%%%%%%%%%%%%%%%%%%
%%%%%%%%%%%%%%%%%%%%%%%%%%%%%%%%%%%%%%%%%%%%%%%%%%%%%%%%%%%%%
In the next case $K\sin\varphi=-0.5Q$ (arc $C_1B_2D_1$ in Fig.
\ref{BZNew}), as in the previous one, in the general directions
one can observe two single lines which form a split doublet in the
resonant direction $B_2$.  However at the final points $C_1$ and
$D_1$ the arc touches the high symmetry lines.  Here the low
frequency single line vanishes (there is no scattering at $k=0$)
while the high frequency line transforms into a doublet because of
an inter-band resonance ({\bf ii}) in one of the arrays (Fig.
\ref{Kll_03}, right panel).

\begin{figure}[htb]
\begin{center}
\includegraphics[width=120mm,height=45mm,angle=0,]{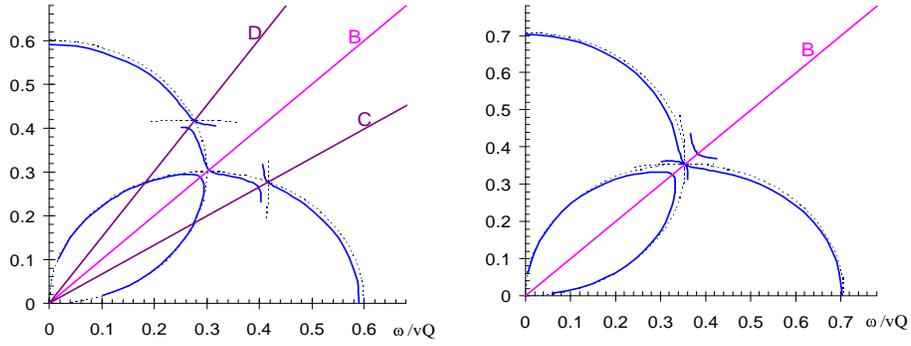}
\caption{{\bf{Left panel}}: Positions of the scattering peaks for
$|k/Q|=0.6.$ Two doublets at the BZ boundaries (points $C_2$ and
$D_2$) in Fig. \ref{BZNew}) are shifted from the high symmetry
directions. {\bf{Right panel}}: Positions of the scattering peaks
for $|k/Q|=\sqrt{2}/2.$ Resonance triplet corresponding to point
$B_4$ in Fig. \ref{BZNew} (two of four frequencies remain
degenerate in our approximation). In IR experiments only one of
the triplet components is seen.}
 \label{Kll_06}
\end{center}
\end{figure}

Further increase of the arc radius $K\sin\varphi=-0.6Q$ (arc
$C_2B_3D_2$ in Fig.  \ref{BZNew}) leads to appearance of two
points $D_2$ and $C_2$ where the arc intersects with high symmetry
lines (BZ boundaries).  Each of these points generates an
inter-band resonance doublet, which coexists with the low
frequency single peak (Fig.\ref{Kll_06}, left panel).

%%%%%%%%%%%%%%%%%%%%%%%%%%%%%%%%%%%%%%%%%%%%%%%%%%%%%%%%%%%%%
%%%%%%%%%%%%%%%%%%%%%%%%%%%%%%%%%%%%%%%%%%%%%%%%%%%%%%%%%%%%%%

In the case $K\sin\varphi=-\sqrt{2}Q/2$ the corresponding arc
includes the BZ corner $B_4.$ This is the point of a double
inter-array and inter-band resonance ({\bf v}).  Moreover, here,
the two quartets described above coincide. Therefore, there are
three lines in Fig. \ref{Kll_06}, right panel. The low-frequency
line as its high-frequency partner is two-fold degenerate while
the central line is four-fold degenerate. We emphasize that each
quartet manifests itself in three lines, contrary to the IR
absorption, where selection rules make two of them invisible.

\begin{figure}[htb]
\begin{center}
\includegraphics[width=54mm,height=45mm,angle=0,]{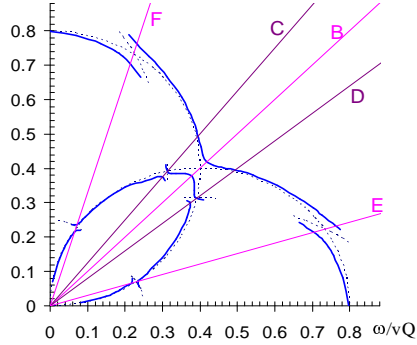}
\caption{Positions of the scattering peaks for $|k/Q|=0.8.$ Two
pairs of doublets appear corresponding to excitation of two pairs
of plasmons in the two arrays (points $E$ and $F$ in Fig.
\ref{BZNew}).}
 \label{Kll_08}
\end{center}
\end{figure}

The last case $K\sin\varphi=-0.8Q$ demonstrates one more
possibility related to inter-band resonance simultaneously in two
arrays ({\bf iii}).  Each point $E$ and $F$ generates (in the
corresponding direction) two doublets describing the inter-band
splitting in different arrays (see Fig.  \ref{Kll_08}).

%%%%%%%%%%%%%%%%%%%%%%%%%%%%%%%%%%%%%%%%%%%%%%%%%%%%%%%%%%%%%
%%%%%%%%%%%%%%%%%%%%%%%%%%%%%%%%%%%%%%%%%%%%%%%%%%%%%%%%%%%%%%

%%%%%%%%%%%%%%%%%%%%%%%%%%%%%%%%%%%%%%%%%%%%%%%%%%%%%%%%%%%%%%
\section{Conclusions}
 \label{subsec:UV-Conc}
%%%%%%%%%%%%%%%%%%%%%%%%%%%%%%%%%%%%%%%%%%%%%%%%%%%%%%

In conclusion, we studied inelastic UV Raman scattering on QCB. We
derived an effective Hamiltonian for QCB-light interaction which
is expressed via the same Bose variables that the QCB itself. With
the help of this Hamiltonian we calculated the differential
scattering cross section as a function of detector orientation and
scattered frequency. Scanning these parameters, one can observe a
set of sharp peaks in the scattering spectrum. The number of peaks
and their positions strongly depend on the direction of the
scattered wave vector.  This results in a dimensional crossover.
It manifests itself in the splitting of the peak frequencies and
therefore in appearance of multiplets (mostly doublets) instead of
single lines in the scattering spectrum.

The sizes of peak splitting are determined by the nature of
interaction which lifts the corresponding degeneracy. In the case
of initial inter-array degeneracy, the splitting is proportional
to the dimensionless interaction strength $\phi\approx 0.007$. The
inter-band splitting is proportional to the square of this
parameter. For chosen QCB parameters it is smaller than the
inter-array interaction strength in spite of an additional large
multiplier $a/r_0$. In all cases, the splitting increases with
increasing the interaction in QCB crosses. The peak amplitudes are
proportional to the Luttinger parameter $g$ in a single nanotube.
Therefore strong electron-electron interaction suppresses the
peaks.

The effectiveness of UV scattering is related to the possibility
of changing continuously the excited plasmon frequency. Due to
other selection rules, some lines which are invisible in the IR
absorption spectrum, become observable in the UV scattering. UV
scattering spectroscopy enables one to restore parameters
describing both interaction in QCB crosses and electron-electron
interaction in a single QCB constituents.

{\em Our studies of optical properties of QCB (this Chapter and
Chapters \ref{sec:Infrared} and \ref{sec:Damp}) show that these
nanoobjects possess a unique combination of optical spectra.
Firstly, they are active in IR and UV frequency range.  Secondly,
they may be observed in various kinds of optical processes, namely
direct and indirect absorption, diffraction, energy loss
transmission, and Raman-like spectroscopy.}

%\newpage

%%%%%%%%%%%%%%%%%%%%%%%%%%%%%%%%%%%%%%%%%%%%%%%%%%%%%%%%%%%%%%%
\appendix
%%%%%%%%%%%%%%%%%%%%%%%%%%%%%%%%%%%%%%%%%%%%%%%%%%%%%%%%%%%%%%%

%%%%%%%%%%%%%%%%%%%%%%%%%%%%%%%%%%%%%%%%%%%%%%%%%%%%%%%%%%%%%%%
\chapter{Empty Super-Chain}
 \label{append:Empty}
%%%%%%%%%%%%%%%%%%%%%%%%%%%%%%%%%%%%%%%%%%%%%%%%%%%%%%%%%%%%%%%

Here we construct eigenfunctions, spectrum, and quasi-particle
operators for an ``empty super-chain'' - quantum wire in an
infinitely weak periodic potential with period $a$. Excitations in
an initial wire are described as plane waves $L^{-1/2}\exp (ikx)$
with wave number $k=2\pi n/L,$ with integer $n,$ and dispersion
law $\omega(k)=v|k|$ (the array number is temporarily omitted).
The following orthogonality relations are valid
\begin{eqnarray*}
    &&\int_{-L/2}^{L/2}\psi_{k}^{*}(x)\psi_{k'}(x)dx=\delta_{k,k'},
    \ \ \ \ \ \ \
    \sum_{k}\psi_{k}^{*}(x)\psi_{k}(x')=\delta_{L}(x-x'),
\end{eqnarray*}
where $\delta_{L}$ stands for  periodic delta-function
$$\delta_{L}(x-x') \equiv \sum_{n}\delta (x-x'-nL).$$

An ``empty super-chain'' is characterized by a space period $a$
and a corresponding reciprocal lattice wave number $Q=2\pi/a.$
Each excitation in such a super-chain is described by its
quasi-wavenumber $q$ and a band number $s$ ($s=1,2,\ldots$) that
are related to the corresponding wave number $k$ by the following
relation,
$$
    k=q+Q(-1)^{s-1}\left[\frac{s}{2}\right]
    {\mbox{ sign }}q.
$$
Here, square brackets denote an integral part of a number. The
corresponding wave function $\psi_{s,q}(x)$ has the Bloch-type
structure,
\begin{equation}
    \psi_{s,q}(x)=\frac{1}{\sqrt{L}}
    e^{iqx}
    u_{s,q}(x),
    \label{WaveFunc}
\end{equation}
and satisfies the orthogonality relations
\begin{eqnarray*}
  &&\int_{-L/2}^{L/2}\psi_{s,q}^{*}(x)\psi_{s',q'}(x)dx=\delta_{s,s'}
 \sum_{m}\delta_{q+mQ,q'},
    %\delta_{Q;q,q'},
  \ \ \
  \sum_{s,q}\psi_{s,q}^{*}(x)\psi_{s,q}(x')=\delta_{L}(x-x'),
\end{eqnarray*}
%where
%$$\delta_{Q;q,q'}=\sum_{n}\delta_{q+nQ,q'}.$$
Within the first BZ, $-Q/2\leq q<Q/2,$ Bloch amplitude and
dispersion law $\omega_{s}$ have the following form
\begin{eqnarray}
    u_{s,q}(x)=
    \exp
    \left\{
    iQx(-1)^{s-1}\left[\frac{s}{2}\right]
    {\mbox{ sign }}q
    \right \},
    \ \ \ \ \
    \omega_{s}(q) =  vQ\left(
    \left[\frac{s}{2}\right]
    +(-1)^{s-1}\frac{|q|}{Q}
    \right).
    \label{WaveFunc1}
\end{eqnarray}
Taking into account that both Bloch amplitude $u_{s,q}(x)$ and
dispersion law $\omega_{s}(q)$ are periodic functions of $q$ with
period $Q,$ one obtains general equations for the Bloch amplitude
\begin{eqnarray*}
        u_{s,q}(x) & = &
    \displaystyle{
    \sum_{n=-\infty}^{\infty}
    \frac{\sin\xi_{n}}{\xi_{n}}}
    \cos[(2s-1)\xi_{n}]
    \exp\left(-4i\xi_{n}\frac{q}{Q}\right),
    \ \ \ \ \ \
    4\xi_{n}=Q(x-na),
\end{eqnarray*}
and dispersion law $\omega_{s}(q)$
\begin{eqnarray*}
    \frac{\omega_{s}(q)}{{v}{Q}} & = &
    \frac{2s-1}{4}+
    \displaystyle{\sum_{n=1}^{\infty}}
    \frac{2 (-1)^{s}}{\pi^{2}(2n+1)^{2}}
        \cos \frac{2\pi (2n+1)q}{Q}.
\end{eqnarray*}
The relations between quasiparticle operators for a free wire,
$c_{k},$ for momentum $k\neq nQ/2$ with $n$ integer, and those for
an empty super-chain, $C_{s,q},$ for quasimomentum $q$ from the
first BZ, $-Q/2<q<Q/2,$ look as
\begin{eqnarray*}
    c_{k}=C_{s,q}{\mbox{sign}}k,
    \ \ \ \ \ \
    s=1+\left[\frac{2|k|}{Q}\right],
    \ \ \ \ \
    q=Q\left(\left\{\frac{k}{Q}
    +\frac{1}{2}\right\}-\frac{1}{2}\right)
\end{eqnarray*}
\begin{eqnarray*}
    C_{s,q}=(-1)^{\nu}c_{k},
    \ \ \ \ \
    k=q+(-1)^{\nu}Q
    \left[\frac{s}{2}\right],
    \ \ \ \ \
    \nu=s+1+\left[\frac{2q}{Q}\right],
\end{eqnarray*}
where curely brackets denote a fractional part of a number. For
obtaining these relations we used the following expression
\begin{eqnarray*}
    \int_{-L/2}^{L/2}\psi_{k}^{*}(x)\psi_{s,q}(x)dx=
    \delta_{s,s(q)}{\rm{sign}}k\sum_{m}\delta_{q+mQ,k},
    \ \ \ \ \ \ \ \ \ \
    s(q)=1+\left[\frac{2|q|}{Q}\right],
\end{eqnarray*}
for the transition amplitude $\langle k|s,q\rangle $. In case when
$k=nQ/2$ with $n$ integer, hybridization of the neighboring bands
should be taken into account. This modifies the above relations by
the following way
\begin{eqnarray*}
    &&c_{nQ/2}=\theta(n)\left[\alpha_{n}C_{n,q_{n}}+
    \beta_{n}C_{n+1,q_{n}}\right]
    +\theta(-n)\left[\beta^{*}_{-n}C_{-n,q_{n}}-
    \alpha^{*}_{-n}C_{-n+1,q_{n}}\right],
    \\
    &&q_{n}=Q\left(\left\{\frac{n+1}{2}\right\}-
    \frac{1}{2}\right),
%\end{eqnarray*}
%\begin{eqnarray*}
    \ \
    C_{s,q_{s}} =\alpha^{*}_{s}c_{sQ/2}+
    \beta_{s}c_{-sQ/2},
    \ \
    C_{s+1,q_{s}} =\beta^{*}_{s}c_{sQ/2}-
    \alpha_{s}c_{-sQ/2},
\end{eqnarray*}
where $\alpha$, $\beta$ are hybridization coefficients.
Corresponding relations between wave functions follow immediately
from these formulas.

Alternatively, each mode with the quasi-wavenumber $q$ in the
energy band $s$ (reduced BZ description) can be described by the
wave number $k=q+mQ$ (extended BZ description), where
$m=(-1)^{s}{\rm{sign}}(q)[s/2]$. Within this scheme, Bloch
amplitude and dispersion law $\omega(k)$ have the following simple
form $u_{k}(x)=e^{ikx}$, $\omega(k)=v|k|$.

To write down any of these formulas for a specific array, one
should add the array index $j$ to the wave function $\psi$, Bloch
amplitude $u$, coordinate $x$, quasimomentum $q$, and to the
periods $a$ and $Q$ of the super-chain in real and reciprocal
space.

%\newpage

%%%%%%%%%%%%%%%%%%%%%%%%%%%%%%%%%%%%%%%%%%%%%%%%
\chapter{Spectrum and Correlators of Square QCB}
 \label{append:DoublSpectr&Corr}
%%%%%%%%%%%%%%%%%%%%%%%%%%%%%%%%%%%%%%%%%%%%%%%%

%%%%%%%%%%%%%%%%%%%%%%%%%%%%%%%%%%%%%%%%%%%%%%%%
\section{Square QCB Spectrum}
 \label{append:DoublSpectr}
%%%%%%%%%%%%%%%%%%%%%%%%%%%%%%%%%%%%%%%%%%%%%%%%

Here we obtain analytical expressions for dispersion laws and wave
functions of QCB. For quasimomenta far from the BZ boundaries, the
energy spectrum of the first band can be calculated explicitly.
Assuming that $\omega^2\ll\omega_s^2(q_j)$, $s=2,3,4,\ldots$, we
omit $\omega^2$ in all terms in the r.h.s.  of Eq. (\ref{F_j})
except the first one, $s=1$. As a result the secular equation
(\ref{F_j}) reads
$$
  \prod\limits_{j=1}^{2}
  \left(
       \frac{R_0}{a}
       \frac{\zeta_1^2(q_j)\omega^2}
            {\omega^2(q_j)-\omega^2}+
       F_0
  \right)
  =\frac{1}{\varepsilon},
$$
where $\omega^2(q)=\omega_1^2(q)$. The solutions of this equation
have the form:
\begin{eqnarray}
  {\omega}_{{\nu}1{\bf{q}}}^{2}=
  \frac{\tilde\omega_1^{2}({\bf q})+
        \tilde\omega_2^{2}({\bf q})}
       {2}\pm
       \left[
            \left(
            \frac{\tilde\omega_1^{2}({\bf q})-
                  \tilde\omega_2^{2}({\bf q})}
                 {2}
            \right)^{2}+
            \phi_{\bf{q}}
       \right]^{1/2}.
  &&
  \label{omega-12}
\end{eqnarray}
Here $\phi_{\bf{q}}=\phi^2\zeta^2_1(q_1)\zeta^2_1(q_2)
\omega^2(q_1)\omega^2(q_2)$, $\nu=+,-$ is the branch number,
$\widetilde\omega_{j}({\bf q})$ is determined as
\begin{equation}
    \widetilde\omega_{1}^2({\bf q})=
              \omega^{2}(q_{1})
              \frac{1-\varepsilon{F}_0
                                 (F_0-{\varphi}^{2}(q_2))}
                   {1-\varepsilon(F_0-{\varphi}^{2}(q_1))
                                 (F_0-{\varphi}^{2}(q_2))},
    \ \ \ \ \
    \varphi^2(q_j)=\frac{R_0}{a}\zeta_1^2(q_j)
    \label{RenFr}
\end{equation}
for $j=1.$ Expression for $\widetilde\omega_{2}^2({\bf q})$ can be
obtained by permutation $1\leftrightarrow 2$. The expression
inside the parentheses on the r.h.s. of Eq. (\ref{RenFr})
describes the contributions to $F_0$ from higher bands. Therefore
$\widetilde\omega_j^2({\bf q})$ is the $j$-th array frequency
renormalized by the interaction with higher bands. In principle,
contribution of higher bands may turn the interaction to be
strong. However, for the specific case of carbon nanotubes, one
stays far from the critical value $\varepsilon_c$ (see estimates
at the end of subsection \ref{subsubsec:Approx}). Therefore the
interaction with higher bands is weak in almost all the BZ except
its boundaries.

The resonance line equation is $\omega(q_1)= \omega(q_2)$. Out of
this line the branch number is in fact the array number and the
renormalized frequencies are frequencies of a boson propagating
along one of the arrays slightly modified by interactions with the
complementary array. In case when $\omega(q_{1})>\omega(q_{2}),$
one obtains
\begin{equation}
  {\omega}_{+,1{\bf{q}}}^{2}\approx
  \omega^2(q_1)
  \left(1-\varepsilon F_0 \varphi^2(q_1)\right),
  \ \ \ \ \
  {\omega}_{-,1{\bf{q}}}^{2}\approx
  \omega^2(q_2)
  \left(1-\varepsilon F_0 \varphi^2(q_2)\right).
  \label{omega-SecOrd}
\end{equation}
In the opposite case one should replace indices
$+\leftrightarrow-$.

Consider the frequency correction in the latter equation in more
details.  The correction term can be approximately estimated as
$\omega^2(q_j)S(q_j)$ with
\begin{equation}
   S(q) =
   \varepsilon{F}_0\varphi^2(q) =
   \varepsilon
   \frac{R_0}{a}
   \zeta_{1}^{2}(q)
   \int d\xi\zeta^2(\xi).
    \label{Sq}
\end{equation}
Due to the short-range character of the interaction, the matrix
elements $\zeta_1(q)\sim1$ vary slowly with the quasimomentum
$q\le{Q}$. Therefore, the r.h.s. in Eq.(\ref{Sq}) can be roughly
estimated as $S(q)\sim\varepsilon{R_0}/{a}=0.1{R_0}/a\ll 1$.
%\begin{equation}
%    S(q_{1})\sim \varepsilon\frac{R_{0}}{a}=0.1
%    \frac{R_{0}}{a} \ll 1.
%    \label{est1}
%\end{equation}
One should also remember that the energy spectrum of nanotube
remains one-dimensional only for frequencies smaller than some
$\omega_{m}.$ Therefore, an external cutoff arises at
$s\sim{a}k_m$ where $k_m\sim\omega_m/v$. As a results one gets an
estimate $S(q)\sim \varepsilon k_{m}R_0^2/a$.
%\begin{equation}
%    S(q_{1})\sim \varepsilon \frac{R_{0}}{a}k_{m}R_{0}.
%    \label{est2}
%\end{equation}
Hence, one could hope to gain additional power of the small
interaction radius. However, for nanotubes, $k_{m}$ is of the
order of $1/R_{0}$ (see Refs. \cite{Ando,Egger}) and both
estimates coincide. For quasimomenta close to the BZ center, the
coefficient $S(q)$ can be calculated exactly.  For exponential
form of $\zeta(\xi)\propto\exp(-|\xi|)$, one obtains,
$S(0)=0.14R_0/a$. Thus, the correction term in
Eq.(\ref{omega-SecOrd}) is really small.

The eigenstates of the system are described by renormalized field
operators. Within the first band they have the form
\begin{eqnarray}
 \tilde\theta_{11{\bf{q}}} & = &
 \alpha_{1{\bf{q}}}
 \left(
      u_{\bf{q}}\theta_{11{\bf{q}}}-
      v_{\bf{q}}\theta_{21{\bf{q}}}
 \right)-
 \sum\limits_{s=2}^{\infty}
 \left(
      \phi_{1s{\bf{q}}}u_{\bf{q}}\theta_{2s{\bf{q}}}+
      \phi_{2s{\bf{q}}}v_{\bf{q}}\theta_{1s{\bf{q}}}
 \right),
 \label{tilde-teta1}
 \\
 \tilde\theta_{21{\bf{q}}} & = &
 \alpha_{2{\bf{q}}}
 \left(
      v_{\bf{q}}\theta_{11{\bf{q}}}+
      u_{\bf{q}}\theta_{21{\bf{q}}}
 \right)-
 \sum\limits_{s=2}^{\infty}
 \left(
      \phi_{1s{\bf{q}}}v_{\bf{q}}\theta_{2s{\bf{q}}}+
      \phi_{2s{\bf{q}}}u_{\bf{q}}\theta_{1s{\bf{q}}}
 \right).
 \label{tilde-teta2}
\end{eqnarray}
Here the coefficients $u_{\bf{q}}$ and $v_{\bf{q}}$ describe
mixing between the modes with different array indices, within the
first band,
\begin{eqnarray}
  u_{\bf{q}}=
  \left(
       \frac{
             \sqrt{\Delta_{\bf{q}}^{2}+\phi_{1{\bf{q}}}^{2}}+
             \Delta_{\bf{q}}
            }
           {2\sqrt{\Delta_{\bf{q}}^{2}+\phi_{1{\bf{q}}}^{2}}}
  \right)^{1/2},
  \ \ \ \ \
  v_{\bf{q}}=
  \left(
      \frac{
            \sqrt{\Delta_{\bf{q}}^{2}+\phi_{1{\bf{q}}}^{2}}-
            \Delta_{\bf{q}}
           }
           {2\sqrt{\Delta_{\bf{q}}^{2}+\phi_{1{\bf{q}}}^{2}}}
  \right)^{1/2},
 \label{uv}
\end{eqnarray}
and
$$\Delta_{\bf{q}}=(\omega^2(q_2)-\omega^2(q_1))/2,
  \ \ \ \ \
  \phi_{1{\bf{q}}}=\sqrt{\varepsilon}\zeta_1(q_1)\zeta_1(q_2)
  \omega_1(q_1)\omega_1(q_2),$$
$\varepsilon=\phi{a/R_0}$.
%\begin{eqnarray}
%  \Delta_{\bf{q}}=
%  \frac{\omega_1^2(q_2)-\omega_1^2(q_1)}{2},
%  \ \ \ \ \ \
%  \phi_{1{\bf{q}}}=
%  \sqrt{\varepsilon}
%  \zeta_1(q_1)\zeta_1(q_2)\omega_1(q_1)\omega_1(q_2).
%  \label{phi1}
%\end{eqnarray}
The parameters $\phi_{js{\bf{q}}}$, $s=2,3,\ldots,$ in Eqs.
(\ref{tilde-teta1}), (\ref{tilde-teta2}) correspond to inter-band
mixing,
$$\phi_{1s{\bf{q}}}=\phi\zeta_1(q_1)\zeta_s(q_2)\omega_1(q_1)/\omega_s(q_2),$$
%$$
% \phi_{1s{\bf{q}}}=\sqrt{\varepsilon}
% \frac{R_0}{a}\zeta_1(q_1)\zeta_s(q_2)
% \frac{\omega_1(q_1)}{\omega_s(q_2)},
%$$
and the coefficients $\alpha_{1{\bf q}}$ take into account
corrections from the higher bands
$$
 \alpha^2_{1{\bf{q}}}=
 1-
 \sum\limits_{s=2}^{\infty}
 \left(
      \phi_{1s{\bf{q}}}^{2}u_{\bf{q}}^{2}+
      \phi_{2s{\bf{q}}}^{2}v_{\bf{q}}^{2}
 \right).
$$
Expressions for $\phi_{2s{\bf{q}}}$ and $\alpha_{2{\bf{q}}}$ can
be obtained by permutation ${1}\leftrightarrow{2}$.

Equations (\ref{omega-12}), (\ref{tilde-teta1}) and
(\ref{tilde-teta2}) solve the problem of QCB energy spectrum
within the first band and away from the BZ boundaries.  However,
due to smallness of the interaction, the general expressions for
the eigenstates of QCB for arbitrary energy band can be obtained.
For quasimomenta far from the high symmetry lines $k_j=\pm{Q}/2$,
the energy spectrum can be calculated explicitly. Let us consider
the $s$-th energy band and assume that
$$
 |\omega^2-\omega_{s'}^2(q_j)|\ll\phi^2\omega_{s'}^2(q_j),
 \ \ \ \ \ s'\ne s.
$$
Moreover, we consider the wave vector lying away from the resonant
lines $k_1=\pm k_2+mQ$ with $m$ integer. Then, in zeroth
approximation, the QCB eigenstates coincide with ``empty QCB''
eigenstates $|1,s_1,q_1\rangle$ and $|2,s_2,q_2\rangle$, i.e.,
belong to a given array and energy band. (Here we represent the
wave vector ${\bf{k}}$ as ${\bf{k}}={\bf{q}}+{\bf{m}}$, where the
quasi-momentum ${\bf{q}}$ belongs to the first BZ, whereas
${\bf{m}}={\bf{m}}_1+{\bf{m}}_2$ is reciprocal super-lattice
vector, ${\bf{m}}_j=m_jQ{\bf{e}}_j$, $j=1,2$, $m_j$ is integer.
The band numbers $s_j$ are $s_j=1+[2|k_j|/Q]$.) The inter-array
interaction mixes the modes propagating along both arrays. The
eigenstates of QCB are described by renormalized field operators
\begin{eqnarray}
 \widetilde\theta_{1,s,{\bf{q}}}
 &=&
 \alpha_{1,s,{\bf{q}}}
 \theta_{1,s,{\bf{q}}}+
 \phi
 \sum\limits_{s'=1}^{\infty}
   \frac{\zeta_{s'}(q_2)\omega_{s'}(q_2)
         \zeta_{s}(q_1)\omega_{s}(q_1)}
        {\omega_{s'}^2(q_2)-\omega_{s}^2(q_1)}
   \theta_{2,s',{\bf{q}}},
% \nonumber\\
% \tilde\theta_{2,s,{\bf{q}}}
% &=&
% \alpha_{2,s,{\bf{q}}}
% \theta_{2,s,{\bf{q}}}+
% \phi
% \sum\limits_{s'=1}^{\infty}
%   \frac{\zeta_{1s'}(q_1)\omega_{1s'}(q_1)
%         \zeta_{2s}(q_2)\omega_{2s}(q_2)}
%        {\omega_{1s'}^2(q_1)-\omega_{2s}^2(q_2)}
%   \theta_{1,s',{\bf{q}}},
 \label{Re-plasmon}
\end{eqnarray}
where
\begin{eqnarray*}
 \alpha_{1,s,{\bf{q}}}^2
 &=& 1-
 \phi^2
 \sum\limits_{s'=1}^{\infty}
   \left(\frac{\zeta_{s'}(q_2)\omega_{s'}(q_2)
               \zeta_{s}(q_1)\omega_{s}(q_1)}
              {\omega_{s'}^2(q_2)-\omega_{s}^2(q_1)}
   \right)^2.
% \\
% \alpha_{2,s,{\bf{q}}}^2
% &=& 1-
% \phi^2
% \sum\limits_{s'=1}^{\infty}
%   \left(\frac{\zeta_{1s'}(q_1)\omega_{1s'}(q_1)
%               \zeta_{2s}(q_2)\omega_{2s}(q_2)}
%              {\omega_{1s'}^2(q_1)-\omega_{2s}^2(q_2)}
%   \right)^2.
\end{eqnarray*}
The corresponding eigenfrequency is a slightly modified frequency
of bosons propagating along the first arrays,
\begin{eqnarray}
    \widetilde\omega_{1,s,{\bf q}}^2
    &=&
    \omega_{s}^{2}(q_{1})-
    \phi^2
    \sum_{s'=1}^{\infty}
    \frac{\zeta_{s'}^2(q_2)\omega_{s'}^2(q_2)
          \zeta_{s }^2(q_1)\omega_{s}^2(q_1)}
         {\omega_{s'}^2(q_2)-\omega_{s}^2(q_1)},
%    \nonumber\\
%    \tilde\omega_{2,s,{\bf q}}^2
%    &=&
%    \omega_{2s}^{2}(q_2)-
%    \phi^2
%    \sum_{s'=1}^{\infty}
%    \frac{\zeta_{1s'}^2(q_1)\omega_{1s'}^2(q_1)
%          \zeta_{2s}^2(q_2)\omega_{2s}^2(q_2)}
%         {\omega_{1s'}^2(q_1)-\omega_{2s}^2(q_2)}.
    \label{RenFr-s}
\end{eqnarray}
Expressions for $\widetilde\theta_{2,s,{\bf{q}}}$,
$\alpha_{2,s,{\bf{q}}}^2$, and $\widetilde\omega_{2,s,{\bf q}}^2$
can be obtained by permutation $1\leftrightarrow2$.

As was mentioned in subsection \ref{subsubsec:Spectr}, description
of QCB plasmons in terms of array plasmons fails at the points
lying at two specific groups of lines in reciprocal space.
Nevertheless, QCB plasmons even at these lines can be considered
as  finite combinations of array plasmons generated by a more
complicated way. Below we consider various types of the
interference of array plasmons in QCB.

The first group of specific lines is formed by the high symmetry
lines $k_j=p_jQ/2$ with $p_j$ an integer (the corresponding
quasi-momentum has a component $q_j=0$ for $p_j$ even or $Q/2$ for
$p_j$ odd, the reciprocal lattice vector is
${\bf{m}}_j=[|p_j|/2]$). In an ``empty lattice'' approximation,
these lines are degeneracy lines which separate $s_j$-th and
$(s_j+1)$-th bands of the $j$-th array ($s_j=|p_j|$). Inter-array
interaction mixes the array states, lifts the (inter-band)
degeneracy and splits the corresponding frequencies. As a result,
the QCB plasmons related to a point ${\bf{k}}$ lying at a high
symmetry line (e.g. all $C,D$ points in Fig. \ref{BZNew}), are
built from the array plasmons associated not only with the points
${\bf{k}}_{1,2}$ but also with the symmetric (with respect to
coordinate axes of a reciprocal space) points
${\bf{k}}_{\overline{1}}=(-k_1,q_2)$ and
${\bf{k}}_{\overline{2}}=(q_1,-k_2)$. (Indeed, if the condition
$k_j=p_jQ/2$ is fulfilled, $-k_j=k_j-p_jQ$, and then both these
wave-numbers correspond to the same quasi-wave-number $q_j$ in the
first BZ.)

Consider for definiteness the case $j=1$ and assume first that the
ratio $2k_2/Q$ is non-integer. In this case, the point
${\bf{k}}={\bf{q}}+{\bf{m}}$ generates three QCB plasmons. The
first of them is plasmon $\widetilde\theta_{2,s_2,{\bf{q}}}$
(\ref{Re-plasmon}) propagating along the second array with
quasimomentum $q_2$ and frequency $\omega_{2,s_2,{\bf{q}}}$
(\ref{RenFr-s}). The two others
\begin{eqnarray}
   \widetilde\theta_{1,s_1,g,{\bf{q}}}
   &=&
   \frac{1}{\sqrt{2}}
   \left(
        \theta_{1, s_1, {\bf{q}}}+
        \theta_{1,s_1+1,{\bf{q}}}
   \right),
   \label{states_1}\\
   \widetilde\theta_{1,s_1,u,{\bf{q}}}
   &=&
   \frac{1}{\sqrt{2}}
   \left(
        \theta_{1, s_1, {\bf{q}}}-
        \theta_{1,s_1+1,{\bf{q}}}
   \right)+
   \sqrt{2}\phi
   \sum\limits_{s'=1}^{\infty}
     \frac{\zeta_{s'}(q_2)\omega_{s'}(q_2)
           \zeta_{s_1}(q_1)\omega_{s_1}(q_1)}
          {\omega_{s'}^2(q_2)-\omega_{s_1}^2(q_1)}
     \theta_{2,s',{\bf{q}}}
   \nonumber
\end{eqnarray}
are even or odd superpositions of the $1$-st array states from the
two zones $s_1$ and $s_1+1$ with eigenfrequencies
\begin{eqnarray}
    \tilde\omega_{1,s_1,g,{\bf{q}}}^2=\omega_{s_1}^2(q_1),
    \ \
    \tilde\omega_{1,s_1,u,{\bf q}}^2
    =
    \omega_{s_1}^{2}(q_1)-
    2\phi^2
    \sum_{s'=1}^{\infty}
    \frac{\zeta_{s'}^2(q_2)\omega_{s'}^2(q_2)
          \zeta_{s_1}^2(q_1)\omega_{s_1}^2(q_1)}
         {\omega_{s'}^2(q_2)-\omega_{s_1}^2(q_1)}.&&
    \label{freq_b_1}
\end{eqnarray}
The case $j=2$ is described similarly after change
$1\leftrightarrow2$.

The second set is formed by the resonant lines defined by the
equation $k_1+rk_2=nQ,$ where $r=\pm1$ and $n$ are two integer
parameters determining the line. The corresponding quasi-momentum
satisfy the equality $q_1+rq_2=0$, then it lies on the diagonal of
the first BZ (see Fig. \ref{sp-sq}). In an ``empty lattice''
approximation, these lines are also degeneracy lines. However this
degeneracy has more complicated nature. Here there is a dual point
${\bf{\overline{k}}}=(-rk_2,-rk_1)$ which satisfies the equalities
$-rk_2=k_1-nQ$ and $-rk_1=-k_2+rnQ$, and therefore this point
corresponds to the {\em same} quasimomentum ${\bf{q}}$ as the
point ${\bf{k}}$. As a result, the array plasmons associated with
the points ${\bf{k}}$ and ${\bf{\overline{k}}}$ are involved in
the resonance. Inter-array interaction mixes degenerate modes,
lifts the degeneracy, and splits degenerate frequencies. As a
result, the QCB plasmons related to a point ${\bf{k}}$ at the
resonant line (e.g. all $B,F,E$ points in Fig. \ref{BZNew}), are
built from the array plasmons associated not only with the points
${\bf{k}}_{1,2}$ but also with the symmetric (with respect to one
of the bisector lines of the coordinate system of a reciprocal
space) points ${\overline{\bf{k}}_1}=(-rk_1,-rq_2)$ and
${\overline{\bf{k}}_2}=(-rq_1,-rk_2).$ Moreover, if the point
${\bf{k}}$ lies at an intersection of two resonant lines (the
point $W$ in Fig. \ref{sp-sq}), the QCB plasmons include also
array states associated with the
${\overline{\bf{k}}}_{\overline{j}}$ points.

In the general case, the point ${\bf{k}}={\bf{q}}+{\bf{m}}$ lies
only on one of the resonant lines $q_1={\pm}q_2$ away from the
high symmetry lines. Here the point ${\bf{k}}$ generates two pairs
of even and odd superpositions of the first and the second array
states (two doublets). The first of them,
$\widetilde\theta_{g/u,s_1,{\bf{q}}}$, lies in the energy band
$s_1=1+\left[2|k_1|/Q\right]$,
\begin{eqnarray}
 &&\tilde\theta_{g/u,s_1,{\bf{q}}}=
   \theta_{g/u,s_1,{\bf{q}}}+
   \phi
   \sum\limits_{s'{\ne}s_1}
   \frac{\zeta_{s'}(q_2)\omega_{s'}(q_2)
         \zeta_{s_1}(q_1)\omega_{s_1}(q_1)}
        {\omega_{s'}^2(q_2)-\omega_{s_1}^2(q_1)}
   \theta_{g/u,s',{\bf{q}}},
 \label{first_1}\\
 &&\theta_{g/u,s_1,{\bf{q}}}=
   \frac{1}{\sqrt{2}}
   \left(
        \theta_{1,s_1,{\bf{q}}}
        \pm
        \theta_{2,s_1,{\bf{q}}}
   \right),
   \ \ \ \ \ j=1,2.
   \nonumber
\end{eqnarray}
The corresponding eigenfrequencies form two doublets
\begin{eqnarray}
   \omega_{g/u,s_1,{\bf{q}}}^2=
   \left(1\pm\phi\right)\omega_{s_1}^2(q_1)-
   \phi^2
   \sum\limits_{s'{\ne}s_1}
   \frac{\zeta_{s' }^2(q_2)\omega_{s' }^2(q_2)
         \zeta_{s_1}^2(q_1)\omega_{s_1}^2(q_1)}
        {\omega_{s'}^2(q_2)-\omega_{s_1}^2(q_1)}.
    \label{freq_a_1}
\end{eqnarray}
The eigenfunctions and eigenfrequencies of another doublet are
obtained similarly by replacing the indices $1\leftrightarrow2$.

The crossing points of the resonant lines can also be divided into
two groups. The first group is formed by the crossing points of
only two high symmetry lines (points $X_1$ and $X_2$ in Fig.
\ref{BZ2}). Here, each of the generated vectors ${\bf{k}}_1$ and
${\bf{k}}_2$ corresponds to a pair of array eigenstates belonging
to two adjacent $1D$ bands. Thus an {\em inter-band} mixing is
significant in both arrays. Consider for definiteness the point
$X_1$. In this case the wave vector ${\bf{k}}$ (${\bf{k}}_j$) of
the extended BZ can be represented as ${\bf{k}}={\bf{q}}+{\bf{m}}$
(${\bf{k}}_j={\bf{q}}+{\bf{m}}_j$), with ${\bf{q}}=(Q/2,0)$. As a
result, in ``empty lattice'' approximation, we have two sets of
degeneracy lines. The first one of them corresponds to odd and
next even bands of the first array ($q_1=Q/2$), whereas the second
one describes degeneracy of even and next odd bands of the second
array ($q_2=0$). The corresponding QCB plasmons are even or odd
combinations of the $j$-th array plasmons ($j=1,2$). For $j=1$, we
have,
\begin{eqnarray}
   \tilde\theta_{1,s_1,g,{\bf{q}}}=
   \theta_{1,s_1,g,{\bf{q}}},
   \ \
   \tilde\theta_{1,s_1,u,{\bf{q}}}=
   \theta_{1,s_1,u,{\bf{q}}}+
   2\phi
   \sum\limits_{n=1}^{\infty}
       \frac{\zeta_{2n}(q_2)\omega_{2n}(q_2)
             \zeta_{s_1}(q_1)\omega_{s_1}(q_1)}
            {\omega_{2n}^2(q_2)-
            \omega_{s_1}^2(q_1)}
       \theta_{2,2n,u,{\bf{q}}},
 \label{high_cross}
\end{eqnarray}
where
$$
 \theta_{j,s,g/u,{\bf{q}}}=
 \frac{1}{\sqrt{2}}
 (\theta_{j,s,{\bf{q}}}\pm
 \theta_{j,s+1,{\bf{q}}}).
$$
The corresponding eigenfrequencies are
\begin{eqnarray}
    \tilde\omega_{1,s_1,g,{\bf{q}}}^2=\omega_{s_1}^2(q_1),
    \ \
    \tilde\omega_{1,s_1,u,{\bf{q}}}^2=
    \omega_{s_1}^{2}(q_1)-
    4\phi^2
    \sum_{n=1}^{\infty}
    \frac{\zeta_{2n}^2(q_2)\omega_{2n}^2(q_2)
          \zeta_{s_1}^2(q_1)\omega_{s_1}^2(q_1)}
         {\omega_{2n}^2(q_2)-\omega_{s_1}^2(q_2)}.
    &&
    \label{freq_b}
\end{eqnarray}
The case $j=2$ is described similarly after change
$1\leftrightarrow2$ and $2n\to2n+1$ in Eqs. (\ref{high_cross}) and
(\ref{freq_b}).

The second group consists of the crosses of two resonant lines
(points $\Gamma$ and $W$ in Fig. \ref{sp-sq}). These points are
also the crosses of two high symmetry lines (this is always true).
Here the QCB plasmons, generated by the point
${\bf{k}}={\bf{q}}+{\bf{m}}$ with ${\bf{q}}\equiv(q,q)=(0,0)$
($\Gamma$ point) or $(Q_0/2,Q_0/2)$ ($W$ point), form two
quartets. The first quartet is really generated by the point
${\bf{k}}_1.$ It consists of four symmetrized combinations of
single-array states
\begin{eqnarray}
  \widetilde\theta_{s_1,g,\nu,{\bf{q}}}=
    \theta_{s_1,g,\nu,{\bf{q}}},
  \ \
  \widetilde\theta_{s_1,u,\nu,{\bf{q}}}=
    \theta_{s_1,u,\nu,{\bf{q}}}-
    2\phi
    \sum\limits_{s'{\ne}s_1}
        \frac{\zeta_{s' }(q_2)\omega_{s' }(q_2)
              \zeta_{s_1}(q_1)\omega_{s_1}(q_1)}
             {\omega_{s'}^2(q_2)-\omega_{s_1}^2(q_1)}
   \theta_{s',u,\nu,{\bf{q}}},
   \label{eigenfun-quartet}
\end{eqnarray}
where $\nu=g,u$; the band numbers $s_1$ and $s'$ are even for
$q=0$ and odd for $q=Q/2$;
\begin{eqnarray*}
  \theta_{s,g/u,g,{\bf{q}}}
  &=&
  \frac{1}{2}
  \sum_{j=1,2}
  \Big(
      \theta_{j, s, {\bf{q}}}\pm
      \theta_{1,s+1,{\bf{q}}}
  \Big);
  \ \ \ \ \
  \theta_{s,g/u,u,{\bf{q}}}
  =
  \frac{1}{2}
  \sum_{j=1,2}(-1)^j
  \Big(
      \theta_{j, s, {\bf{q}}}\pm
      \theta_{j,s+1,{\bf{q}}}
  \Big).
\end{eqnarray*}
QCB eigenstates of the second quartet are generated by the point
${\bf{k}}_2.$ They are obtained from the equations
(\ref{eigenfun-quartet}) by replacing $1\leftrightarrow2$. Even
array eigenstates are degenerate with the frequencies
\begin{equation}
    \omega_{s_j,g,g/u,{\bf{q}}}^2=
    \omega_{s_j}(q_j)^2,
    \label{evar}
\end{equation}
while the odd array eigenstates are split
\begin{equation}
    \omega_{s_1,u,g/u,{\bf{q}}}^2=
    (1\pm2\phi)\omega_{s_1}^2(q_1)-
    4\phi^2
    \sum\limits_{s'{\ne}s_1}
        \frac{\zeta_{s' }^2(q_2)\omega_{s' }^2(q_2)
              \zeta_{s_1}^2(q_1)\omega_{s_1}^2(q_1)}
             {\omega_{s'}^2(q_2)-\omega_{s_1}^2(q_1)}.
    \label{odar}
\end{equation}

Equations (\ref{states_1}) - (\ref{odar}) exhaust all cases of QCB
plasmons, generated by the point ${\bf{k}}_2$ lying on a specific
line, via the interference of array plasmons.

%\newpage

%%%%%%%%%%%%%%%%%%%%%%%%%
\section{AC Conductivity}
 \label{append:conduct}
%%%%%%%%%%%%%%%%%%%%%%%%%%

For interacting wires, where $\zeta_{js}(q_j)\neq{0},$ the
correlator (\ref{CurrCorr}) may be easily calculated after
diagonalization of the Hamiltonian (\ref{TotHam2}) by means of the
transformations considered in Appendix \ref{append:DoublSpectr}.
For the first band, for example, these are the transformations
(\ref{tilde-teta1}) and (\ref{tilde-teta2}). As a result, one has:
\begin{eqnarray*}
  \left\langle\left[
                 {J}_{11{ \bf{q}}}(t),
                 {J}_{11{\bf{q}}}^{\dag}(0)
  \right]\right\rangle
  &=&
  -2ivg\left(u_{\bf{q}}^{2}
  {\omega}_{+,1{\bf{q}}}
  \sin({\omega}_{+,1{\bf{q}}}t)
  +v_{\bf{q}}^{2}
  {\omega}_{-,1{\bf{q}}}
  \sin({\omega}_{-,1{\bf{q}}}t)\right),
  \\
  \left\langle\left[
                 {J}_{11{ \bf{q}}}(t),
                 {J}_{21{\bf{q}}}^{\dag}(0)
  \right]\right\rangle
  &=&
  -2ivgu_{\bf{q}}v_{\bf{q}}
  \left(
     {\omega}_{-,1{\bf{q}}}
     \sin({\omega}_{-,1{\bf{q}}}t)-
     {\omega}_{+,1{\bf{q}}}
     \sin({\omega}_{+,1{\bf{q}}}t)
  \right).
\end{eqnarray*}
where $u_{\bf{q}}$ and $v_{\bf{q}}$ are defined in Eqs.(\ref{uv}).
Then, for the optical absorption ${\sigma}'$ one obtains
\begin{eqnarray}
 {\sigma}'_{11}({\bf{q}},\omega)
 &=&
 {\pi}{v}{g}
 \Bigl[
      u_{\bf{q}}^{2}
      \delta\left(
                {\omega}-
                {\tilde\omega}_{+,1{\bf{q}}}
            \right)+
      v_{\bf{q}}^{2}
      \delta\left(
                 {\omega}-
                 {\tilde\omega}_{-,1{\bf{q}}}
            \right)
      \Bigr]
 \label{opt_l}
 \\
 {\sigma}'_{12}({\bf{q}},\omega)
 &=&
 {\pi}{v}{g}
 u_{\bf{q}}v_{\bf{q}}
 \Bigl[
      \delta\left(
                 {\omega}-
                 {\tilde\omega}_{-,1{\bf{q}}}
            \right)-
      \delta\left(
                 {\omega}-
                 {\tilde\omega}_{+,1{\bf{q}}}
            \right)
 \Bigr].
 \label{opt_t}
\end{eqnarray}
For quasimomentum ${\bf{q}}$ away from the resonant coupling line,
$u_{\bf{q}}^{2}\approx{1}$ and
$v_{\bf{q}}^{2}\sim\phi_{1\bf{q}}^{2}$ for $\Delta_{\bf{q}}>0$
($v_{\bf{q}}^{2}\approx{1}$ and
$u_{\bf{q}}^{2}\sim\phi_{1\bf{q}}^{2}$ for $\Delta_{\bf{q}}<0$).
Then the longitudinal optical absorption (\ref{opt_l}) (i.e. the
absorption within a given set of wires) has its main peak at the
frequency ${\omega}_{+,1{\bf{q}}}\approx{v\vert{q_1}\vert}$ for
$\Delta_{\bf{q}}>0$ (or
${\omega}_{-,1{\bf{q}}}\approx{v\vert{q_1}\vert}$ for
$\Delta_{\bf{q}}<0$), corresponding to the first band of the
pertinent array, and an additional weak peak at the frequency
${\omega}_{-,1{\bf{q}}}\approx{v\vert{q_2}\vert}$, corresponding
to the first band of a complementary array.  It contains also a
set of weak peaks at frequencies $\omega_{2,s{\bf{q}}}\approx
[s/2]vQ$ ($s=2,3,\ldots$) corresponding to the contribution from
higher bands of the complementary array (in Eq.(\ref{opt_l}) these
peaks are omitted).  At the same time, a second observable becomes
relevant, namely, the transverse optical absorption (\ref{opt_t}).
It is proportional to the (small) interaction strength and has two
peaks at frequencies ${\omega}_{+,1{\bf{q}}}$ and
${\omega}_{-,1{\bf{q}}}$ in the first bands of both sets of wires.

If the quasimomentum ${\bf{q}}$ belongs to the resonant coupling
line $\Delta_{\bf{q}}=0$, then
$u_{\bf{q}}^{2}=v_{\bf{q}}^{2}=1/2$. In this case the longitudinal
optical absorption (\ref{opt_l}) has a split double peak at
frequencies ${\omega}_{+,1{\bf{q}}}$ and ${\omega}_{-,1{\bf{q}}}$,
instead of a single main peak. The transverse optical absorption
(\ref{opt_t}), similarly to the non-resonant case (\ref{opt_t}),
has a split double peak at frequencies ${\omega}_{+,1{\bf{q}}}$
and ${\omega}_{-,1{\bf{q}}}$, but its amplitude is now of the
order of unity. For $\left\vert{\bf{q}}\right\vert\to 0$
Eq.(\ref{opt_l}) reduces to that for an array of noninteracting
wires (\ref{Drude_peak}), and the transverse optical conductivity
(\ref{opt_t}) vanishes.

The imaginary part of the {\it ac} conductivity
${\sigma}''_{jj'}({\bf{q}},\omega)$ is calculated within the same
approach. Its longitudinal component equals
\begin{eqnarray*}
 {\sigma}''_{11}({\bf{q}},\omega)
 &=&
 \displaystyle{\frac{2vg}{\omega}
 \left[
      \frac{u_{\bf{q}}^{2}{\omega}_{+,1{\bf{q}}}^{2}}
           {{\omega}_{+,1{\bf{q}}}^{2}-\omega^2}+
            \frac{v_{\bf{q}}^{2}{\omega}_{-,1{\bf{q}}}^{2}}
                 {{\omega}_{-,1{\bf{q}}}^{2}-\omega^2}
            \right]}.
\end{eqnarray*}
Beside the standard pole at zero frequency, the imaginary part has
poles at the resonance frequencies $\omega_{+,1{\bf q}}$,
$\omega_{-,1{\bf q}}$, and an additional series of high band
satellites (omitted here).  For quasimomenta far from the resonant
lines, only the first pole is well pronounced while amplitude of
the second one as well as amplitudes of all other satellites is
small.  At the resonant lines, the amplitudes of both poles
mentioned above are equal. The corresponding expression for
${\sigma}'_{22}({\bf{q}},\omega)$ can be obtained by replacement
$1\leftrightarrow 2$.

The transverse component of the imaginary part of the {\it ac}
conductivity has the form:
\begin{eqnarray*}
{\sigma}'_{12}({\bf{q}},\omega)=
        \displaystyle{
        \frac{2vg}{\omega}u_{{\bf q}}v_{{\bf q}}
        \left[
             \frac{{\omega}_{-,1{\bf{q}}}^{2}}
                  {{\omega}^{2}-{\omega}_{-,1{\bf{q}}}^{2}}-
             \frac{{\omega}_{+,1{\bf{q}}}^{2}}
                  {{\omega}^{2}-{\omega}_{+,1{\bf{q}}}^{2}}
        \right]}.
\end{eqnarray*}
It always contains two poles and vanishes for noninteracting
wires. For quasimomenta far from the resonance lines the
transverse component is small while at these lines its amplitude
is of the order of unity.

%\newpage

%%%%%%%%%%%%%%%%%%%%%%%%%%%%%%%%%%%%
 \chapter{Tilted QCB}
 \label{append-tilted}
%%%%%%%%%%%%%%%%%%%%%%%%%%%%%%%%%%%%

%%%%%%%%%%%%%%%%%%%%%%%%%%%%%%%%%%%%%%%%%%%%
\section{Geometry, Notions and Hamiltonian}
 \label{sec:tilt-Hamilt}
%%%%%%%%%%%%%%%%%%%%%%%%%%%%%%%%

A tilted QCB is a $2D$ grid, formed by two periodically crossed
arrays of $1D$ quantum wires or carbon nanotubes. Like in a square
QCB, arrays are labelled by indices $j=1,2$ and wires within the
first (second) array are labelled by an integer index $n_2$
($n_1$). The arrays are oriented along the unit vectors ${\bf
e}_{1,2}$ with an angle $\varphi$ between them. The periods of a
crossbars along these directions are $a_1$ and $a_2,$ and the
corresponding basic vectors are ${\bf a}_j=a_j{\bf e}_j$. In
experimentally realizable setups, QCB is a cross-structure of
suspended single-wall carbon nanotubes lying in two parallel
planes separated by an inter-plane distance $d$. Nevertheless,
some generic properties of QCB may be described under the
assumption that QCB is a genuine $2D$ system. We choose coordinate
system so that the axes $x_j$ and corresponding basic unit vectors
${\bf{e}}_j$ are oriented along the $j$-th array. The basic
vectors of the reciprocal superlattice for a square QCB are
$Q_{1,2}{\bf{g}}_{1,2},$ $Q_j=2\pi/a_j$ so that an arbitrary
reciprocal superlattice vector ${\bf m}$ is a sum ${\bf m}={\bf
m}_{1}+ {\bf m}_{2},$ where ${\bf m}_{j}=m_{j}Q_j{\bf g}_{j},$
($m_j$ integer). Here ${\bf g}_{1,2}$ are the vectors of the
reciprocal superlattice satisfying the standard orthogonality
relations $({\bf e}_i\cdot {\bf g}_j)=\delta_{ij}$. The first BZ
is $|q_{1,2}|\leq Q_{1,2}/2.$

A single wire of $j$-th array is characterized by its radius
$r_j$, length $L_j$, and LL interaction parameter $g_j$. The
minimal nanotube radius is $0.35$~nm\cite{Louie}, maximal nanotube
length is $L=1$~mm, and the LL parameter is estimated as $0.3$
\cite{Egger}. In typical experimental setup\cite{Rueckes} the
characteristic lengths mentioned above have the following values
\begin{eqnarray*}
       d\approx 2 ~{\rm nm},\phantom{aa}
       L_j\approx 0.1 ~{\rm mm},
    \label{lengths}
\end{eqnarray*}
so that the inequalities
\begin{eqnarray*}
    r_{1,2}\ll d\ll a_{1,2}\ll L_{1,2}
    \label{ineq}
\end{eqnarray*}
are satisfied.

The QCB Hamiltonian
\begin{equation}
    H=H_{1}+H_{2}+H_{12}.
    \label{HamiltTot}
\end{equation}
consists of three terms. The first two of them describes LL in the
first and second arrays
\begin{eqnarray*}
    H_1 &=&
         \frac{\hbar v_1}{2}\sum_{n_2}
              \int\limits_{-L_1/2}^{L_1/2}dx_1
              \biggl\{
                   g_1\pi_1^2\left(x_1,n_2a_2\right)+
                   \frac{1}{g_1}
                   \left(
                        \partial_{x_1}
                        \theta_1
                        \left(x_1,n_2a_2\right)
                   \right)^2
         \biggr\},
    \\
    H_2 &=&
         \frac{\hbar v_2}{2}\sum_{n_1}
              \int\limits_{-L_2/2}^{L_2/2}dx_2
              \biggl\{
                   g_2\pi_2^2\left(n_1a_1,x_2\right)+
                   \frac{1}{g_2}
                   \left(
                        \partial_{x_2}
                        \theta_2
                        \left(n_1a_1,x_2\right)
                   \right)^2
         \biggr\}.
\end{eqnarray*}
The inter-array interaction is described by the last term in
Eq.(\ref{HamiltTot})
\begin{eqnarray}
   H_{12} &=&
   \frac{2e^{2}}{d}
   \sum\limits_{n_1,n_2}
   \int dx_1 dx_2
   \zeta\left(\frac{x_1-n_1a_1}{r_1}\right)
   \zeta\left(\frac{n_2a_2-x_2}{r_2}\right)
   \times\nonumber\\&&\times
   \partial_{x_1}\theta_1(x_1,n_2a_2)
   \partial_{x_2}\theta_2(n_1a_1,x_2).
   \label{Inter-tilted}
\end{eqnarray}
It results from a short-range contact capacitive coupling in the
crosses of the bars. The dimensionless envelope function
(introduced phenomenologically) $\zeta(\xi_j)$ describes
re-distribution of a charge in a tube $j$ induced by the
interaction with tube $i.$ This function is of order unity for
$|\xi|\sim 1$ and vanishes outside this region so that the
dimensionless integral
\begin{eqnarray*}
    \int\zeta(\xi)e^{ikr_j\xi}d\xi\sim 1
    \label{barzeta}
\end{eqnarray*}
is of order unity for all $|k|$ smaller than a certain ultraviolet
cutoff.

The QCB Hamiltonian (\ref{HamiltTot}) is a quadratic form in terms
of the field operators, so it can be diagonalized exactly.  Such a
procedure is rather cumbersome. However, due to the separability
of the interaction (\ref{Inter-tilted}) the spectrum can be
described analytically.

%%%%%%%%%%%%%%%%%%%%%%%%%%%%%%%%%%%%
\section{Spectrum}
 \label{sec-tilt-spectr}
%%%%%%%%%%%%%%%%%%%%%%%%%%%%%%%%%%%%

\begin{figure}[htb]
\centering
\includegraphics[width=55mm,height=55mm,angle=0,]{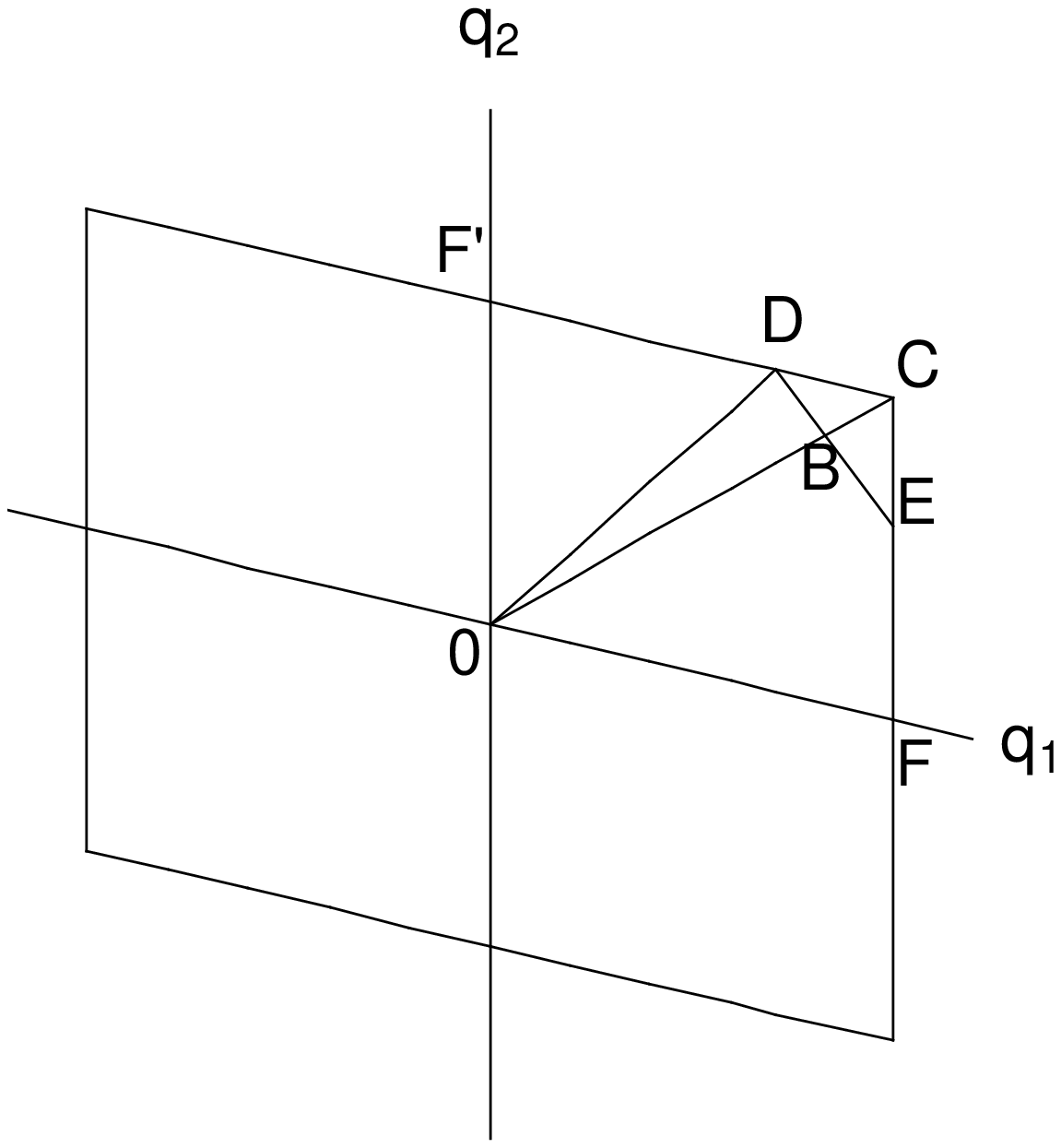}
\caption{BZ of a tilted QCB.} \label{BZ3}
\end{figure}

Now we consider the spectrum of a generic double QCB. The
resonance condition (\ref{res}) is fulfilled not at the BZ
diagonal but at the resonant polygonal line. Its part $ODE,$ lying
in the first quarter of the BZ, is displayed in Fig. \ref{BZ3}
(all figures of this Section correspond to the specific values
$v_2Q_2=1,\phantom{aa}v_1Q_1=1.4$). This results in qualitative
modifications of the spectrum that are related first of all to the
appearance of two points $D$ and $E$ of the three-fold degeneracy
for a titled QCB (Fig. \ref{BZ3}) instead of a single point $W$ of
four-fold degeneracy for a square QCB (Fig. \ref{sp-sq}).
%%%%%%%%%%%%%%%%%%%%%
%%%%%%%%%%%%%%%%%%%

\begin{figure}[htb]
\centering
\includegraphics[width=70mm,height=60mm,angle=0,]{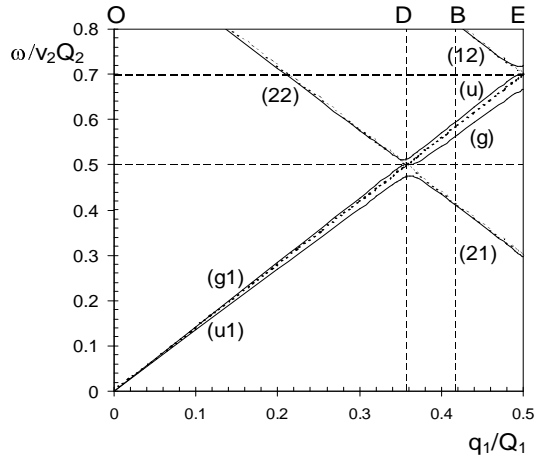}
\caption{The energy spectrum of a tilted QCB (solid lines) and
noninteracting arrays (dashed lines) for quasimomenta on the
resonant line of the BZ (line $ODE$ in Fig. \ref{BZ3}).}
\label{sp-res}
\end{figure}

We start with the resonant line $ODE$ (Fig. \ref{BZ2}). The
dispersion curves at its $OD$ part and the symmetry properties of
the corresponding eigenstates are similar to those at the $OC$
resonant line for the square QCB (Fig. \ref{sp-sq}). The only
difference is that instead of the four-fold degeneracy at the BZ
corner $C$ of the square QCB, there is a three-fold degeneracy at
the point $D$ lying at the BZ boundary. A completely new situation
takes place at the $DE$ line, where two other modes $(1,1)$ and
$(2,2),$ corresponding to different arrays and different bands,
are degenerate. The interaction lifts this degeneracy and the two
middle lines in Fig. \ref{sp-res} describe even $(g)$ and odd
$(u)$ combinations of these modes. The even mode corresponds to
the lowest frequency and the odd mode corresponds to the higher
one. At the point $E$ one meets another type of a three-fold
degeneracy described in more detail in the next paragraph.

\begin{figure}[htb]
\centering
\includegraphics[width=70mm,height=60mm,angle=0]{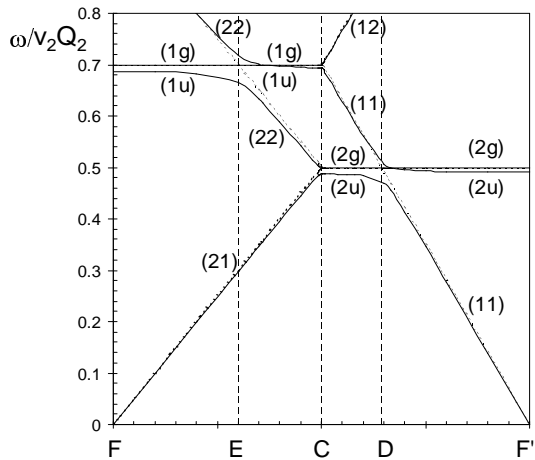}
\caption{The energy spectrum of a tilted QCB (solid lines) and
noninteracting arrays (dashed lines) for quasimomenta at the BZ
boundary (line $FCF'$ in the Fig. \ref{BZ3}).}
 \label{sp-boun}
\end{figure}

Dispersion curves corresponding to quasi momenta lying at the BZ
boundary $q_1=Q_1/2,$ $0\leq{q}_2\leq{Q}_2/2$ ($FC$ line in Fig.
\ref{BZ3}) and $q_{2}=Q_2/2,$ $0\leq{q}_{1}\leq{Q}_{1}/2$ ($CF'$
line in Fig. \ref{BZ3}), are displayed in Fig. \ref{sp-boun}. The
lowest and highest curves in the $FE$ part of the latter figure,
describe two waves propagating along the second array. They are
nearly linear, and deviations from linearity are observed only
near the point $E$ where the interaction has a resonant character.
Two modes propagating along the first array, in zero
approximation, are degenerate with an unperturbed frequency
$\omega=0.7.$ The interaction lifts the degeneracy.  The lowest of
the two middle curves corresponds to $(1,u)$ boson, and the upper
of one describes $(1,g)$ boson.  Note that $(1,g)$ boson conserves
its unperturbed frequency $\omega=0.7$.  The latter fact is
related to the symmetry $\zeta_j(\xi)=\zeta_j(-\xi)$ of the
separable interaction (\ref{zeta}). At the point $E$, the two
modes propagating along the first array and the mode propagating
along the second array in the second band are degenerate. The
interaction lifts the degeneracy, and, as a result, the $(1,u)$
and $(2,2)$ waves are strongly mixed and the eigen-modes are their
even (highest frequency) and odd (lowest frequency) combinations,
and the $(1,g)$ mode (middle level).
%%%%%%%%%%%%%%%%%%%%%%%%
%%%%%%%%%%%%%%%%%%%%%%%%%%
%%%%%%%%%%%%%%%%%%%%%%%%
%%%%%%%%%%%%%%%%%%%%%%%%%%
%%%%%%%%%%%%%%%%%%%%%%%%%
\begin{figure}[htb]
\centering
\includegraphics[width=65mm,height=60mm,angle=0,]{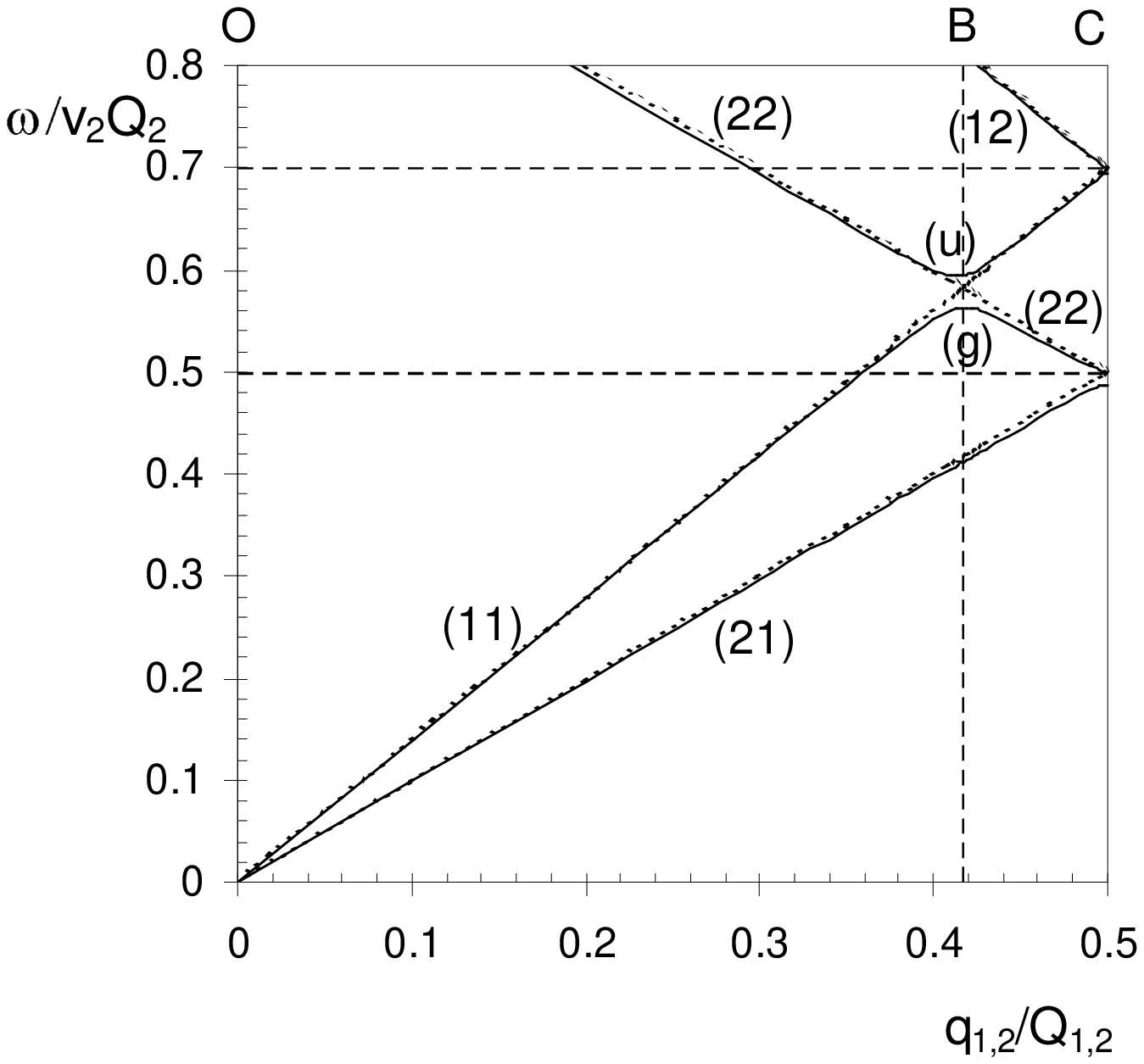}
\caption{The energy spectrum of a titled QCB (solid lines) and
noninteracting arrays (dashed lines) for quasimomenta on the BZ
diagonal (line $OC$ in Fig. \ref{BZ3}).} \label{sp-diag}
\end{figure}
%%%%%%%%%%%%%%%%%%%%%%%%%%%%%
%%%%%%%%%%%%%%%%%%%%%%%%%%%%%%%%%%%%%%%
\begin{figure}[htb]
\centering
\includegraphics[width=70mm,height=60mm,angle=0,]{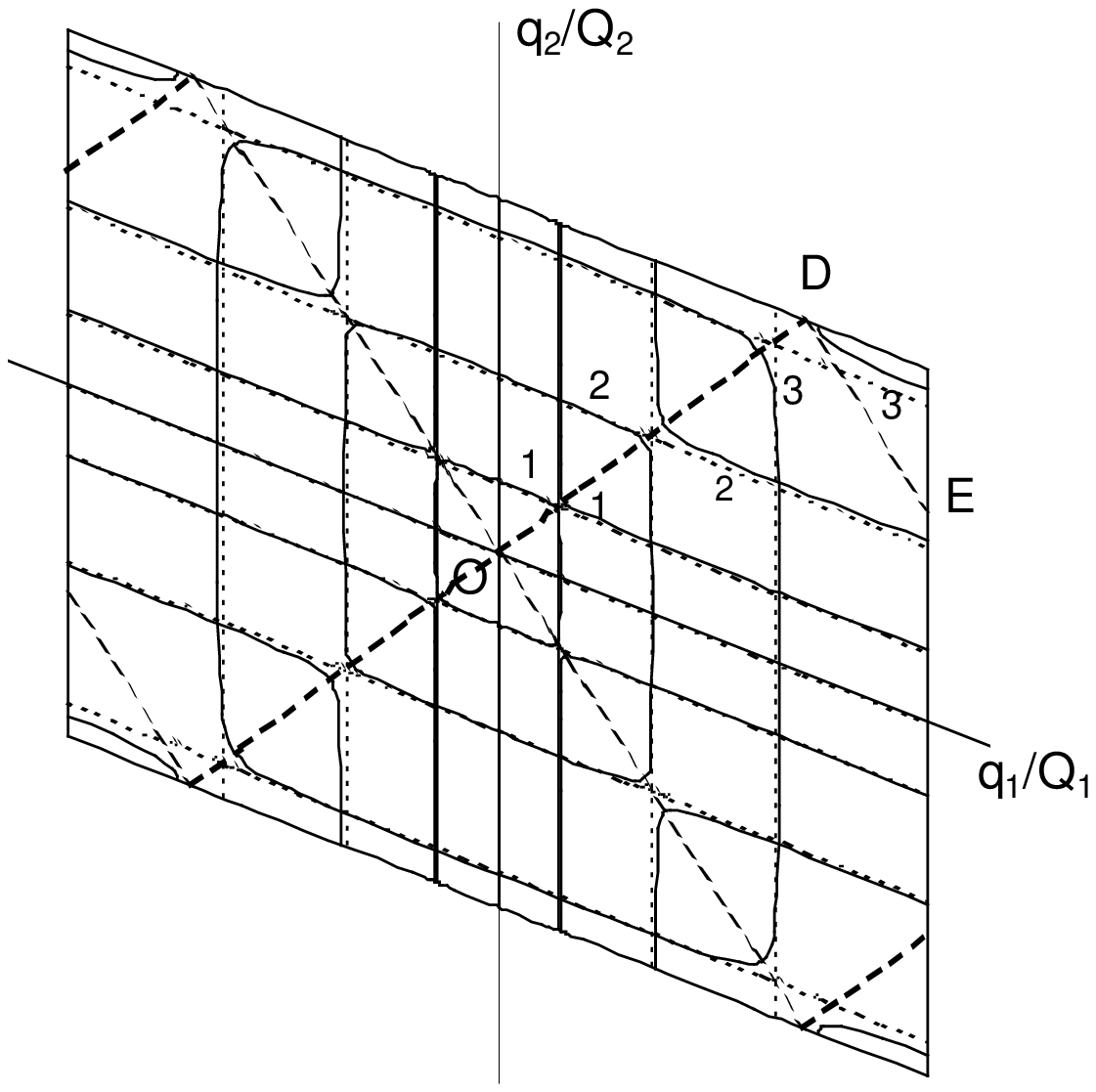}
\caption{Lines of equal frequency for a tilted QCB (solid lines)
and noninteracting arrays (dashed lines). Lines $1,2,3$ correspond
to frequencies $\omega_{1}=0.1$, $\omega_{2}=0.25$,
$\omega_{3}=0.45$.}
 \label{IsoEn1}
\end{figure}
%%%%%%%%%%%%%%%%%%%%%%%%%%%%%%%%%%%%
%%%%%%%%%%%%%%%%%%%%%%%%%%%%%%%%%%%%%%%
\begin{figure}[htb]
\centering
\includegraphics[width=70mm,height=60mm,angle=0,]{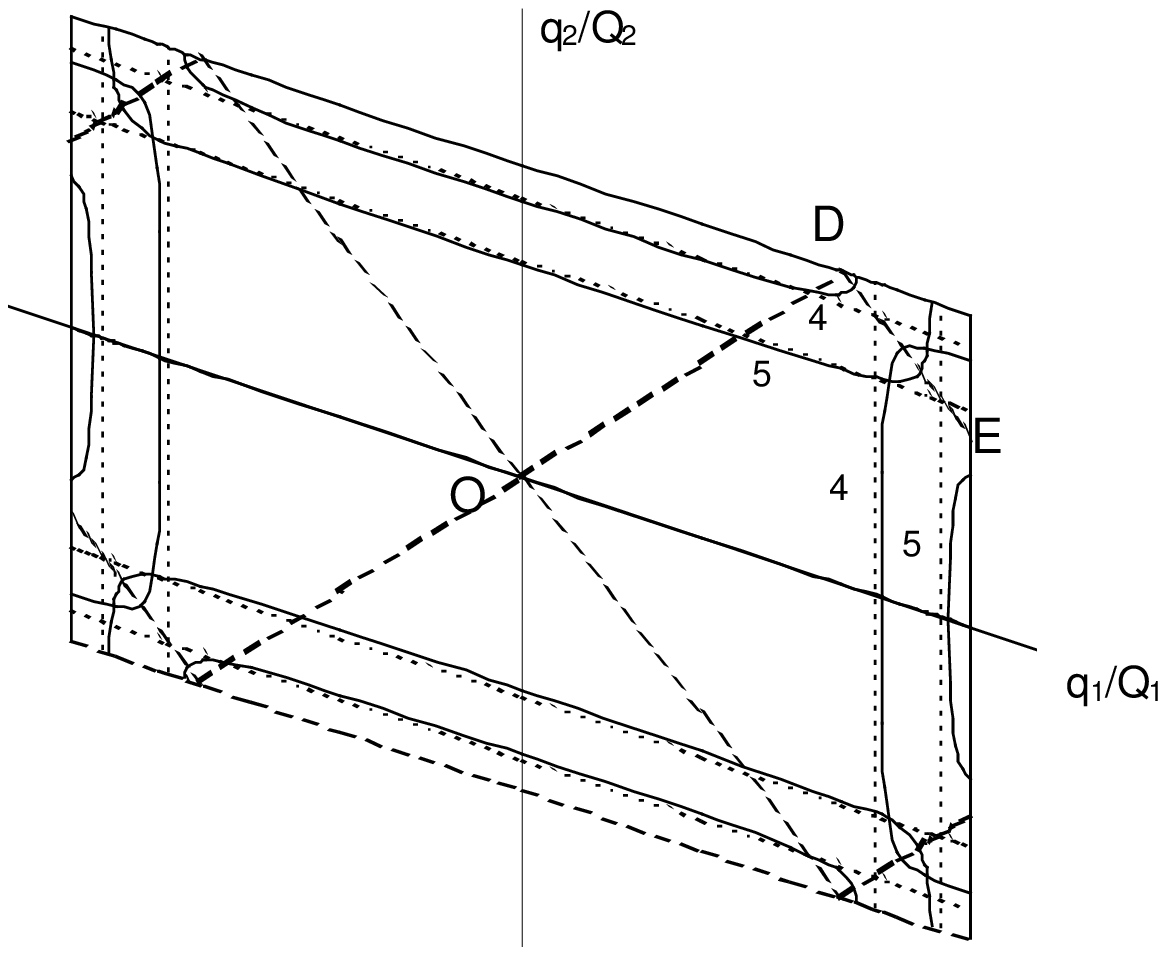}
\caption{Lines of equal frequency for a tilted QCB (solid lines)
and noninteracting arrays (dashed lines). Lines $4,5$ correspond
to frequencies $\omega_{4}=0.55$, $\omega_{5}=0.65.$}
\label{IsoEn2}
\end{figure}
%%%%%%%%%%%%%%%%%%%%%%%%%%%%%%%%%%%%

There are two separate degeneracies within each array at the
corner $C$ of a titled QCB BZ. Both of them are related to
inter-band mixing conserving array index.  The spectral behavior
along the $CF'$ boundary of the BZ is similar to that considered
above but in the vicinity of the point $D$ of three-fold
degeneracy.  Here, two modes propagating along the second array in
the separable potential approximation (\ref{separ}) remain
degenerate.  This degeneracy is lifted only if deviation from
separability is accounted for.

The diagonal $OC$ of a tilted QCB BZ represents a new type of
generic line, that crosses a resonant line (Fig. \ref{sp-diag}).
Here the spectrum mostly conserves its initial systematics, i.e.
belongs to a given array, and mostly depends on a given
quasimomentum component. However, at the crossing point $B$, the
modes $(1,1)$ and $(2,2),$ corresponding to both different arrays
and bands, become degenerate (two middle dashed lines in Fig.
\ref{sp-diag}). Interaction between the wires lifts the
degeneracy. The eigenstates of QCB have a definite parity with
respect to transposition of these two modes. The lowest and upper
of the two middle lines correspond to even ($g$) and odd ($u$)
mode, respectively.

Like in a square QCB, bosons with quasimomenta close to the
resonant lines are strongly mixed bare $1D$ bosons.  These
excitations are essentially two-dimensional, and therefore lines
of equal energy in the vicinity of the resonant lines are modified
by the $2D$ interaction (see Figs. \ref{IsoEn1} and \ref{IsoEn2}).
Deviations from $1D$ behavior occur only in this small part of the
BZ. For $\omega < 0.5 v_2Q_2$ the lines of equal energy within BZ
consist of closed line around the BZ center and four open lines
(within the extended bands scheme these lines are certainly
closed) around the BZ corners (lines 1, 2, 3 in Fig.
\ref{IsoEn1}). At the line $OD$ in BZ, the modes of QCB are
strongly coupled bare bosons propagating along both arrays in the
first band.

For $0.5 v_2Q_2 <\omega < 0.5 v_1Q_1$ (lines 4, 5 in Fig.
\ref{IsoEn2}) the topology of lines of equal energy is modified.
In this case, lines of equal energy within the BZ consist of four
open lines. The splitting of lines at the direction $DE$
corresponds to strong coupling of modes propagating along the
first array in the first band with those propagating along the
second array in the second band.

%\newpage

%%%%%%%%%%%%%%%%%%%%%%%%%%%%%%%%%%%%%%%%%%%%%%%%
\chapter{Triple QCB}
 \label{subsec:Triple}
%%%%%%%%%%%%%%%%%%%%%%%%%%%%%%%%%%%%%%%%%%%%%%%%

%%%%%%%%%%%%%%%%%%%%%%%%%%%%%%%%%%%%%%%%%%%%
\section{Notions and Hamiltonian}
\label{subsubsec:NotHam}
%%%%%%%%%%%%%%%%%%%%%%%%%%%%%%%%%%%%%%%%%%%%

\begin{figure}[htb]
\centering
\includegraphics[width=48mm,height=44mm,angle=0,]{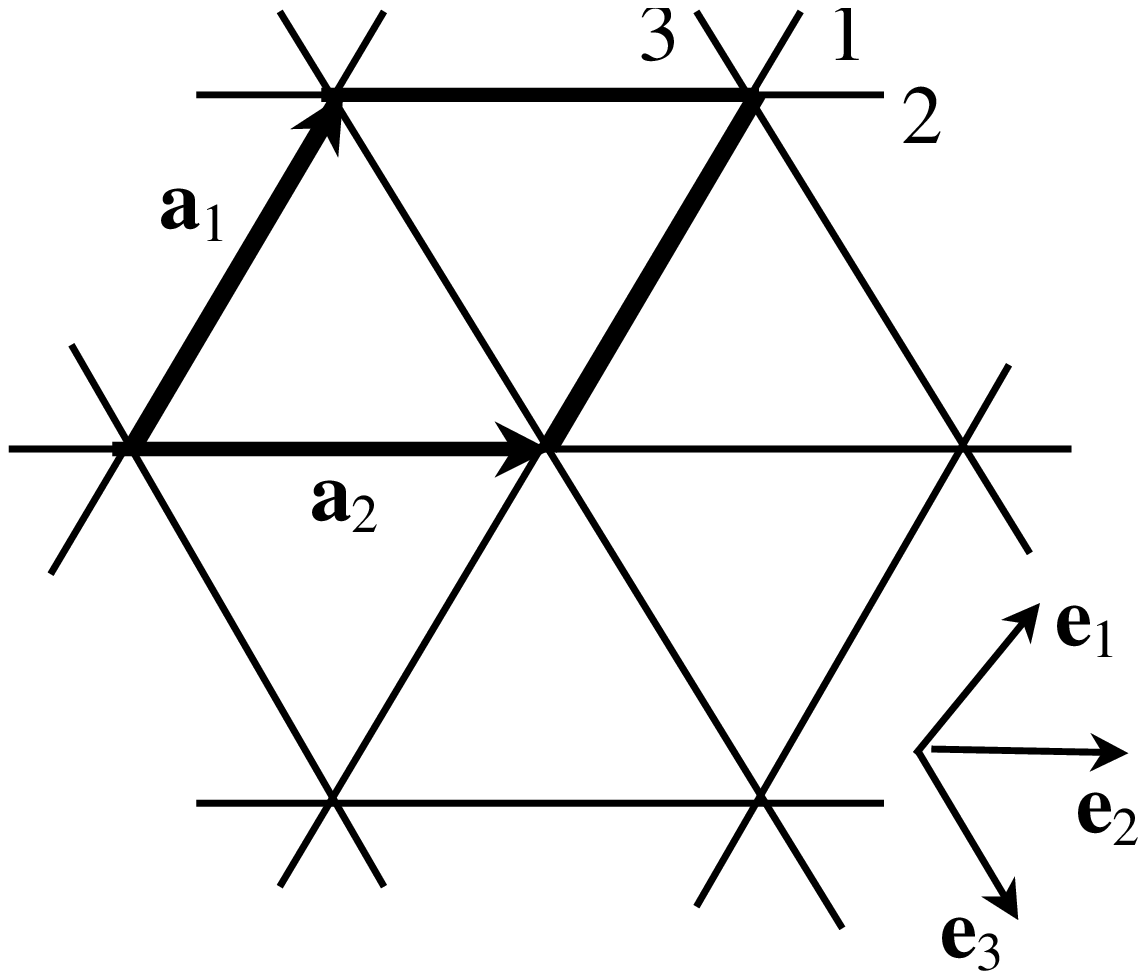}
\caption{Triple QCB.} \label{Bar4}
\end{figure}

Triple quantum bars is a $2D$ periodic grid with $m=3$, formed by
three periodically crossed arrays $j=1,2,3$ of $1D$ quantum wires.
These arrays are placed on three planes parallel to the $XY$ plane
and separated by an inter-plane distances $d.$ The upper and the
lower arrays correspond to $j=1,2,$ while the middle array has
number $j=3.$ All wires in all arrays are identical.  They have
the same length $L,$ Fermi velocity $v$ and Luttinger parameter
$g.$ The arrays are oriented along the $2D$ unit vectors
\begin{eqnarray*}
    {\bf{e}}_{1}=\left(\frac{1}{2}, \frac{\sqrt{3}}{2}\right),
    \ \ \ \ {\bf{e}}_2=(1,0),\ \ \ \
    {\bf{e}}_{3}={\bf{e}}_2-{\bf{e}}_{1}.\nonumber
\end{eqnarray*}
The periods of QCB along these directions are equal, $a_{j}=a,$ so
we deal with a regular triangular lattice.  In what follows we
choose ${\bf a}_{1,2}=a{\bf e}_{1,2}$ as the basic vectors of a
superlattice (see Fig. \ref{Bar4}).

The wires within the $j$-th array are enumerated with the integers
$n_j.$ Define $2D$ coordinates along the $n_{j}$-th wire ${\bf
r}_{j}$ as ${\bf r}_{j}=x_{j}{\bf{e}}_{j}+n_{j}a{\bf{e}}_{3}$ for
upper and lower arrays ($j=1,2$) and ${\bf
r}_{3}=x_{3}{\bf{e}}_{3}+n_{3}a{\bf{e}}_{1}$ for the middle array.
Here $x_{j}$ are $1D$ continuous coordinates along the wire. The
system of three non-interacting arrays is described by the
Hamiltonian
\begin{eqnarray*}
    H_{0}=H_{1}+H_{2}+H_{3},
\end{eqnarray*}
where
\begin{eqnarray}
 H_1 & = & \frac{{\hbar}v}{2}\sum\limits_{n_1}\int dx_1
 \left[
      g{\pi}_1^2(x_1{\bf{e}}_1+n_1a{\bf{e}}_3)+
      \frac{1}{g}
      \left(
           \partial_{x_1}{\theta}_1(x_1{\bf{e}}_1+n_1a{\bf{e}}_3)
      \right)^2
 \right],
 \label{H1}
 \\
 H_2 & = & \frac{{\hbar}v}{2}\sum\limits_{n_2}\int dx_2
 \left[
      g{\pi}_2^2(x_2{\bf{e}}_2+n_2a{\bf{e}}_3)+
      \frac{1}{g}
      \left(
           \partial_{x_2}{\theta}_2(x_2{\bf{e}}_2+n_2a{\bf{e}}_3)
      \right)^2
 \right],
 \label{H2}
 \\
 H_3 & = & \frac{{\hbar}v}{2}\sum\limits_{n_3}\int dx_3
 \left[
      g{\pi}_3^2(x_3{\bf{e}}_3+n_3a{\bf{e}}_1)+
      \frac{1}{g}
      \left(
           \partial_{x_3}{\theta}_3(x_3{\bf{e}}_3+n_3a{\bf{e}}_1)
      \right)^2
 \right],
 \label{H3}
\end{eqnarray}
and $\pi_{j}$ and $\partial_{x_{j}}\theta_{j}$ are canonically
conjugate fields describing LL within the $j$-th array.

%%%%%%%%%%%%%%%%%%%%%
%%%%%%%%%%%%%%%%%%%
%%%%%%%%%%%%%%%%%%%%%
\begin{figure}[htb]
\centering
\includegraphics[width=50mm,height=53mm,angle=0,]{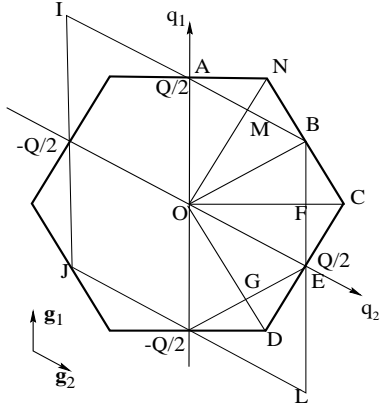}
\caption{Elementary cell $BIJL$ of the reciprocal lattice and the
BZ hexagon of the triple QCB.} \label{BZ4}
\end{figure}
%%%%%%%%%%%%%%%%%%%

Interaction between the excitations in different wires of adjacent
arrays $j,j'$ is concentrated near the crossing points with
coordinates $n_j{\bf a}_j+n_{j'}{\bf a}_{j'}$.  It is actually
Coulomb interaction screened on a distance $r_0$ along each wire
which is described by Hamiltonian
\begin{eqnarray*}
    H_{int}=H_{13}+H_{23},
\end{eqnarray*}
where
\begin{eqnarray}
 H_{j3} &=&  V_0\sum\limits_{n_j,n_3}\int dx_jdx_3
 \Phi\left(
          \frac{x_j-n_3a}{r_0}{\bf{e}}_j-
          \frac{x_3-n_ja}{r_0}{\bf{e}}_3
 \right)
 \times\nonumber\\&&\times
 \partial_{x_j}{\theta}_j(x_j{\bf{e}}_j+n_ja{\bf{e}}_3)
 \partial_{x_3}{\theta}_3(n_3a{\bf{e}}_j+x_3{\bf{e}}_3).
 \label{H13}
\end{eqnarray}
Here the effective coupling strength $V_{0}$ is defined by
Eq.(\ref{strength}), the dimensionless interaction $\Phi$ is
separable
\begin{equation}
 \Phi(\xi_{j}{\bf{e}}_{j}+\xi_{3}{\bf{e}}_{3})=
 \zeta(\xi_{j})\zeta(\xi_{3}),
 \ \ \ \ \ j=1,2,
 \label{separab}
\end{equation}
and $\zeta(\xi)$ is a dimensionless charge fluctuation in the
$j$-th wire (see Eq. (\ref{zeta})).

Such interaction imposes a super-periodicity on the energy
spectrum of initially one dimensional quantum wires, and the
eigenstates of this superlattice are characterized by a $2D$
quasimomentum ${\bf q}=q_1{\bf g}_1+q_2{\bf g}_2 \equiv(q_1,q_2)$.
Here ${\bf g}_{1,2}$ are the unit vectors of the reciprocal
superlattice satisfying the standard orthogonality relations
$({\bf e}_i\cdot {\bf g}_j)=\delta_{ij}, \ \ j=1,2.$ The
corresponding basic vectors of the reciprocal superlattice have
the form $Q(m_1{\bf g}_1 + m_2{\bf g}_2)$, where $Q=2\pi/a$ and
$m_{1,2}$ are integers. In Fig. \ref{BZ4} the elementary cell
$BIJL$ of the reciprocal lattice is displayed together with the
hexagon of the Wigner-Seitz cell that we choose as the BZ of the
triple QCB.

To study the energy spectrum and the eigenstates of the total
Hamiltonian
\begin{equation}
    H=H_{0}+H_{int},
    \label{H3t}
 \end{equation}
we define the Fourier components of the field operators
\begin{eqnarray}
 {\theta}_1(x_1{\bf{e}}_1+n_1a{\bf{e}}_3)
 &=&
 \frac{1}{\sqrt{{N}{L}}}
 \sum\limits_{{s},{\bf{q}}}
 \theta_{1s{\bf{q}}}e^{i(q_1x_1+q_3n_1a)}
 u_{s,q_1}(x_1),
 \label{Fuorier1}
 \\
 {\theta}_2(x_2{\bf{e}}_2+n_2a{\bf{e}}_3)
 &=&
 \frac{1}{\sqrt{{N}{L}}}
 \sum\limits_{{s},{\bf{q}}}
 \theta_{2s{\bf{q}}}e^{i(q_2x_2+q_3n_2a)}
 u_{s,q_2}(x_2),
 \label{Fuorier2}
 \\
 {\theta}_3(x_3{\bf{e}}_3+n_3a{\bf{e}}_1)
 &=&
 \frac{1}{\sqrt{{N}{L}}}
 \sum\limits_{{s},{\bf{q}}}
 \theta_{3s{\bf{q}}}e^{i(q_3x_3+q_1n_3a)}
 u_{s,q_3}(x_3).
 \label{Fuorier3}
\end{eqnarray}
Here
$${\bf q}=q_{1}{\bf g}_{1}+q_{2} {\bf g}_{2},\ \ \
q_{3}=q_{2}-q_{1},$$ and $N=L/a$ is the dimensionless length of a
wire.  In the ${\bf q}$ representation, the Hamiltonians $H_{j}$
(Eqs. (\ref{H1})-(\ref{H3})) and $H_{j3}$ (Eq. (\ref{H13})) can be
written as
\begin{eqnarray*}
 H_j  &=&  \frac{{\hbar}{v}{g}}{2}\sum\limits_{{s},{\bf{q}}}
           \pi_{js{\bf{q}}}^{+}\pi_{js{\bf{q}}}
           +\frac{\hbar}{{2}{v}{g}}\sum\limits_{{s},{\bf{q}}}
           {\omega}_{s}^{2}(q_j)
           \theta_{js{\bf{q}}}^{+}\theta_{js{\bf{q}}},
 \ \ \ \ \ \ \ \ \ \ \ \ \ \ \ \ \ \ \ \ j=1,2,3,
 \\
 H_{j3}  &=&  \frac{V_0r_0^2}{{2}{v}{g}}\sum\limits_{s,s',{\bf{q}}}
 \phi_{s}(q_3)\phi_{s'}(q_j)\omega_{s}(q_3)\omega_{s'}(q_j)
 \left(
      \theta_{3s{\bf{q}}}^{+}
      \theta_{js'{\bf{q}}}
      +h.c.
 \right), \ \ \ \ j=1,2,
\end{eqnarray*}
where
$$
\omega_s(q)=v\left(\left[\frac{s}{2}\right]Q+
\left(-1\right)^{s-1}|q|\right),
 \ \ \ \ \ Q=\frac{2{\pi}}{a},
$$
Thus the total Hamiltonian (\ref{H3t}) describes a system of
coupled harmonic oscillators, and can be diagonalized exactly like
in the case of double QCB.

%%%%%%%%%%%%%%%%%%%%%%%%%%%%%%%%%%%%%%%%%%%%%%%%%%%%%%%%%%%%%%%%%%
\section{Spectrum}
 \label{subsubsec:SpecTriple}
%%%%%%%%%%%%%%%%%%%%%%%%%%%%%%%%%%%%%%%%%%%%%%%%%%%%%%%%%%%%%%%%%%
Separability of the interaction (\ref{separab}) allows one to
derive analytical equations for the spectrum of the total
Hamiltonian (\ref{H3t}). Here we describe the behavior of the
spectrum and the corresponding states along some specific lines of
the reciprocal space.

To diagonalize the Hamiltonian(\ref{H3t}), we write down equations
of motion
\begin{eqnarray}
 \left[
      \omega_s^2(q_{j})-\omega^2
 \right]\theta_{js{\bf{q}}}
 +\sqrt{\varepsilon}\phi_s(q_j)\omega_s(q_j)
 \frac{r_0}{a}
 \sum\limits_{s'}
 \phi_{s'}(q_3)\omega_{s'}(q_3)
 \theta_{3s'{\bf{q}}}
 &=& 0,
 \label{EqMotion1-2}
 \\
 \left[
      \omega_s^2(q_{3})-\omega^2
 \right]\theta_{3s{\bf{q}}}
 +\sqrt{\varepsilon}\phi_s(q_3)\omega_s(q_3)
 \frac{r_0}{a}
 \sum\limits_{j,s'}
 \phi_{s'}(q_j)\omega_{s'}(q_j)
 \theta_{js'{\bf{q}}}
 &=& 0.
 \label{EqMotion3}
\end{eqnarray}
Here $j=1,2$, and $\varepsilon$ is defined by Eq.(\ref{epsilon1}).
The solutions of the set of equations (\ref{EqMotion1-2}) -
(\ref{EqMotion3}) have the form:
\begin{eqnarray*}
 \theta_{js{\bf{q}}} &=& A_j
 \frac{\phi_s(q_j)\omega_s(q_j)}
      {\omega_s^2(q_j)-\omega^2},
 \ \ \ \ \ j=1,2,3.
\end{eqnarray*}
Substituting this equation into Eqs.(\ref{EqMotion1-2}) and
(\ref{EqMotion3}), we have three equations for the constants
$A_j$:
\begin{eqnarray*}
 A_1+A_3{\sqrt{\varepsilon}}F_{q_3}(\omega^2)=0,
 \ \ \ \
 A_2+A_3{\sqrt{\varepsilon}}F_{q_3}(\omega^2)=0,
 \ \ \ \
 A_3+\sum\limits_{j=1,2}A_j{\sqrt{\varepsilon}}F_{q_j}(\omega^2)=0,
\end{eqnarray*}
where
$$
 F_q(\omega^2)=\frac{r_0}{a}\sum\limits_{s}
 \frac{\phi_s^2(q)\omega_s^2(q)}{\omega_s^2(q)-\omega^2}.
$$
Dispersion relations can be obtained from the solvability
condition for this set of equations
$$
 \varepsilon F_{q_3}(\omega^2)
 \left(
      F_{q_1}(\omega^2)+
      F_{q_2}(\omega^2)
 \right)=1.
$$
The function $F_{q_{s}}(\omega^2)$ has a set of poles at
$\omega^2=\omega_{s}^{2}(q)$, $s=1,2,3,\ldots$ .  For
$\omega^{2}<\omega_{s}^{2}(q)$, i.e. within the interval
$[0,\omega_{1}^{2}(q)]$,  $F_{q_{s}}(\omega^2)$ is positive
increasing function. Its minimal value $F$ on the interval is
reached at $\omega^2=0$ and does not depend on the quasi-momentum
$q$
$$
 F_q(0) =\frac{r_0}{a}\sum\limits_{s}
 \phi_{s}^{2}(q)=\int d\xi \zeta_j^2(\xi)\equiv F.
$$
If the parameter $\varepsilon\equiv\eta^{2}$ is smaller than the
critical value
$$
  \varepsilon_c=
  \frac{1}{2F^{2}},
$$
then all the solutions $\omega^{2}$ of the characteristic equation
are positive. When $\varepsilon$ increases, the lowest QCB mode
softens and its square frequency vanishes \textit{in the whole BZ}
at $\varepsilon=\varepsilon_{c}$.  For the exponential interaction
model $\zeta(\xi)=\exp(-|\xi|),$ one obtains $\varepsilon_c\approx
1$.

The high symmetry of the triple QCB leads to a number of lines
where inter-array or inter-band resonant interaction occurs:
\emph{all} lines in Fig. \ref{BZ4} possess some resonant
properties. These lines may be classified as follows:

On the Bragg lines where one of the three array wave-numbers
$q_{j}$ is a multiple integer of $Q/2,$ there is a strong
intra-band mixing of modes of the $j$-th array. In Fig. \ref{BZ4},
these lines are the boundaries of the elementary cell of the
reciprocal lattice $IJLB,$ axes $q_{1}$ and $q_{2}$, lines $OB$
and $EH.$ In particular, along the lines $OA$ ($q_{2}=0$) and $OB$
($q_3=0$) two modes corresponding to the second and third bands
and to the second ($OA$) or third ($OB$) array are mixed. Along
the line $AB$ ($q_{1}=Q/2$) the same mixing occurs between $(1,1)$
and $(1,2)$ modes.  Moreover, the resonant mixing of different
arrays within the same band occurs along the medians $OA,$ $OB,$
etc. There are two types of such resonance. The first one (e.g.,
$OA$ line) is the resonance between neighboring arrays
($q_1=-q_3$) and therefore it is of the main order with respect to
interaction. The second one (e.g., $OB$ line) is the resonance
between remote arrays ($q_1=q_2$) and it is one order smaller.

The second family consists of resonant lines formed by the BZ
hexagon boundaries and diagonals.  Thus, the diagonal $OC$
realizes a first order resonance between the first and the third
arrays $q_{1}=q_{3},$ and the BZ boundaries $HD$ and $AN$
correspond to the same resonance up to an umklapp process
($q_{1}=q_{3}-Q$ and $q_{1}=q_{3}+Q$ respectively).  Along the
diagonal $OD$ and the BZ boundary $NC$ a second order resonance
takes place with resonance conditions $q_{2}=-q_{1}$ and
$q_{2}=-q_{1}+Q$ respectively.

In the reciprocal space of the triple QCB there are four different
types of crossing points. Two of them include the bases of BZ
medians (e.g., points $A,$ $B,$ $E$ and so on). Here one deals
with the four-fold degeneracy of the modes corresponding to the
first order resonance between the neighboring arrays (e.g., point
$A,$ $\omega_{1,s}=\omega_{3,s'},$ $s,s'=1,2$), or to the second
order resonance between remote arrays (like point $B,$
$\omega_{1,s}=\omega_{2,s'},$ $s,s'=1,2$).  One more family
consists of crossing points of the BZ diagonals and the lines
connecting the bases of its medians (points $M,$ $F,$ $G$ and so
on).  Here one deals with three types of two-fold degeneracy
simultaneously.  For example, at the point $M$ two separate pairs
of modes corresponding to neighboring arrays $(2,1),$ $(3,1),$ and
$(2,2),$ $(3,2),$ are degenerate, as well as two modes
corresponding to the first array, $(1,1),$ $(1,2).$ Finally the BZ
hexagon vertices form the most interesting group of points where
the three-fold degeneracy between modes corresponding to all three
arrays takes place. The typical example of such a point is the
vertex $C$ where the resonance condition
$q_{1}=-q_{2}+Q=q_{3}=Q/3$ is satisfied.

%%%%%%%%%%%%%%%%%%%%%%%%%%%%%%%%%%%%%%
\begin{figure}[htb]
\centering
\includegraphics[width=75mm,height=60mm,angle=0,]{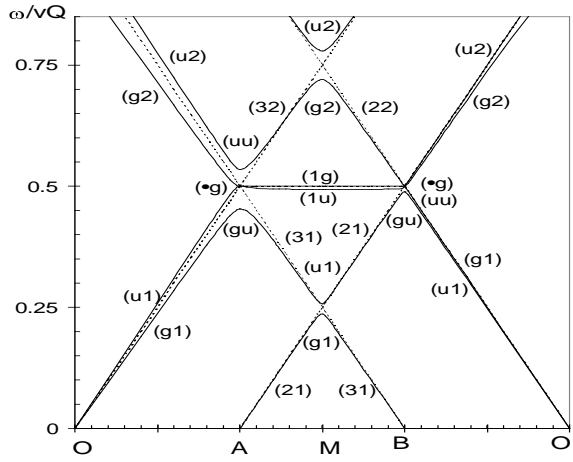}
\caption{Dispersion curves at the $OAMBO$ polygon of BZ.}
\label{Tri1}
\end{figure}
%%%%%%%%%%%%%%%%%%%

Almost all these peculiarities of the triple QCB spectrum can be
illustrated in Fig. \ref{Tri1} where the dispersion curves along
the closed line $OABO$ are displayed.  We emphasize once more that
in the infrared limit $\omega,{\bf q}\to 0$ triple QCB like double
QCB preserves the characteristic LL properties of the initial
arrays.

%%%%%%%%%%%%%%%%%%%%%%%%%%%%%%%%%%%%%%%%%%%%%%%%%%%%%%%%%%%%%%%%%%
\section{Observables}
 \label{subsubsec:Observ3}
%%%%%%%%%%%%%%%%%%%%%%%%%%%%%%%%%%%%%%%%%%%%%%%%%%%%%%%%%%%%%%%%%%

The structure of the energy spectrum analyzed above strongly
influences optical and transport properties of the triple QCB. As
in the case of double QCB (part \ref{subsubsec:Observ2}), one
expects to observe four peaks of the optical absorption near the
points $A,B,E,H$ of the four-fold degeneracy.  Then, specific
features of space correlators like those considered in
\ref{subsubsec:SpecTriple} can be observed.  But the most
pronounced manifestation of a triangular symmetry of the triple
QCB are its Rabi oscillations.

Consider the vicinity of the point $C$ of the BZ of three-fold
degeneracy, mixing all three arrays,
$$
 q_1=q_3=-q_2+Q=\frac{Q}{3},
 \ \ \ \ \
 \omega_{11}(Q/3)=\omega_{21}(2Q/3)=\omega_{31}(Q/3)\equiv\omega_0.
$$
Equations of motion at this point in the resonance approximation
read
\begin{eqnarray*}
  &&\left[\frac{d^2}{dt^2}+\omega_{0}^{2}\right]
  \theta_{j}+
  \sqrt{\varepsilon}\phi^2\omega_{0}^2\theta_{3} = 0,
  \ \ \ \
  \left[\frac{d^2}{dt^2}+\omega_{0}^{2}\right]
  \theta_{3}+
  \sqrt{\varepsilon}\phi^2\omega_{0}^2
  \left(\theta_{1}+\theta_{2}\right) = 0,
\end{eqnarray*}
where $\theta_{j}\equiv\theta_{j{\bf q}}.$ General solution of
this system looks as
\begin{eqnarray*}
    \left(
    \begin{array}{c}
          \theta_{1}(t)\\
          \theta_{2}(t)\\
          \theta_{3}(t)
    \end{array}\right)=
    \theta_{0}
    \left(
    \begin{array}{c}
          1\\-1\\0
    \end{array}
    \right)
    e^{i\omega_{0}t}+
    \theta_{+}
    \left(
    \begin{array}{c}
          1\\1\\\sqrt{2}
    \end{array}
    \right)
    e^{i\omega_{+}t}+
    \theta_{-}
    \left(
    \begin{array}{c}
          1\\1\\-\sqrt{2}
    \end{array}
    \right)
    e^{i\omega_{-}t},
\end{eqnarray*}
where one of the eigenfrequencies coincides with $\omega_{0},$
while the other two are
$$\omega_{\pm}=\omega_{0}\sqrt{1\pm\sqrt{2}\phi^{2}},$$
and $\theta_{0,\pm}$ are the corresponding amplitudes.

\begin{figure}[htb]
\centering
\includegraphics[width=75mm,height=105mm,angle=0,]{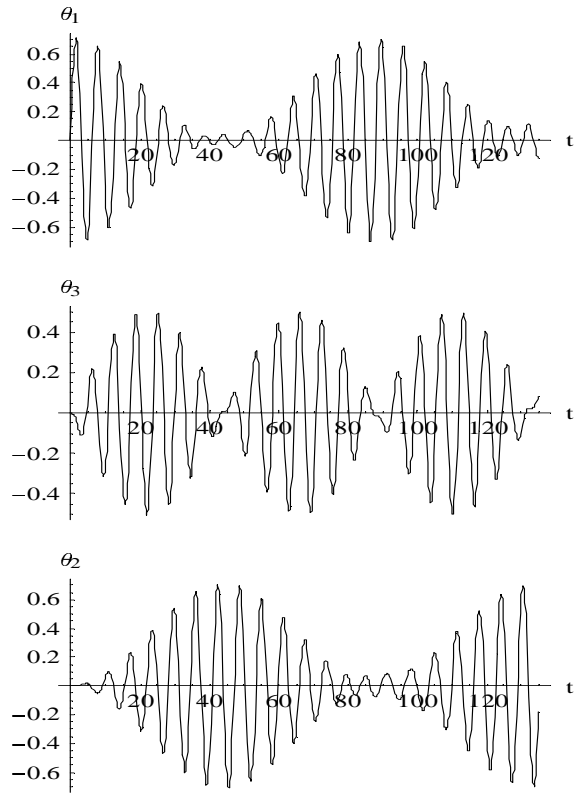}
\caption{Periodic energy transfer between three arrays at the
triple resonant point $C$ of the BZ.} \label{ROTr}
\end{figure}

Choosing initial conditions ${\theta}_{1}(0)=i\theta_0,$
${\dot{\theta}}_{1}(0)=\omega_0\theta_0,$
${\theta}_{2}(0)={\theta}_{3}(0)=0,$
${\dot{\theta}}_{2}(0)={\dot{\theta}}_{3}(0)=0,$ we obtain for the
field amplitudes at the coordinate origin
\begin{eqnarray*}
  \theta_1(0,0;t) &=& \frac{\theta_0}{4}
  \left[
       \frac{\omega_0}{\omega_+}\sin(\omega_+t)+
       \frac{\omega_0}{\omega_-}\sin(\omega_-t)
  \right]+
  \frac{\theta_0}{2}\sin(\omega_0t),
  \nonumber\\
  \theta_2(0,0;t) &=& \frac{\theta_0}{4}
  \left[
       \frac{\omega_0}{\omega_+}\sin(\omega_+t)+
       \frac{\omega_0}{\omega_-}\sin(\omega_-t)
  \right]-
  \frac{\theta_0}{2}\sin(\omega_0t),
  \nonumber\\
  \theta_3(0,0;t) &=& \frac{\theta_0}{2\sqrt{2}}
  \left[
       \frac{\omega_0}{\omega_+}\sin(\omega_+t)-
       \frac{\omega_0}{\omega_-}\sin(\omega_-t)
  \right].
\end{eqnarray*}
In the limiting case $\varepsilon\ll 1$ these formulas lead to the
following time dependence of the field operators in the coordinate
origin in real space
\begin{eqnarray}
  \theta_1(0,0;t) &=& \theta_0
  \sin(\omega_0t)
  \cos^2\left(\frac{\sqrt{2\varepsilon}\phi^2}{4}\omega_0t\right),
  \nonumber\\
  \theta_2(0,0;t) &=&\theta_0
  \cos(\omega_0t)
  \sin^2\left(\frac{\sqrt{2\varepsilon}\phi^2}{4}\omega_0t\right),
  \label{sol3}\\
  \theta_3(0,0;t) &=& \theta_0
  \sin(\omega_0t)
  \cos\left(\frac{\sqrt{2\varepsilon}\phi^2}{2}\omega_0t\right).
  \nonumber
\end{eqnarray}
The field operators of all three arrays demonstrate fast
oscillations with the resonant frequency $\omega_{0}$ modulated by
a slow frequency.  It is the same for the two remote arrays, and
doubled for the intermediate array.  These beatings are
synchronized in a sense that zero intensity on the intermediate
array always coincides with the same intensity on one of the
remote arrays.  At these moments all the energy is concentrated
solely within one of the remote arrays. These peculiar Rabi
oscillations are displayed in Fig. \ref{ROTr}.

%%%%%%%%%%%%%%%%%%%%%%%%%%%%%%%%%%%%%%%%%%%%%%%%%%%%%%%%%%%
%%%%%%%%%%%%%%%%%%%

%\newpage

%%%%%%%%%%%%%%%%%%%%%%%%%%%%%%%%%%%%%%%%%%%%%%%%%%%%%%%%%%%%%%%%%%%%%%%%%%
\chapter{Derivation of QCB-Light Interaction Hamiltonian}
 \label{append:Inter}
%%%%%%%%%%%%%%%%%%%%%%%%%%%%%%%%%%%%%%%%%%%%%%%%%%%%%%%%%%%%%%%%%%%%%%%%%%
Light scattering on QCB is described by equations
(\ref{H-int-initial}) - (\ref{Pol_Matr1}).  In this Appendix we
briefly explain the main steps which lead to such a description.

\noindent \textbf{1.  Nanotube-light interaction, Eq.
(\ref{H-int-initial}).} We start with a consideration of a single
nanotube of the first array, interacting with an external
electromagnetic field.  The characteristic time of all nanotube
energies, including Coulomb interaction, is of order of an inverse
plasmon frequency.  The scattering process occurs during much
shorter time interval, which is of the order of an inverse photon
frequency. Hence, Coulomb interaction is irrelevant for the
scattering processes. This enables us to restrict ourselves by a
kinetic part of the nanotube Hamiltonian. Within the ${\bf k}-{\bf
p}$ approximation, this part is given by \cite{Ando}
\begin{eqnarray}
  h_{kin} &=& v_F\int\frac{dx_{1}d\gamma}{2\pi}
  \psi_{\alpha}^{\dag}(x_{1},r_{0},\gamma)
  \hbar{\bf{k}}\cdot
  {\mbox{$\boldsymbol\sigma$}}_{\alpha,\alpha'}
  \psi_{\alpha'}(x_{1},r_{0},\gamma).
  \label{H-kinetic-1}
\end{eqnarray}
{In the presence of a magnetic field, one should add}
$e\textbf{A}/c$ {to the electron momentum operator} $\hbar {\bf
k}.$ \textit{An additional part of the Hamiltonian}
(\ref{H-kinetic-1}) {exactly coincides with the nanotube-light
interaction Hamiltonian}(\ref{H-int-initial}).

The Hamiltonian (\ref{H-kinetic-1}) is diagonalized by a two step
canonical transformation. The first one is Fourier transformation
\begin{equation}
    \psi_{\alpha}(x_{1},\gamma)=
     \frac{1}{\sqrt{L}}
     \sum_{k,m}c_{\alpha,k,m}e^{ikx_{1}+im\gamma},
    \label{Fourier}
\end{equation}
where the orbital moment $m$ is restricted by the condition
$|m|\leq m_0=[\pi r_0/a_0]$ due to the finite number of honeycomb
cells along the nanotube perimeter ($a_{0}\sqrt{3}$ is the lattice
constant).  The second rotation is defined as
\begin{eqnarray*}
  c_{p,k,m}=\frac{1}{\sqrt{2}}
      \left(
       {c}_{Akm}+p
       {e}^{i\phi_m}
       {c}_{Bkm}
      \right),\ \ \  p=\pm,
  \label{c-AB--c-pm}
\end{eqnarray*}
where
$$
 \cos\phi_m=\frac{kr_0}{\sqrt{(kr_0)^2+m^2}},
 \ \ \ \ \
 \sin\phi_m=\frac{m}{\sqrt{(kr_0)^2+m^2}}.
$$
As a result, the Hamiltonian takes the form
\begin{equation}\label{kin1}
H_{kin}=\sum_{p,k,m}\hbar \omega_p(k,m)c^\dag_{pkm}c_{pkm}
\end{equation}
with eigenfrequencies
\begin{equation}
    \omega_p(k,m)=pv_F \sqrt{k^2+\frac{m^2}{r_0^2}}.
    \label{eigenenergies}
\end{equation}

\noindent \textbf{2.  QCB-light interaction, Eq.
(\ref{Eff_Int_QCB}).} Substituting into Eq.  (\ref{H-int-initial})
the scalar product ${\bf{A}}\cdot{\boldsymbol\sigma}$ in the form
\begin{eqnarray*}
 \left({\bf A}\cdot{\mbox{$\boldsymbol\sigma$}}\right)
 &=&
 \left(
      \begin{array}{cc}
      0     & A^{-}\\
      A^{+} & 0
      \end{array}
 \right),
 \label{sigma*A}
\end{eqnarray*}
where $A^{\pm}=A_1\pm iA_{\gamma},$ we can write the
nanotube-light interaction as
\begin{equation}
 H_{nl} = \frac{ev_F}{ c}\int\frac{dx_1d\gamma}{2\pi}
         \Big(
          \psi_A^{\dag}
          A^{-}
          \psi_B
          +h.c.
         \Big).
         \nonumber
 \label{H-int-pm}
\end{equation}

We are interested in an effective QCB-light interaction
Hamiltonian obtained in second order of perturbation theory which
describes transitions between initial states $|i\rangle$ and final
ones $|f\rangle.$ Initial states are one-photon states of the
electromagnetic field and the electron ground state of the
nanotube whereas final states consist of one photon and an
electron above the Fermi level. It will be seen later that just
these states form a one-plasmon array state. The energy of the
incident photon $E=\hbar{ck}$ is much higher than the excitation
energies of the nanotube. Therefore, absorption of the incident
photon and radiation of the scattered photon occur practically
without retardation. All this results in the interaction
Hamiltonian
\begin{eqnarray}
  h_{int} &=&
       \frac{e^2}{\hbar c}
       \left(\frac{v_F}{c}\right)^2
       \int \frac{dx_1d\gamma}{2\pi}
       \int\frac{dx'_1d\gamma'}{2\pi}
       \sum_{|{v}\rangle}
       \frac{1}{k}
       \Big(
       \psi^{\dag}_{A}(x'_1,\gamma')
       A^{-}(x'_1,0,0)
       \times\nonumber\\&&
       |{v}\rangle
       \langle{v}|
       \psi_B(x'_1,\gamma')
       \psi^{\dag}_B(x_1,\gamma)
       |{v}\rangle
       \langle{v}|
       A^{+}(x_1,0,0)
       \psi_A(x_1,\gamma)+
       {A}\leftrightarrow{B}\Big),
  \label{H-eff-int-1}
\end{eqnarray}
which corresponds to the diagram shown in Fig. \ref{X-Scatt}b
(there is no photons in an intermediate state).

Consider now the matrix element $\langle{v}|
\psi_{\alpha}(x'_1,\gamma') \psi^{\dag}_{\alpha}(x_1,\gamma)
|{v}\rangle$ which enters this Hamiltonian. Due to our choice of
initial and final states, only diagonal elements with respect to
both virtual states $|v\rangle$ and sublattice indices $\alpha$
survive. In the $c_{\alpha,k,m}$ representation (\ref{Fourier}),
they have the form
\begin{eqnarray*}
 &&\langle{v}|
       \psi_{\alpha}(x'_1,\gamma')
       \psi^{\dag}_{\alpha}(x_1,\gamma)
 |{v}\rangle=\frac{1}{L}\sum_{k,k'}\sum_{m,m'}
    \langle{v}|
     c_{\alpha,k,m}
     c^{\dag}_{\alpha,k',m'}
    |{v}\rangle
    e^{ikx_1-ik'x'_1+im\gamma-im'\gamma'}.
 \label{n-alpha}
\end{eqnarray*}
Internal matrix elements on the r.h.s. of the latter equation are
$$
    \langle{v}|
     c_{\alpha,k,m}
     c^{\dag}_{\alpha,k',m'}
    |{v}\rangle
    =\frac{1}{2}\delta_{k,m}\delta_{k',m'}
$$
(here the symmetry property $n\left(E_{-}(k,m)\right)+
n\left(E_{+}(k,m)\right)=1$ is used). Therefore
$$
 \langle{v}|
 \psi_{\alpha}(x'_1,\gamma')
 \psi^{\dag}_{\alpha}(x_1,\gamma)
 |{v}\rangle
 =
 \pi\delta(x_1-x'_1)
 S(\gamma-\gamma'),
$$
where
\begin{equation}
    S(\gamma)=\frac{1}{2\pi}\sum_{m=-m_0}^{m_0}e^{im\gamma}.
    \nonumber
    \label{gamma}
\end{equation}

Thus, the interaction (\ref{H-eff-int-1})  takes the form
\begin{eqnarray*}
       h_{int} =
       \frac{e^2}{\hbar c}
       \left(\frac{v_F}{c}\right)^2
       \int\frac{dx_1d\gamma d\gamma'}{4\pi k}
       S(\gamma-\gamma')
       A^{-}(x_1,0,0)
       A^{+}(x_1,0,0)\sum_{\alpha}
       \psi^{\dag}_{\alpha}(x_1,\gamma')
       \psi_{\alpha}(x_1,\gamma).
\end{eqnarray*}
The field dependent factor here is
\begin{eqnarray}
       A^{-}(x_1,0,0)A^{+}(x_1,0,0)=
       A_1^2(x_1,0,0)+
       A^2(x_1,0,0)\sin(\gamma_{A}-\gamma')
       \sin(\gamma_A-\gamma),
    \label{field}
\end{eqnarray}
and $(A_{1},A,\gamma)$ are cylindrical components of the vector
potential ${\bf A}$ (\ref{VecPot}). Taking into account the
angular dependence of the field (\ref{field}), we can omit it in
the electron operators. Indeed, according to Eq.
(\ref{eigenenergies}) (see also Fig. \ref{CN-sp-2D}) in an
energy-momentum region where we work, the $m=1$ spectral band with
nonzero orbital moment is separated from that with $m=0$ by an
energy of order of $\hbar v_{F}/r_{0},$ which is much higher than
the QCB plasmon energy. Keeping only the zero moment field
operators which form the electron density operator
$$
 \sum_{\alpha}\psi^{\dag}_{\alpha}(x_1)\psi_{\alpha}(x_1)=
 \rho(x_1)\equiv \sqrt{2}\partial_{x_1}\theta(x_1),
$$
and integrating over $\gamma,\gamma',$ we obtain an interaction
Hamiltonian in form
\begin{equation}
 h_{int} =\frac{\sqrt{2}}{4k}
 \frac{e^2}{\hbar c}
 \left(\frac{v_F}{c}\right)^2
 \int dx_1
 \partial_{x_1}\theta(x_1)
 {\bf{A}}_1^2(x_1,0,0),
 \label{Effect-int-3}
\end{equation}
 where
\begin{equation}
 {\bf{A}}_1={\bf{A}}+
 (\sqrt{2}-1)A_1{\bf{e}}_1.
 \nonumber
 \label{A-1}
\end{equation}
{Straightforward generalization of this expression to the QCB case
leads exactly to the Hamiltonian} (\ref{Eff_Int_QCB}).

\noindent \textbf{3.  Polarization matrix, Eq.
(\ref{Pol_Matr1}).} To study the scattering process, we should
modify the last expression for the interaction Hamiltonian.  To
proceed further, we define Fourier transforms $\theta_{j, {\bf
Q}_j}$ of the bosonic fields
\begin{eqnarray}
  \theta_1(x_1,n_2a) &=&
  \frac{1}{\sqrt{{N}{L}}}
  \sum_{{\bf k}_1}
  \theta_{1,-{\bf k}_1}
  e^{-ik_1x_1-iq_2n_2a},
  \nonumber\\
  \theta_2(n_1a,x_2) &=&
  \frac{1}{\sqrt{{N}{L}}}
  \sum_{{\bf k}_2}
  \theta_{2,-{\bf k_2}}
  e^{-iq_1n_1a-ik_2x_2}.
  \label{theta-Fourier}
\end{eqnarray}
Here $N=L/a$ is the number of QCB cells in both directions.  The
electromagnetic field also can be expanded in a sum of harmonics
with a wave vector ${\bf k}$ and polarization $\lambda=||,\bot,$
\begin{eqnarray}
 {\bf{A}}({\bf{r}}) &=&
 \sum_{{\bf{K}}\lambda}
 {\bf{n}}_{{\bf{K}}\lambda}A_{{\bf{K}}\lambda}
 e^{i{\bf{Kr}}}.
 \label{A-Fourier}
\end{eqnarray}
The polarization vectors
\begin{eqnarray*}
{\bf{n}}_{{\bf{K}}||} &=&
 \frac{iK_2{\bf{e}}_1}{\sqrt{K_1^2+K_2^2}}-
 \frac{iK_1{\bf{e}}_2}{\sqrt{K_1^2+K_2^2}},
 \\
{\bf{n}}_{{\bf{K}}\bot} &=&
 \frac{K_1K_3{\bf{e}}_1}{K\sqrt{K_1^2+K_2^2}}+
 \frac{K_2K_3{\bf{e}}_1}{K\sqrt{K_1^2+K_2^2}}-
 \frac{\sqrt{K_1^2+K_2^2}}{K}{\bf{e}}_3,
\end{eqnarray*}
are normalized, $\left|{\bf{n}}_{{\bf{K}}\lambda}\right|=1,$ and
satisfy the orthogonality conditions,
${\bf{n}}_{{\bf{K}}\lambda}\cdot{\bf{K}}=
 {\bf{n}}_{{\bf{K}}||}\cdot{\bf{n}}_{{\bf{K}}\bot}=0.$ The
field operators $A_{{\bf{K}}\lambda}$ satisfy the condition
$A^{\dag}_{{\bf{K}}\lambda}=A_{-{\bf{K}}\lambda}$, so that
${\bf{A}}^{\dag}({\bf{r}})={\bf{A}}({\bf{r}})$.

Substituting equations (\ref{theta-Fourier}), and
(\ref{A-Fourier}) into the Hamiltonian (\ref{Eff_Int_QCB}), we
obtain
\begin{eqnarray}
&&H_{int}=
 -i\frac{\sqrt{2NL}}{4}
 \frac{e^2}{\hbar c}
 \left(\frac{v_F}{c}\right)^2
 \sum_{{\bf{K}},{\bf{K}'},{\bf{k}}}\sum_{j,\lambda,\lambda'}
 P_{j;\lambda',\lambda}
 \left(\frac{{\bf K}'}{K'},
       \frac{{\bf K}}{K}
 \right)
 \frac{k_j}{K}
 A^{\dag}_{{\bf K}',\lambda'}
 A_{{\bf{K}},\lambda}
 \theta_{j,-{\bf k}_j}.
 \label{Eff-Int-k}
\end{eqnarray}
Here
\begin{equation}
    \label{polariz}
    P_{j;\lambda',\lambda}
    \left(
         \frac{{\bf K}'}{K'},
         \frac{{\bf K}}{K}
    \right)=
  ({\mbox{$\boldsymbol\kappa$}}^*_{j,{\bf K}',\lambda'}
  \cdot
  {{\mbox{$\boldsymbol\kappa$}}}_{j,{\bf K},\lambda})
\end{equation}
is the polarization matrix, ${\bf k}={\bf q}+{\bf m},$ and
\begin{eqnarray*}
 {{\mbox{$\boldsymbol\kappa$}}}_{j,{\bf{K}},\lambda}
  &=&
  {\bf{n}}_{{\bf{K}}\lambda}+
  \left(\sqrt{2}-1\right)
  \left({\bf{n}}_{{\bf{K}}\lambda}\cdot{\bf{e}}_j\right){\bf{e}}_j.
 \label{ej}
\end{eqnarray*}
{In the case of normal incidence,} Eqs. (\ref{polariz}),
(\ref{ej}) {result in form} (\ref{Pol_Matr1}) {of polarization
matrix.}

\noindent \textbf{4.  Scattering Hamiltonian, Eq.
(\ref{Fin_Int}).} In the next step, we express the Fourier
transforms of the Bose fields $\theta$ via creation ($a^{\dag}$)
and annihilation ($a$) operators of the array plasmons
$$
  \theta_{j,-{\bf k}_j} =
  \sqrt{\frac{g}{2|k_j|}}
  \left(
       a_{j,-{\bf k}_j}+
       a^{\dag}_{j,{\bf k}_j}
  \right).
$$
The electromagnetic field amplitudes $A_{{\bf K},\lambda}$ should
also be expressed via photon creation ($c^{\dag}$) and
annihilation ($c$) operators
\begin{eqnarray*}
 A_{{\bf{K}},\lambda}(t) &=&
 \sqrt{
      \frac{\hbar{c}}{2VK}}
      \left(
           c_{{\bf{K}},\lambda}+
           c_{-{\bf{K}},\lambda}^{\dag}
      \right).
      \label{A-opers}
\end{eqnarray*}
Substituting these expansions into Eq. (\ref{Eff-Int-k}) we obtain
the final form of the effective interaction. \emph{In the case of
normal incidence}, \emph{this interaction is written as}
\begin{eqnarray}
 H_{int} =&&
 \frac{i\sqrt{gNL}}{V}
 \left(\frac{ev_F}{2cK}\right)^2
 \sum_{{\bf{K}},{\bf{K}'}}\sum_{j,\lambda,\lambda'}
 h_{\lambda',\lambda}({\bf k}')
 c^{\dag}_{{\bf{K'}},{\lambda}'}
 c_{{\bf{K}},\lambda},
 \label{Eff-Int-ck-1}
\end{eqnarray}
{where} $h_{\lambda',\lambda}({\bf K}')$ {is the Hamiltonian}
(\ref{Fin_Int}), {where} $\lambda_f,\lambda_i,{\bf K}$ {are
replaced by} $\lambda',\lambda,{\bf K}'.$

\noindent \textbf{5.  Scattering cross section, Eqs.
(\ref{eq:cross1}) - (\ref{Fin_Cross}).} Standard procedure based
on the Fermi golden rule leads to the following expression of the
differential scattering cross section per unit QCB square
\begin{equation}
 \frac{1}{L^2}\frac{d\sigma}{d\omega do}=\frac{1}{\pi}
 \left(\frac{Vk_f}{Lc\hbar} \right)^2
    \overline{\big|\big\langle f\big|H_{int}\big|i\big \rangle\big|^2}
      \delta\left(\frac{\varepsilon_i-\varepsilon_f}{\hbar}\right)
 \label{cross-def}
\end{equation}
(here bar denotes averaging with respect to polarization of both
incident light and scattered quanta).  Choose an initial ket-state
$\big | i \big \rangle$ such that it contains an incident photon
with momentum ${\bf K}_i,$ frequency $\Omega_i=cK_i,$ and
polarization $\lambda_i,$ and does not contain any QCB plasmon.
This state can be written as
$\big|i\big\rangle=|{\bf{K}}_i\rangle_l\bigotimes|0\rangle_p,$
where $| 0 \rangle_p$ is the plasmon vacuum,
$|{\bf{K}}_i\rangle_l=c^{\dag}_{{\bf K}_i,\lambda_i}|0\rangle_l,$
and $|0\rangle_l$ is the photon vacuum. A final bra-state
$\big\langle f\big|$ contains a scattered photon with momentum
${\bf K}_f,$ frequency $\Omega_f=cK_f,$ and polarization
$\lambda_f.$ It contains also a QCB plasmon $P$ with the frequency
$\omega_P$ (its quantum numbers will be specified below).  The
final state is written as $ \big \langle f \big | = \langle P
|\bigotimes \langle {\bf K}_f |,$ where $\langle{\bf K}_f
|=\left(c^{\dag}_{{\bf K}_f,\lambda_f}|0\rangle_l\right)^{\dag}.$

A matrix element of the interaction which enters Eq.
(\ref{cross-def}), is
\begin{equation}
    \label{MatrEl}
    \big\langle f\big|H_{int}\big|i\big \rangle=
    \langle P|\overline{H}_{int}|0\rangle_p,
\end{equation}
where
\begin{eqnarray}
    \overline{H}_{int}
    &\equiv&
    \left\langle{\bf K}_f,\lambda_f\left|
    H_{int}
    \right|{\bf K}_i,\lambda_i\right\rangle
    =
    -\frac{i\sqrt{gNL}}{V}
    \left(\frac{ev_F}{2cK}\right)^2
    \sum_{j,{\bf k}}
    \frac{k_j}{\sqrt{|k_j|}}
    \delta_{{\bf K}^{\|}_f,{\bf K}^{\|}_i-{\bf k}}
    \times\nonumber\\&&\times
    P_{j;\lambda_f,\lambda_i}
    \left(
         \frac{{\bf K}_f}{K_f},
         \frac{{\bf K}_i}{K_i}
    \right)
    \left(
         a_{j,-{\bf k}_j}+
         a^{\dag}_{j,{\bf k}_j}
    \right),
    \label{matel}
\end{eqnarray}
where ${\bf{K}}^{\|}_{i,f}={\bf{K}}_{i,f}-
({\bf{K}}_{i,f}\cdot{\bf{e}}_3){\bf{e}}_3$. {In the case of normal
incidence} Eqs. (\ref{cross-def}) - (\ref{matel}) {are equivalent
to} Eqs. (\ref{Eff_Int_QCB}) - (\ref{Fin_Int}).

%\newpage

%%%%%%%%%%%%%%%%%%%%%%%%%%%%%%%%%%%%%%%%%%%%%%%%%%%%%%%%%%%%%%%%%%%%

%%%%%%%%%%%%%%%%%%%%%%%%%%%%%%%%%%%%%%%%%%%%%%%%%%%%%%%%%%%%%%%%%%%%

\end{document}